# Laser Plasma Physics

## FORCES AND THE NONLINEARITY PRINCIPLE

### Heinrich Hora

Emeritus Professor
University of New South Wales, Sydney, Australia

Dedicated to the Memory of
**Edward Teller**

and with sincere thanks to my co-authors who have contributed
to the progress in this field

# TABLE OF CONTENTS

# PREFACE TO THE SECOND EDITION

The second edition of this book can well be based on all what was said in the preface to the first edition about summarizing the solid knowledge. On this knowledge, the later plasma theory was based since the beginning in 1845 by Kelvin with forces in electrodynamics on electrically neutral materials with no electic charges if there is a dielectric response. This force of ponderomotion is basically nonlinear by combining mechanical macroscopic forces with quadratic terms of the force quantities of electric and magnetic fields. It has now been clarified that this is interwoven even with clouds of electrons as in the Boreham experiment or in the Umstadter experiment and as it is visible in the Particle in Cell (PIC) computations where electric double layers appear as Debye sheath in the Target Normal Sheath Acceleration (TNSA), though there is some difference to optical properties in the explicitly clarified forces defined by laser fields in the skin layer of plasma boundaries.

This all is related to the measurements of ultrahigh acceleration of plasma blocks by Sauerbrey (1996) and Földes et al (2000) as it was theoretically-numerically predicted since 1978 and as it is accessible since the turning point in laser technology by the work of Mourou and collaborators since 1986 with the CPA (Chirped Pulse Amplification).

The second edition tries to present several specific views to the very broad stream of developments during the 15 years since the first edition has been published. From the very numerous results of these developments one trace is mostly followed up with respect to the application of the ultrahigh accelerated plasma blocks driven by the nonlinear force applying to laser ignition of controlled nuclear fusion. For this purpose, the two appendices of the first edition had to be integrated into the main text. It had to be clarified what a difference it is for laser driven fusion with nanosecond (ns) pulses using thermal processes for heating, compressing and igniting fusion fuel; this is in absolute contrast to the picosecond (ps) and shorter laser pulse interaction for exclusion of the complex systems of thermal processes and how to arrive at non-thermal direct conversion of laser pulse energy into macroscopic mechanical energy of motion by the ultrahigh acceleration.

The result is rather interesting because it is opening a way how fusion is possible beyond the usual fuel of deuterium-tritium (DT). There is now an access for burning protons with boron-11 (HB11) where the generation of very dangerous nuclear radiation is not more than from burning coal, and this can be ignored. Finally the new access to produce magnetic fields in the range of 10 kiloTesla may lead to conditions for developing power stations where ps laser pulses with few dozens of petawatt (PW) power are needed. These pulses are not too far in the future to be developed also for many other important basic applications apart from fusion.

After the first edition covered a lot of settled physics of the past – last not least by opening a new dimension of research by nonlinearity from a discussion with Richard Feynman – the new aspects for application on fusion energy are now a first steps with the plasma block acceleration. This refers to the attention on research for a wide field of unexplored phenomena as next steps. In this sense the second edition has well changed the character of this book. While these explorations are connected to usuall settled "classic" knowledge including aspects from a discussion with Richard Feynman (Section 6.3) there is a number of new points now needing clarification. We may offer some well fascinating aspects in general as well as for the future of energy production and the access to solutions.

<div style="text-align: right;">Heinrich Hora<br>Sydney, June 2014</div>

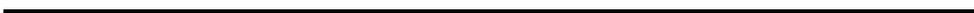

# PREFACE TO THE FIRST EDITION

Many renowned physicists speak very articulately about the "ponderomotive potential." It is now used in a wider context of high intensity laser interaction with plasma, or for acceleration of electrons by lasers, or in free electron lasers and many other related fields. There is now a growing interest in this topic. However, taking the desire for a clear definition into account, we see there are historical difficulties regarding the definition of *ponderomotive force*. One may a little less precisely say that it relates to *electrodynamic forces* acting on plasmas or on free electrical charges, which are basically different from the Coulomb force. These forces are essentially nonlinear and in a general sense are called *nonlinear forces*.

The study of these forces was necessary with the advent of the laser for understanding the very strange observations of interaction with plasma. This fascinating new field has a much broader basic importance than the high intensity laser interaction with materials and plasmas. It was necessary to underline the confusion with the correct definition, insufficiencies in applications and their exact solution for the highly advancing field of laser physics. Diverging opinions of authorities were not helpful. The more or less settled view about this is being presented here.

The following very abbreviated reviews on hydrodynamics and electrodynamics may be an enlightening path for the advanced student as well as for the expert with many years of experience to see how the theoretical and experimental studies of the nonlinear (ponderomotive) force led to the new foundation and completion of magnetohydrodynamics. Some colleagues have mentioned to me that they now understood Maxwell's theory better. It should be mentioned that the waves in inhomogeneous media are presented here in a more general way (within all shortness) than in Ginzburg's best-selling book.

But this is not the main mission. The reader should know that the work he does in understanding this text mostly in the area of classical physics will result in a very global view about physics in general. The thrilling story of the highly unexpected phenomena of laser–plasma interaction with the generation of GeV ions after relativistic self-focusing, with a stochastically pulsating (stuttering) interaction within a 20 ps scale, electron acceleration, fusion energy, laser ablation of materials in technology, etc., led to the codiscovery of the longitudinal components of laser beams in a vacuum. What was so surprising is that the mechanical motion of electrons in the intense laser fields depends drastically on very minor changes of these fields. This teaches us that nonlinear physics needs the highest possible degree of accuracy, much higher than that required by linear physics.

This accuracy principle of nonlinearity implies that in the future, completely undreamable phenomena and applications will be derived in all fields of physics if only sufficient accuracy, for example, with the most advanced computations, will be performed. We know now what we have to watch next: we have to be aware that not only curiosities but fully unexpected phenomena will be revealed. This is the new message for a rich future of physics. It was impossible to suppress some remarks against the views of a saturation and "an end" of physics as expressed in some way by Stephen Hawking or Carl Friedrich von Weizsäcker. The work on the nonlinear (ponderomotive) force and the related phenomena of *driving* plasmas or particles by *lasers* led unavoidably through this new gate of physics as a further promise for a bright future.

The initial version of the text[1] suffered indeed from a large number of misprints which have been removed, such that this rather unusual and not trivial text may now be absorbed better by the reader. As an appendix, two recent publications have been added as a first step to demonstrate directions to the reader where the applications are following from the rather compactly presented and interwoven development of this new chapter of physics.

Support for the preparation of this text by the Department of Theoretical Physics of the University of New South Wales in Sydney, Australia, and by the new University of Applied Sciences in Deggendorf, Germany, and encouragement by Präsident Prof. Dr. Reinhard Höpfl are gratefully acknowledged.

Heinrich Hora
Sydney, March 2000

---

[1] H. Hora, *Nonlinear Force and Ponderomotion*, Institute of Laser Engineering, Osaka University, 1996 ISBN 4-9900502-1-5.

# CHAPTER 1

# Introduction to the Ponderomotive Processes and Overview about Related Phenomena

It was fully beyond any understanding and a shocking surprise when Roland Sauerbrey (1996) measured ultrahigh acceleration of macroscopic objects of about solid state density. These accelerations were above 10.000 times higher than ever measured before in a laboratory. His convincing measurement at laser interaction with a target was easily seen by the Doppler-effect from blue shift of the spectral lines in the reflected light. Sauerbrey mentioned this most extraordinary observation in the abstract drastically in a direct way and not only hidden in a long paper. Why was no exceptional attention given to this fact? One problem was that extremely high quality lasers were needed as discovered and clarified by Jie Zhang (1998), see third paragraph of Section 8.

The acceleration was in the order of one billion times one billion times higher than the gravitational acceleration *g* on the Earth. It was well known from laser interaction with targets that very high accelerations of dense blocks of materials can be produced when laser pulses of the largest laser on Earth at the National Ignition Facility NIF in Livermore/California could heat a metal surface and ablate it from the generated plasma by an acceleration million times million times faster than *g*. This respectable result was achieved with very short laser pulses of one thousands of a millionth time of a second (nanosecond ns).

However, with thousand times shorter high power pulses, a basically different mechanism was observed. This new situation appeared following the development of the discovery by Gerard Mourou and associates (Strickland et al. 1985; Mourou 1994; Perry et al. 1994; Mourou et al. 1998; Mourou et al 2013) where lasers with powers in the range of petawatt (PW) could be produced with a pulse duration more than thousand times shorter than with nanosecond pulses, with picoseconds (ps) and now as short as attoseconds (as = $10^{-18}$s, see Krausz et al. 2009). In this case the heating and gasdynamic pressure processes of the thermal ablation could be ignored (determining the interaction with ns pulses) and in contrast, a non-thermal direct conversion of laser pulse energy was to go immediately into mechanical energy of motion of the ablated dense plasma blocks.

The way to these results in physics was paved by a large number of single steps which are in the focus of this book. After the discovery of the gravitation force by Newton as a force in the mechanical motion of masses was seen since knowing the precise elliptical planetary motion by Kepler, the subsequent force in electricity by Coulomb was a crucial step. But a next important step was the discovery by William Thomson (1845) – later Lord Kelvin – to show how electrically uncharged bodies can be moved by electric fields as ponderomotion. This force was formulated as a quadratic expression of the force quantities of electrostatic fields representing a nonlinear relation. Indeed, this *nonlinear-type of forces* had later to be combined with magnetic and temporal non-constant processes by the stress tensor of Maxwell using his electrodynamics.

The main advantage of the extremely fast laser-plasma interaction for ignition of controlled nuclear fusion for power generation (see sections 9 and 10) is related to overcome the enormous problems of complex systems. This was discovered and treated by Robert May (1972; 2011) – now Lord May of Oxford - focusing on physics of thermal processes. This is in contrast to the microscopic world of atoms and as it was formulated also in 1952 by Edward Teller (2001). The pioneering work of Lord May consists in the universal application of these solutions not only in physics, but also to zoology with the population of animals, also to medical problems as infectious deceases, or to economics including the banking systems and the financial crisis (Haldrane et al. 2011).

This opportunity to reduce the complex problems by extremely short time interactions is the special advantage of the here specifically presented new scheme. Lasers permit the driving of reactions through side-on ignition of fusion at very short times of uncompressed nuclear fusion fuel using nonlinear force acceleration of plasma blocks.

Teller (2001) summarized the problem in the following way: "Research on controlled fusion means dealing with the hydrodynamics of a plasma. I had a thorough respect for the fearsome nature of hydrodynamics, where every little volume does its own thing. Plasma does not consist of molecules, like a gas, but of ions—heavy slow moving positive ions—and light fast moving electrons those, in turn, creat and are coupled with electric and magnetic fields. For each little volume of plasma, several questions have to be answered: How many positive ions? How many electrons? How fast does each move on the average? What is the electric force, and what is the magnetic force acting on them?

"Mathematicians can predict the flow of matter as the volumes involved move in an orderly way. But even hydrodynamics of air was (and to some extend is—see wheather forecasts) beyond the grasp of mathematics. Theoretician of the nineteenth century proved that flying was impossible! In the twentieth century, they retreated to the statement that flying is impossible unless the air flow is confused and disordered (turbulent). Hydrodynamics as a science remains uncharted water.

"The same complications occure in planning a thermonuclear explosion. But an explosion occurs in a so short time that many of the complicated phenomena have no chance to develop. Even so it took a decade from Fermi's first suggestion of a thermonuclear reaction to the point (which occurred after the first full-scale demonstration of fusion) that the theoretical calculations of the explosions were reasonably complete. I had no doubt that demonstrating controlled fusion would be even more difficult".

This problem with plasma physics and hydrodynamics is still not fully solved now 60 years later, though a lot has been learnt. Only a research with thorough depth can lead us forward, as seen from the eminent achievements of Lord May of Oxford who ingeniously changed the initial insufficient approaches form the nineteenth century in theoretical physics now to master complex systems. His question is "will a large complex system be stable?" (May 1972) or to consider "systemic risk in banking systems" (Haldrane et al 2011).

The new situation for a here presented laser fusion scheme is, as Teller mentioned before, that the interaction of petawatt-picoseond laser pulses is fast enough "occurring in so short time" to avoid the usually appearing complications. This is the reason why this was considered to be revolutionary or as Steven Haan from the biggest NIF laser for fusion at the Lawrence Livermore National Laboratory said in an interview to the Royal Society of Chemistry (Haan 2011) that "this has the potential to be the best route to fusion energy".

Puting these very general aspects aside, the questions of Kelvin's ponderomotion well influenced the key development of laser physics and the exploration of forces in irradiated targets to extraordinary new streams of applications including the problems of generation of clean a-neutronic nuclear fusion for possible energy generation where even the problems of intermediary neutron generation is avoided (Tahir et al 1997) to eliminate the still existing problem connected with the usual fusion of heavy and superheavy hydrogen, deuterium and tritium DT.

The introductory overview of this Chapter should explain in an abbreviated way the historical background of the ponderomotive force—extensively discussed in the second last century—which has nearly been forgotten since then, until laser-produced plasma required a reconsideration. This renaissance included a lot of conscious or unconscious misunderstandings and incorrect expressions; therefore, we are sketching here first some of the crucial points in order to explain that there is a much more general description necessary for the *electrodynamic forces to single particles or plasmas produced by the laser fields* which sometimes have extremely high values and are sometimes of durations in the range of fs (femtoseconds).

Furthermore, there will be a rough description of some phenomena that are interesting in laser–plasma interaction and are related to the electrodynamic forces, which since their introduction were called *nonlinear forces*. This refers to the discussion in the following chapters that are more general, including the special ponderomotive forces depending not only on static but on time averaged high-frequency, stationary or transient fields with or without dissipation (thermalizing by collisional absorption). Indeed this all is restricted to plasmas including their properties of dispersion.

The *ponderomotive force* was discovered by Kelvin (1845). It is a force density $f_K$ produced in a dielectric medium with a refractive index $n$ ($n^2=\varepsilon$ is the dielectric constant) causing a polarization $P$, given by the scalar product with the tensor $\nabla E$ (where $E$ is the electric field)

$$f_K = \mathbf{P}\cdot\nabla\mathbf{E}\text{(in SI units)} = \mathbf{P}\cdot\nabla\mathbf{E}/(4\pi)\text{(in cgs units)}. \quad (1\text{-}1)$$

The vector symbols are used in the following way. The scalar product between two vectors *a* and *b* is given by *a* ·*b*, the vector product by *a*×*b*, and the dyadic (undefined) product leading to the tensor *ab* is given without any sign between the vectors.

In order to drastically underline the controversial (or at least complex) situation: One referee of a leading journal argued that Eq. (1–1) contains all that is going to be discussed in this book and none of this needs to be published because it is not new and everything is known about Eq. (1–1). I plan to convince the reader that there is more to be said, at least for plasmas and electrons.

The formulation of Kelvin (1–1) should now be rewritten. Remembering the definition of the polarization

$$\mathbf{P} = (n^2-1)\mathbf{E}/(4\pi), \quad (1\text{-}2)$$

Eq. (1–1) can be rewritten as

$$f_K = \frac{1}{4\pi}\left(\frac{n^2-1}{2}\right)\nabla \mathbf{E}^2 - \frac{1}{4\pi}(n^2-1)\mathbf{E}\times(\nabla\times\mathbf{E}). \tag{1-3}$$

In electrostatics one can express the electric field **E** by the gradient of a potential $\psi$ which means that this is free of curls

$$\mathbf{E} = -\nabla\psi; \qquad \nabla\times\mathbf{E} = 0. \tag{1-4}$$

This means that the last term in Eq. (1–3) is zero in electrostatics and one simply has the ponderomotive force expressed by the gradient of $\mathbf{E}\cdot\mathbf{E}$, the well-known *electrostriction, or electrostrictive force*. One may note that in the case that the rotation of *E* (i.e., $\nabla\times E$) is different from zero as given for non-electrostatic cases or for the high frequency case by the Maxwellian equations, even the historic ponderomotive force of Kelvin is more complicated than simply the gradient of the square of the electric field.

One may remember the following cases of the classical electrostatic ponderomotive force of Kelvin, Eq. (1–1). Figure 1–1(a) shows the Coulomb force acting on a charged particle within the homogeneous field of two condenser plates. If in Fig. 1–1(b) the charged particle is substituted by a sheet of dielectric material, no force will act at it since there is no excess charge and the tensor $\nabla E$ for a force [Eq. (1–1)] is zero. In Fig. 1–1(c) we consider a metallic sphere with positive charge producing a radial electric field in the air. If a dielectric material with $\varepsilon > 1$ is within this electric field it will be pushed by the ponderomotive force [first term of Eq. (1–3)] towards the metal, i.e., towards increasing $E^2$ by electrostriction, in absevne of any Coulomb force.

If the dielectric constant of the material is less than that of the surrounding medium, i.e., if there is a liquid with high dielectric constant, Fig. 1–2, then the material will be pushed towards lower values of $E^2$. If there are more positively charged metal spheres (Fig. 1–2) or if there is a tetrahedral geometry, the electric field will result in minimum between the spheres and the material will then be pushed into the minimum and perhaps be squeezed together with another material of the same lower dielectric constant than the surrounding liquid. This is an important technique for merging living cells and forcing them to improved genetic properties or to study cancer problems (Coster et al. 1995).

Being aware of the restrictions when mixing electrostatics with high frequency fields (limited strictly to nontransient time averaged cases) we now follow a consideration to see the special case of the *ponderomotive potential*. We expressively note that this is a very special limitation. For plasma irradiated by an electromagnetic wave of the frequency $\omega$, as will be shown in more detail, the refractive index *n* is given by the plasma frequency $\omega_p$

$$n^2 = 1 - \omega_P^2/\omega^2; \qquad \omega_P^2 = 4\pi e n_e/m, \tag{1-5}$$

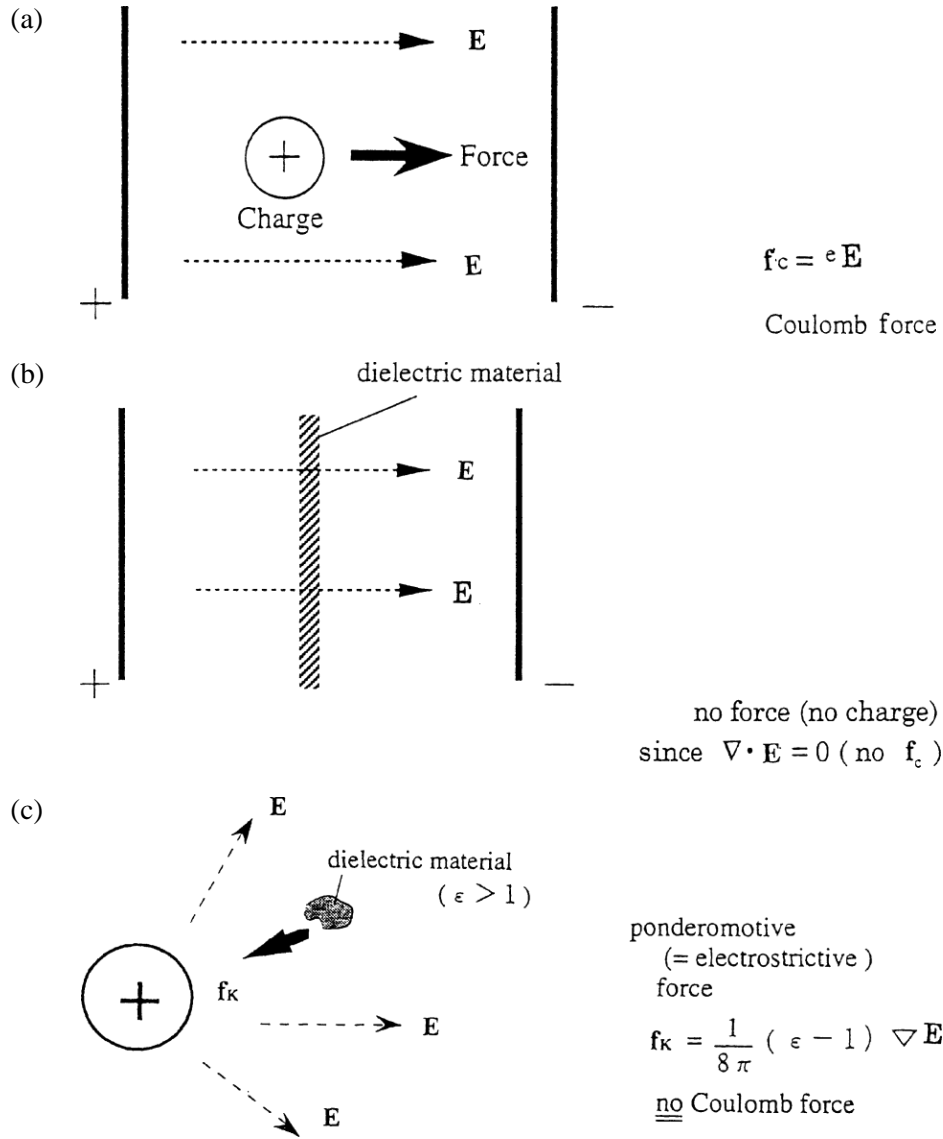

**Figure 1–1.** Electrostatic forces (Coulomb or ponderomotive) to materials in electric fields.

where $e$ is the charge, $n_e$ is the density, and $m$ is the mass of the electron. The (static) ponderomotive force of Kelvin acting at an electron fluid of density $n_e$, is expressed by the force density

$$f_K = mn_e \frac{d}{dt}v_e = \frac{n^2-1}{8\pi}\nabla \mathbf{E}^2, \tag{1-6}$$

where $v_e$ is the velocity of the electrons. Substituting the refractive index $n$ from

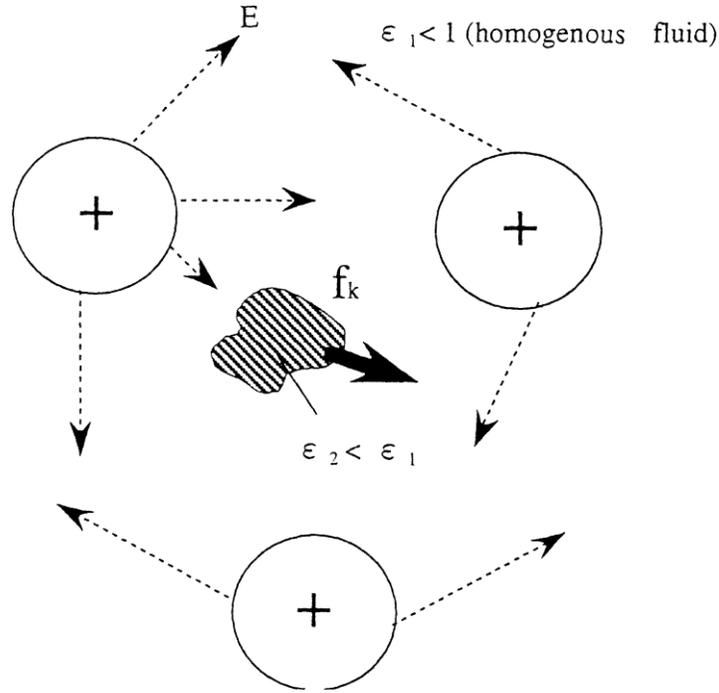

**Figure 1–2.** Dielectric material with a refractive index less than the surroundings within a minimum-$E$ field being pushed to the minimum.

Eq. (1–5) one can cancel the electron density $n_e$ and arrive at the equation of motion given by the force to one single electron

$$f_{Ke} = m\frac{d}{dt}v_e = -\frac{e^2}{2\omega^2 m}\nabla \mathbf{E}^2. \tag{1-7}$$

This force which can be expressed by the gradient of a scalar is [like $E$ in electrostatics, Eq. (1–4)] curlfree

$$\nabla \times f_{Ke} = 0, \tag{1-8}$$

and the force can be expressed by the gradient of a mechanical potential

$$f_{Ke} = -\nabla\phi; \quad \varphi = \frac{e^2}{2\omega^2 m}\mathbf{E}^2 \tag{1-9}$$

called the *ponderomotive potential*.

If the force in Eq. (1–7) is acting on an electron along a path between a point $r_1$ and a point $r_2$, the energy $E$ gained or lost is independent of the path chosen for the integration (conservative force) given by

$$E = \int_{r_1}^{r_2} f_{Ke} \cdot dr = -\frac{e^2}{2\omega^2 m}\left[\mathbf{E}(r_2)^2 - \mathbf{E}(r_1)^2\right]. \qquad (1\text{-}10)$$

If the ponderomotive potential is zero at $\mathbf{r}_2$,

$$\phi(r_2) = 0,$$

the energy gained by the electron is given by the ponderomotive potential where it starts. Under stationary conditions and if there is no emission or absorption of radiation (no Poynting flux) this can be applied to a cw-laser beam where an electron is generated in its center, e.g., by ionization from an atom. The electron is then emitted from the beam along the gradient of $\mathbf{E}^2$, i.e., in the radial direction of the beam, and it will gain the energy given by the ponderomotive potential [as measured first by Boreham et al. (1979)] where the laser intensity was high enough that the energy needed for the ionization could be neglected.

We shall see in the following that the calculation of the electron motion under these conditions is much more complex than a quiver drift. This usually arrives at the same global results as given by the ponderomotive potential.

It is therefore a point of caution to clarify whether the stationary conditions are fulfilled and the last term in Eq. (1–3) of the Kelvin ponderomotive force can be neglected or not. The questions of dissipation and of time dependent interaction are then a further point of attention.

It is interesting to note that the Helmholtz (1881) formulation of the ponderomotive force (Pavlov 1978) arrives for plasmas at the same expression as Kelvin's formulation. If the dielectric response given by the refractive index of solids is used, a discrepancy remains. This old controversy is in a similar way given for the question of what the correct relativistic description is for the propagation of light in media, the Abraham–Minkowski controversy. A transparent solution was possible for plasmas from reproducing Fresnel's formulas of reflectivity (Hora 1974a): the photons behave half like Abraham and half like Minkowski predicted (Klima and Petrzilka 1972; implicitly given by Hora (1969) see Hora 1991: Section 9.4) and as was clarified by Nowak (1983). It may be noticed that the problem for condensed matter seems to have been solved by Peierls (1976) in mutual confirmation and clarifying questions of the Schwarz-Hora effect (Hora et al. 2013).

It is indeed important to note that the ponderomotive potential acts on a free electron in the same way as an electric voltage $U$ or the electrostatic potential $\psi$, Eq. (1–4), acts on an electron. The ponderomotive force is therefore similar with its quadratic $\mathbf{E}^2$-gradient as is the electric field $\mathbf{E}$ in the Coulomb force $\mathbf{f}_C$ acting on an electron

$$\mathbf{f}_C = e\mathbf{E} = e\nabla\psi. \qquad (1\text{-}11)$$

One only has to be aware that the electrodynamic force (apart from the electrostatic case) is more complex than the ponderomotive force, Eq. (1–7) to the electron and only the used of all components of Maxwell's stress tensor arrives at the correct solutions (Cicchitelli et al. 1990).

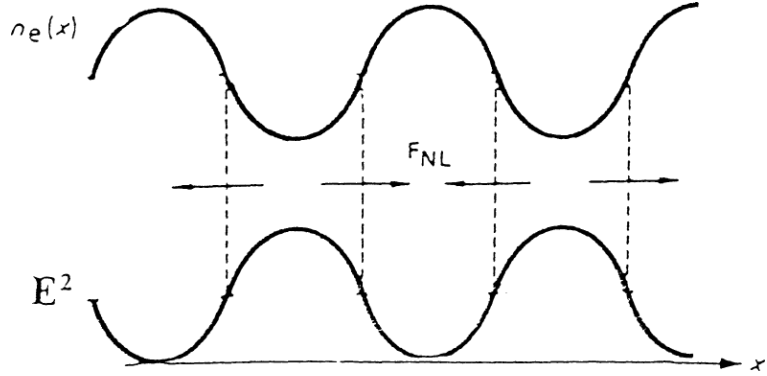

**Figure 1–3.** $E^2$ for a standing wave field where the nonlinear (ponderomotive) force pushes the electrons into the nodes with a resulting electron density (dashed curve) as shown by Weibel (1957, 1958).

Historically it is interesting to note that the ponderomotive force of the kind illustrated in Eq. (1–7) after its discussion in electrostatics—now without dielectric media—was first used for the time-dependent high frequency case by Erich Weibel (1957, 1958) for the standing wavefield for microwaves (Fig. 1–3, however much more complicated in details, anticipating Fig. 7-2 calculated only after a number of iterations: Cicchitelli et al. 1990a), followed by Gapunov and Miller (1958) and Boot et al. (1959). Weibel discovered that electrons will be driven into the nodes of the standing wavefield. Stationary laser beams interacting with electrons without dielectric effects were first discussed by Kibble (1966). The essential source of the dielectric effects to the electrodynamic forces at laser interaction with plasmas was first formulated by Hora et al. (1967) and generalized as a nonlinear force, where the macroscopic theory of the equation of motion in plasmas had to be modified by adding nonlinear terms in order to achieve momentum conserveation (Hora 1969).

Despite all the complications for the hydrodynamic theory the result is rather easy to understand. If light in one dimension is moving into a plasma with monotonously increasing electron density $n_e$ (Fig. 1–4), the averaged electric field squared $E^2$ of the laser light coming from vacuum with an amplitude $E_v$ is being swelled up due to the refractive index $n$ decreasing from the vacuum value to a minimum value at the critical density [although $n$ in (1–5) is modified by absorption].

The resulting ponderomotive or nonlinear force $f_{NL}$ is given by the negative gradient of $E^2$ tearing the plasma into an ablative part and into a compressive interior. The compression corresponds to an ordinary radiation pressure, which, however, is increased by the swelling $1/n$. After detailed analysis, the Z-times ionized plasma ions emitted into the vacuum achieve an amount of energy

$$\varepsilon_i = Z\frac{e^2}{2\omega^2 m}\mathbf{E}_v^2\left[\frac{1}{|n|_{min}} - 1\right] \tag{1-12}$$

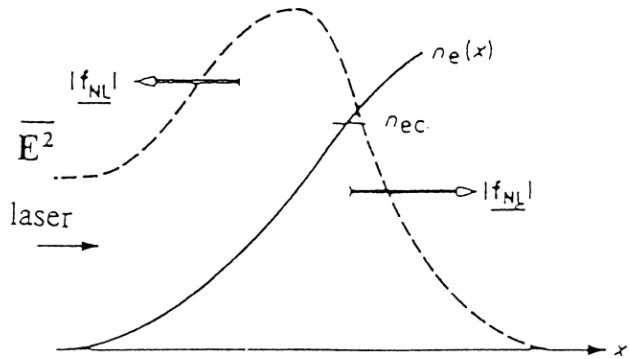

**Figure 1–4.** Electron density $n_e$ depending on the plasma depth $x$ with laser light incident from the vacuum side where $n_e$ is zero. The averaged laser field $\overline{\boldsymbol{E}^2}$ grows over its vacuum value $\boldsymbol{E}_v^2$ due to the decreasing refractive index $n$ to a maximum value near the critical electron density $n_{ec}$. The negative gradients of the $\boldsymbol{E}^2$-field are the nonlinear forces which drive the plasma corona towards the vacuum (ablation) and give a compression reaction to the plasma interor (Hora et al. 1967; Hora 1991).

which is exactly given by the change of the ponderomotive potential between the plasma interior and the vacuum. This linear separation of the ion energy was observed since the very first days of laser plasma interaction experiments and is characteristic of the action of the nonlinear (ponderomotive) force*. Any gas dynamic mechanism with thermal equilibrium of the ions cannot result in such a separation of the ions by their charge number $Z$.

The need to look into the details of these processes as explained in the following can be seen from an example where one calculates the interaction of initially resting electrons with "pancakes" of photons, e.g., 800 nm wave length laser pulses of 18 fs duration (5.4 micrometer long) and of 50 $\mu$m focus diameter (Barty 1996), Fig. 1–5. The single electron interaction of the photon pancake of a few PW power will accelerate a single electron to energies in the GeV range (Häuser et al. 1994). It turns out that the maximum energies of the electrons are nearly as high as if the electron was accelerated within the ponderomotive potential. It is rather curious what the ponderomotive potential is able to reproduce although it would not have been clear at the beginning how the thin pancake of photons would do this. Indeed this all depends on further conditions (Häuser et al. 1994) to be explained. The critical point in this is that the purely electrostatic ponderomotive force of Kelvin (1–1) has been extended to plasma for which in some sense no electrostatics exist and the high frequency refractive index was used formally only to get a result from Eq.1–1).

---

*This is the reason why the processes treated in this book cannot simplifying be called "Radiation Pressure Acceleration RPA" because there is a detailed inclusion of the dielectric response involved given by Maxwell's stress tensor. An example is shown where all components of the tensor are needed to arrive at the correct nonlinear solution (Cicchitelli et al. 2000) to agree with the experiment. This turns out to be a principle reasons in contrast to linear physics (Section 6.3) .

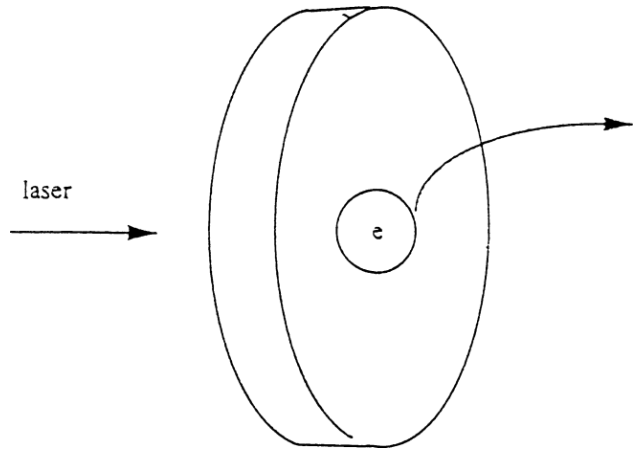

**Figure 1–5.** A 18 fs petawatt laser pulse focused to 50 μm diameter is like a pancake of photons. Hitting resting electons will relativistically accelerate them as a free wave accelerator to GeV energies (Häuser et al. 1994; Barty 1996).

The ponderomotive force—generalized now for the non-electrostatic properties of the plasma—is related to the Lorentz force which is usually considered a typical relativistic result. We shall show how the Lorentz force can be derived from nonrelativistic Maxwell's theory which indeed has the connection to Maxwell's stress tensor (see Section 3.2). Therefore, the plasma generalized ponderomotive force, the Lorentz force, the Maxwellian stress tensor and the new aspects of plasma theory created by the very high laser intensity interaction with plasmas are interwoven. They are considered here in their full complexity in view of the basic background. The reference to Chris Barty (1996) was anticipating Mourou's discovery of the initially mentioned generation of Petawatt-picosecond laser pulses (Mourou 1994; Miley 1994; Perry et al. 1994), see Fig. 8-1 (Mourou 2011).

The following description of the electrodynamic laser plasma interaction by the nonlinear (ponderomotive) forces is given in two steps. First the macroscopic hydrodynamic plasma theory is discussed and then the cases of single particle motion.

In order to illustrate the deeper meaning of the nonlinear force and ponderomotion, a review of the basic framework of hydrodynamics and of electrodynamics will be given as would be appropriate for a graduate course where the student should have detailed knowledge of theoretical physics on hydrodynamics and on electrodynamics. We concentrate then on an elaboration and review of the main conceptual background of these fields while the details of the derivations are referred back to the earlier courses. Some details of the mathematical derivations and proofs of the basic questions relevant to the nonlinear force and ponderomotion are documented in an earlier reference (Hora 1991).

Coming back to Kelvins discovery of ponderomotion in electrostatics, it should not be overlooked that it was not trivial but it was historical understanding that after Coulomb discovered the electric forces between charges that the electric forces are possible also between uncharged neutral bodies with a dielectric response showing the quadratc (nonlinear) relation.

We shall mostly focus on the application of the dielectric and not immediate charge related force on energy problems for clean fusion but we have to be aware what extreme frame is spanned. The upper end refers to accelerations at the surface of black holes with Hawking and Unruh radiation at intensities of pair production in vacuum (Stait-Gardner 2006, Eliezer et al. 2002: see there Sect. 1.5) where electrons can not longer be fermions (Hora et al 1961). The lower end of intensities is how the dielectric generated force at nondestructive intensities in living tissues can contribute in medicine for healing (Liebert et al. 2014).

The motivation for this discovery by Kelvin was somehow guided at his time from observations. One could see the pictures of diagrams of the flowing of liquids and similar pictures of the electric fields or of static magnetic fields or from the vector distribution of magnetic fields. It was tried to understand whether there were relations. We shall see that the relation between mechanical and electrical phenomena is linear only with the Coulomb force. The general relation is basically nonlinear as seen from Maxwell's stress tensor including the derivation of the Lorentz-force being based on (Section 3.2) in a way that this opened a route to Einstein's relativity theory. But it is more: in some generalized sense, the understanding of nonlinearity opens a basically new door or a dimension for physics exploration as it will be explained from discussions with Feynman (Section 6.3) and reflecting the position of Hawking that theoretical physics is not at the end but nonlinearity is opening a door for a new dimension of knowledge.

# CHAPTER 2
# Elementary Plasma Properties and Hydrodynamics

## 2.1 DEFINITION OF PLASMA

Reflecting the physics problems and the mix of presumptions of electrostatics with electrodynamics described in the preceding section, there may be questions as to what extent the earlier definition of plasma is still valid: *Plasma is a physical state of high electrical conductivity and mostly gaseous mechanical properties at high temperature*.

This is to be compared with the usual definition that "plasma is the fourth state of aggregation: solid, liquid, gas, and plasma". This latter definition conflicts with the fact that the plasma state is possible in solids and liquids (metals and semiconductors). This brings us back to the physics before the middle of the 19th century, especially before Maxwell's theory. The theory of electricity was then based on electrostatics [Coulomb's law, electric fields (definition of force **E** or through the quantity of charge density **D**) and the connection to magnetic fields **H** or **B** given by Ampére's and Faraday's laws]. The media were polarizable bodies (solid, liquid or gas) used in condensers or in electrophorus. Metals were strange: one needed metals to show the Ampére current which produced the magnetic field **H** around the conductor, or in Farday's closed wire loop within which the temporal change of the enclosed magnetic field **B** produced a voltage. But what was the dielectric constant of a metal? In electrostatics, one could not define this easily. A metal between two condenser plates discharged the plates and no polarization appeared from the dielectric insulators.

An ingenious step was taken by Hendrik Antoon Lorentz who tried to describe the metals as an ensemble of electrons and ions (after the electron was realized by Lenard at the end of the nineteenth century) where the dielectric constant in between is that of a vacuum. This plasma-like description in the Lorentz theory of metals could indeed explain a number of phenomena. In an electrostatic sense one cannot define a dielectric constant. When including the electrodynamics of the high frequency fields, a dielectric constant and a refractive index $n$ given for the radian frequency $\omega$ of the interaction could be defined by Eq. (1–5). This was not reached by Lorentz since the plasma frequency $\omega_p$ was discovered several years later by Langmuir. How close Lorentz was with this including the Debye length of a plasma discovered later is remarkable. The tragedy is that Lorentz, otherwise, was so prominent just to derive the optical properties for high frequency of dielectrics with the refractive index $n$ from Maxwell's theory with a splendid reproduction of the reflection laws which Fresnel derived before purely empirically.

After Maxwell discovered from his wave equations that electromagetic quantities do not propagate infinitely fast but with the limited speed of light *c*, Lorentz went a step further, asking what happened in electrodynamics if all phenomena were limited to velocities less than *c*. This was the discovery of the Lorentz transform for relativity. The next step, taken by Einstein in 1905, was to draw the consequences for mechanics discovering time dilatation, mass variation at high speeds and the relation $E = mc^2$.

We see that metals and with them semiconductors and ionized gases with electric conductivity in magnetic fields do not directly fit into the picture of the earlier electrodynamics. It is no surprise that the study of the electrical conductivity of electrolytes arrived at one of the first substantial formulations for plasma by S.R. Milner (1912, 1913) and later by Peter Debye and his school, such that Denis Gabor (1953) speaks about the "Debye–Milner theory".

It is therefore no surprise that the electrostatic ponderomotive force of Kelvin underwent many modifications by plasma theory much later under the extreme conditions of laser interaction.

In view of all this, one should not define the plasma simply as a fourth state of matter. Perhaps the definition given above, reflecting the views of the Lorentz theory of metals, comes closer to the facts which are also illuminated by the non-linear (ponderomotive) force aspects.

## 2.2 Elementary Plasma Parameters

One important parameter is the *plasma frequency* $\omega_p$ which determines the oscillations of the electron gas between the ions in a plasma. Electrons in a homogeneous plasma with a density $n_e$ are removed from their equilibrium position by any process, e.g. incident electromagnetic waves, as shown in Fig. 2–1. By simple geometry we find the relative change of the electron density given by

$$\frac{dn_e}{n_e} = -\frac{d\xi}{dx} . \qquad (2\text{-}1)$$

The electric field **E** produced by this change is given by the differential change $dn_e$ of the electron density according to Gauss' law, and using Eq. (2–1)

$$\frac{d\mathbf{E}}{dX} = -4\pi e dn_e = +4\pi e n_e \frac{d\xi}{dx} . \qquad (2\text{-}2)$$

This field produces a force on each electron which can be used in a Newton equation of motion, describing the temporal dependence of the disturbance length $\xi(t)$

$$m\frac{d^2\xi}{dt^2} = -e\mathbf{E} = -4\pi n_e e^2 \xi(t). \qquad (2\text{-}3)$$

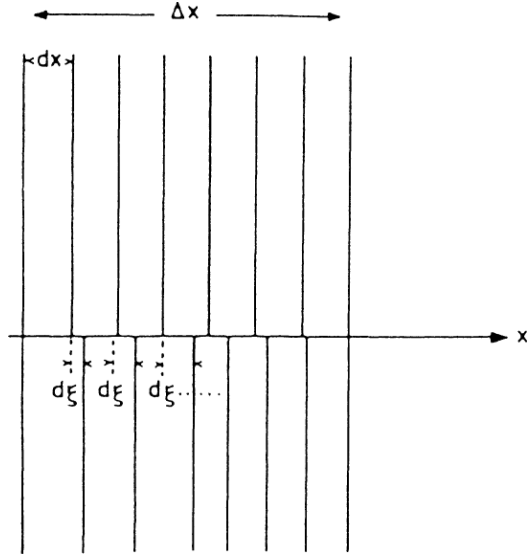

**Figure 2–1.** Electron gas between the ions in a homogeneous plasma (upper part) are removed from their equilibrium position by a length $d\xi$.

This second order differential equation can be solved by the ansatz $\xi(t) = \text{const} \cdot \exp(i\omega_p t)$ and describes an oscillation with a frequency

$$\omega_p^2 = \frac{4\pi e^2 n_e}{m} \ ; \qquad \omega_p = 5.65 \times 10^4 \sqrt{n_e} \ , \tag{2-4}$$

which is called the plasma frequency. The units used are the Gaussian magnetic cgs system where the electron charge is $e = 4.803 \times 10^{-10}$ cm$^{3/2}$s$^{-1}$g$^{1-2}$ and the electron mass is $m = 0.90109 \times 10^{-27}$ g. The electron density $n_e$ is given in particles per cm$^3$. This plasma frequency was discovered by Langmuir in order to explain why radio waves were reflected at the ionosphere like light is reflected by metals. Another plasma parameter is the *Debye length* $\lambda_D$ which is given as a kind of wavelength to the electrostatic Langmuir oscillations of the plasma frequency. But instead of the product of wavelength × frequency being a wave speed, we now have to take the product of wavelength $\lambda_D$ times the radian frequency ($2\pi$-times the frequency) where the wave velocity is identical with the averaged thermal velocity $v_e$ of the electrons. This is related to the electron temperature $T$ by the Boltzmann constant $K = 1.38 \times 10^{-16}$ erg/degree through $mv_e^2/29 = KT/2$ per degree of freedom. So, we find then the Debye length as

$$\lambda_D = \frac{v_e}{\omega_p} = \left[\frac{KT}{4\pi n_e e^2}\right]^{1/2} \ ; \tag{2-5}$$

$$\lambda_D (\text{cm}) = 6.9 \left[ \frac{T(\text{K})}{n_e(\text{cm}^{-3})} \right]^{1/2} = 743 \left[ \frac{T(eV)}{n_e(cm^{-3})} \right]^{1/2}. \tag{2-5a}$$

In the first part of Eq. (2-5a) the temperature is given in degree Kelvin and in the second part it is given in electron volts with the relation $T$ (Kelvin) $= 1.16 \times 10^4$ (eV).

Debye found this length not in such a way but from the statistical thermal fluctuations of the electrons in their microscopic state between the ions. The missing $2\pi$ in relation to Langmuir's plasma oscillation may have caused some headache for Debye because he was always very critical of wavelength and frequency as essentially related to a wave equation. This way of thinking was fundamental for the creation of the Schrödinger equation of quantum mechanics. The essence of the Nobel prize-winning PhD thesis of de Broglie was that the product of the momentum of an electron and a hypothetical wavelength has to be Planck's constant $h$. When Schrödinger presented this result in an appraising seminar in Berlin in 1924, Debye simply asked: if there is a wave length where is then the wave equation? Schrödinger's bitterness at not being able to give an answer forced him to find the solution: his equation. As shown before (Hora 1991: Appendix A), a logically different derivation of the Schrödinger equation directly from the Hamilton equation of classical mechanics is given.

Another meaning of the Debye length in Eq. (2–5), differing also from Debye's thermostatistical derivation, is presented here for the physicist looking into laser produced plasmas. Figure 2–2 shows a plasma of a temperature $T$ and electron density $n_e$ at the right-hand side expanding into a vacuum. The equally energetic but much lighter electrons escape faster than the ions into the vacuum such that a high concentration of ions in the plasma surface is generated. Their positive charge (as a double layer producing an exit potential of $3kT/2$ for the electrons) prevents additional electrons from leaving to the vacuum. Going through the derivation one arrives at a thickness of the ion layer given by the Debye length.

An important quantity in plasma (we are for simplicity always talking about fully ionized plasmas) is the *collision frequency* $\nu$ between the electrons and ions. This process can be described with very elementary mathematics. We shall show later that this is in rough agreement with the more complex theory of second quantization.

If an electron is moving (Fig. 2–3) towards an ion of positive charge (the ion may be a nucleus with $Z$ protons) at a lateral separation $r_0$ at large distance, (called the impact parameter), it will be attracted by a Coulomb force **f** where **r** is the distance-vector between the electron and the ion

$$f = \frac{-Ze^2 \mathbf{r}}{r^3}. \tag{2-6}$$

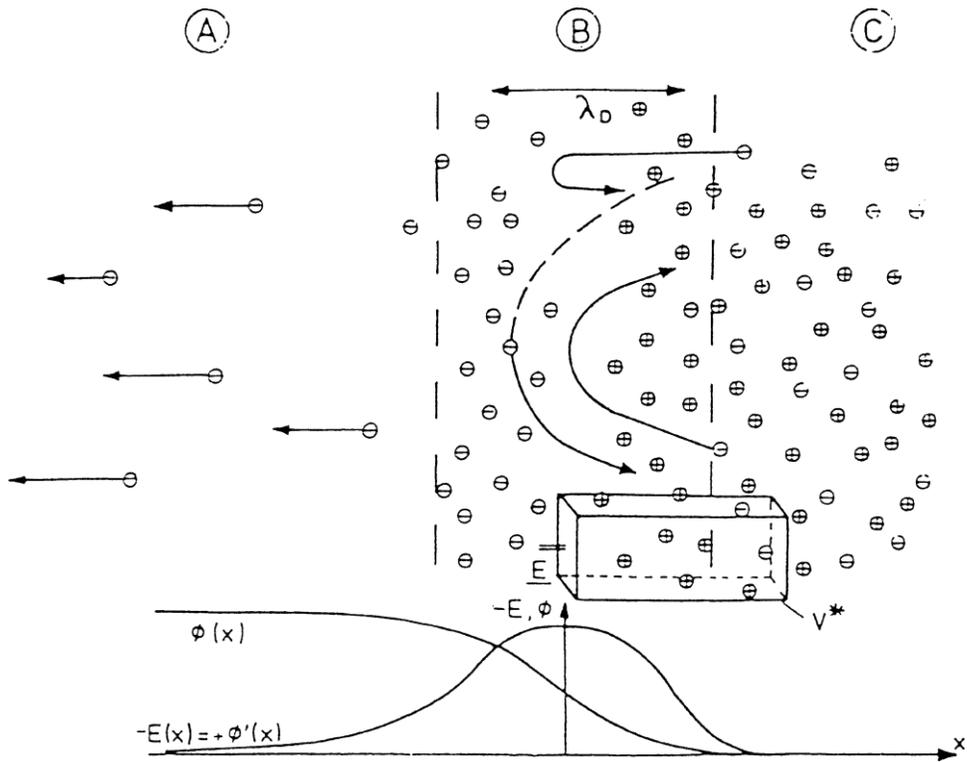

**Figure 2–2.** Between the vacuum range A and the space charge neutral plasma interior C, the plasma surface sheath is depleted by the escape of fast electrons until such a strong space charge is built up that the following fast electrons from the plasma C are electrostatically returned into C. The electric field **E**(x), due to the space charge separation in B, and its potential $\varphi$ are given.

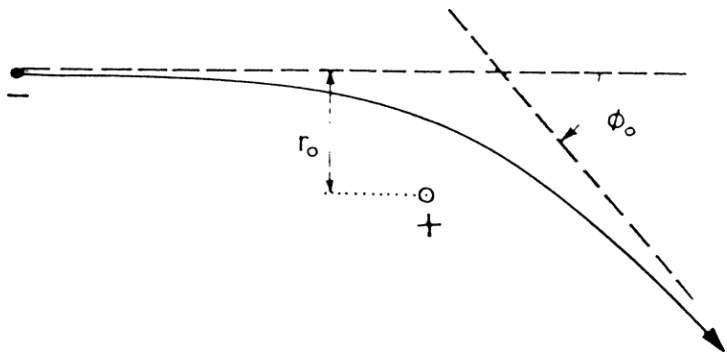

**Figure 2–3.** An electron (–) having an impact parameter $r_0$ moves around a positively charged ion along a hyperbolic curve due to Coulomb attraction if the ion mass is assumed to be sufficiently large.

The main interaction occurs during a time

$$t = \frac{r_0}{v} \qquad (2\text{-}7)$$

if the electron has a velocity $v$. During the interaction time $t$, a change of the electron momentum occurs, resulting in an angle $\varphi_0$ (Fig. 2–3) given by the value

$$\Delta(mv) = |ft| = \frac{Ze^2}{r_0 v}. \qquad (2\text{-}8)$$

If we restrict attention to only 90° collisions, then the relation results

$$\Delta(mv) \square |ft| \square \frac{Ze^2}{r_0 v}; \quad r_0 = \frac{Ze^2}{mv^2}, \qquad (2\text{-}9)$$

defining the value $r_0$ of the impact parameter for the Coulomb collision. There was a long discussion about small angle collision in the past, but we can simply use the fact, well-proven experimentally, that the 90° scattering results in a collision frequency which is rather close to the values observed experimentally.

The Coulomb collision cross section is given by

$$\bar{p} = \pi r_0^2 = \frac{z^2 \pi e^4}{m^2 v^4}, \qquad (2\text{-}10)$$

which results in the collision frequency in a plasma of an ion density $n_i$

$$v_{ei} = n_i \bar{p} v = \frac{Z n_e \pi e^4}{m^2 v^3}. \qquad (2\text{-}11)$$

Expressing the electron velocity $v$ by the electron temperature $T$ arrives at

$$v_{ei} = \frac{Z n_e \pi e^4 3^{-3/2}}{m^{1/2} (KT)^{3/2}}. \qquad (2\text{-}12)$$

This determines the Ohmic conductivity of a plasma as

$$\sigma = \frac{3(KT)}{2Z\sqrt{m/3\pi e^2}}, \qquad (2\text{-}12a)$$

which is independent of the plasma density and has a value equal to that of the best metallic conductors when the plasma temperature is about one million degrees.

For the modification by small angle scattering and for the quantum generalization of the collision frequency at higher plasma temperatures to produce anomalous resistivity (see Sections 2.5 and 2.6) we simply add here the results.

The Spitzer (1962) formula for the electron ion collision frequency [instead of Eq. (2–12)] is

$$v_e = \frac{n_e}{(KT)^{3/2}} \frac{Z\pi^{3/2} e^4 \ln\Lambda}{m^{1/2} 2^{5/2} \gamma_e(Z)}, \qquad (2\text{-}13)$$

where the factor $\gamma_e$ describes the correction due to electron–electron collisions which value is between 0.5 for $Z = 1$ monotonously growing to 1 for high $Z$. The Coulomb logarithm in Eq. (2–13) is the ratio between the Debye length [Eq. (2–5)] and the impact parameter for 90° scattering [Eq. (2–9)]

$$\Lambda = \frac{\lambda_D}{r_0(90°)} = \frac{3}{2Ze^3}\left(\frac{K^3 T^3}{\pi n_e}\right)^{1/2}$$

resulting in the numerical values

$$v_{ei} = 8.513\times 10^{-7} \frac{Z n_e}{\gamma_e(Z) T_e^{3/2}} \ln\left(1.55\times 10^{10}\frac{T_e^{1/2}}{Z n_e^{1/2}}\right).$$

The quantum correction of the collisions appears when the impact parameter for the 90° scattering (Fig. 2–3) can no longer be described by classical point mechanics but where the de Broglie wavelength is larger than the impact parameter and a quantum mechanical diffraction of an electron by the ion is valid (Hans Bethe 1934, Hora 1981a). The impact parameter then changes to

$$r_0 = \frac{r_{\text{Bohr}}}{2Z}\frac{1}{(1+4T/T^*)^{1/2}-1}$$

$$= \begin{cases} \dfrac{r_{\text{Bohr}} T^*}{4ZT} = \dfrac{Ze^2}{3kT} & \text{if } T < T^*, \\ \dfrac{r_{\text{Bohr}}}{4Z}\sqrt{\dfrac{T^*}{T}} = \dfrac{h}{2\sqrt{3kT/m}} & \text{if } T > T^*, \end{cases} \qquad (2\text{-}13a)$$

where the plasma temperature for the transition is

$$T^* = \frac{4Z^2 mc^2 \alpha^2}{3K} = 4.176\times 10^5 Z^2 \ (\text{K})$$

using the fine structure constant $\alpha = 2\pi e^2/(hc)$ with Planck's constant $h$, and the Bohr radius $r_{\text{Bohr}} = h^2/(4\pi^2 me^2)$.

The collision frequency has then to be generalized to

$$v_{ei} = \frac{\pi r^2_{Bohr} n_e}{4Z^3} \sqrt{\frac{3kT}{m}} \left[ \left(1 + 4T/T^*\right)^{1/2} - 1 \right]^{-2}$$

$$= \begin{cases} \dfrac{Z\pi e^4}{3^{3/2} m^{1/2}} \dfrac{n_e}{(kT)^{3/2}} = v_{ei} & \text{if } T \ll T^*, \\[2ex] \dfrac{\pi h^2}{3^{1/2} Z m^{3/2}} \dfrac{n_e}{(kT)^{3/2}} = v_{ei} \dfrac{T}{T^*} & \text{if } T \gg T^*, \end{cases} \quad (2\text{-}13b)$$

using the 90° scattering formula [see the identical value with (2–12) for temperatures below $T^* = 35.6 Z^2$ eV)]. The electrical conductivity is then generalized to the formula

$$\sigma = \frac{en_e}{2mv_{ei}} = \frac{eZ^3}{\sqrt{3mkT} r_{Bohr}} \left[ \left(1 + 4T/T^*\right)^{1/2} - 1 \right]^2$$

$$= \begin{cases} \dfrac{3^{3/2} (kT)^{3/2}}{2Z m^{1/2} e^2} = \sigma_{cl} & \text{if } T \ll T^*, \\[2ex] T^* \sigma_{cl}/T & \text{if } T \gg T^*, \end{cases}$$

and the diffusion of plasma across magnetic fields changes in the same way. This was drastically confirmed by measurements in stellarators (Grieger et al. 1981) working with deuterium at a temperature $T = 800$ eV. The diffusion across the magnetic field was 20 times larger than the classical value. This behavior was reproduced by using the factor $800/T^* = 22$. From these evaluations and comparingwith experimental results we concluded that the Tokamak works only because of this quantum correction of the collisions (anomalous resistivity) and plasmon effects (Boreham et al. 1995).

## 2.3 Hydrodynamics

High temperature plasmas follow the hydromechanical properties of gases. We review here some basic equations of hydrodynamics and refer to the detailed derivations (Hora 1991: Chapter 4). In the following subsection the basic concept of the microscopic plasma theory (Hora 1991: Chapter 5) is explained on which hydro- dynamics also can be based.

Leonhard Euler extended Newton's single particle mechanics to the entity of a fluid or gas. Instead of having Newton's equation of motion where mass time acceleration is the force, Euler had to formulate all to the mass density and the force density **f** in the hydrodynamic system. The mass density $\rho$ in a plasma is expressed by the dependence on the spatial coordiantes $x$, $y$, and $z$, and on the time $t$

$$\rho(x, y, z, t) = m_i n_i (x, y, z, t) + m n_e (x, y, z, t), \quad (2\text{-}14)$$

where $m_i$ is the ion mass, $n_i$ is the ion density, $m$ is the electron (rest) mass and $n_e$

is the electron density.

Mass times acceleration is then the mass density $\rho$ times the acceleration in the velocity field **v** of the fluid:

$$\rho \frac{d\mathbf{v}}{dt} = -\nabla p + \eta \nabla^2 \mathbf{v} \tag{2-15}$$

where the force density on the right-hand side is given by the negative gradient of the pressure $p$

$$p = \frac{3}{2} n_e K T_e + \frac{3}{2} n_i K T_i \approx \frac{3}{2}(1+Z) n_i K T \tag{2-16}$$

Pressures (force per area) are equal to energy density (force per volume) and the gradient of this energy density is the force density as shown in Eq. (2–15). This is mainly the Euler equation; the second term in Eq. (2–15), the Navier–Stokes term, was added later to express the friction losses in the fluid or gas due to the viscosity $\eta$. In most cases of a one fluid model for plasmas this viscosity may be neglected, however, not in the following two-fluid models

The total differentiation of the velocity **v** in Eq. (2–15) has to be performed by partial differentiation to all coordinates.

$$\rho \frac{d\mathbf{v}}{dt} = \rho \frac{\partial \mathbf{v}}{\partial t} + \rho \frac{\partial \mathbf{v}}{\partial x}\frac{dx}{dt} + \rho \frac{\partial \mathbf{v}}{\partial y}\frac{dy}{dt} + \rho \frac{\partial \mathbf{v}}{\partial z}\frac{dz}{dt}. \tag{2-17}$$

Remembering the meaning of the velocity components one can then write the Euler equation of a plasma in the following vectorial form

$$\rho \frac{\partial \mathbf{v}}{\partial t} + \rho \mathbf{v} \cdot \nabla \mathbf{v} = -\nabla p. \tag{2-18}$$

This *equation of motion* corresponds to the conservation of momentum.

The equation of conservation of the mass is expressed in hydrodynamics by amount of density $\rho$ which is lost or gained when a fixed volume $V$ (Fig. 2–4) is moved substantially in space. This is given by geometric relations expressed by

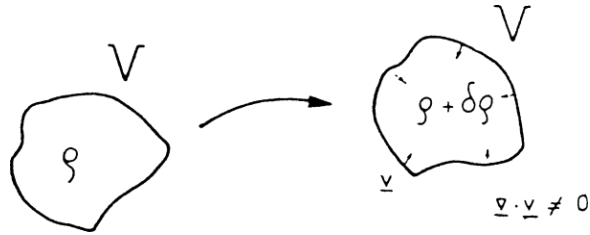

**Figure 2–4.** Change of density in a fixed volume $V$ when moving within a fluid resulting in the equation of continuity.

vectors and using Gauss' theorem (which was derived at this occasion). The result is (see Section 4.3) that

$$\frac{\partial \rho}{\partial t} + \rho \nabla \cdot \mathbf{v} = 0. \tag{2-19}$$

This *equation of continuity* has to be satisfied in hydrodynamics.

In hydrodynamics the equation of *energy conservation* has to be satisfied. This can immediately be understood by writing

$$\frac{\partial}{\partial t}\frac{\rho}{2}\mathbf{v}^2 = \frac{\partial}{\partial t}\frac{3}{2}n_i KT(1+Z) - \frac{3}{2}n_i(1+Z)\nabla\cdot(\kappa_T \nabla T) + W. \quad (2\text{-}20)$$

The local change (partial differentiation) of the kinetic energy density of the fluid (left-hand side) has to be compensated by the change of the internal energy (first term on the right-hand side) which expresses the pressure as the thermokinetic energy density (Eliezer et al. 1986: Chapter 2) modified by a second term. This second term describes any energy loss or gain from thermal conductivity ($\kappa_T$ is the coefficient of thermal conduction) plus a term *W* describing the local gain or loss of energy density by other than thermal processes, e.g. radiation emission or absorption of laser energy. This term needs special attention and involves the complete calculation of the temporally- and spatially-changing laser field in the irradiated plasma including penetrating and reflected waves.

With the three equations, the vector equation of motion (2–18), the scalar equation of continuity (2–19), and the scalar equation of energy conservation (2–20), we can solve completely the temporal and spatial behavior of the three quantities: density $\rho$, velocity **v**, and temperature *T*, if the initial and boundary values for the motion are defined.

We now consider an example of hydrodynamic motion from the field of laser-produced plasmas (Engelhardt et al. 1970). In Fig. 2–5 we see the side-on framing camera pictures taken from a free-falling aluminium sphere irradiated from the left-hand side by a laser pulse of 3.4 J at different times after the irradia- tion. Two groups of plasma could be identified by microscopic analysis of the photographs. There was a spherically symmetric core of plasma expanding slowly and a further half-moon-like asymmetric plasma expanding with high velocity against the laser light with ion energies of more than 10 keV. The expansion velocity of the core which contained about 95% of the plasma corresponded to reasonable temper- atures of a few tens of eV. The hydrodynamic computation of the time dependence of the inner core could be followed up, e.g., by the diagram of Fig. 2–6.

What is surprising is that the one-sided irradiation of the target re- sulted in the highly symmetric thermal distribution of the radiation to the whole thermokinetically-expanding inner sphere. This was visible evidence that the plasma expansion followed the self-similarity model of hydrodynamics (for a de- tailed explanation, see Hora 1991: Chapter 5). This could also be confirmed by comparing the expected energy deposition from the laser into the plasma core from the self-similarity theory (Basov and Krokhin 1964; Dawson 1964; Hora 1964, 1971) with the measurements as shown in Fig. 2–7.

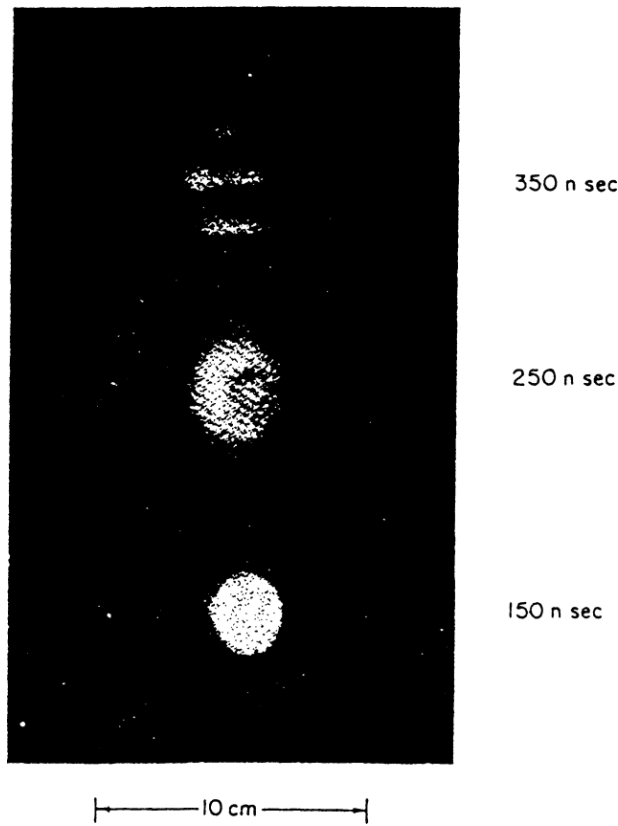

**Figure 2–5.** Side-on framing camera picture of a plasma produced from an aluminum sphere of 80 µm radius at the time indicated after irradiation by a 30 ns ruby laser pulse focused to 0.4 mm diameter. The second frame shows the outer part of a rapidly expanding plasma and an inner spherical thermally expanding part [Engelhardt et al. (1970).]

The self-similarity model is a rare case where the otherwise very complex numerical solution of the three hydrodynamic equations (2–18), (2–19) and (2–20) for the conservation of momentum, mass and energy can be achieved analytically. If one has the initial conditions of a spherical gas (or plasma) with radial density profile of Gaussian shape (Fig. 2–8), with a locally constant initial temperature and with an initial velocity profile depending linearly on the radius, then the plasma will expand or compress (depending on the direction of the initial velocity) while always keeping a Gaussian profile with varying width and varying central maximum but constant overall integral. The temperature will also be locally constant at each time step, varying with time as given by adiabatic compression or expansion. This is well-known from astronomy (Heckmann 1942) or from the expansion of the universe with the present blackbody radiation-temperature of three Kelvin de-grees. This was tacitly used for laser produced plasmas where the assumption of a box-like density profile instead of a Gaussian could be mathematically justified in good approximation (Hora 1971a).

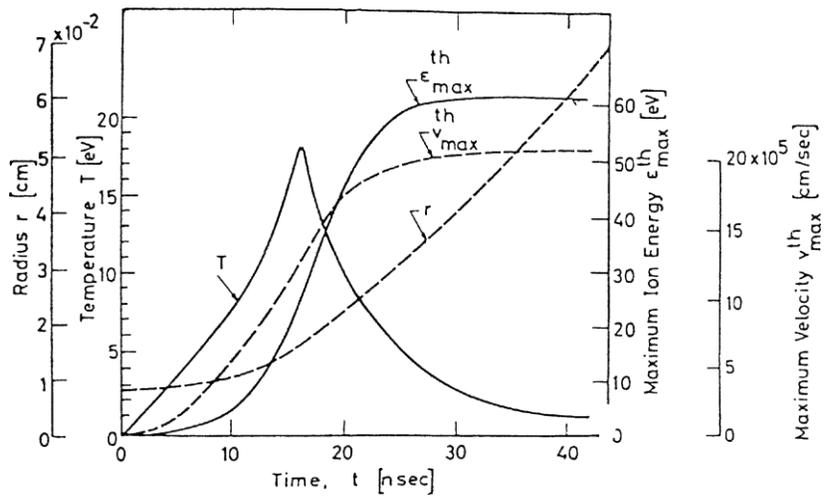

**Figure 2–6.** Example of the numerical computation of the time history of radius, temperature, maximum velocity and maximum thermal energy of the aluminium plasma of initial radius 80 μm at irradiation of a 3.4 J, 16 ns ruby laser pulse. The iterative solution included the time dependence of the interaction cross section.

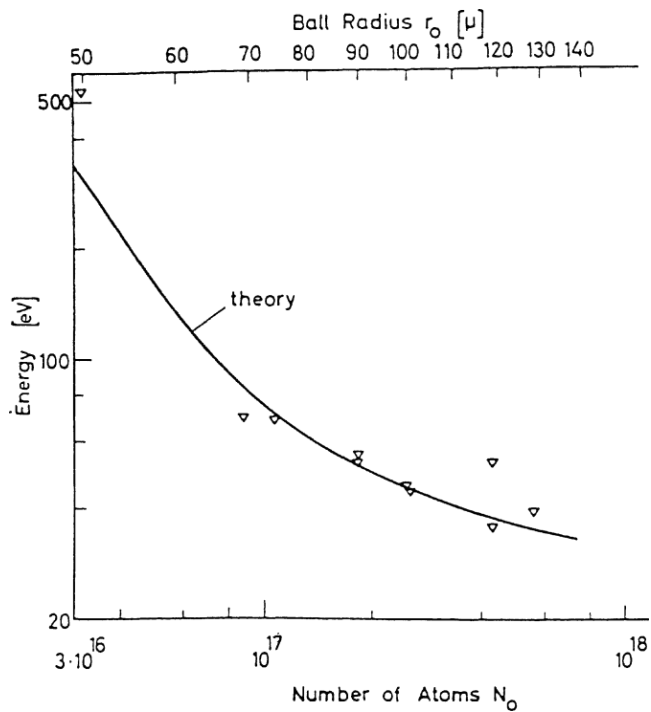

**Figure 2–7.** Measured maximum ion energies of plasmas produced from aluminium balls of varying ball radius with irradiation by laser pulses of about 70 MW and 30 ns pulse length (∇) compared with theoretical values, based on the self-similarity model (curve).

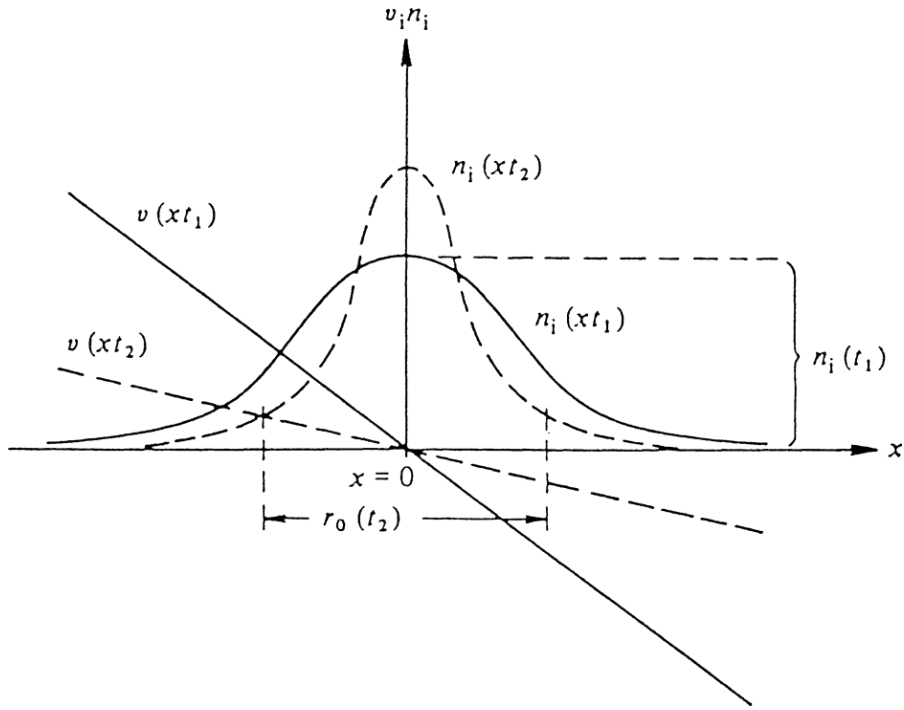

**Figure 2–8.** Self-similarity model for the ideal adiabatic compression of a plasma whose initial density profile is Gaussian. The (instantaneous) temperature profile is spatially constant and the velocity profile is linear on the radius of the plasma sphere (Hora 1971a).

It is interesting to consider the *further case of numerical solution of the hydrodynamic equations* for the plane wave laser irradiating from a frozen hydrogen layer perpendicularly. The laser intensity was chosen at a level where the nonlinear force effects considered later were not yet acting but only the thermokinetic pressure $p$. The driving term $W$ in Eq. (2–20) described how the laser radiation is propagating through the plasma corona until the critical plasma density. Beyond this the light cannot penetrate and decays within the skin depth as in the case of a metal. The time- and space-dependent optical absorption was described using the classical absorption constant determined by the collision frequency, Eq. (2–12), through the electron density $n_e$ and the plasma temperature $T$.

The result (Mulser 1970) is seen in Fig. 2–9. While the electron density in the solid hydrogen is $6 \times 10^{22}$ cm$^{-3}$, the light can penetrate only to densities around $10^{21}$ cm$^{-3}$ within the plasma corona which contains plasma of about 1 million degrees Kelvin blown off (ablated) into the vacuum with velocities of more than $10^7$ cm/s. Though no light is touching the high density parts of the solid hydrogen block there is thermal conduction to heat the plasma below the corona to temperatures up to several 10,000 K. The momentum of the ablated plasma in the corona results automatically in a recoil to the plasma below the corona and compresses this conductivity heated plasma to densities four times solid density. This plasma heating and compression by ablation was the first publication [together with the result of Rehm (1970)] which showed laser-produced plasma ablation and compression of plasma, as expected from shock wave studies (Guderley 1942).

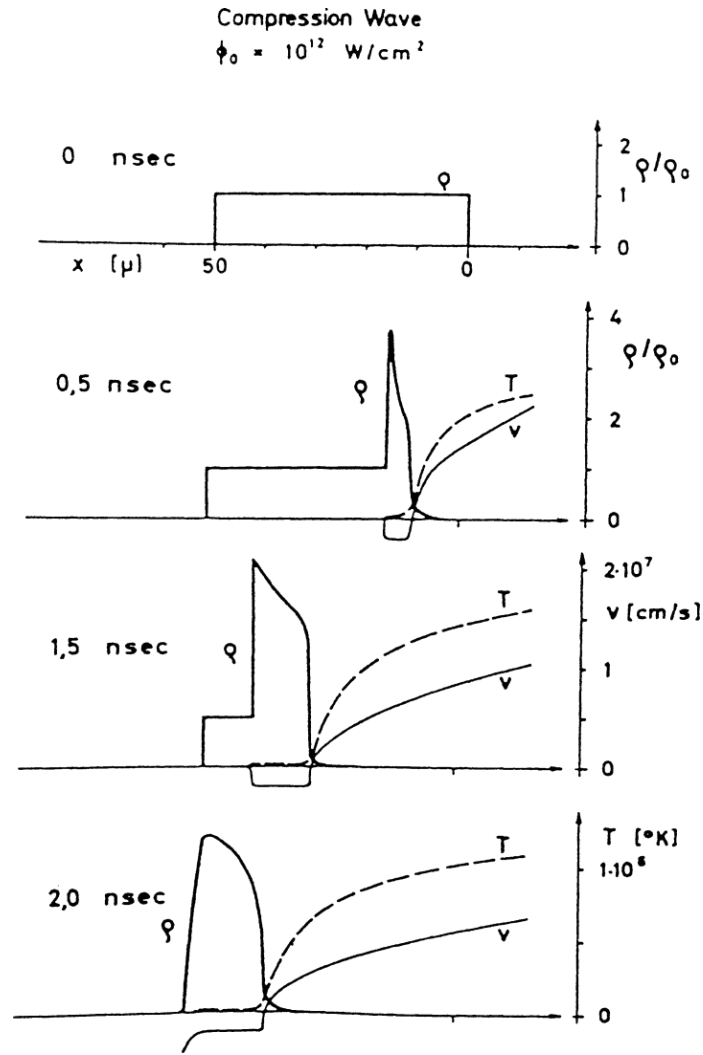

**Figure 2–9.** One-dimensional numerical solution of the hydrodynamic equations of conservation for laser light with a steplike intensity of $10^{12}$ W/cm$^2$ incident on a 50 μm thick slab of solid hydrogen (density $\rho_0$). The resulting density $\rho = n_i m_i$, velocity $v$, and temperature $T$ are shown for times $t = 0$, 0.5, and 2.0 ns. After Mulser (1970).

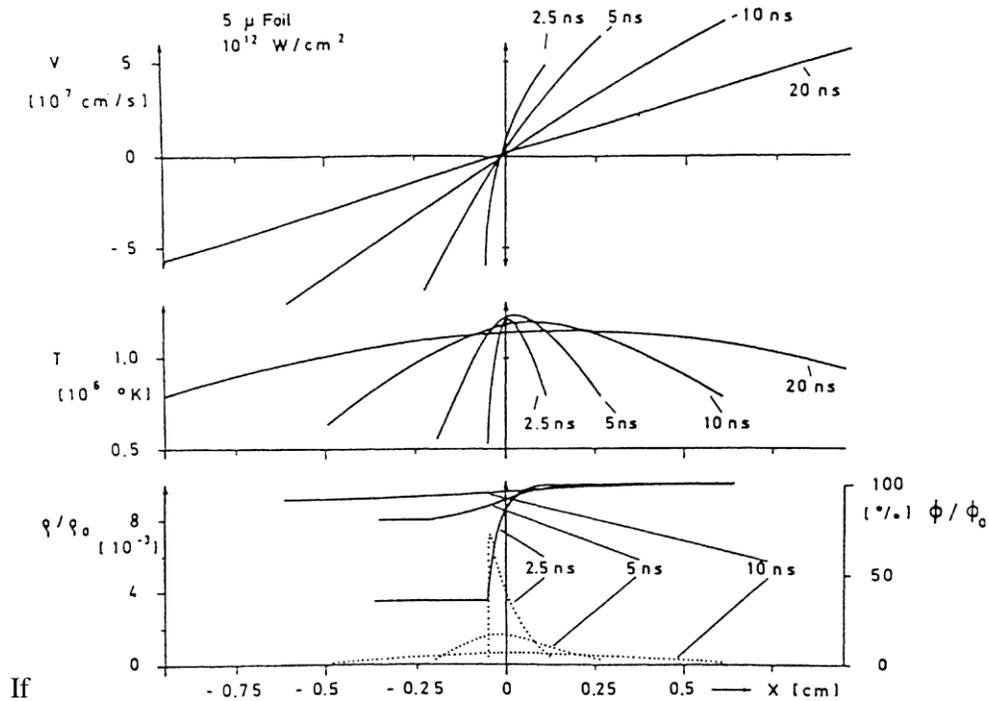

**Figure 2–10.** One-dimensional numerical solution of laser plasma interaction for a hydrogen foil of 5 μm thickness. A linear velocity profile and Gaussian density profile resulted at 5 ns and later. After Mulser (1970).

the target is spherical and if a special time dependence of the "tailored" laser pulses is used, densities of laser compressed polyethylene spheres more than 2000 g/cm$^3$ have been produced (Azechi et al. 1991).

When the irradiated hydrogen is a very thin foil (Fig. 2–10), the plasma density profile develops on time more and more symmetrically—despite the one-sided irradiation—and approximates automatically to a Gaussian density profile as given by the self-similarity solution (Hora 1971a; Phipps et al. 1998, 1993).

It is interesting to note that the measurement of the burn-through time for the hydrogen layers (Sigel 1970) is ten times shorter than that calculated from hydrodynamics (Mulser 1970). And even the hydrodynamic calculation should have resulted in a longer burn-through time if the reduction of the thermal conductivity would have been included due to the electric double layer. *The electric double layer* between the hot corona and the cold dense plasma causes a depletion of electrons (see Fig. 2–2) and thermal conduction is then performed by the ions, giving a thermal conductivity which is reduced by the square root of the mass ratio of ions to electrons (Cicchitelli et al. 1984), Fig. 2–11.

Obviously, the interaction of the laser beam with the hydrogen layers in Sigel's experiment did not occur by the way that the laser beam interacted like an ideal plane wave front. The beam had hot spots and other mechanisms occurred [such as ponderomotive self-focusing (Hora 1969a), or anomalous quantum-collisions; Eq. (2-13b)] so that the laser light spread much faster than the plane hydrodynamic front moved into the solid hydrogen. The measured faster burn-through time could be followed up by observing self-focusing filaments, showing that the laser produced plasma moved from the end of the foil back towards the laser light (Hohla 1970). These processes may explain why the laser light spread so quickly and uniformly into the aluminum spheres with one sided illumination (Engelhardt et al. 1970), Fig. 2–5, and why the inner core temperature of these plasmas between 20 and 80 eV agreed so well with the predictions of the self-similarity model.[1]

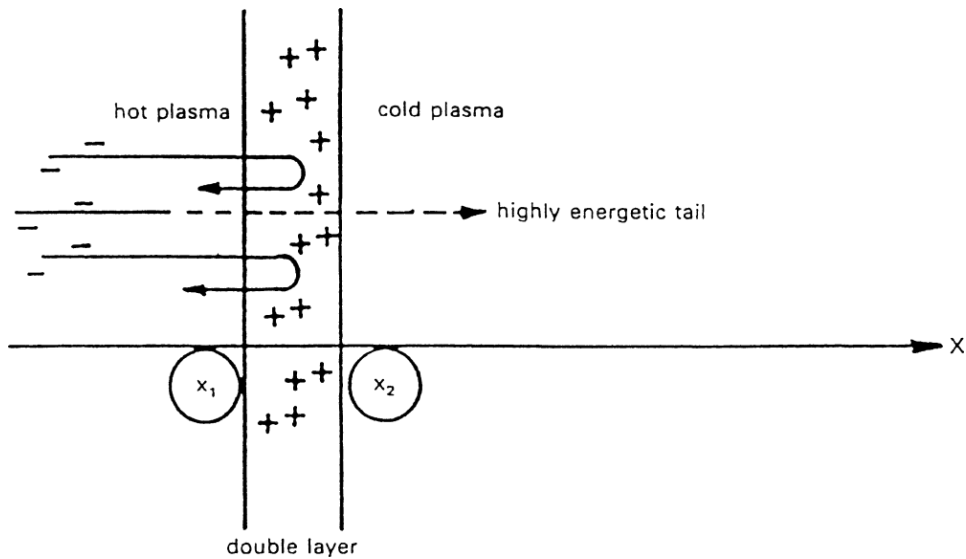

**Figure 2–11.** The positive charge of the double layer between the hot and the cold plasma causes a return of the electrons to the hot plasma with the exception of the electrons from the energetic tail of the energy distribution.

In contrast, the ions in the outer half moon like plasma of Fig. 2–5 do not at all follow hydrodynamics, and their energies up to 20 keV were just the reason to look into non-thermodynamic interaction mechanisms leading to the nonlin- ear (ponderomotive) forces discussed later. Here we summarize the results from Fig. 2–5.

The keV ion energy did not at all depend on the size of the irradiated aluminium ball (Fig. 2–12) where the cases with the same laser power and pulse duration had to be selected. Further, the maximum keV ion energy increased superlinearly with an exponent 1.5 on the laser power for selected cases of the same pulse duration (Fig. 2–13) while any thermal processes can only be sublinear (theories arrive at exponents between 0.3 and 0.6). The keV ions are therefore definitely of a nonlinear origin.

---

[1] The agreement of the laser-driven compression dynamics of deuterium–tritium fusion plasmas with the self-similarity model in recent experiments for the highest neutron gains and compression to 2000 times the solid density was rather a surprise. The facts are reported in Section 8-10.

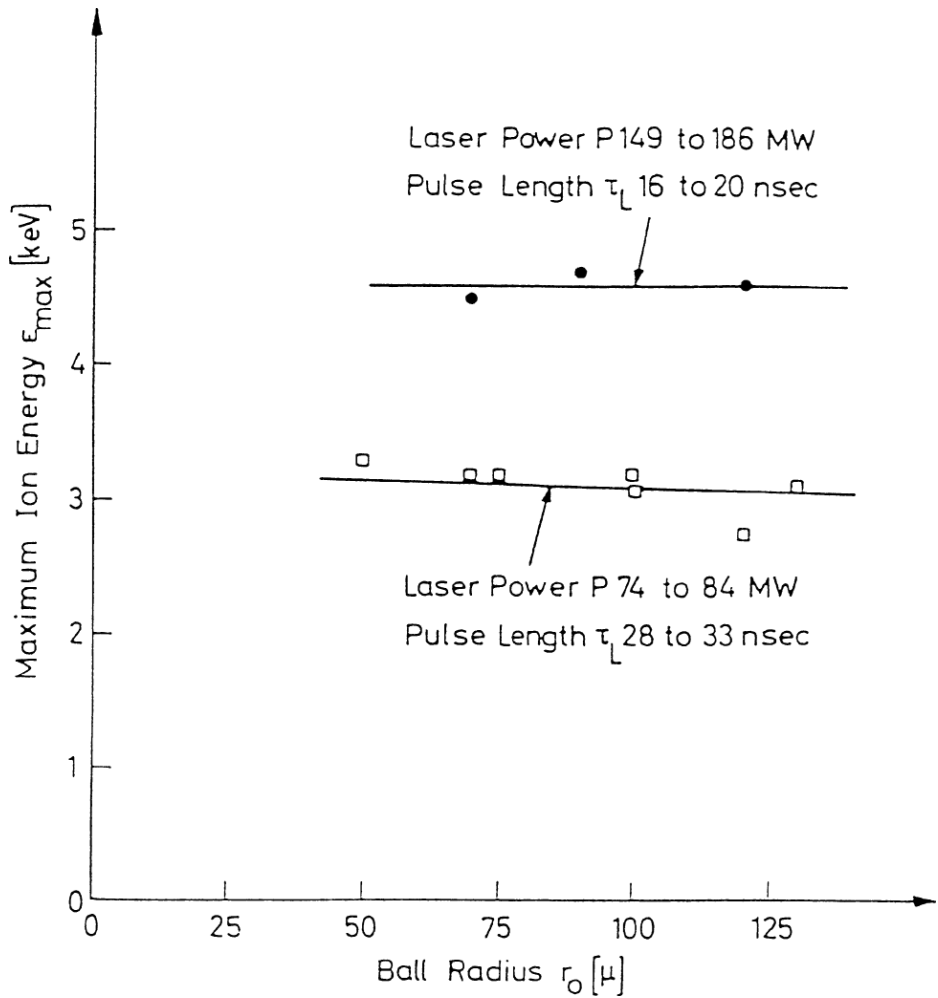

**Figure 2–12.** Maximum ion energy of the keV ions of the half moons in the experiments (Engelhardt et al. 1970) which are independent of the size of the irradiated aluminium ball.

Another important curiosity is the fact that the measured keV (or later MeV) ions are separated by their charge number $Z$ where ion energy $\varepsilon_i$ linear with $Z$ ($\varepsilon_i \propto Z$), a fact which in no way can be explained thermally. Thermal equilibrium results in the same ion energies whatever charge number $Z$ these have. To demonstrate this we see the result of Ehler (1975) from carbon targets irradiated with $CO_2$ lasers of different intensity. Side-on spectroscopic pictures immediately showed (Boland et al. 1968) how the plasma cloud was separated into groups of plasma with different ionization (see the later Fig. 5–2).

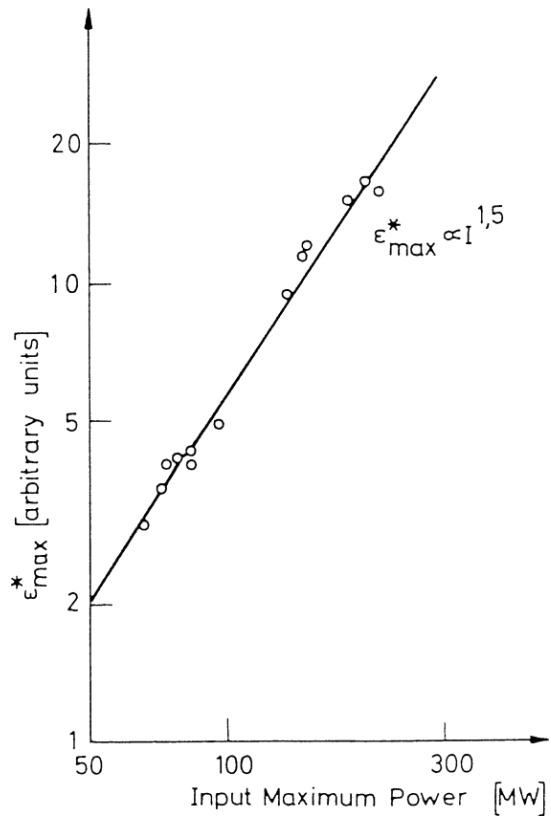

**Figure 2–13.** Superlinear increase of the maximum energy of the keV ions in the half moons of Fig. 2–5 (Hora et al. 1971b).

It can be concluded that the main part of the keV ions cannot be generated by the ambipolar electric fields in the plasma surface: Taking the thickness of the Debye layer, Fig. 2–2, times the surface of the plasma results in less than $10^9$ ions in this volume for the ambipolar acceleration. The measured number of ions of each Z-group is more than $10^{13}$, therefore ambipolar acceleration is not dominating while for a group of $10^9$ ions with no dependence on the laser polarization was a proof that this may have been due to ambipolar effects (Wägli et al. 1978).

The facts of the very nonthermal and nonlinear behavior of the laser generated plasma with the most unexpected high ion energies of keV, MeV and more were first noted around 1962–1963. What was observed in this very early stage of laser interaction with plasmas by Honig (1962) and by Linlor (1963) were anomalies which could not easily be understood:

(a) Electron emission of more than 1000 times higher current densities than the Langmuir–Child law of space charge limitation permitted for electron emission (Honig 1962), and

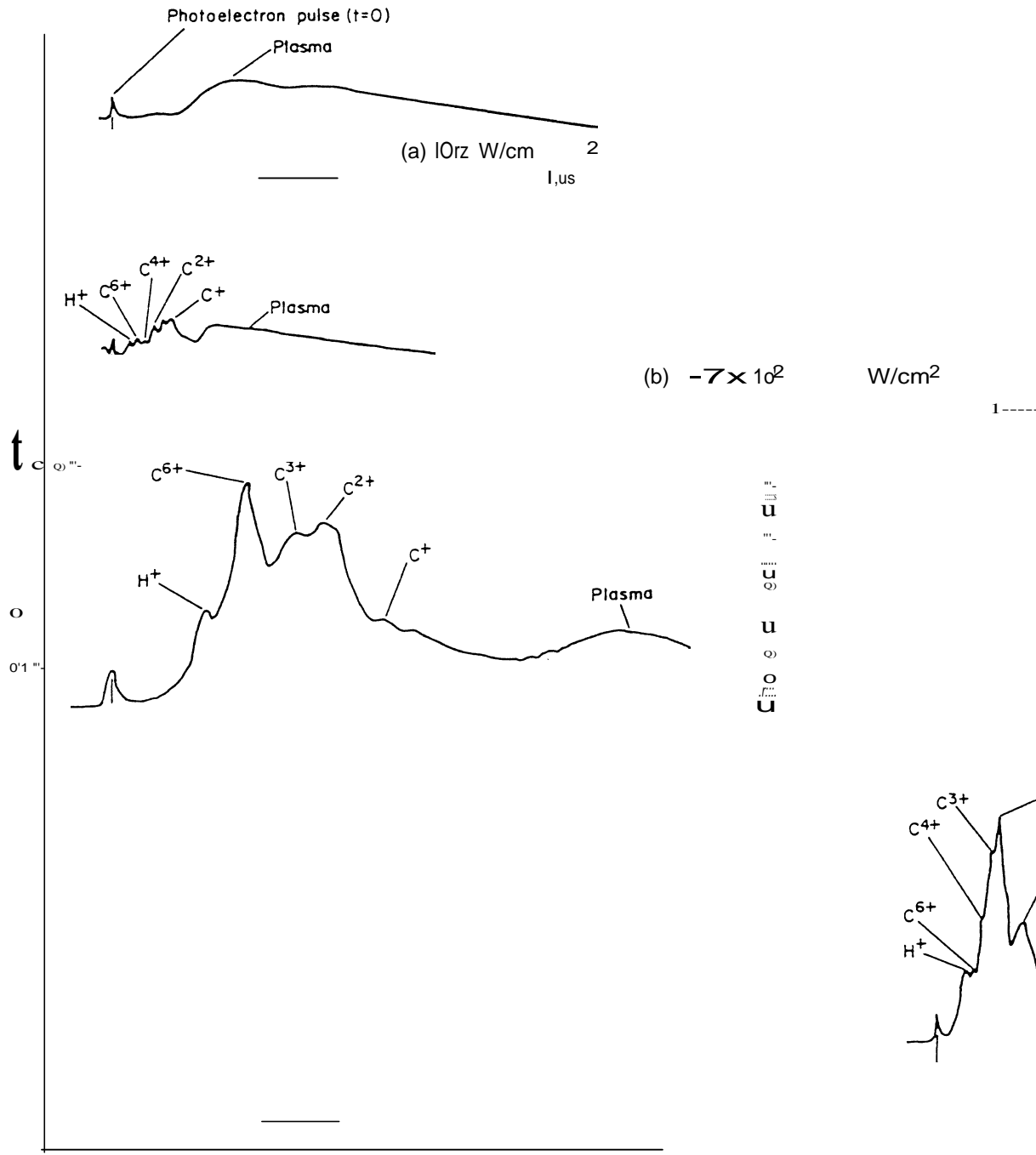

**Figure 2-14.** Time of flight diagrams of the charge collector signals in front of a polyethylene target irradiated by $CO_2$ laser intensities $\phi$ (Ehler 1975).

(b) The generation of 10 keV ions from plasmas for which the temperature was shown to be only a few dozen eV (Linlor 1963). The transferred momentum showed a similar jump (Metz 1973; Hora 1973b).

The only condition was that the laser power $P$ had to be above about 1 MW. Below 1 MW, the plasma behaved in a fully classical manner; following thermodynamic kinetics (Ready 1971).

The same was seen in Fig. 2–5: the half moon like pattern corresponded to aluminium ions of more than 10 keV. From a very large number of observations of the kind of Fig. 2–5, one could determine that these keV ions followed highly nonlinear properties (Hora 1971b) and was one of the motivations to look into the nonlinear laser–plasma interaction by the direct electrodynamic forces to the plasma (Hora et al. 1967).

## 2.4 Concept of Microscopic Plasma Theory

Hydrodynamic plasma theory, where additional forces have to be included in the equation of motion (2–18), e.g., the Lorentz force, can be derived from a microscopic theory similar to Lorentz's attempt to describe the metals from an ensemble of a large number of electrons and ions. This is in contrast to the purely phenomenological hydrodynamics based on Euler's equation of motion, Gauss's clarification of continuity and energy conservation elaborated from the classical gas laws.

The model considers a large number of electrons and ions giving them (classically!) numbers $n$ from 1 to $N$. Each of these particles follows an equation of motion where the acceleration of the $n$-th particle with the mass $m_n$ is given by the forces from all the other particles 1 to $N$ (with exception of $n$) given by their position coordinates $(x, y, z)$ and their velocities (temporal derivatives of the coordinates). We have then a set of $3N$ differential equations

$$m_n \frac{d^2 x_n}{dt^2} = f_{xn}\left(x_i, ..., z_N, \frac{\partial x_1}{\partial t}, ..., \frac{\partial z_n}{\partial t}\right); \quad n = 1, ..., N,$$

$$m_n \frac{d^2 y_n}{dt^2} = f_{yn}\left(x_i, ..., z_N, \frac{\partial x_1}{\partial t}, ..., \frac{\partial z_n}{\partial t}\right); \quad n = 1, ..., N, \quad (2\text{-}21)$$

$$m_n \frac{d^2 z_n}{dt^2} = f_{zn}\left(x_i, ..., z_N, \frac{\partial x_1}{\partial t}, ..., \frac{\partial z_n}{\partial t}\right); \quad n = 1, ..., N,$$

This system of equations is treated by computers to describe plasmas with $N$ up to one million and more. The forces $f$ are simplified and only the interaction of the particles with their immediate neighbors are covered. While a considerable number of plasma phenomena like oscillations, instabilities, etc., can be perfectly described, others where the Coulomb collision frequency is important cannot. Even with the biggest available computers, the long range Coulomb collision could only be covered approximately and the theoretical predictions in these cases are not perfect.

The next step for a microscopic description of plasmas is to summarize the large number of particles by introducing a distribution function (which loses some information). To understand a distribution function one may start from a number $N$ of elements $a_i$ ($i = 1$ to $N$) of which an average value $M$ has to be found

$$M = \frac{1}{N}\sum_{i=1}^{N} a_i. \tag{2-22}$$

Instead of this easy procedure one may bundle elements of equal value, where $f_i$ describes the number of the elements of each bundle. With this weighting of numbers one can then perform the same averaging task as in (2–22)

$$M = \frac{\sum_{i=1}^{n} f_i a_i}{\sum_{i=1}^{n} f_i}. \tag{2-23}$$

The numbers $f$ describe the distribution of all the elements and this can also be expressed for a continuous (instead of discrete) manifold by a distribution function $f(x)$

$$\langle a \rangle = \frac{\int f(x) a(x) \, dx}{\int f(x) \, dx} \tag{2-24}$$

in order to get the expectation value $a$, the weighted average value of the manifold $a(x)$. Up to this point all is fully deterministic. The fact that these methods are also in gambling and probability theory does not mean that the results are uncertain.

Our ensemble of $N$ particles can be described by a distribution function $f$ depending on the coordinates $x_i$ and the velocity components $v_i$ ($i = 1, 2, 3$) giving the number of particles of these (6-dimensional) coordinates which are located in the differential element $dx_1\, dx_2\, dx_3\, dv_1\, dv_2\, dv_3 = d^3x\, d^3v$

$$f(x_1, x_2, x_3, v_1, v_2, v_3, t) d^3x d^3v. \tag{2-25}$$

One has then to realise that instead of talking about quantities $Q$ directly one looks into the distribution function and then derives the expectation (average) values of $Q$ from the procedures given by (2–24)

$$\overline{Q(\mathbf{r},t)} = \frac{\iiint_{-\infty}^{+\infty} Q(\mathbf{r},\mathbf{w},t) f(\mathbf{r},\mathbf{w},t) d^3 w}{\iiiint\!\!\iiint_{-\infty}^{+\infty} f(\mathbf{r},\mathbf{w},t) d^3 w d^3 x dt} \tag{2-26}$$

The denominator is equal unity for a normalized distribution function $f$. If $Q = 1$, one obtains the particle density $n_e$ for electrons or for ions of the plasma. If $Q$ is the (microscopic) velocity $\mathbf{w}$ one arrives at the macroscopic plasma velocity $\mathbf{v}$ (or drift velocity) while the difference $\mathbf{u}$ is the random (thermal) velocity $\mathbf{w} = \mathbf{v} + \mathbf{u}$ $u$ can be a Maxwellian distribution (Hora 1991; Appendix 2) but can just as well describe any other non-equilibrium distribution.

The advantage of the description with the distribution function $f$, called kinetic theory, consists in the fact that non-equilibrium states and their relaxation can be described. This is not possible with simple one fluid macroscopic hydrodynamics where temperatures $T$ are defined from an equilibrium state. The intriguing property of the distribution function $f$ consists in the fact that there is a very simple way for deriving the macroscopic hydrodynamics with- out needing the phenomenological pictures of Euler, Gauss or the classical gas laws. This is given from the (logically very simple) equation that any temporal change of $f$ has to be described by the collisions between the particles

$$\frac{d}{dt}f = \left(\frac{\partial f}{\partial t}\right)_c. \tag{2-27}$$

The intriguing property of $f$ can be seen when integrating Eq. (2–30) by $d^3w$ over the whole velocity space. The integration over Boltzmann's collisional term is zero and the other terms immediately result in the macroscopic hydrodynamic equation of continuity (2–19).

$$\frac{d}{dt}f = \frac{\partial}{\partial t}f + \frac{\partial}{\partial x_1}f\frac{\partial x_1}{\partial t} + \frac{\partial}{\partial x_2}f\frac{\partial x_2}{\partial t} + \frac{\partial}{\partial x_3}f\frac{\partial x_3}{\partial t}$$
$$+ \frac{\partial}{\partial w_1}f\frac{\partial w_1}{\partial t} + \frac{\partial}{\partial w_2}f\frac{\partial w_2}{\partial t} + \frac{\partial}{\partial w_3}f\frac{\partial w_3}{\partial t}. \tag{2-28}$$

Using the nabla operator for the spatial differentiation and the nabla operator with an index $w$ for the differentiation towards the velocity components and remembering that the force $\mathbf{F}$ can be expressed according to Newton's law by

$$\mathbf{F} = m\frac{\partial}{\partial t}\mathbf{w}, \tag{2-29}$$

we finally arrive at the usual formulation of Boltzmann's equation

$$\frac{\partial f}{\partial t} + \mathbf{w}\cdot\nabla f + \frac{\mathbf{F}}{m}\nabla_w f = \left(\frac{\partial f}{\partial t}\right)_c. \tag{2-30}$$

The intriguing property of $f$ can be seen when integrating Eq. (2–30) by $d^3w$ over the whole velocity space. The integration over Boltzmann's collisional term is zero and the other terms immediately result in the macroscopic hydrodynamic equation of continuity (2–19). Multiplying Eq. (2–30) by $m\mathbf{w}$ and integrating over the whole velocity space arrives at the macroscopic equation of motion (2–18). Multiplying Eq. (2–30) with $mw^2$ and integrating over the whole velocity space results in the hydro-dynamic equation of energy conservation (2–20). The details of these integrations need certain simplifying interpretations but are straightforward, as can be seen in Chapter 3 of Hora (1991). All of these remarks should enable one to reflect briefly on how the macroscopic theory can be based on microscopic derivation while the very abstract distribution function and the Boltzmann equation has very easily un- derstandable property in that their momentum weighted velocity space integrations lead to these macroscopic equations.

As can be seen (Hora 1991; Chapter 3), there were a number of simplifications in the derivation of the macroscopic equations. The advantage of the kinetic theory with the Boltzmann equation consists in the fact that the general properties for non-equilibrium and of numerous nonlinear properties are included in a much more sophisticated way than in the macroscopic theory. This is the reason why plasma instabilities and other complex phenomena are treated in a more preferable manner by the kinetic theory. However, the Boltzmann collision term in Eq. (2–30), right-hand side, is still a difficulty. Such approximations as the Krook term or the Fokker–Planck term are well known. Nevertheless, a complete solution does not exist yet. Generally, the Boltzmann collision term consists of more than one thousand (!) terms such that cybernetic methods have to be used to order them in an algebraic way (Drska 1987).

# CHAPTER 3
# Electrodynamics and Plasma

## 3.1 Maxwell's Equations

The theory of electrodynamic phenomena is presented in textbooks in very different ways. Most of the books still use the historical method from the time before 1864 when James Clark Maxwell introduced his equations. His work arrived at a closed field of knowledge and made such revolutionary predictions as the existence of electromagnetic waves and the fundamental meaning of the speed of light. The old-fashioned books are not wrong but the reader misses the very easy overview Maxwell achieved.

The pre-Maxwellian presentation starts with the electrostatics from the Coulomb force, then goes to magnetostatics, then to the chapter of slowly changing magnetic fields and then to a new and separate chapter on the electromagnetic waves. This is not wrong. But a better way is, after the student has some basic knowledge of electric charges, condensers, electric currents, and so on, to explain the central meaning and properties of the Maxwell equations, using all students' knowledge of vector algebra and vector calculus and to derive from there the properties of electromagnetic waves together with the equations of electrostatics, magnetostatics, etc., as special cases of Maxwell's equations.

Before describing Maxwell's equations, the following elementary definitions have to be given. Electrodynamics begins with Newton. For thousands of years people have observed that when rubbed, amber lifts up pieces of paper or other materials. Only since Newton revealed that there is a force from the amber (Greek: elektra) acting on the paper against the gravitational force of the earth, could Coulomb describe the (attractive or repulsive) electrostatic forces between two charges $q_1$ and $q_2$ (with unequal or equal sign, respectively) over a distance $r$, given by

$$|\mathbf{F}| = \frac{q_1 q_2}{r^2} c_0 \tag{3-1}$$

Describing the force in the dimension of dyn = g cm s$^{-2}$, r in cm, and using a constant $c_0 = 1$, the dimension of the charge is $[q] = g^{1/2}$ cm$^{3/2}$ s$^{-2}$. In the following we shall use mostly this (Gaussian) cgs system because it is common in plasma physics. The charge of an electron is then $4.805 \times 10^{-10}$ cgs units. If the unit of the charge is the Coulomb (C), [the charge going through the electrolyte of silver nitrate which discharges 1.118 mg silver], the charge of an electron is $1.602 \times 10^{-19}$ C. With these SI units [meters, kilograms, seconds and Coulombs], the constant $c_0$ in Eq. (3–1) is

$$c_0 = \frac{1}{4\pi\varepsilon_0} \tag{3-2}$$

where $\varepsilon_0$ is a constant which will be given in Eq. (3–8).

From the Coulomb force (3–1) one derives the *definition of the electric field strength* **E** as the force per probe charge $q_2$ which another charge $q_1$ or a number of such charges or a continuous distribution of such electric charges is producing at the point with the coordinates $x$, $y$, $z$, in space

$$\mathbf{E} = \frac{\mathbf{F}}{q2} \left[ \frac{\text{kgm}}{\text{s}^2\text{C}} \right]. \tag{3-3}$$

In such a temporally constant electrostatic field **E**, the line integral between two points $A$ and $B$

$$\int_A^B \mathbf{E} \cdot d\mathbf{s} = U \left[ \frac{\text{kgm}}{\text{s}^2\text{C}} \right] \tag{3-4}$$

gives an independent value $U$, the electrical voltage, where the path between $A$ and $B$ can be chosen arbitrarily. This means that **E** is called conservative and can be expressed by a potential [see Eq. (1–4)].

This **E** is the *force* definition of the electric field. One can also define the electrostatic field by a *quantity* definition **D** describing the amount of electrical charge which is inside of a volume $V_0$ which has a (closed) surface $S_0$. The amount of charge $q$ inside of $S_0$ is given by

$$\oiint \mathbf{D} \cdot d^2\mathbf{f} = q, \tag{3-5}$$

which, following Gauss' theorem of vector analysis, can be expressed by the electrical charge density $\rho$ within a volume integral over $V_0$

$$\oiint \mathbf{D} \cdot d^2\mathbf{f} = \iiint \nabla \cdot \mathbf{D}\, d^3\tau = \iiint \rho(x, y, z)\, d^3\tau = q. \tag{3-5a}$$

Since this definition has to be independent from the special volume chosen, it follows that

$$\nabla \cdot \mathbf{D} = \rho \tag{3-6}$$

with the dimension of the electrical *flux density* **D** (also called *dielectric displacement*) and the charge density $\rho$

$$[\mathbf{D}] = \text{Cm}^{-2} \tag{3-7a}$$

$$[\rho] = \text{Cm}^{-3} \ . \tag{3-7b}$$

The relation between the two definitions is given by

$$\mathbf{D} = \varepsilon\varepsilon_0 \mathbf{E} \ , \tag{3-8}$$

where the constant $\varepsilon_0$ is given by

$$\varepsilon_0 = \frac{1}{36\pi \times 10^9} \left[ \frac{\text{s}^2\text{C}^2}{\text{kgm}^3} \right] \tag{3-9}$$

called the dielectric constant of the vacuum. The constant $\varepsilon$ is the dimensionless *relative dielectic constant* which is larger than one for polarizable media and may even be a tensor (for crystals). It is less than one in plasmas or metals, there however only by a high frequency definition, while it is undefined in electrostatics for materials with an electric conductivity.

For very high fields, the relation (3–8) is due to a nonlinear extension

$$\mathbf{D} = \varepsilon_0 \varepsilon \mathbf{E} \left( 1 + \varepsilon_1 \mathbf{E}^2 + \ldots \right), \tag{3-9a}$$

where the constant $\varepsilon_1$ was measured for the electrostatic case more than 100 years ago for various dielectrics.

Magnetic fields are basically connected with flowing electrical charges. The electrical current $I$ is given by the electrical current density $\mathbf{j}$ integrated over the whole cross section $A$ through which the current is flowing. The current is driven by a voltage $U$. The connection between these is given by the electrical conductivity $\sigma$ (Ohm's law)

$$\mathbf{j} = \sigma \mathbf{E} \ . \tag{3-10}$$

For the magnetic fields there is also a force and a quantity definition $\mathbf{B}$ and $\mathbf{H}$ in analogy to $\mathbf{E}$ and $\mathbf{D}$ with the relation

$$\mathbf{B} = \mu\mu_0 \mathbf{H} \ , \tag{3-11}$$

where $\mu$ is a dimensionless material constant, called relative magnetic permeability, and the quantity

$$\mu_0 = \frac{4\pi}{10^7} \left[ \frac{\text{kgm}^2}{\text{C}^2} \right] \tag{3-12}$$

is the permeability of the vacuum. Since the magnetic field so far as we know at present, has no magnetic monopoles as sources [contrary to the source of the electric field given by the charge density, Eq. (3–6)] the relation is defined according to Eq. (3–6)

$$\nabla \cdot \mathbf{B} = 0 \ . \tag{3-13}$$

This relation is basic and goes back to Heinrich Hertz. Some confusion as to the meaning of the choices [it would have been better to use $\nabla \cdot H = 0$ instead of Eq. (3–13) though there is numerically nearly no difference] for the magnetic quantities as it has been described by Sommerfeld (1955).

After giving these definitions we can introduce the Maxwell equations. These are the generalization of the following two laws:

(a) The Ampére law that an electric current $I$ in a conductor produces a magnetic field $\mathbf{H}$ around the conductor such that the product of $\mathbf{H}$ times the path for a closed circumference around the conductor (Fig. 3–1) is constant whatever path is used. If the path around the wire is a concentric cycle of radius $r$ perpendicular to the wire, this law reduces to the simple formula

$$2\pi r |\mathbf{H}| = I . \tag{3-14}$$

Using any bent path with the appropriate spatial dependence of the vector field $\mathbf{H}$ one arrives at the following vector formulation of Ampére's law

$$\oint \mathbf{H} \cdot d\mathbf{r} = I = \int \int \mathbf{j} \cdot d^2\mathbf{f} . \tag{3-15}$$

(b) Faraday's induction law measures the temporal variation of the magnetic flux within a metallic loop that produces a ring voltage in the loop measured by a voltage $U$ if one opens the wire and connects it to a voltmeter (Fig. 3–2). Again, with the simplified geometry that the loop is a circle of radius $r$, this results in

$$-\frac{\partial}{\partial t} \pi r^2 |\mathbf{B}| = U = 2\pi r |\mathbf{E}| , \tag{3-16}$$

and for a generally bent loop in space and with a general magnetic field changing in space one arrives at the vector formulation

$$\oint \mathbf{E} \cdot d\mathbf{r} = -\frac{\partial}{\partial t} \int \int \mathbf{B} \cdot d^2\mathbf{f} . \tag{3-17}$$

Maxwell performed the following changes: He first assumed that the relations (3–15) and (3–17) will be valid everywhere in space and therefore at places where the conducting loop or the current $I$ are not fixed in conducting wires. The further very ingenious assumption was that he added to the current density in Am- pére's law the *displacement current* given by the temporal derivative of the $\mathbf{D}$ field,

$$\mathbf{M} = \mathbf{j} + \frac{\partial}{\partial t} \mathbf{D} , \tag{3-18}$$

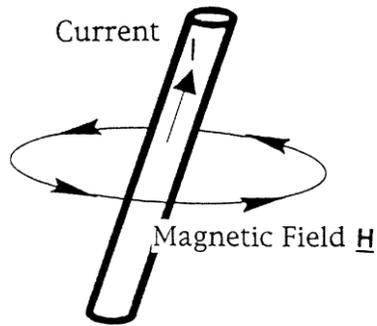

**Figure 3–1.** Ampére's law of generation of a magnetic field **H** by an electric current $I$.

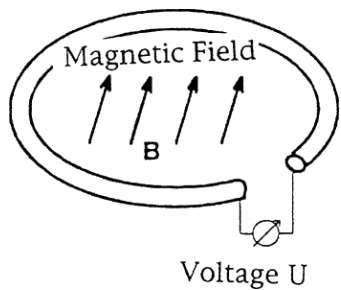

**Figure 3–2.** Faraday's law where the temporal change of the flux of a magnetic field **B** through a closed loop of wire produced a ring voltage $U$.

resulting in the corrected current density $M$ in Eq. (3–15). Rewriting the closed loop integrals in Eqs. (3–15) and (3–17) into surface integrals following Stoke's theorem of vector analysis

$$\oint \mathbf{H} \cdot d\mathbf{r} = \iint \nabla \times \mathbf{H} \cdot d^2\mathbf{f};$$
$$\oint \mathbf{E} \cdot d\mathbf{r} = \iint \nabla \times \mathbf{E} \cdot d^2\mathbf{f},$$
(3-19)

one can then express the vector equation in differential form instead of the integral form since any area over which it is to be integrated has to be valid:

$$\nabla \times \mathbf{E} = -\frac{\partial}{\partial t}\mathbf{B} \; ;$$
(3-20)

$$\nabla \times \mathbf{H} = \mathbf{j} + \frac{\partial}{\partial t}\mathbf{D} \; ,$$
(3-21)

These are the Maxwellian equations with the additional conditions (3–8), (3–11), (3–6), (3–13) and (3–10).

Using Eqs. (3–8) and (3–11), we can substitute the vectors **B** and **D** in Eqs. (3–20) and (3–21) and arrive at

$$\nabla \times \mathbf{E} = -\frac{\partial}{\partial t} \mu \mu_0 \mathbf{H} \tag{3-22}$$

$$\nabla \times \mathbf{H} = \mathbf{j} + \frac{\partial}{\partial t} \varepsilon \varepsilon_0 \mathbf{E}. \tag{3-23}$$

The categories of all phenomena of electrodynamics can then be seen very simply. Electrostatics is the case where all time derivatives are zero and no magnetic field exists. From Maxwell's equations there remains only $\nabla \times \mathbf{E} = 0$ and $\nabla \cdot \mathbf{D} = \rho$. From there the fundamental equations (1–4) of electrostatics with the potential equation (Laplace or Poisson equations) follow automatically for deriving the results of electrostatics including condensers electric double layers, etc. Magnetostatics follow in a similar way with no time derivative, and $\mathbf{E} = 0$ but with a distribution of stationary currents **j**. From all these equations remaining from the Maxwell equations, including that of a vector potential, one can calculate all the phenomena of magnetism. The next step of slowly temporally varying fields permits the description of the technology of electrical machines where the speed of light is not involved, and even the skin effect can be described (Hora 1994).

For compleness we write these equations in the cgs system (Hora 1991: Section 6.3)

$$\nabla \times \mathbf{E} = -\frac{1}{c} \frac{\partial \mu \mathbf{H}}{\partial t} \tag{3-24}$$

$$\nabla \times \mathbf{H} = -\frac{1}{c} \frac{\partial \varepsilon \mathbf{E}}{\partial t} + \frac{4\pi}{c} \mathbf{j} \tag{3-25}$$

$$\mathbf{B} = \mu \mathbf{H} \tag{3-26}$$

$$\mathbf{D} = \varepsilon \mathbf{E} \tag{3-27}$$

$$\nabla \cdot \mathbf{D} = 4\pi \rho \tag{3-28}$$

$$\nabla \cdot \mathbf{B} = 0 \tag{3-29}$$

$$\nabla \cdot \mathbf{B} = 0 \tag{3-30}$$

where *c* is the velocity of light in vacuum.

Continuing in SI units, we eliminate one of the two vector equations [3–22 and 3–23] by applying the operator "$\nabla \times$" to Eq. (3–22) and to Eq. (3–23) the operator $(\partial/\partial t)$ and obtain:

$$\nabla \times (\nabla \times \mathbf{E}) = -\mu_0 \frac{\partial}{\partial t} \nabla \times \mathbf{H}; \qquad (3\text{-}31)$$

$$\frac{\partial}{\partial t} \nabla \times \mathbf{H} = \varepsilon\varepsilon_0 \frac{\partial^2}{\partial t^2} \mathbf{E} + \frac{\partial}{\partial t} \mathbf{j}, \qquad (3\text{-}32)$$

assuming that we have no magnetic response ($\mu = 1$). From these two equations we can eliminate $\mathbf{H}$. Remembering the vector theorem

$$\nabla \times (\nabla \times \mathbf{E}) = \nabla\nabla \cdot \mathbf{E} - \nabla^2 \mathbf{E} \qquad (3\text{-}33)$$

and using the special case of a situation where no space charges exist ($\sigma = 0$, therefore $\nabla \cdot E = 0$) and no electric current appears ($j = 0$), we arrive at

$$\nabla^2 \mathbf{E} = \varepsilon\varepsilon_0\mu_0 \frac{\partial^2}{\partial t^2} \mathbf{E}. \qquad (3\text{-}34)$$

This is a wave equation with a velocity $c$ of the wave propagation

$$c = \frac{1}{\sqrt{\varepsilon_0\mu_0}} \frac{1}{\sqrt{\frac{1}{36\pi \times 10^9} \frac{4\pi}{10^7}}} = 3 \times 10^8 \, \frac{\text{m}}{\text{s}}. \qquad (3\text{-}35)$$

This is the Maxwell relation showing that any change of the electromagnetic quantities $\mathbf{E}$ and $\mathbf{H}$ are spreading in space with a wave velocity which is equal to the velocity of light. Since the velocity of light is not exactly the value given in Eq. (3–35), the following corrections to the constants have to be given

$$\varepsilon_0 = 8.8541878 \times 10^{-12} \, \frac{\text{s}^2\text{C}^2}{\text{kgm}^3}; \qquad (3\text{-}36\text{a})$$

$$\mu_0 = 4\pi \times 10^{-7} = 12.566370614 \times 10^{-7} \, \frac{\text{kgm}}{\text{C}^2}. \qquad (3\text{-}36\text{b})$$

If $\varepsilon$ is different from 1 (vacuum value) and spatially constant as $\varepsilon = n^2$, we have a medium with an optical refractive index $n$.

In the same way, by elimination of $\mathbf{E}$ one can derive a wave equation for the magnetic vector

$$\nabla^2 \mathbf{H} - \left(\frac{n}{c}\right)^2 \frac{\partial^2}{\partial t^2} \mathbf{H} = 0. \qquad (3\text{-}37)$$

We now discuss the plane wave solutions of the wave equations (3–34) and (3–37). Without giving up generality, we discuss waves expanding into the $x$-

direction

$$\frac{\partial}{\partial y} = \frac{\partial}{\partial z} = 0,$$

and for the charge-free vacuum space follows

$$\nabla \cdot \mathbf{E} = 0; \quad \mathrm{div}\mathbf{E} = \frac{\partial}{\partial x}E_x + \frac{\partial}{\partial y}E_y + \frac{\partial}{\partial z}E_z = 0.$$

This shows that the wave propagating into the *x*-direction has a vanishing *x*-component

$$\frac{\partial}{\partial x}E_x = 0 \quad \left(\text{and} \quad \frac{\partial^2}{\partial t^2}E_x = 0\right) \Rightarrow E_x = \text{const.} \tag{3-38}$$

The wave therefore has only transverse components $E_y$ and $E_z$ which combine to the general elliptically polarized wave. The wave is linearly polarized if $E_z$ is zero. A plane wave solution with this linear polarization and a frequency $\omega$ is

$$\mathbf{E} = \mathbf{e}_y E_{y0} \mathrm{e}^{i(kx-\omega t)}, \tag{3-39}$$

where the abbreviation for the *wave number k* has been used

$$k = \frac{\omega}{c}. \tag{3-40}$$

In order to calculate the magnetic vector one can start from the Maxwell equation

$$\mu_0 \frac{\partial \mathbf{H}}{\partial t} = \nabla \times \mathbf{E},$$

and one arrives with (3–39) at

$$\nabla \times \mathbf{E} \begin{vmatrix} \mathbf{e}_x & \mathbf{e}_y & \mathbf{e}_z \\ \frac{\partial}{\partial x} & 0 & 0 \\ 0 & E_y & 0 \end{vmatrix} = \mathbf{e}_z \frac{\partial}{\partial x}E_y; \quad -\nabla \times \mathbf{E} = -\mathbf{e}_z ikE_{y0}\mathrm{e}^{i(kx-\omega t)};$$

$$\mathbf{H} = \mathbf{e}_z \frac{k}{\omega}\mu_0 E_{y0}\mathrm{e}^{i(kx-\omega t)} \frac{1}{c\mu_0}$$

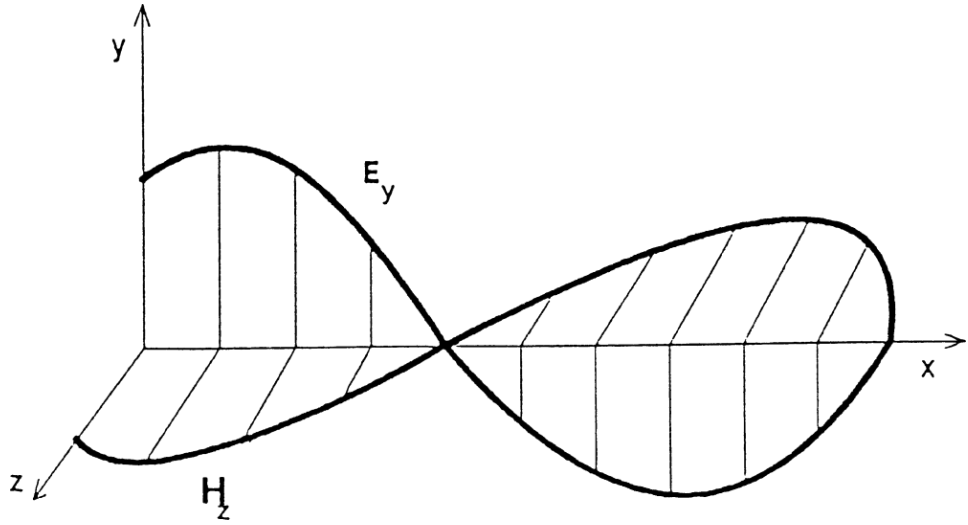

**Figure 3–3.** Linearly polarized electromagnetic wave propagating into the *x* direction showing equally phased transversal **E** and **H** vectors, Eqs. (3–39) and (3–41).

where the last equation was the result of temporal integration. Remembering Eq. (3–35) we finally arrive at

$$\mathbf{H} = \mathbf{e}_z H_z = \mathbf{e}_z \sqrt{\frac{\varepsilon_0}{\mu_0}} E_{y0} e^{i(kx-\omega t)} \tag{3-41}$$

The plane wave has therefore purely transverse wave solutions (Fig. 3–3) with no phase difference between **E** and **H**.

## 3.2 Derivation of the Lorentz Force

We shall now derive the relations of energy densities from Maxwell's equations in the usual way and use this to derive the Lorentz force in a direct way. The well-known derivation of the Lorentz force (Sommerfeld 1955) is based on the Coulomb force for relativistic motions and the Lorentz force automatically appears from the Lorentz transform. In many other textbooks, the Lorentz force simply falls from the sky without any explanation. Other authors say this is an empirically given force and do not worry about any derivation.

Scalar multiplication of the first Maxwell equation (3–20) from the left-hand side with **H** and the second equation (3–21) with **E** and adding the resulting equations we arrive at

$$\mathbf{H} \cdot \dot{\mathbf{B}} + \mathbf{E} \cdot \dot{\mathbf{D}} + \mathbf{E} \cdot \mathbf{j} = \mathbf{E} \cdot \nabla \times \mathbf{H} - \mathbf{H} \cdot \nabla \times \mathbf{E} \tag{3-42}$$

In Eq. (3–42), the dot in an abbreviation for partial temporal differentiation

$$\dot{\mathbf{B}} = \frac{\partial}{\partial t}\mathbf{B}; \quad \dot{\mathbf{D}} = \frac{\partial}{\partial t}\mathbf{D}; \quad \dot{\mathbf{H}} = \frac{\partial}{\partial t}\mathbf{H}; \quad \dot{\mathbf{E}} = \frac{\partial}{\partial t}\mathbf{E}. \tag{3-43}$$

If $\varepsilon$ and $\mu$ are not dependent on time we can write

$$\mathbf{H} \cdot \dot{\mathbf{B}} = \mathbf{H} \cdot \mu\mu_0 \dot{\mathbf{H}} = \frac{1}{2}\mu\mu_0 \frac{\partial}{\partial t}\mathbf{H}^2 \tag{3.45}$$

$$\mathbf{E} \cdot \dot{\mathbf{D}} = \mathbf{E} \cdot \varepsilon\varepsilon_0 \dot{\mathbf{E}} = \frac{1}{2}\varepsilon\varepsilon_0 \frac{\partial}{\partial t}\mathbf{E}^2 \tag{3.45}$$

The right-hand side of Eq. (3–42) can be rewritten by taking into account that only vectors on the right-hand side of a nabla symbol are to be differentiated and that a change of the sign appears by changing the sequence of the vectors in the right-hand side of Eq. (3–43), to arrive at

$$\mathbf{E} \cdot \nabla \times \mathbf{H} - \mathbf{H} \cdot \nabla \times \mathbf{E} = -\nabla \cdot \mathbf{E} \times \mathbf{H}$$
$$= -\nabla \cdot \mathbf{S} \tag{3-46}$$

where the vector $\mathbf{S} = \mathbf{E} \times \mathbf{H}$ is called the *Poynting vector*.

Summarizing the substitutions (3–44), (3–45), and (3–46) and using Ohm's law (3–10) we can then write Eq. (3–42) as

$$\frac{\partial}{\partial t}\varepsilon\varepsilon_0 \mathbf{E} \cdot \mathbf{E}/2 + \frac{\partial}{\partial t}\mu\mu_0 \mathbf{H} \cdot \mathbf{H}/2 + \sigma\mathbf{E}^2 = -\nabla \cdot \mathbf{S} \tag{3-47}$$

The meaning of the first term is the temporal change of the electrical field energy density, the second term is the temporal change of the magnetic field density and the third term is generation of heat by Ohmic conductivity per time (Ohmic power generation). The meaning of the Poynting vector can be seen from its dimension

$$[\mathbf{S}] = \frac{\text{kgm}}{\text{s}^2\text{C}} \frac{\text{C}}{\text{m s}} = \frac{\text{joule}}{\text{m}^2\text{s}} = \frac{\text{watt}}{\text{m}^2} \tag{3-48}$$

as power emitted per area. Equation (3–47) therefore represents the law of electromagnetic field conservation: Any change of electrical or magnetic field energy including Ohmic losses by thermal energy has to be compensated by the flux of electromagnetic field energy given by the divergence of the Poynting vector.

The time average of the Poynting vector is the intensity $I$ of an electromagnetic wave

$$I = \overline{|\mathbf{S}|} = \overline{|\mathbf{E}||\mathbf{H}|} = E_0^2 \frac{1}{2}\sqrt{\frac{\varepsilon_0}{\mu_0}}, \tag{3-49}$$

using the result (3–39) and (3–41). The amplitude $E_y$ is expressed here by $E_0$ and is given from Eq. (3–49) in SI units as

$$E_0 = \sqrt{2 \cdot 377}\sqrt{I} = 27.45\sqrt{I} \qquad (3\text{-}50)$$

The same equation is valid for the usual dimension in laser physics of $[E_0]=$ V/cm and $[I]=$ W/cm$^2$.

For an easier connection with plasma physics we rewrite the energy law (3–47) in Gaussian cgs units

$$\frac{\partial}{\partial t}\left[\varepsilon\mathbf{E}^2 + \mu\mathbf{H}^2\right]\frac{1}{8\pi} + \sigma\mathbf{E}^2\frac{1}{4\pi} + \nabla\cdot\mathbf{S} = 0 \qquad (3\text{-}51)$$

The *derivation of the Lorentz force* relates the exchange of electromagnetic field energy into mechanical energy of motion. Therefore we have no change of energy into (Ohmic) heat and no energy loss or gain by electromagnetic radiation:

$$\sigma = 0; \quad \mathbf{S} = 0 \qquad (3\text{-}52)$$

These are adiabatic conditions, where the electromagnetic field energy can only go into mechanical energy of motion given by a force density $\mathbf{f}$. We assume vacuum conditions, i.e., $\varepsilon = 1$ and $\mu = 1$. The energy which goes from the electromagnetic field into mechanical energy is that of

$$\frac{1}{8\pi}\left(\mathbf{E}^2 + \mathbf{H}^2\right) \qquad (3\text{-}53)$$

Any force density due to this energy density is given by the gradient. As a first step we calculate a force density with an additional term

$$f_1 = \frac{1}{8\pi}\nabla\left(\mathbf{E}^2 + \mathbf{H}^2\right) - \frac{1}{4\pi}\mathbf{E}\nabla\cdot\mathbf{E}. \qquad (3\text{-}54)$$

We further assume that the following approximation is possible when spatially deriving the following vector expressions

$$\nabla\cdot(\mathbf{E}\mathbf{E} + \mathbf{H}\mathbf{H}) = \nabla(\mathbf{E}\cdot\mathbf{E} + \mathbf{H}\cdot\mathbf{H}) \qquad (3\text{-}55)$$

which is equivalent to the approximate relation using the unity tensor $\underline{\mathbf{1}}$ (Eq. 4–49)

$$\frac{1}{4\pi}\nabla\cdot\left[\mathbf{E}\mathbf{E} + \mathbf{H}\mathbf{H} - \frac{1}{2}\left(\mathbf{E}^2 + \mathbf{H}^2\right)\underline{\mathbf{1}}\right] \approx \frac{1}{8\pi}\nabla\left(\mathbf{E}^2 + \mathbf{H}^2\right) \qquad (3\text{-}56)$$

With this relation and using the zero Poynting term we arrive from Eq. (3–54) at

$$\begin{aligned}f_1 &= \frac{1}{4\pi}\nabla\cdot\left[\mathbf{E}\mathbf{E} + \mathbf{H}\mathbf{H} - \frac{1}{2}\left(\mathbf{E}^2 + \mathbf{H}^2\right)\underline{\mathbf{1}}\right]\\ &\quad -\frac{1}{4\pi}\mathbf{E}\nabla\cdot\mathbf{E} - \frac{1}{4\pi c}\frac{\partial}{\partial t}\mathbf{E}\times\mathbf{H};\end{aligned} \qquad (3\text{-}57)$$

$$f_1 = -\frac{1}{8\pi}\nabla \cdot \mathbf{1}\left(\mathbf{E}^2 + \mathbf{H}^2\right) - \frac{1}{4\pi}\mathbf{E}\nabla \cdot \mathbf{E} - \frac{1}{4\pi^c}\left(\frac{\partial}{\partial t}\mathbf{E}\right) \times \mathbf{H} - \frac{1}{4\pi c}\mathbf{E}$$
$$\times\left(\frac{\partial}{\partial t}\mathbf{H}\right) + \frac{1}{4\pi}\left(\mathbf{E}\cdot\nabla\mathbf{E} + \mathbf{E}\nabla\cdot\mathbf{E}\right) + \frac{1}{4\pi}\mathbf{H}\cdot\nabla\mathbf{H} \qquad (3\text{-}58)$$

where the last two terms result from the product law of differentiation

$$\nabla \cdot \mathbf{EE} = \mathbf{E}\cdot\nabla\mathbf{E} + \mathbf{E}\nabla\cdot\mathbf{E};$$
$$\nabla \cdot \mathbf{HH} = \mathbf{H}\cdot\nabla\mathbf{H} + \mathbf{H}\nabla\cdot\mathbf{H},$$

and for the fact that

$$\nabla \cdot \mathbf{H} = 0$$

is in vacuum.

In Eq. (3–58), the terms $\mathbf{E}\ \nabla\cdot\mathbf{E}/4\pi$ cancel and using the relation

$$\left(\nabla \times \overset{\downarrow}{\mathbf{H}}\right) \times \mathbf{H} = \mathbf{H}\cdot\nabla\mathbf{H} - \frac{1}{2}\nabla\mathbf{H}^2, \qquad (3\text{-}59)$$

where the arrow indicates that the differentiation refers only to this vector and not to the other vector. This is the reason why the factor 1∕2 appears in Eq. (3–59). We can do the same for terms with $\mathbf{E}$ and the result (3–58) can be summarized to

$$f_1 = \frac{1}{4\pi}\left(\nabla \times \overset{\downarrow}{\mathbf{H}}\right) \times \mathbf{H} + \frac{1}{4\pi}\left(\nabla \times \overset{\downarrow}{\mathbf{H}}\right) \times \mathbf{E} - \frac{1}{4\pi c}\left(\frac{\partial}{\partial t}\mathbf{E}\right) \times \mathbf{H}$$
$$-\frac{1}{4\pi c}\mathbf{E} \times \frac{\partial}{\partial t}\mathbf{H} \qquad (3\text{-}60)$$
$$= \frac{1}{4\pi}\left[\nabla \times \overset{\downarrow}{\mathbf{H}} - \frac{1}{c}\frac{\partial}{\partial t}\mathbf{E}\right]\times\mathbf{H} + \frac{1}{4\pi}\left[\nabla \times \mathbf{E} + \frac{1}{c}\frac{\partial}{\partial t}\mathbf{H}\right]\times\mathbf{E}.$$

The first square bracket is, according the second Maxwell equation (3–25), equal to $4\pi\mathbf{j}/c$, and the second square bracket is zero because of the first Maxwell equation (3–24). Therefore we obtain

$$f_1 = \frac{1}{c}\mathbf{j}\times\mathbf{H} \qquad (3\text{-}61)$$

Remembering the meaning of the added term in Eq. (3–54) [$4\pi\rho = \nabla\cdot\mathbf{E}$ as given by Eq. (3–28) and (3–27)] we finally find for our initial adiabatic transfer of electromagnetic field energy into mechanical energy (Eq. 3–53)

$$f = \frac{1}{4\pi}\nabla\left(\mathbf{E}^2 + \mathbf{H}^2\right) = \frac{1}{c}\mathbf{j}\times H + \rho\mathbf{E} \qquad (3\text{-}62)$$

The mechanical force density is given by the Lorentz force and the further term expressing the electrostatic force by the Coulomb law.

Can we say that the ponderomotive force next to the Coulomb force in Eq. (3–62) is the same as the Lorentz force? Since the force density leading to the Lorentz force is related to the $\nabla \mathbf{E}^2$ in Eq. (3–54) or (3–62), this is electrostriction as known from the ponderomotive force, Eq. (1–3), or from the magnetostrictive force $\nabla H^2$. Considering Eq. (3–62) it is evident that the Lorentz force cannot generally be identified with the ponderomotive force. Looking at Fig. 1–1 one cannot simply say that the attraction of the dielectric material [without a v and without a B, Eq. (3–61)] by the ponderomotive force is a Lorentz force! Here we see how difficult it is to define a plasma and what exception of all solid materials with good electric conductivity like metals or semiconductors are against the dielectric insulators in view of electrodynamics.

With respect to the approximation (3–55) it can be shown easily from the tensor relations and their differentiation by writing out all components that in homogeneous media (as assumed here for the adiabatic conditions), the approximation is fulfilled exactly. The case of inhomogeneous media has not yet been evaluated along the lines of the derivation of the Lorentz force given in this subsection.

It is worthwhile to realize that the derivations in this subsection are based on the fields of a charge density $\rho$, on a field of current densities $\mathbf{j}$, on vacuum conditions with $\varepsilon = 1$ and $\mu = 1$, on electric and magnetic fields and on Maxwell's equations only, and *not on any force*. By considering how electromagnetic energy may convert into mechanical energy in a volume under adiabatic conditions did we arrive automatically at the Lorentz force and at the Coulomb force.

## 3.3 SCHLÜTER'S TWO-FLUID EQUATIONS AND OPTICAL PROPERTIES OF PLASMA

After learning about the basics of electrodynamics we are going to combine this with hydrodynamics in order to describe plasmas. One may begin with Euler's hydrodynamic equation (2–15) including the Navier–Stokes term of viscosity. For plasmas, the right-hand side with the force densities has then to be extended to the electrondynamic forces, the Coulomb force and the Lorentz force. The Coulomb force is mostly discarded since the interior of homogeneous plasmas (like uniform metals) have no internal electric fields (contrary to the inhomogeneous plasmas, a fact that is mostly overlooked). With this one fluid description it was possible for Hannes Alfvén (1942, 1942a) to derive the magnetohydrodynamic waves, the Alfvén waves which were experimentally confirmed later and now belong to the trivial repertoire of a plasma physicist. Alfvén received the Nobel Prize for this discovery. It should also be noted that the Nobel Prize was given to Irving Langmuir for his discovery of the plasma frequency, Eq. (2–4) and to Peter Debye for his discovery of the Debye length, Eq. (2–5).

Another important step in plasma theory was Schlüter's (1950) introduction of the hydrodynamic two-fluid model. This described the electron fluid of a plasma separately by the following Euler equation

$$mn_e \frac{dv_e}{dt} = -n_e e\mathbf{E} - n_e \frac{e}{c} \mathbf{v}_e \times \mathbf{H} - \nabla \frac{3}{2} n_e KT_e \quad (3\text{-}63)$$
$$+ mn_e \nu_{ei}(\mathbf{v}_i - \mathbf{v}_e) + \mathbf{K}_e$$

and by the Euler equation for the ion fluid of the plasma by

$$m_i n_i \frac{d\mathbf{v}_i}{dt} = Zn_i e\mathbf{E} + n_i \frac{Ze}{c} \mathbf{v}_i \times \mathbf{H} - \nabla \frac{3}{2} n_i KT_i \quad (3\text{-}64)$$
$$- mn_e \nu_{ei}(\mathbf{v}_i - \mathbf{v}_e) + \mathbf{K}_i$$

as equations of motion for the electron velocity $\mathbf{v}_e$, electron mass $m$ and electron density $n_e$, for the ion velocity $\mathbf{v}_i$, ion mass $m_i$ and the ion density $n_i$. If there are unionized neutral atoms in between one may add a third equation (Alfvén and Fälthammar 1973), but this is our interest here where we focus on fully ionised plasmas.

The forces on the right-hand side of the Euler equations (3–63) and (3–64) are the Coulomb forces due to the electron charge e and the electrical field E, and the Lorentz force where the electron current density is expressed by je = neve and the ion current density with the index "i", the thermokinetic pressure terms are determined by the temperature of the electrons Te and of the ions Ti , and the viscosity terms are determined by the electron–ion collision frequency vei . This latter, apart from Spitzer's (1962) modifications by logarithmic expressions, is given by Eq. (2–11). The final terms Ke and Ki are further force densities due to gravitation, centrifugal forces, Coriolis forces, etc., which will not be discussed in the following.

By adding the Euler equation (3–63) and (3–64) and using the abbreviations for the net velocity

$$\mathbf{v} = \frac{m_i \mathbf{v}_i + Zm v_e}{m_i + Zm} \quad (3\text{-}65)$$

of the whole plasma and the current density

$$\mathbf{j} = e(n_i \mathbf{v}_i - n\mathbf{v}_e) \quad (3\text{-}66)$$

and assuming space charge neutrality, one arrives at the *equation of motion of the whole plasma*

$$f = m_i n_i \frac{dv}{dt} = -\nabla p + \frac{1}{c} \mathbf{j} \times \mathbf{H} + \frac{1}{4\pi}\left(\frac{\omega p}{\omega}\right)^2 \mathbf{E} \cdot \nabla \mathbf{E} \quad (3\text{-}67)$$

Here, the following transcription of the last term, the Schlüter term, was used

$$\frac{1}{4\pi}\left(\frac{\omega p}{\omega}\right)^2 \mathbf{E}\cdot\nabla\mathbf{E} = \mathbf{j}\cdot\nabla\frac{1}{n_c}\mathbf{j}\frac{m}{e^2} \qquad (3\text{-}68)$$

which is possible only for the high-frequency case of the fields and currents with a frequency $\omega$.

By subtracting the Euler equation for the electrons and ions one arrives at the *diffusion equation* or *generalized Ohm's law*

$$\frac{m}{e^2 n_c}\left(\frac{d\mathbf{j}}{dt}+\nu\mathbf{j}\right) = \mathbf{E} + \frac{1}{c}\mathbf{v}\times\mathbf{H} + \frac{1}{en_c c}\mathbf{j}\times\mathbf{H} + \frac{c}{en_c}\frac{\nabla p}{1+1/Z}. \qquad (3\text{-}69)$$

The detailed derivation of these steps is rather complex. The only publication about a derivation since (Hora 1981) is given in Hora 1991: Appendix C where some details complementing Schlüter's case were added due to later knowledge of laser interaction with plasmas.

The first term on the right-hand side of Eq. (3–69) shows immediately the relation to Ohm's law, the second term is the Hall term, the third the Lorentz term and the last the diffusion term showing the ambipolar generation of an electric field due to the gradient of pressure $p$ in the plasma. For the cases of laser plasma interaction with relativistic intensities, all of these terms can be neglected in Ohm's law, although the ambipolar diffusion will be discussed later in another context. With these simplifications and recalling the expression of the plasma frequency $\omega_p$, Eq. (2–4), we arrive from (3–69) at

$$\frac{d\mathbf{j}}{dt} + \nu\mathbf{j} = \frac{\omega_p^2}{4\pi}\mathbf{E} \qquad (3\text{-}70)$$

Here we have the current density we need in Maxwell's equations from Eqs. (3–24) and (3–25).

The logical problem arises from the fact that the Euler equations (3–63) and (3-64) follow the presumptions of Lorentz's electron theory of metals, having charge densities and current densities in vacuum, i.e. having $\varepsilon = \mu = 1$, therefore the Maxwellian equations

$$\nabla\times\mathbf{E} = -\frac{1}{c}\frac{\partial}{\partial t}\mathbf{H} \qquad (3\text{-}71)$$

$$\nabla\times\mathbf{H} = \frac{4\pi}{c}\mathbf{j} + \frac{1}{c}\frac{\partial}{\partial t}\mathbf{E} \qquad (3\text{-}72)$$

Nevertheless in the following steps a refractive index (corresponding to a dielectric constant) will appear that we must mention here explicitly.

Using monochromatic oscillations with the radian frequency $\omega$ for the field quantities $\mathbf{E}$, $\mathbf{H}$, and $\mathbf{j}$,

$$\begin{aligned} \mathbf{E} &= \mathbf{E}_r \exp(i\omega t); \\ \mathbf{H} &= \mathbf{H}_r \exp(i\omega t); \\ \mathbf{j} &= \mathbf{j}_r \exp(i\omega t), \end{aligned} \qquad (3\text{-}73)$$

with the amplitudes with index $r$ depending only on spatial coordinates $x$, $y$, and $z$, we find from integration of Eq. (3–70)

$$\mathbf{j} = \frac{\omega_p^2}{4\pi i \omega (1 - i\nu/\omega)} \mathbf{E} \qquad (3\text{-}74)$$

Substituting this into Maxwell's equations (3–71) and (3–72) we arrive at

$$\nabla \times \mathbf{E}_r = -\frac{i\omega}{c} \mathbf{H}_r \qquad (3\text{-}75)$$

$$\nabla \times \mathbf{H}_r = -\frac{i\omega_p^2}{c\omega(1 - i\nu/\omega)} \mathbf{E}_r + \frac{1}{c}\omega \mathbf{E}_r \qquad (3\text{-}76)$$

Using steps similar to those we used in Section 3.1 to derive a wave equation from Maxwell's equations, we apply the operation $\nabla \times$ on Eq. (3–76) for a substitution of Eq. (3–75) into the resulting equation to obtain

$$\nabla^2 \mathbf{H}_r + \frac{\omega^2 n^2}{c^2} \mathbf{H}_r - i\frac{\omega}{c} \mathbf{E}_r \times \nabla \mathbf{n}^2 = 0 \qquad (3\text{-}77)$$

where the abbreviation

$$\varepsilon = \mathbf{n}^2 = 1 - \frac{\omega_p^2}{\omega^2 (1 - i\nu/\omega)} \qquad (3\text{-}78)$$

was used. Substituting $\mathbf{E}_r$ from Eq. (3–76) into the last equation and remembering the harmonic time dependence of the vectors, Eq. (3–73) linked with $\mathbf{E}$ and $\mathbf{H}$ by integration, we arrive at the general wave equation for the magnetic field

$$\nabla^2 \mathbf{H} - \frac{\mathbf{n}^2}{c^2}\frac{\partial^2}{\partial t^2}\mathbf{H} - \frac{1}{\mathbf{n}^2}(\nabla \times \mathbf{H}) \times \nabla \mathbf{n}^2 = 0 \qquad (3\text{-}79)$$

or, after expanding the triple cross product,

$$\nabla^2 \mathbf{H} - \frac{\mathbf{n}^2}{c^2}\frac{\partial^2}{\partial t^2}\mathbf{H} + 2(\nabla \mathbf{H}) \cdot \nabla \ln \mathbf{n} - 2(\nabla \ln \mathbf{n}) \cdot \nabla \mathbf{H} = 0 \qquad (3\text{-}80)$$

With the abbreviation $\mathbf{n}^2$ [Eq. (3–78)] we arrived at a dielectric response of the plasma with a complex refractive index $\mathbf{n}$. In view of the logical problem with the Lorentz assumptions for the plasma with vacuum properties of the dielectric constant and the magnetic permeability of the vacuum between the electrons and ions of the plasma, it sounds very strange that we now have a refractive index and a (high-frequency) dielectric constant $\varepsilon$. The fact that the real part of Eq. (3–78) immediately explains Langmuir's total reflection of radio waves at the ionosphere encourages us to go ahead with the result. The fact that the refractive index in laser produced plasmas agrees with Eq. (3–78) is a further argument. The first derivation of this dielectric response of a plasma from the two-fluid equations was given by Lüst (1953).

Eq. (3–80) is a very complicated wave equation. The corresponding equation for $\mathbf{E}$ is

$$\nabla^2 \mathbf{E} + 2(\nabla \mathbf{E}) \cdot \nabla \ln \mathbf{n} - \frac{1}{c^2}\left[\mathbf{n}^2 + 2\left(\frac{c}{\omega}\right)^2 \nabla^2 \ln \mathbf{n}\right]\frac{\partial^2}{\partial t^2}\mathbf{E} = 0 \quad (3\text{-}81)$$

where it had to be taken into account when differentiating in space that the refractive index is spatially varying (though a temporal independence was implied). Otherwise the wave equations (3–80) and (3–81) would be further complicated. We shall come back to these general (and mathematically very complicated) wave equations in the next subsection when we discuss waves in inhomogeneous plasmas.

For homogeneous plasma, where $\nabla \mathbf{n} = 0$, the wave equations are very simple as given in Eqs. (3–34) and (3–37). This results in the ideal purely transverse plane waves (or spherical waves, see Hora 1994) as shown in Fig. 3–3. The solutions only have the refractive index $\mathbf{n}$ (constant) and have for collisionless plasma the simple refraction properties of transparent media. If there are plasma collisions we have damped electromagnetic waves, and the absorption constant can be evaluated.

The property of wave damping can be seen from the solution of the Maxwellian equations with the complex refractive index $\mathbf{n}$ in Eq. (3–81) which for the homogeneous medium is

$$\nabla^2 \mathbf{E} + \frac{\omega^2 \mathbf{n}^2}{c^2}\mathbf{E} = 0 \quad (3\text{-}82)$$

The solution for the linearly polarized plane wave moving into the $x$ direction, compared with the undamped case of Eq. (3–39), is

$$\mathbf{E} = \mathbf{e}_y E_y e^{-\frac{\bar{K}}{2}x} e^{-i\left[\omega t - \frac{\omega}{c}nx\right]} \quad (3\text{-}83)$$

where the absorption coefficient $\kappa$ defines the absorption constant

$$\bar{K} = \frac{2\omega}{c}\kappa \quad (3\text{-}84)$$

defining the exponential decrease of the intensity (given by Poynting Vector) of the electromagnetic radiation when propagating along a depth x. the electromagnetic radiation when propagating along a depth $x$.

The real part $n$ and the imaginary part $\kappa$ of the complex refractive index **n** is given by

$$\mathbf{n} = n + ik = \mathrm{Re}(\mathbf{n}) + i\mathrm{Im}(\mathbf{n}) = \left[1 - \frac{\omega_p^2}{\omega^2(1 + iv/\omega)}\right]^{1/2} \tag{3-85}$$

resulting in

$$n = \frac{1}{\sqrt{2}}\left[\sqrt{\left(1 - \frac{\omega_p^2}{\omega^2 + v^2}\right)^2 + \left(\frac{v}{\omega}\frac{\omega_p^2}{\omega^2 + v^2}\right)^2} + \left(1 - \frac{\omega_p^2}{\omega^2 + v^2}\right)\right]^{1/2} \tag{3-86}$$

and

$$\kappa = \frac{1}{\sqrt{2}}\left[\sqrt{\left(1 - \frac{\omega_p^2}{\omega^2 + v^2}\right)^2 + \left(\frac{v}{\omega}\frac{\omega_p^2}{\omega^2 + v^2}\right)^2} - \left(1 - \frac{\omega_p^2}{\omega^2 + v^2}\right)\right]^{1/2}. \tag{3-87}$$

The magnetic field derived for the solution of **E** from the Maxwellian equations as before in the same way as for the collisionless case, Eq. (3–41), is from Eq. (3–83)

$$\mathbf{H} = \mathbf{e}_z E_y \sqrt{\frac{\varepsilon_0}{\mu_0}} e^{-\frac{\bar{K}}{2}x} e^{-i\left[\omega t - \frac{\omega}{c}\mathrm{Re}(\bar{\mathbf{n}})x\right]} \left[\mathrm{Re}(\mathbf{n}) + i\mathrm{Im}(\mathbf{n})\right] \tag{3-88}$$

$$= \mathbf{e}_z \sqrt{\frac{\varepsilon_0}{\mu_0}} |\mathbf{E}| e^{i\phi} |\mathbf{n}| \tag{3-89}$$

where a phase $\varphi$ between $E$ and $H$ appeared (contrary to the undamped wave) of

$$\phi = \frac{\mathrm{Im}(\mathbf{n})}{\mathrm{Re}(\mathbf{n})} = \frac{\kappa}{n}. \tag{3-90}$$

It is interesting to consider the refractive index for the collisionless case [which we used in advance in Eq. (1–5)] from Eq. (3–85) for $v = 0$,

$$\mathbf{n} = n = (1 - \omega_p^2/\omega^2)^{1/2} \quad (\text{if } v = 0). \tag{3-91}$$

One immediately sees here if the electron density ne is growing to produce a plasma frequency (2–4), which is then reaching $\omega = \omega_p$, the refractive index is then zero, which results in total reflection according to Langmuir's conclusion of total reflection. The special electron density in this case is from Eq. (2–4), the "critical" density

$$n_{ec} = \frac{m\omega^2}{4\pi e^2} \quad (\omega = \omega_p). \tag{3.92}$$

For neodymium glass laser radiation with a wavelength of 1.053 μm the cut-off density of the electrons is nec of 1.0 × 1021 cm−3 . For the same laser we can calculate the absorption constant (3–84) from the imaginary part of the refractive index, Eq. (3–87), depending on the electron density ne as given in the plasma frequency (2–4) and using the collision frequency according to Eq. (2–12) including Spitzer's correction and depending on the electron temperature via the collision frequency (2–12). We then receive the dependence on the plasma temperature shown in Fig. 3–4 with plasma density as parameter.

We see that there is a pole in the function near the cut-off density of $10^{21}$ cm$^{-3}$ below which we have low absorption which is well known in low-density plasmas. Above the critical cut-off density, there is a very strong absorption comparable to the metallic absorption which leads to very high reflection (just below total reflection). At high temperatures one is close to the pole with very high (partial) derivatives of the absorption constant on the density.

The evaluation of the real part of the complex refractive index is shown in Fig. 3–5. Again we see the pole near the critical density. At densities below the critical value, the refractive index is nearly independent of the temperature and has values below unity, or very close to it for densities less than one tenth of the critical value. One may note that the metallic behaviour leads to very low real parts of the refractive index as an expression of the skin effect.

In both Figs. 3–4 and 3–5 the curves end at low temperatures. These are the points where the Spitzer collision frequencies are invalid because of the conditions of very high density plasmas. The lines are dashed in the ranges where the classical approximation becomes invalid due to the Fermi–Dirac statistics of the electrons.

The question was how it was possible that the ordinary direct-current collision frequency *v* could explain the high-frequency absorption of laser radiation in the plasma, another curiosity with respect to the initial assumption of the Lorentz theory for the two-fluid model of the plasma. First, it also agrees with the electrical conductivities of metals (see, e.g., Hora 1994). But what was much more interesting was that the absorption constants agreed with the quantum electrodynamical derivation of Gaunt (1930) using Dirac's second quantization. Some minor differences of order unity exist for the low density range only: the collective effects at high densities were not covered by the initial inverse bremsstrahlung theory of Gaunt. From this it is evident that the most primitive 90' collision model of Fig. 2–3 comes to the same result as the most sophisticated quantum mechanics. A modification of the collision frequency at high plasma temperatures due to the quantum modification for the anomalous absorption should be mentioned. However, there appears to be a more drastic difference between the quantum electrodynamic solution for the collisions (Gaunt 1930) and the quantum modified collisions, Eq. (2–13b) (Hora 1982).

A nonlinear modification of the optical constants is necessary when the quiver energy of the electrons in the laser field

$$\varepsilon_{osc} = \frac{I}{cn_{ec}} \tag{3-93}$$

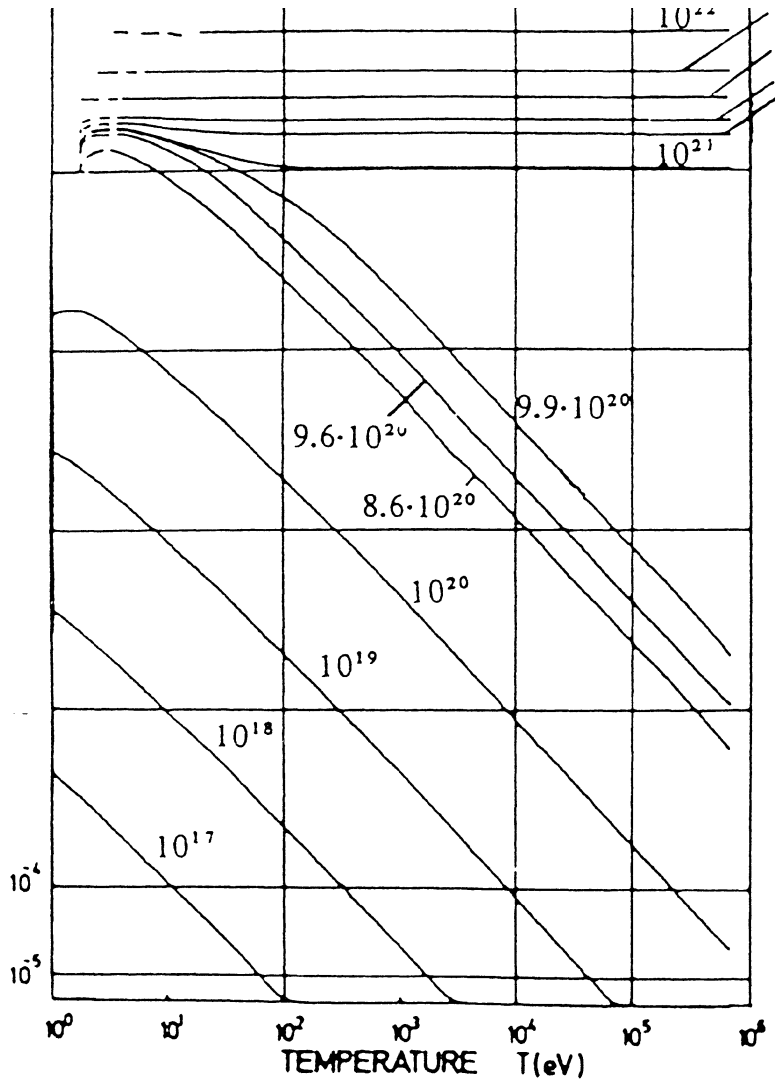

Figure 3-4. Absorption constant, Eq. (3-84), for neodymium glass laser radiation in deuterium plasma of various temperatures and density ne (cm-3) as a parameter (Hora et al. 1970).

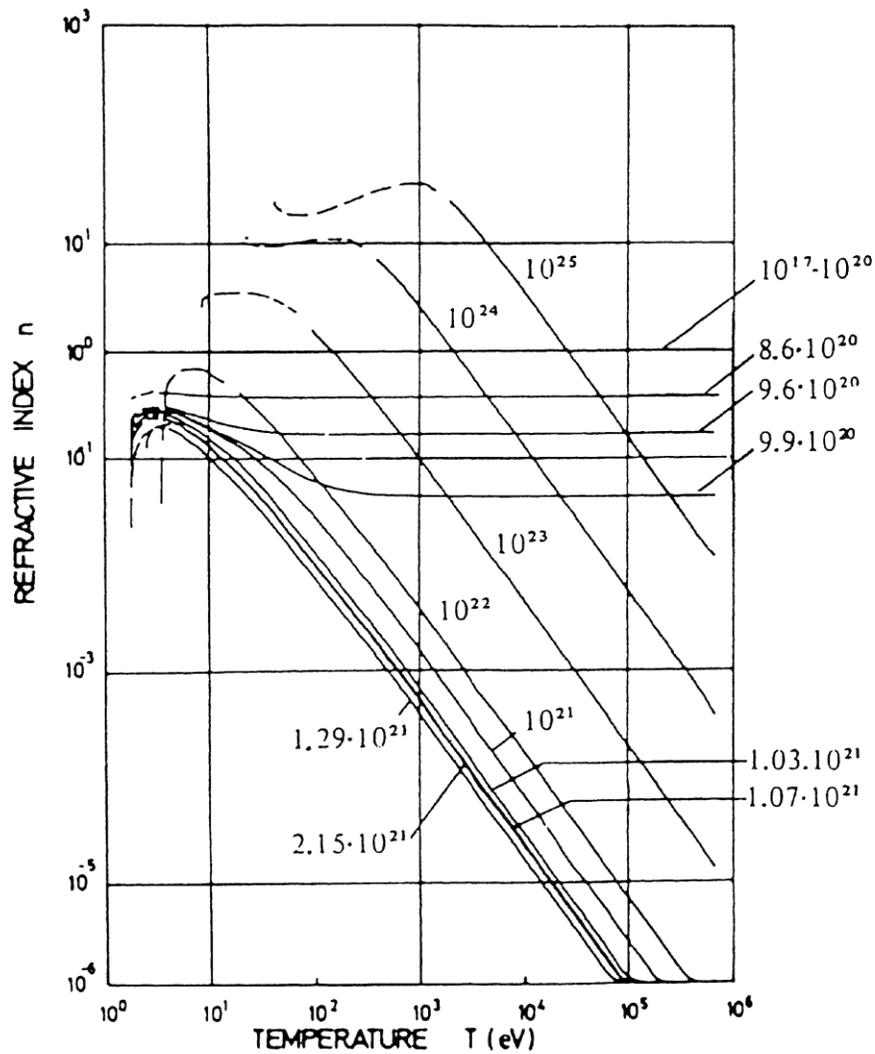

**Figure 3–5.** Real part of the refractive index for the same conditions as in Fig. 3–4 (Hora et al. 1970).

[intensity $I$ has to be given in equivalent dimensions as the quiver energy, Eq. (3–50)] is equal to or larger than the thermal energy $kT_e$ of the electrons. One then has simply to substitute the temperature in the collision frequency by the effective temperature

$$kT_{\text{eff}} = \frac{2}{3}\varepsilon_{\text{osc}} + kT_e .\qquad(3\text{-}94)$$

At very high laser intensities, one can simply ignore the chaotic thermal energy and work only with the quiver energy of coherent motion of the laser field.

This nonlinear absorption constant agrees rather well with the quantum mechanical derivation of Rand (1964).

If the quiver motion to values where the velocity is approaching the speed of light, the relativistic mass change has to be taken into account. The exact solution of the equation of motion of an electron results in a quiver energy of

$$\varepsilon_{osc} = mc^2 \left\{ \left[ 1 + \frac{A(I)e^2 |\mathbf{E}_r|^2}{m^2 \omega^2 c^2 |n|} \right]^{1/2} - 1 \right\}$$

$$= mc^2 \left\{ \sqrt{1 + 3A(I)\frac{1}{I_r}} - 1 \right\}, \quad (3\text{-}95)$$

where $m$ is the rest mass of the electron. The factor $A$ is unity for circularly polarized light and is growing from 1 to 1.05 with relativistic laser intensities (Eq. 6–76). $I_r$ is the relativistic threshold intensity where the quiver energy of the electron is $mc^2$ corresponding to an electric laser field amplitude $E_r$, such that

$$E_r = \frac{\sqrt{3}m\omega c}{e}; \quad I_r = \frac{3m^2\omega^2 c^3}{8\pi e^2}, \quad (3\text{-}96)$$

approximating $A(I) = 1$, where the values for neodymium glass and for $CO_2$ lasers are

$$E_r = \begin{cases} 5.11 \times 10^{10} \text{ V/cm (Nd)}; \\ 5.17 \times 10^{10} \text{ V/cm } (CO_2); \end{cases}$$

$$I_r = \begin{cases} 3.68 \times 10^{18} \text{ W/cm}^2 \text{ (Nd)}; \\ 3.68 \times 10^{16} \text{ W/cm}^2 (CO_2). \end{cases}$$

The plasma frequency changes then to

$$\omega_p^2 = \frac{4\pi e^2 n_e}{m} \sqrt{1 - v^2/c^2}$$

$$= \frac{4\pi e^2 n_e}{m} \frac{1}{[1 + 3A(I)I/I_r]^{1/2}}, \quad (3\text{-}97)$$

and the cut-off density to

$$n_{ec} = \frac{\omega^2 m}{4\pi e^2} [1 + 3A(I)I/I_r]. \quad (3\text{-}98)$$

These changes of the optical constants due to relativistic effects were first published in connection with pair production at very intense laser fields (Hora 1973; 1973a).

## 3.4 Waves in Inhomogeneous Media and Phase between **E** and **H**

Up to this point we evaluated—as exact solutions of the Maxwell equations—solutions of infinitely spread plane waves mostly of linear polarization for the special case of a vacuum or of a uniform medium with real refractive index *n*, and arrived at the equations for transverse waves for **E**, Eq. (3–39), and for **H**, Eq. (3–41), see Fig. 3–3. If the refractive index contained an imaginary part given by the plasma collision frequency *v*, we arrived at the damped plane waves, Eqs. (3–83) and (3–88), where again the refractive index and the collision frequency were uniform in the whole space.

We now consider a spatially varying refractive index as usual in the highly inhomogeneous media of laser produced plasmas. We do not yet consider the case of temporally varying refractive indices. The description of the wave solutions in inhomogeneous plasmas is considerably affected by the mathematics of ordinary differential equations giving higher functions (Bessel functions, confluent hypergeometric functions, Laguerre functions, Legendre functions, etc.). These functions were discovered and defined by solving a special case of the ordinary, linear differential equation of the second order of the type

$$\frac{\partial^2}{\partial x^2} f(x) + \bar{a}(x) f(x) = 0 \tag{3-99}$$

for a function *f(x)*. All the complications came from the fact that the coefficient *a(x)* was not constant but depended on the variable *x*.

This problem appears if we try to solve the wave equations (3–82) for **E** and (3–80) for **H** even in the simplified form without the term with a logarithmic spatial dependence of the refractive index **n**. Looking for solutions of waves with one frequency $\omega$ only using the product ansatz (3–73) we arrive, for the spatial part of the solutions $\mathbf{E}_r$ and $\mathbf{H}_r$, at

$$\nabla^2 \mathbf{E}_r + \frac{\omega^2}{c^2} \mathbf{n}^2 \mathbf{E}_r = 0; \tag{3-100}$$

$$\nabla^2 \mathbf{H}_r + \frac{\omega^2}{c^2} \mathbf{n}^2 \mathbf{H}_r = 0, \tag{3-101}$$

with the refractive index **n** as given by Eq. (3–78). For the collisionless case with real refractive index, the simplification of the one-dimensional (e.g., *x*-dependent) case results in a differential equation where $E_y$ or $H_z$ stands for *f(x)* in Eq. (3–99). Instead of (3–100) and (3–101) we then have the one-dimensional problem

$$\frac{\partial^2}{\partial x^2} E_y + \frac{\omega^2}{c^2} \mathbf{n}^2(x) E_y = 0; \tag{3-102}$$

$$\frac{\partial^2}{\partial x^2} H_z + \frac{\omega^2}{c^2}\mathbf{n}^2(x)H_z = 0. \tag{3-103}$$

The partial differentiation can then be changed into direct differentiation and these equations are then ordinary differential equations like Eq. (3–99).

It would not help much, if we had to use a special x -dependence of the refractive index [e.g., of n = (1 + 4/x2)] and we ran into the problem of a Bessel differential equation because this special refractive index does not meet the general case of an inhomogeneous plasma. The same is the case if we used the requirement $\mathbf{n}^2 = ax + is$ which results in the Airy differential equations with solutions of the Airy functions (Lindl and Kaw 1971), even if such special cases can be very useful for understanding the physical processes involved as we shall show in the following Rayleigh case.

Fortunately there is a special approximation which can be used within certain limitations to describe the general case of a spatial dependence of the refractive index. This is the WKB approximation, sometimes called the WKBJ (after Wentzel, Kramers, Brilluoin and Jordan) who used this well-known method of the last century for solving wave mechanical problems.

The WKB approximation for solving Eq. (3–102) uses the ansatz

$$E_y = \frac{E_V}{|\mathbf{n}|^{1/2}} \exp\left(i\frac{\omega}{c}\int^x \mathrm{Re}(\mathbf{n})\mathrm{d}\xi - \frac{\omega}{c}\int^x \mathrm{Im}(\mathbf{n})\mathrm{d}\xi\right) \tag{3-104}$$

or

$$E_y = \frac{E_V}{|\mathbf{n}|^{1/2}} \exp(iF); \quad F = i\frac{\omega}{c}\int^x \mathbf{n}\,\mathrm{d}\xi. \tag{3-105}$$

As one can see this ansatz substituted into the Eq. (3–102) satisfies the differential equation if the following conditions of the sufficiently slowly varying refractive index are fulfilled:

$$\frac{\sqrt{3}}{2|\mathbf{n}|}\left|\frac{\partial \mathbf{n}}{\partial x}\right| \Box \frac{\omega}{c}|\mathbf{n}| \quad \text{or} \quad \Theta = \frac{\sqrt{3}}{2}\frac{c}{\omega|\mathbf{n}|^2}\left|\frac{\partial \mathbf{n}}{\partial x}\right| \Box 1 \tag{3-106}$$

and

$$\frac{1}{2|\mathbf{n}|}\left|\frac{\partial^2 \mathbf{n}}{\partial x^2}\right| \Box \frac{\omega^2}{c^2}|\mathbf{n}^2| \quad \text{or} \quad \psi = \frac{1}{2|\mathbf{n}|^3}\frac{c^2}{\omega^2}\left|\frac{\partial^2 \mathbf{n}}{\partial x^2}\right| \Box 1. \tag{3-107}$$

The solution for the electric field is then

$$\mathbf{E} = \mathbf{i}_y \frac{E_V}{|\mathbf{n}|^{1/2}} \cos\left(\frac{\omega}{c}\int^x \mathrm{Re}(\mathbf{n})\mathrm{d}\xi - \omega t\right) \exp\left(\frac{-\bar{k}x}{2}\right), \tag{3-108}$$

if one uses the averaged absorption coefficient

$$\bar{k} = 2\frac{\omega}{xc}\int^x \text{Im}(\mathbf{n})d\xi. \qquad (3\text{-}109)$$

The magnetic field of the wave is determined then again through the Maxwellian equations. One has to substitute the solution (3–108) into one of the Maxwellian equations and then [similar to the evaluation following Eq. (3–40)] obtains from differentiation and integration

$$H_z = -\frac{ic}{2\omega}\frac{E_v}{|\mathbf{n}|^{3/2}}\frac{d\mathbf{n}}{dx}\exp(iF) + E_v\sqrt{|\mathbf{n}|}\exp(iF) \qquad (3\text{-}110)$$

or approximating $\mathbf{n}$ by $|\mathbf{n}|$

$$\mathbf{H} = -\mathbf{i}_z\frac{c}{2\omega}\frac{E_v}{|\mathbf{n}|^{3/2}}\frac{d|\mathbf{n}|}{dx}\sin\left(\frac{\omega}{c}\int^x\text{Re}(\mathbf{n})d\xi - \omega t\right)\exp\left(-\frac{1}{2}kx\right)$$
$$+\mathbf{i}_zE_v|\mathbf{n}|^{1/2}\cos\left(\frac{\omega}{c}\int^x\text{Re}(\mathbf{n})d\xi - \omega t\right)\exp\left(-\frac{1}{2}kx\right). \qquad (3\text{-}111)$$

It is important to note that the amplitude of $\mathbf{E}$, Eq. (3–108), is *increasing* when the wave is penetrating into plasma with decreasing refractive index $\mathbf{n}$, while the amplitude of the magnetic field, Eq. (3–110), is *decreasing* by the same amount, such that for collisionless plasma the Poynting vector $\mathbf{E} \times \mathbf{H}$ expresses a constant energy flux since there is no absorption. What is much more important is that next to the cos-functions in $\mathbf{E}$ and $\mathbf{H}$ there is a sin-function in $\mathbf{H}$ which represents a phase shift between the oscillations especially for collisionless plasma determined by the gradient of the refractive index $d\mathbf{n}/dx$. For the spatially constant refractive index, this factor becomes zero and we have no phase shift as is the case in Fig. 3–3. We shall see that *this phase shift is essential for the generation of the nonlinear (ponderomotive) force in the following section.*

If the plane electromagnetic wave is obliquely incident on the one-dimensionally layered (stratified) plasma, the WKB approximation is still applicable at least until a certain angle of incidence $u_0$ (Fig. 3–6) (for details see Hora 1974, 1991: Section 7.2). The penetration of the wave into the inhomogeneous plasma, then, does not go to the depth x = x0 where the critical density is reached, but recedes earlier. If one uses linear polarized waves with s- and p-polarization, the p-polarization results in some longitudinal component when bending around the propagation vector of the wave.

The p-polarized wave produces an oscillation with an $E_x$ component even at the depth $x_0$ which the propagating wave does not reach. For collisionless plasma, this longitudinal component at the critical density can reach minus infinity (White and Chen 1974), which switches into very large positive values as soon as a tiny absorption—e.g., Landau damping—appears (Hora 1979). This Försterling–Denisov resonance absorption was considered exceedingly important in laser-produced plasmas. However, this is definitely not the case since it is polarization dependent and strong only at certain angles of incidence near 20°. To explain the observation of strong resonance at perpendicular incidence this process cannot be used even if artificial assumptions on partially oblique incidence are included. There is a much more severe resonance possible at perpendicular incidence (Hora et al. 1985; Goldsworthy et al. 1988) (Hora 1991: Sections 11.2 and 11.3).

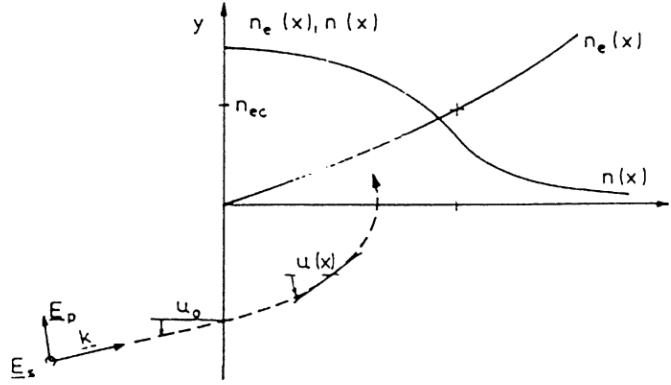

**Figure 3–6.** Linearly polarized plane waves with p- or s-polarization obliquely incident at an angle $u_0$ in vacuum onto a stratified plasma. This angle $u(x)$ varies in the plasma along the $x$-direction.

One may argue that the WKB approximation is uncertain since one may not be able to overlook limitations where the approximations become critical or break down. For this discussion, fortunately, we have a case with an exact solution for comparison which can be easily understood since it results in elementary functions only and does not need the complicated higher functions. This case for electromagnetic waves with varying refractive index was first discussed by Rayleigh and also discussed in a PhD thesis by Schlick (1904) supervised by Max Planck. Using a real refractive index

$$n = \frac{1}{1+ax}, \quad a \geq 0, \tag{3-112}$$

and plane wave solutions propagating into the $x$-direction with linear polarization of **E** into the $y$-direction, we can take the logarithmic terms of the wave equations (3–80) and (3–81) apart from second derivatives of logarithmic functions

$$\frac{\partial^2}{\partial x^2} E_y + \frac{\omega^2}{c^2} n^2 E_y = 0; \tag{3-113}$$

$$\frac{\partial^2}{\partial x^2} H_z - 2\left(\frac{\partial}{\partial x} \ln n\right) \frac{\partial}{\partial x} H_z + \frac{\omega^2}{c^2} n^2 H_z = 0. \tag{3-114}$$

Equation (3–113) is an Euler differential equation which can be solved exactly (see Hora 1991: Section 7.3) to arrive at

$$\mathbf{E} = \mathbf{i}_y \frac{E_y}{n^{1/2}} \exp\left\{\mp \frac{i}{2}\left[\frac{4\omega^2}{c^2\alpha^2} - 1\right]^{1/2} \ln(1+\alpha x) - i\omega t\right\}, \qquad (3\text{-}115)$$

$$\begin{aligned}\mathbf{H} = &\mp\left(\sqrt{1 - \frac{\alpha^2 c^2}{4\omega^2}} - i\frac{\alpha c}{2\omega}\right) \\ &\times \mathbf{i}_z E_y n^{1/2} \exp\left[\mp \frac{i}{2}\sqrt{\frac{4\omega^2}{c^2\alpha^2} - 1} \ln(1+\alpha x) - i\omega t\right].\end{aligned} \qquad (3\text{-}116)$$

For the limit $\alpha$ going to zero (vacuum case) we reproduce the plane waves Eq. (3–39) and (3–41). What is interesting in Eq. (3–115) is the swelling of the electric field amplitude by decreasing refractive index and the inverse behavior for the magnetic field amplitude in (3–116) exactly as seen with the WKB approximation, Eqs. (3–108) and (3–111). Also, the phase between **E** and **H** is given by the complex first factor on the right-hand side of (3–116) whose absolute value is 1 and is constant for the whole wave in the Rayleigh medium. This constant phase will have a special meaning for the Rayleigh case when calculating the nonlinear (ponderomotive) force.

The fact that there is a phase shift results in a reflection at the kink of the refractive index between the vacuum and the interior of the Rayleigh medium. A reflected wave has to be introduced to compensate for the phase shift. This looks strange since there is complete continuity (no step as between air and glass) of the refractive index and nevertheless a reflected wave is generated. For very large values of $α$ this reflectivity can even be total, which is never possible at the edge between two homogeneous media at normal incidence. It also is a more sophisticated proof for Langmuir's high reflection of radio waves at the inhomogeneous ionosphere.

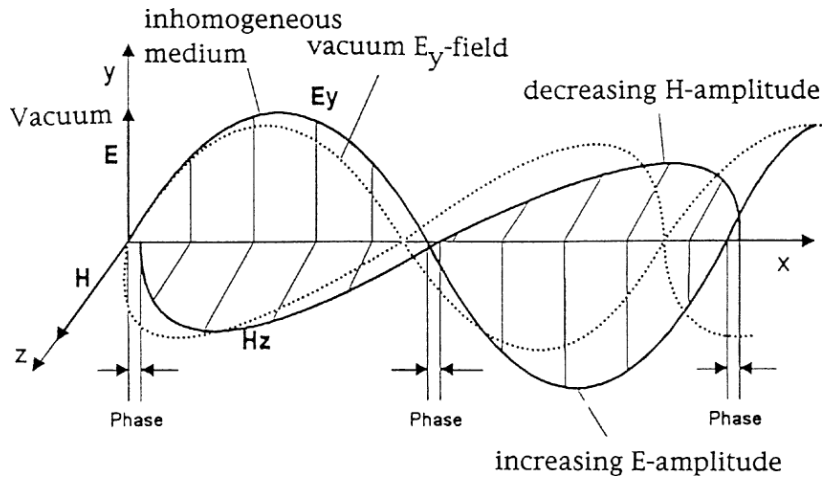

**Figure 3–7.** Exact **E**- and **H**-solution for a wave penetrating into the $+x$ direction within a Rayleigh medium.

Inside an inhomogeneous medium, there are only two linearly independent waves possible, an incoming and an outgoing wave. There is obviously no internal reflection possible, only at the changes from the homogeneous to the Rayleigh medium and vice versa. This was in contrast to the established theory (Ginzburg 1970) as it was pointed out (Hora 1957) to be valid for a general derivation of the non-existence of internal reflection. With an unusual expression of surprise, this was pointed out also by Osterberg (1958) for more general solutions than using Euler's differential equation. This result strongly contradicted the view of the calculations of wave propagation in the ionosphere where one was trying to find an approximate "local reflectivity". How this concept of "local reflectivity" could nevertheless have a meaning theoretically in a perfect way on the background of the exact results (Hora 1957; Osterberg 1958) was explained and clarified against the earlier established theory (Hora 1991: Section 7.3; Lawrence et al. 1980).

# CHAPTER 4
# Hydrodynamic Derivation of the Nonlinear Forces with Ponderomotion

The preceding review of basic plasma physics, hydrodynamics and electrodynamics with special attention to the Lorentz force and the phases between the **E** and the **H** fields of electromagnetic waves and how this was derived, was necessary to understand the nonlinear forces and ponderomotion. If we are not let the problems rest at the numerous open questions on the assumptions about the ponderomotive potential described in Chapter 1, and unless we are to work with this while ignoring the complicated background, we have no choice but to go through these details. For easier reeding of this text, we begin first with a derivation of the nonlinear force for a simplified case of perpendicular incidence and convince ourselves of the clear action of the force. After this we discuss the problems which led to a very general derivation of the hydrodynamic basis of the nonlinear forces which in turn arrived at a clarification of the hydrodynamic two-fluid theory. This is leading to a general derivation of the Maxwell stress tensor in a purely hydrodynamic way while all the other derivations used the elastomechanics.

    The elastomechanical method provided very limited results for slow temporal changes for fluids without dispersion and without dissipation (absorption), while the hydrodynamic method immediately covers the high-frequency case of plasma dispersion and dissipation. Nevertheless this is not the whole story. The problems of single particle motions, the resulting electric double layers and more general numerical evaluations on the basis of a genuine two-fluid model will follow in the next chapter, with the addition of applications for laser accelerators, laser fusion, ion sources and related industrial uses.

## 4.1 SIMPLE DERIVATION OF THE NONLINEAR FORCE FOR PERPENDICULAR INCIDENCE

We consider the forces in a plasma with a one-dimensional inhomogeneity when light is incident perpendicularly (density, temperature and the resulting refractive index depend on the depth coordinate $x$ only). The force density could be calculated from Schlüter's equation (3–67). It is worthwhile to recall that this equation has no electric field force (Coulomb force) which initially was in the Euler equations for each electron and ion fluid which the plasma consists of, while the last term in Eq. (3–67) can be expressed completely by current densities. We identify between the thermokinetic force

$$f_{th} = -\nabla p,\qquad(4\text{-}1)$$

given by the gas dynamic pressure $p$ and call all the remaining forces due to electromagnetic quantities the nonlinear force $\mathbf{f}_{NL}$, where $\mathbf{f}$ is the total force density in the plasma

$$f_{NL} = f - f_{th}.\qquad(4\text{-}2)$$

The last term in Eq. (3–67) was identified by the very sophisticated derivation of Schlüter, and while it did not appear in the derivation from the Boltzmann equation as given by Spitzer (1962). The derivation of the Schlüter term from the Boltz- mann equation should be possible if one watches the numerous simplifications and linearizations in the derivation (Chapter 2.4) but this definitely is not a trivial task.

Schlüter's equations were indeed very successful in calculating numerous plasma problems with nearly static magnetic fields and currents, for example, for theta pinches or for deriving such macroscopic plasma instabilities (Heim et al. 1957) such as the sausage, the kink and other instabilities. When concentrating to the Hall term in the diffusion equation [Ohm's law, Eq. (3–69)] one could derive most of the porperties of the magnetohydrodynamic generators, etc. However, Schlüter's equations could not be used in high-frequency field interactions where interest was arising in 1965 from observations of highly nonlinear nonclassical effects in laser produced plasmas.

For the case of perpendicular incidence of infinite plane electromagnetic waves on stratified plasmas, the last term in Eq. (3–67) is zero. Since $\mathbf{E}$ is transverse and any gradient of components of $\mathbf{E}$ can be only in the $x$-direction; therefore the scalar products in the last term of Eq. (3–67) vanish.

The nonlinear force (4–2) is then simply the Lorentz force

$$f_{NL} = \frac{1}{c}\mathbf{j}\times\mathbf{H}\qquad(4\text{-}3)$$

Inserting the magnetic field and current density $\mathbf{j}$ of the WKB solution (3–111)

$$\mathbf{j} = -\frac{1}{4\pi}\mathbf{e}_y\frac{E_v}{n^{1/2}}\frac{\omega_p^2}{\omega}\sin\left[\frac{\omega}{c}\int^x n(\xi)\,d\xi - \omega t\right]\qquad(4\text{-}4)$$

after using Ohm's law (3–70) for a collisionless plasma ($\nu = 0$)

$$\frac{\partial}{\partial t}\mathbf{j} = \frac{\omega_p^2}{4\pi}\mathbf{E}\qquad(4\text{-}5)$$

we arrive by taking the cross product in Eq. (4–3) at

$$f_{NL} = -\mathbf{e}_x \frac{1}{4\pi c} E_v^2 \frac{\omega_p^2}{\omega} \cos\left[\frac{\omega}{c}\int^x n(\xi)d\xi - \omega t\right]\sin\left[\frac{\omega}{c}\int^x n(\xi)d\xi - \omega t\right]$$
$$+\mathbf{e}_x \frac{\omega_p^2}{8\pi\omega^2} \frac{E_v^2}{n^2} \frac{dn}{dx} \sin^2\left[\frac{\omega}{c}\int^x n(\xi)d\xi - \omega t\right]. \tag{4-6}$$

When time averaging we have to remember the orthogonality of the cos and sin Functions

$$\frac{1}{T}\int^T \cos(...)\sin(...)\,dt = 0; \quad \frac{1}{T}\int^T \sin^2(...)dt = \frac{1}{2}.$$

Then one arrives at the time-averaged value

$$\bar{f}_{NL} = \frac{\mathbf{e}_x}{16\pi} \frac{\omega_p^2}{\omega^2} \frac{E_p^2}{n^2} \frac{dn}{dx} = -\mathbf{e}\frac{1-n^2}{16\pi} \frac{d}{dx} \frac{E_v^2}{n}. \tag{4-7}$$

We have to be aware that this force is different from zero due only to the phase shift [with the sin function of the **H**-vector in Eq. (3–111) for the WKB approximation] which is determined by the gradient of the refractive index

$$\nabla \mathbf{n}$$

Since the WKB approximation (3–109) is essentially

$$\mathbf{E} = \mathbf{E}_v / \mathbf{n}^{1/2} \tag{4-8}$$

where $\mathbf{E}_v$ is the high-frequency field of the electromagnetic wave in vacuum (index 'v') we can write instead of Eq. (4–7)

$$f_{NL} = -\frac{1-\mathbf{n}^2}{16\pi}\nabla\mathbf{E}^2. \tag{4-9}$$

This is formally the ponderomotive force (1–6) where the factor 16 instead of 8 in the denominator is due to the time averaging of the harmonic functions. What then reminds of a Lorentz force? In the electrostatic case of Fig. 1–1(b) there are no velocities nor is there a magnetic field to identify a Lorentz force. Since there was no deeper clarification available, the result of the nonlinear force (4–9) or of (4–7) was called electrostrictive force, j × B -force, ponderomotive force, field gradient force, etc. In our first use (Hora et al. 1967) we underlined the nonlinear (quadratic) character for the high-frequency case and called it the nonlinear force.

As shown in Chapter 1, Eq. (1–9) for a single electron, the nonlinear force density of Eq. (4–9) in a plasma for the special case of perpendicular incidence and stationary irradiation is conservative, i.e., the curl of fNL is zero. Such a con- servative field permits the expression of the force by the gradient of a potential. Expressing the refractive index for the collisionless plasma (ν = 0) by the plasma frequency, Eqs. (3–78) and (2–4), we arrive from Eq. (4–9) at the time averaged nonlinear force

$$f_{NL} = -\frac{\omega_p^2}{16\pi\omega^2}\nabla \mathbf{E}^2 = -\frac{n_e}{4m\omega^2}\nabla \mathbf{E}^2.$$

This can be expressed as

$$f_{NL} = -\frac{n_e}{2}\nabla\left[\frac{e^2}{2m\omega^2}\right]\mathbf{E}^2 = -0.5 n_e \nabla\phi \quad (4\text{-}9a)$$

using the ponderomotive potential $\varphi$ of Eq. (1–9). One may realize that the factor 1/2 in these last equations is the result of time averaging and that the nonlinear force of Eq. (4–9a) is the force density in the plasma acting on all the electrons given by the electron density $n_e$ while the consi-deration of Eq. (1–9) was for a single electron.

## 4.2 PONDEROMOTIVE AND NON-PONDEROMOTIVE TERMS. PREDOMINANCE OF THE NONLINEAR FORCE

It is interesting to note that the result (4–9) also can be reached by starting with the conservation law of electromagnetic energy (3–51) and arriving at the nonlinear force from the gradient of the expression (3–53) which in this case is taken for an inhomogeneous medium contrary to the derivation of the Lorentz force for homogeneous media following Eq. (3–53). We arrive after time averaging, differentiation and using the WKB approximations of (4–8) and $\mathbf{H} = n\mathbf{E}_v^2$,

$$\begin{aligned}\bar{f}_{NL} &= -\mathbf{e}_x \frac{1}{8\pi}\frac{d}{dx}\left(\mathbf{E}^2 + \mathbf{H}^2\right) = -\mathbf{e}_x \frac{E_v^2}{16\pi}\frac{d}{dx}\left(\frac{1}{n} + n\right)\\ &= -\mathbf{e}_x \frac{1}{16\pi}\frac{\omega_p^2}{\omega^2}\frac{d}{dx}\frac{E_v^2}{n},\end{aligned} \quad (4\text{-}10)$$

which is identical to (4-9) or (4-7).

The connection between the result (3–62) for the Lorentz force and nonlinear force (4–10) shows the different sign. This is due to the fact that $\mathbf{f}$ in (3–53) is the force density of the electromagnetic field which results in the counter-reaction to the electrons in the plasma, $\mathbf{f}_{NL}$. It is important to note that in view of Eq. (3–53) the nonlinear force also includes the coulomb force next to the Lorentz force. Many authors—in a less critical way—identify the nonlinear force in plasmas with the ponderomotive force or electrostriction of electrostatics, neglecting—or not being aware of—the Coulomb term.

This is possible in homogeneous plasmas only where any charge density $\rho$ decays within sufficiently short time. However, in inhomogeneous plasmas there is a static charge separation (see the "genuine two-fluid model" in the following Section 6.4) or there are oscillating charge densities as a necessary result to fulfill the law of momentum conservation in laser irradiated plasmas. See the second term on the right hand side of the laser Eq. (4–33). This all indicates that it is necessary to distinguish between Lorentz force, electrostriction, ponderomotive force and the general nonlinear force.

This all was derived for plasmas without collisions. With collisions one arrives at

$$\bar{f}_{NL} = \mathbf{i}_x \frac{E_v^2}{16\pi} \frac{\omega_p^2}{\omega^2} \frac{1}{|\mathbf{n}|^2} \frac{d|\mathbf{n}|}{dx} + \mathbf{i}_x \frac{E_v^2}{16\pi} \frac{\omega_p^2}{\omega^2} \frac{2\omega}{c} \frac{v}{\omega},\qquad(4\text{-}11)$$

(Hora 1969). The last term in (4–11) depends on the collision frequency but it has nothing to do with the thermalization leading to the thermokinetic forces which we previously identified (Eq. 4–2). Therefore it is a term produced purely by the electrodynamic interaction of the electromagnetic wave with the plasma. Since only the first term of Eq. (4–11) has a formal similarity with the ponderomotive force we may concede that the first term in Eq. (4–11) may be called the *ponderomotive term* and the second is the *non-ponderomotive term*. This last term was derived independently and is called the "Stamper term" (Stamper 1977).

Following the result of Eq. (4–10) of the nonlinear force for perpendicular incidence of the electromagnetic radiation on a stratified plasma, we can now consider the total force density in the plasma, Eq. (4–2)

$$f = -\nabla p + \nabla\left(\mathbf{E}^2 + \mathbf{H}^2\right)/8\pi \qquad(4\text{-}12)$$

and can modify Fig. 1–4 by Fig. 4–1 showing the action of the nonlinear force. While the whole block of plasma from the maximum value of the electromagnetic energy density $E = (\mathbf{E}^2 + \mathbf{H}^2)/8\pi$ towards higher depths $x$ is producing the dielectrically increased radiation pressure for compression of the plasma interior, the force in the plasma corona towards the incident laser radiation (coming from the vacuum) is given only by the excess of the force density $E$ over the vacuum value. We shall see from experiments that the swelling of the maximum over the vacuum value can be very large.

When comparing the nonlinear force with the thermokinetic force $\mathbf{f}_{th}$ (4–3) we could simply compare the values behind the $\nabla$ operator but we need to note that an integration constant has to be considered. For the ablating plasma this constant is the value of $E$ in the vacuum since only the excess energy density due to the dielectric swelling in the plasma causes the negative gradient in Eq. (4–10) to accelerate the plasma corona. We then have to compare the pressure $p$ (electron and ion pressure) with the difference of the maximum $\mathbf{E}^2$ over its vacuum value. The pressures are equal at

$$p = \frac{3}{2} n_e KT \left(1 + 1/Z\right) = \frac{\mathbf{E}_v^2}{16\pi} \left[ \left( \frac{1}{|\mathbf{n}|} + |\mathbf{n}| \right) \exp_0 - 1 \right]. \tag{4-13}$$

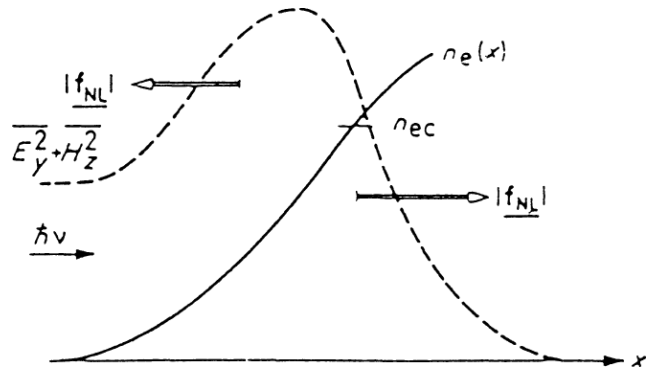

**Figure 4–1**. Action of the nonlinear force fNL as given by the gradient of the electromagnetic energy density E = (E2 + H 2 )/8π [Eq. (4–10)] due to laser light (hv ) propagating from vacuum into a plasma. The electron density ne monotonously increases with x .

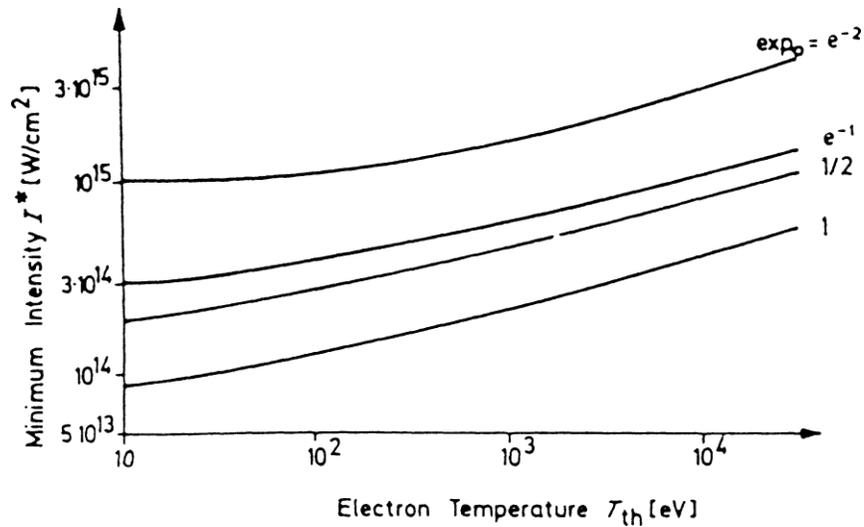

**Figure 4–2.** Threshold neodymium glass laser intensity $I^*$ in a deuterium plasma of given temperature $T_{th}$ above which the nonlinear force exceeds the thermokinetic pressure $p$. The factor $\exp_0$ takes into account to what value the collision absorption may have decreased the electromagnetic field when reaching the critical electron density.

The factor $\exp_0$, see Fig. 4–2, expresses the possible collision damping of the electromagnetic field at the maximum value near the critical density nec. This factor is ≈ 1. As a worst case we consider values for exp0 of 1, 0.5, 1/e and 1/e2 in Fig. 4–2. The maximum of the absolute value of the complex refractive index **n** is given from procedures similar to that of Section 3.3, as previously explained in detail.[2]

The numerical evaluation of the comparison (4–12) where the intensity $I^*$ of a neodymium glass laser exceeds the thermokinetic pressure in a deuterium plasma is shown in Fig. 4–2. We see that for laser intensities around $10^{12}$ W∕cm$^2$ the nonlinear force can be neglected by the thermokinetic force as was done, e.g., by Mulser (1970) in his hydrodynamic computations of laser–plasma interaction. However, at intensities of $10^{16}$ W∕cm$^2$ as used by Shearer, Kidder and Zink (1970) the nonlinear force is clearly predominant and then the gas dynamic action is only a small perturbation.

## 4.3 NUMERICAL AND EXPERIMENTAL RESULTS

After the first results on the nonlinear force (Hora et al. 1967; Hora 1969) the following experiments were explained immediately. There was the observation of a linear increase of the energy of the keV ions on their charge number $Z$ observed in numerous experiments. This was immediately evident from the approximation of ponderomotive potential as explained in Chapter 1. The question why the multi- keV ions were produced apart from thermal plasma (Fig. 2–5) needed an answer by the theory of the ponderomotive self-focusing (Hora 1969a) for which the threshold in the range of MW laser power for ns laser pulses was derived. This laser power was exactly the value above which the usual thermal plasma properties significantly changed in the experiments and the keV ions (Linlor 1963) and the exceedingly large electron emission currents appeared (Honig 1963; Isenor 1964; Opower et al. 1965; Namba et al. 1967; Schwarz 1971). The self-focusing was confirmed experimentally (Korobkin and Alcock 1968) and the depletion process of the plasma at the beam center was measured (Richardson et al. 1971). The beam schrank to such high laser intensities that the nonlinear force acceleration was dominant, and then produced keV ions energetically and linearly separated on the ion charge number $Z$. It was then important to show directly where basic physics about the non-linear force or numerical results could be used directly in order to demonstrate (Hora 1991; p. 180) the details of the mechanism. The first attempts to discount the existence of the force actually supported its existence (Steinhauer and Ahlstrom 1970; Mulser and Green 1970). A positive numerical result was derived by Lindl and Kaw (1971) for a collisionless plasma showing the standing wave field and the force pushing the plasma to its nodes, with a swelling average value indicating the ablation of the plasma corona by the net nonlinear force, Fig. 4–3.

The question was from the beginning (Hora et al. 1967) how the initial density profile of a laser irradiated plasma would develop where the nonlinear forces were predominant. The first clear answer was given by applying the WAZER code by Shearer, Kidder and Zink (1970), as shown in Fig. 4–4. A laser pulse with an intensity and time dependence given in the upper part of Fig. 4–4 is irradiating an initially monotonous deuterium plasma density profile. Due to the nonlinear force, one generates density minima and profile steepening which is characteristic for the nonlinear force interaction. The density minima were later called cavitons and were the most convincing criteria for experiments proving the nonlinear force action.

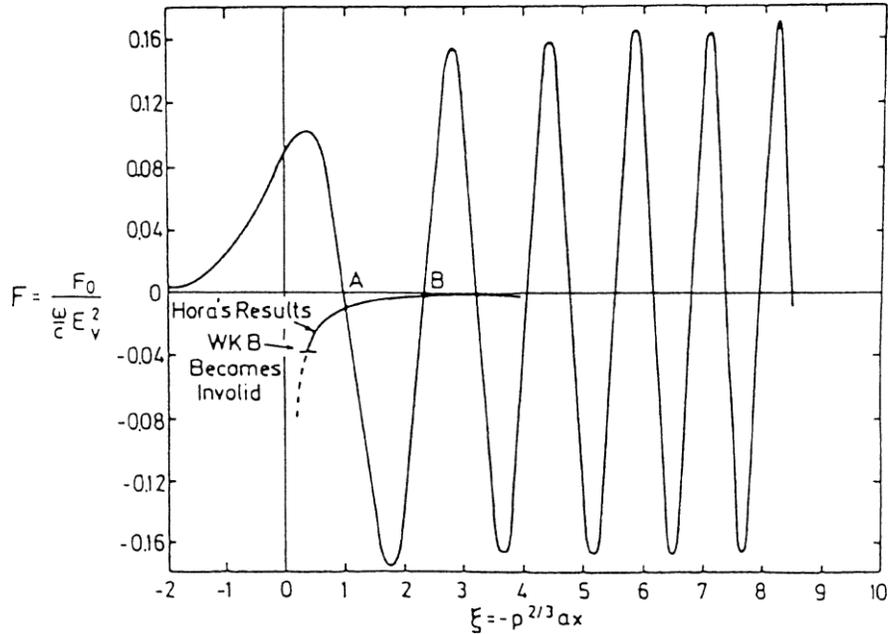

**Figure 4–3.** The nonlinear force for a collisionless plasma of linear density gradients is given. The light (incident from the right) creates a standing electromagnetic wave and the locally oscillating force. The net acceleration of the plasma is given by the double value of the monotonic force close to the abscissa. It is compared with the WKB approximation (Lindl and Kaw 1971).

A numerical study to show both the net acceleration of the plasma corona as well as the standing wave pattern of Lindl and Kaw (1971) with the numerical code developed by Rich Kinsinger (Hora 1975) had the advantage that a detailed solution of the incident and the reflected laser radiation was solved numerically in a complete way including the absorption in the plasma covering the linear collisional absorption and the intensity dependent nonlinear absorption [see Eq. (3–94)]. The initial electron density is a linearly increasing ramp from zero density (vacuum) to the critical density within 50 wavelengths and then increasing more rapidly quickly at supercritical density.

It turns out (Hora 1975) that the light initially penetrates to the critical density where it is then reflected like from a metallic mirror (Fig. 4–5). The standing wave pattern including all the collisional damping then appears only partially. But this is sufficient to push the plasma into the nodes of the standing wave within the first picoseconds of interaction. During this time, the hydrodynamic expansion of a uniformly heated plasma of 100 eV temperature is expanding much less rapidly. Even the push of the corona reaching velocities of order $10^7$ cm/s is too slow to interfere with the plasma being pushed into the nodes of the standing wave (Fig. 4–6).

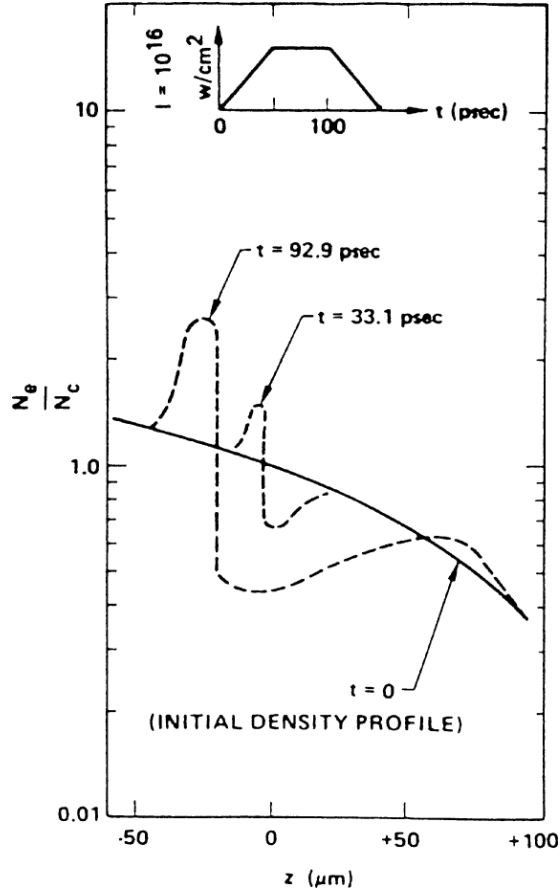

**Figure 4–4.** Density profile at difference times calculated with the dynamic computer program WAZER by Shearer (1971). The low-density maximum is caused by the nonlinear force. The assumed laser pulse is given in the upper part. This was the first numerically discovered density minimum (caviton) and profile steepening due to nonlinear forces. After Shearer, Kidder and Zink (1970).

The density ripple in the standing wave pattern fits ideally a phase reflection grating (von Laue–Bragg grating) which after 2 ps causes a high reflection at the very low-density outermost part of the plasma such that at later time the light no longer penetrates in 50 wavelength thick corona and decays by a factor of 100 before it reaches critical density. This result was very disappointing for laser fusion since it demonstrated that the plasma prevented the laser from depositing its energy into the plasma. This problem has to do with the stuttering interaction of the laser light (later measured by Maddevar et al. 1990) with plasma and the strong phase reflection. This was overcome even later by the methods for smoothing of the laser light (Hora et al. 1992; 1999).

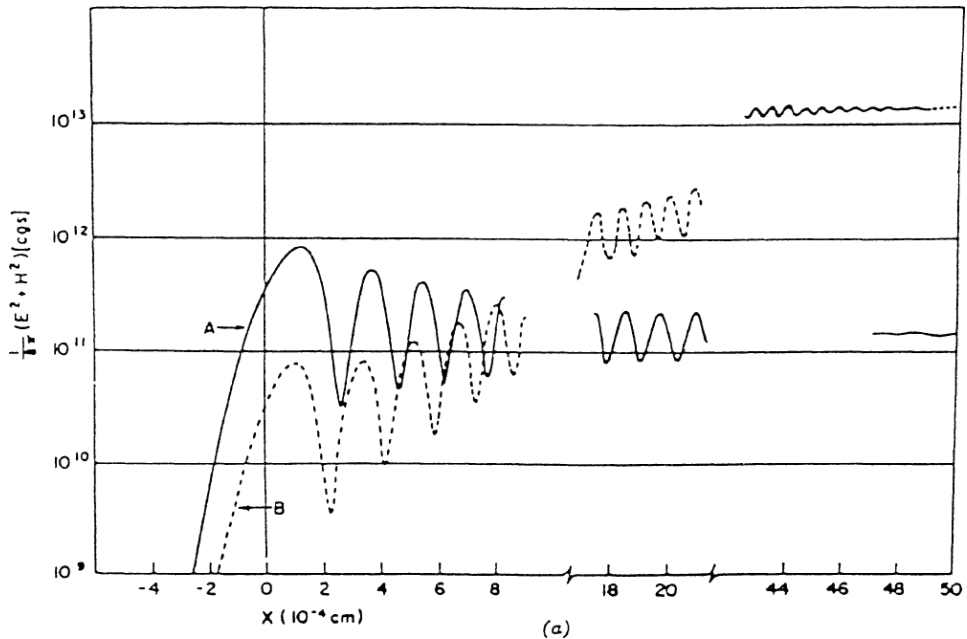

**Figure 4–5**. A laser beam is incident from the right side on a plasma of initial temperature of 100 eV and linear density increasing from zero at x = 50 μm to the cutoff density at x = 0 and then increasing more rapidly. The exact stationary (time-dependent) solution without retardation of the Maxwell equations with a nonlinear refractive index **n**, based on a nonlinear collision frequency Eq. (3–94), results in an oscillation due to the standing wave and dielectric swelling of the amplitude (curve A). At a later time ($2 \times 10^{-12}$ s) the laser intensity is $2 \times 10^{16}$ W/cm$^2$ (curve B), where the relative swelling remains, but the intensity at $x = 0$ is attenuated by the phase reflection of the self-generated von Laue–Bragg-grating.

       The *experimental proof* of the predominance of the nonlinear force action in laser irradiated plasmas could be followed up by looking for the generation of the density minima (cavitons) predicted numerically by Shearer, Kidder and Zink (1970), Fig. 4–4. A first indication was reported from the measurements by Beaudry and Martineau (1973) and a more indirect proof came from a measurement of the ponderomotive self-focusing processes by Marhic (1975). It is important to note that the discovery by Shearer, Kidder and Zink (1970) of the caviton was a straightforward application of hydrodynamics that correctly included the nonlinear forces, and nothing else. There was no concern about parametric

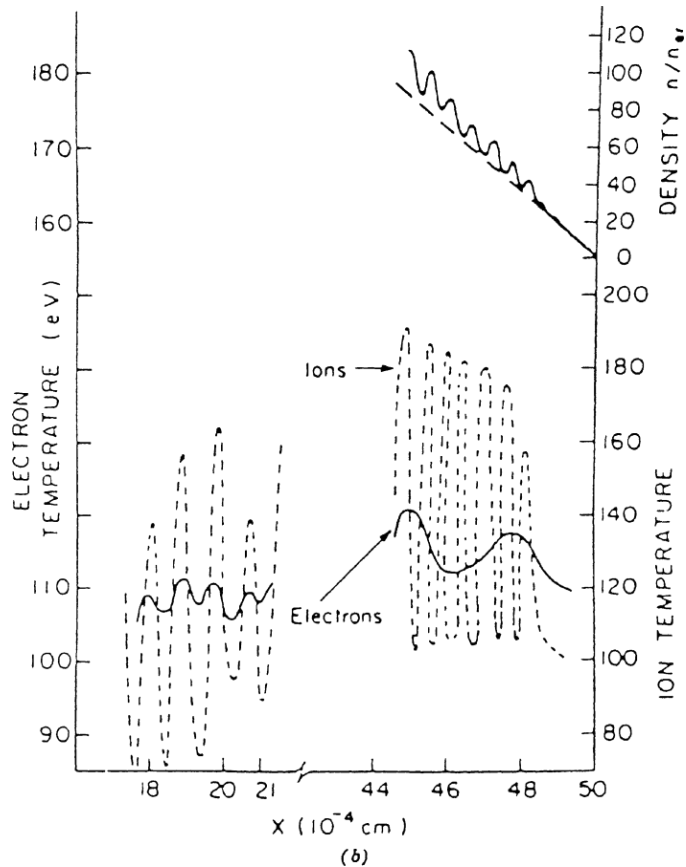

**Figure 4–6.** The initial density (dashed line) and the density along curve B of Fig. 4–5 where a ripple is created by the nonlinear force, pushing plasma towards the nodes of the standing wave. The electron and ion temperatures are increasing following the ripple by dynamic compression at conditions identical to curve.

instabilities (though weakly present as seen from higher harmonic emission) or the Försterling (1950)–Denisov (1957) resonance absorption and other anomalies which, after wasting a very large amount of research capacity, have been realized to be wrong or of less importance.

The first measurement of the caviton was achieved by irradiation of on a plasma by very intense microwaves with electron density beyond the critical value, not by lasers. The longer wavelength and the more comfortable longer time scale made it possible to measure the details of the time dependence of the electromag- netic wave amplitude with the strong swelling above its vacuum value depending on the penetration depth at different time steps [Fig. 4–7(a)]. Then the electron density profiles could be measured for a set of different time steps [Fig. 4–7(c)]. The microwave intensity was not constant but changed sufficiently slowly to keep nearly time constant assumptions but pulsed in order to better demonstrate the swelling and caviton effect. The discussion was correctly based on the nonlinear force effect as described by Lindl and Kaw (1971).

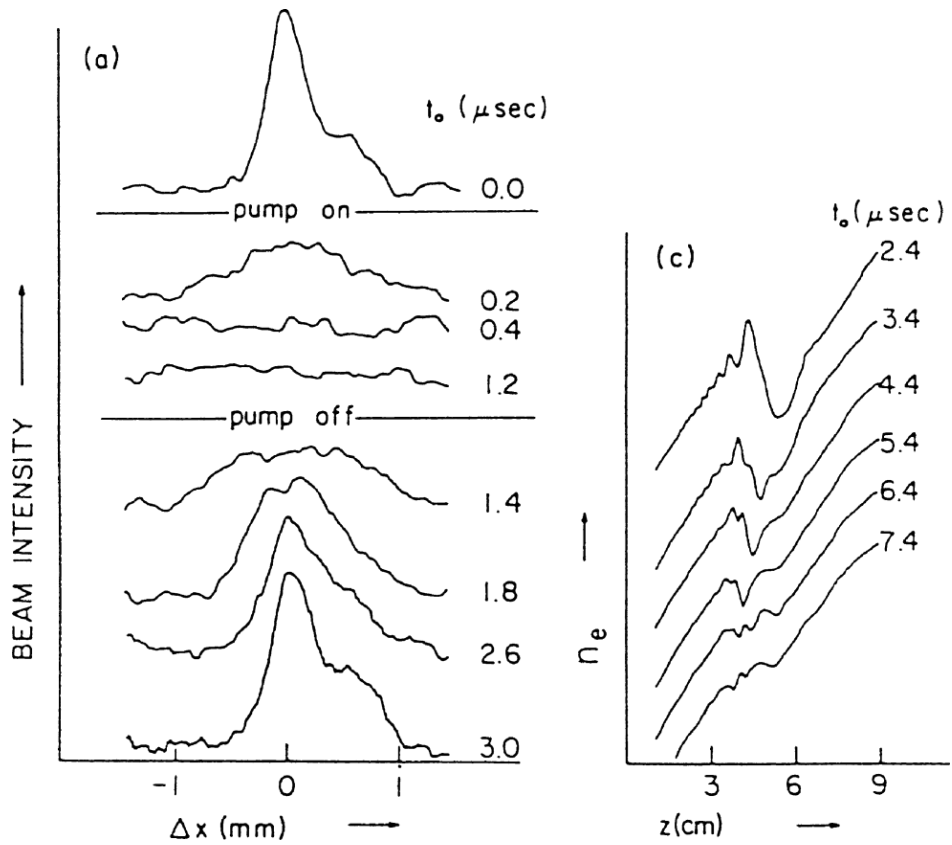

**Figure 4–7.** Electromagnetic energy density or intensity $I$ of microwaves with swelling [Eq. (4–14)] at a depth $x = 0$ where the electron density is critical (a); the density profiles changed after some time $t_0$ of irradiation to generate the caviton (c). After Kim, Stenzel and Wong (1974).

One experiment summarizing these effects was performed (Wong and Stenzel 1975) showing a swelling by a factor S of the incident microwave radiation intensity [as described by the WKB approximation (3–108)]

$$I(x) = \frac{\mathbf{E}_v^2 c}{8\pi \mathbf{n}(x)} = \frac{I_v}{|\mathbf{n}(x)|} = I_v S \qquad (4\text{-}14)$$

using the vacuum intensity $I_v$. It is remarkable that the swelling reached values of up to 600. Figure 4–8 shows electron density profiles not only with a caviton but also with some steepening before the caviton. This radiation trapping process was directly shown from the special numerical analysis and has nothing to do with instabilities.

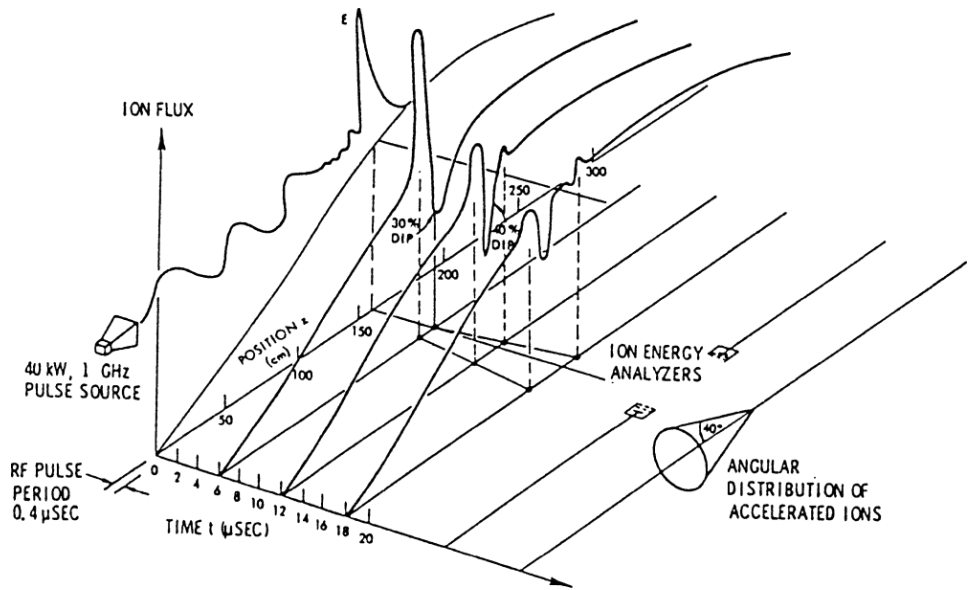

**Figure 4–8.** Intense microwaves irradiating a plasma with monotonously increasing electron density. The swelling of the wave amplitude of **E** at the critical density (shown at time zero) causes the cavitons in the density later (Wong and Stenzel 1975).

With laser produced plasmas it was much more difficult to measure the caviton since the experimental spatial resolution reduces to a few micrometers and the temporal resolution to a few ps duration. The first direct measurement of the caviton was published by Zakharenko et al. (1976), Fig. 4–9, and repeated by Fedoseevs et al. (1977), Fig. 4–10, and in a more dimensional spatial resolution by Azechi et al. (1977), Fig. 4–11.

For completeness, the following numerical results should be mentioned (Lawrence 1978; Hora et al. 1979a). In order to avoid the standing wave patterns of Fig. 4–5 where an initially linearly increasing electron density was irradiated by the laser, and remembering the Rayleigh media, Eq. (3–112), a bi-Rayleigh deuterium plasma density profile of 100 μm thickness for the initial electron distribution was chosen for the computations (Fig. 4–12) for irradiation with neodymium glass lasers. The maximum density was a little below the critical density but because of the (linear and nonlinear) absorption mechanisms the light penetrated only to about 25 μm depth. A typical case of the temporal development of the electromagnetic energy density is shown in Fig. 4–13. As the latest step we see the result of some radiation trapping. This may be of interest in connection with the Wong-Stenzel (1975) experiment, Fig. 4–8.

What was interesting was that during the first 1.5 ps only the plasma corona due to the gradients of the electromagnetic energy density (Fig. 4–14) for various neodymium glass laser intensities showed blocks of plasma moving with velocities (Fig. 4–15) exceeding $10^9$ cm/s against the laser light at laser intensities of $10^{18}$ W/cm$^2$ (close to the relativistic limit). The blocks of plasma interior for compression showed nearly the same high velocities. The initial electron and ion temperatures in the bi-Rayleigh blocks of the assumed values of 100 eV or 1000 eV were changed by up to one order of magnitude. Finally, we can check the essentially nonlinear property of the interaction by computing the transfer of kinetic energy to the ablated plasma corona (Fig. 4–16).

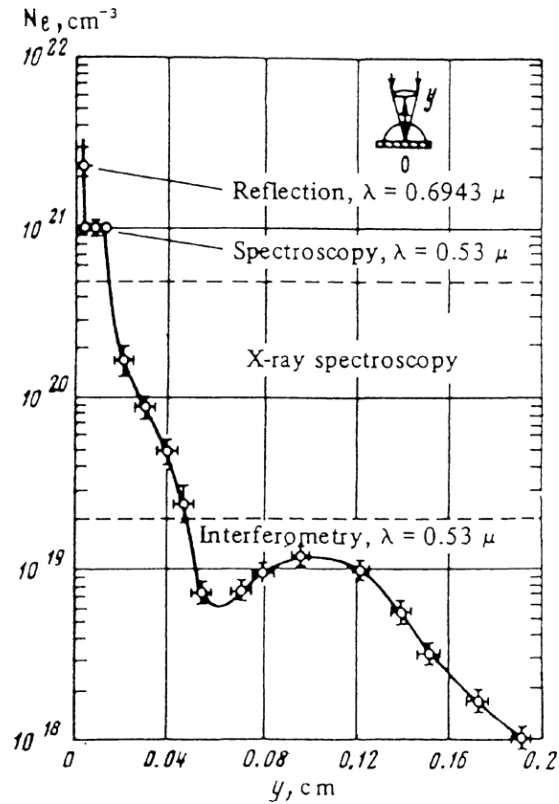

**Figure 4–9.** Density profile with a nonlinear force produced caviton at laser irradiation (Zakharenko et al. 1976).

## 4.4 Problem at Oblique Incidence

The reader will have understood by now that the electrodynamic forces in a plasma are not so easily described by simply saying "ponderomotive potential" and thinking of Eq. (1–9). The macroscopic plasma hydrodynamics and electrodynamics is much more sophisticated. The results with the simple derivation of the nonlinear force for perpendicular incidence of the radiation on the plasma (Section 4.1) are complex enough, leading to the ponderomotive potential, Eq. (4–9a). Its reality has now been demonstrated convincingly from numerous numerical and experimental results quoted.

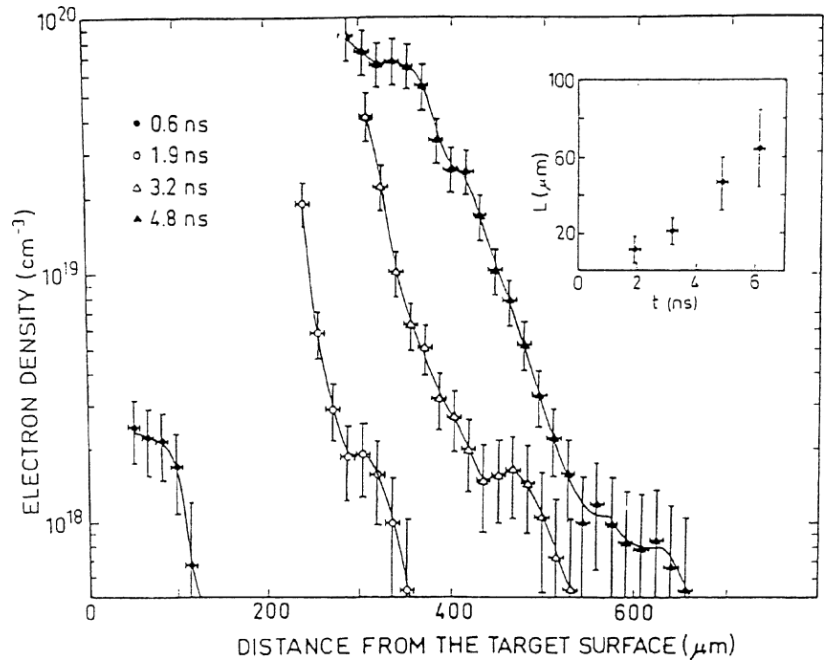

**Figure 4–10.** Density profiles with nonlinear force produced cavitons in a laser irradiated plasma (Fedoseev et al. 1977).

Nevertheless the reader will have to be prepared that a much more complex approach is necessary for the general case of oblique incidence. A number of further nonlinear terms had to be found for the Schlüter equation of motion (3–67) in order to arrive at the necessary momentum balance.

Surprisingly, this required the introduction of electric fields within the plasma which it was hoped would be eliminated forever by Schlüter's equation of motion, since it was "intuitively clear" to plasma physicists familiar with homogeneous media, that there is never an internal electric field present if outside fields are ignored. However, most plasmas are ignored inhomogeneous unlike uniform metals. This intuitive assumption is one of the notorious deficiencies of plasma theory and why it led to disagreement with experiments. It was none other than Hannes Alfvén (1971, 1981) with his minority group (Fälthammar 1988; Hershkovitz 1985; Peratt 1991) who pointed out the interior electric fields in inhomogeneous plasmas based on his knowledge of cosmic plasmas or laboratory plasma devices. The difficulty of accepting these facts can be seen from the remark of a leading plasma theoretician like Kulsrud (1983) that "Alfvén's electric fields which are intuitively not clear." This is the reason why the development of the nonlinear force in the more general purpose than perpendicular incidence is of importance for plasma theory.

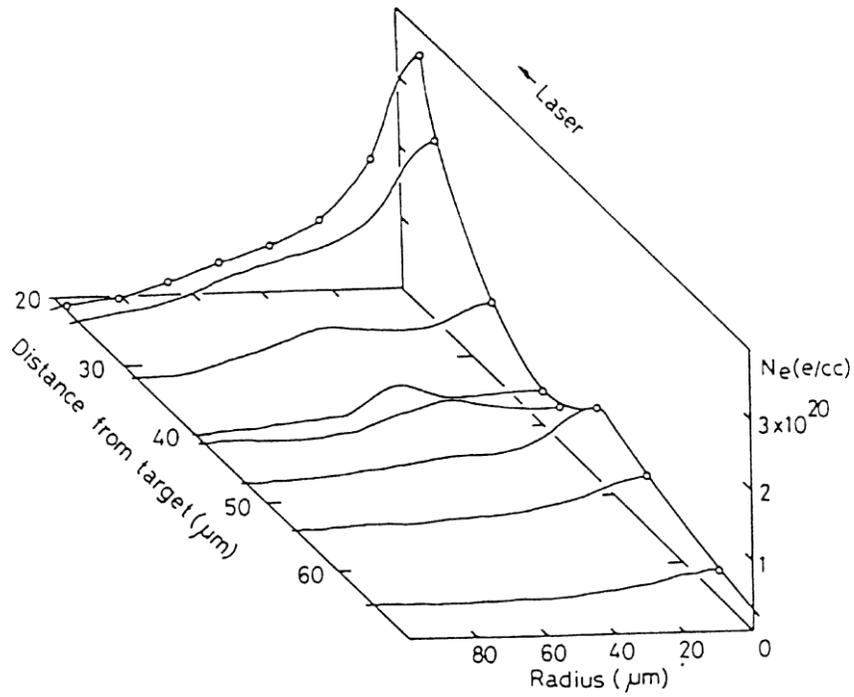

**Figure 4–11.** Two-dimensional resolved density profile measurement at a time step of 370 ps at laser irradiation showing the nonlinear force produced caviton (Azechi et al. 1977).

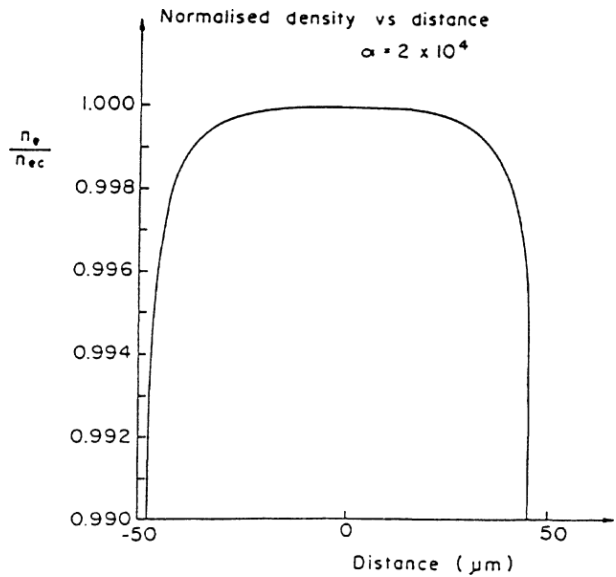

**Figure 4–12.** Initial electron density with a bi-Rayleigh profile of 100 wavelength thickness for very low reflection at irradiation by neodymium glass laser radiation.

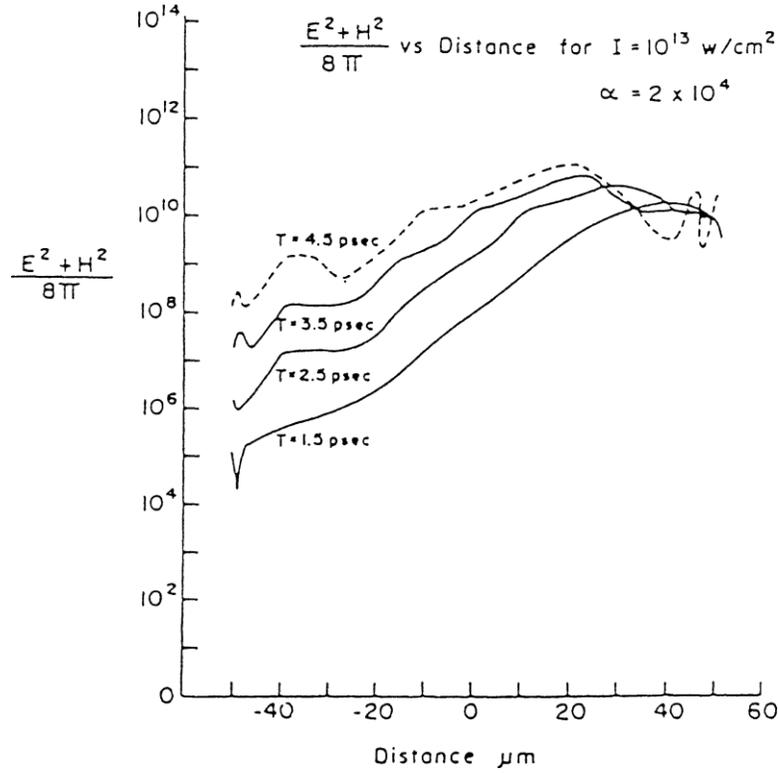

**Figure 4–13.** Laser field energy density at irradiation of an initial bi-Rayleigh deuterium density profile (Fig. 4–12) of initially 100 eV temperature and its temporal dynamic development.

The steps to this development were to look into the processes when laser light is obliquely incident at an angle $u_0$ (see Fig. 3–6) on a stratified plasma. Starting from the WKB approximation for this case the electromagnetic field in the plasma could be calculated and the resulting force densities achieved. It was evident that the equation of motion following Spitzer (1962)

$$f = m_i n_i \frac{\partial \mathbf{v}}{\partial t} + m_i n_i \mathbf{v} \cdot \nabla \mathbf{v} = -\nabla p + \frac{1}{c} \mathbf{j} \times \mathbf{H} \quad \text{(Spitzer)}, \qquad (4\text{-}15)$$

should not be used, since there was the more general equation derived by Schlüter (1950) with the nonlinear term, see Eqs. (3–67) and (3–68). This expression determining the force density in a plasma apart from the thermokinetic term (4–1) solely from current densities and magnetic fields (Schlüter 1950) is:

$$f = m_i n_i \frac{\partial \mathbf{v}}{\partial t} = -\nabla p + \frac{1}{c} \mathbf{j} \times \mathbf{H} + \mathbf{j} \cdot \nabla \frac{1}{n_c} \mathbf{j} \frac{m}{e^2}. \qquad (4\text{-}16a)$$

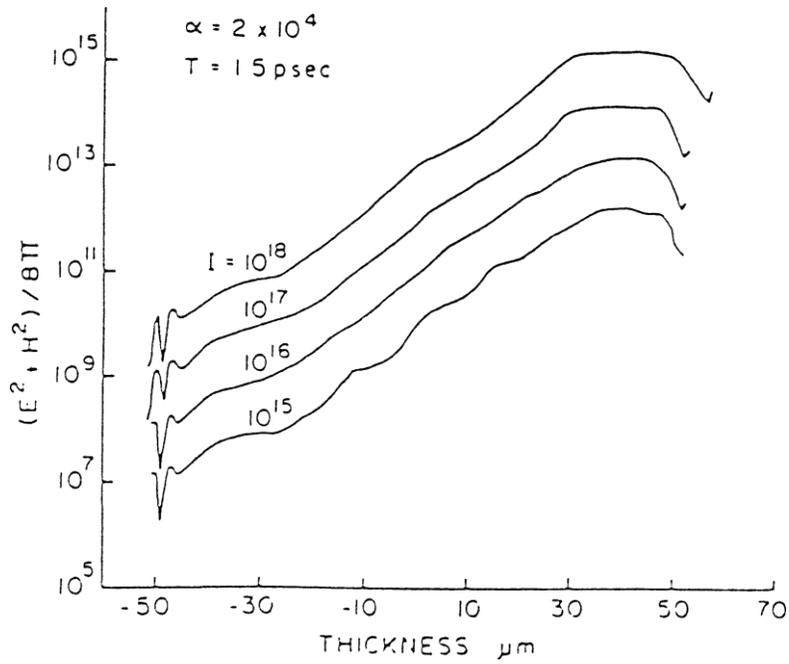

**Figure 4–14.** Electromagnetic energy density in the plasma with initial conditions as in Fig. 4–14 1.5 ps after laser irradiation with various intensities $I$ in W/cm.

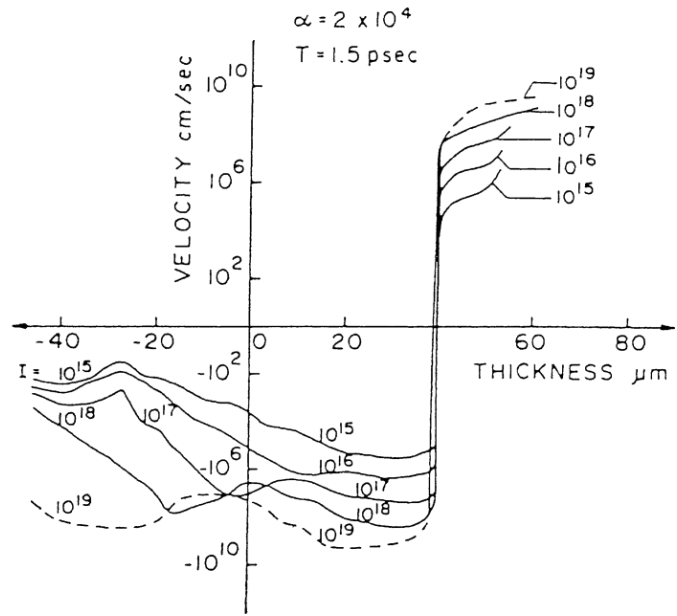

**Figure 4–15.** Resulting velocity profiles of the plasma for the irradiations in Fig. 4–14.

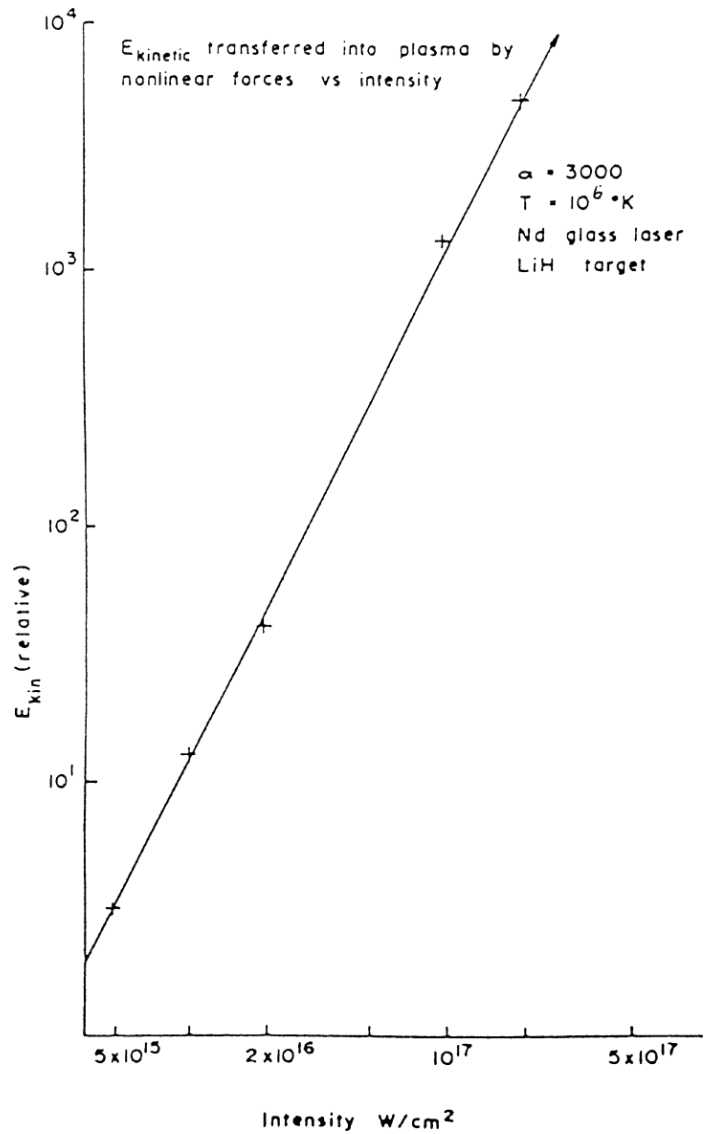

**Figure 4–16.** Kinetic energy transferred to the plasma for irradiation of an initial 100 wavelength thick bi-Rayleigh profile of an LiH plasma of 100 eV initial temperature (Lawrence 1978).

This could be rewritten (Hora 1969) for the special case of monochromatic interactions as

$$f = -\nabla p + \frac{1}{c}\mathbf{j} \times \mathbf{H} - \frac{1}{4\pi}\frac{\omega_p^2}{\omega^2}\mathbf{E}\cdot\nabla\mathbf{E} \quad \text{(Schlüter)}. \tag{4-16b}$$

We note at this stage that the last term in Eq. (4–16b) for high-frequency fields of plasma is formally identical with Kelvin's electrostatic ponderomotive force (1–1) using (3–77) and (1–2). Here the ponderomotive force appears as a different entity next to the Lorentz force while there are other cases where the ponderomotive force (ignoring the assumptions of static and oscillating fields) is identical with the Lorentz force. This case should be a warning how mistaken it can be to mix cases with different presumptions with each other, even though there is a formal identity.

When putting the WKB solutions for the laser field in the plasma at oblique incidence (Hora 1974) into Eq. (4–16), the last term is not vanishing as in the case of perpen- dicular incidence. For the special case of a collisionless plasma we find that for s-polarization (**E**-vector perpendicular to the plane of incidence) the forces are perpendicular to the plane of the plasma (see plasma plume A in Fig. 4–17). In the case of the p-polarization (Fig. 4–17), the result is that the force should have a non-vanishing time averaged component along the direction of the plasma surface such that the plasma plume would expand obliquely (case B in Fig. 4–17).

This result arises from of working with these solutions and Schlüter's equation of motion. It seems to be reasonable that there is a difference between the two different cases of polarization as is well-known from the formulas of reflectivity (Hora 1974a). Measurements were performed which showed clearly from side-on photos that the plasma plume for the case of p-polarization was the oblique case of B in Fig. 4–17 (Büchl et al. 1967). Schlüter stated that this result was not acceptable for the theory of collisionless plasma under stationary (non-transient) conditions. Since there is *no recoil (momentum transfer) possible* along the plasma surface if there are no collisions, in all cases there has to be a plasma acceleration *always perpendicular to the plasma surface*.

The fact that there were clear photographs for the oblique plasma plume raised a question of whether the transient mechanisms of switching on or off of the laser light, or the absorption due to collisions caused the oblique emission.

In order to find the answer for the non-transient collisionless case with no oblique but exclusively perpendicular emission of plasma from the plasma surface, one has to go back to the theory. At this time Krokhin (1967) indicated one should treat the problem on the basis of the book by Landau and Lifshitz (1966). This refers to the Maxwell stress tensor, but checking the reference, it revealed that it did not fit at all the requirements of a plasma. Landau and Lifshitz presented the most advanced derivation of the Maxwell stress tensor describing the stresses in an electromagnetic field by analogy with the stresses in an elastic medium. This is the way this was derived in all known treatments.

The most advanced result by Landau and Lifshitz (1966) arrived at the case where the electromagnetic field at least could be slowly varying and was not longer restricted to static cases, as before. Further, the treatment was done for dielectric media with a dielectric constant larger than one, and the media could be liquid. It was clearly established that the media must not have any optical absorption (dissipation) or dispersion (wavelength dependent dielectric constant or optical absorption constant).

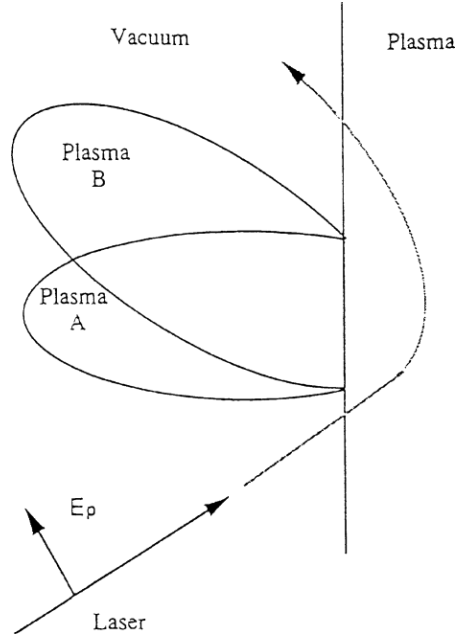

**Figure 4–17.** Scheme for electromagnetic radiation of p-polarization obliquely incident on a stratified plasma. The emitted plasma plume has an emission direction perpendicular to the plasma surface (A) or may have an oblique direction (B).

This was not applicable for plasma at the very high laser frequencies and the very fast acceleration mechanisms. The plasmas have a dispersion [frequency dependence of the optical constant, see. Eq. (3–78)] and we were interested also in the possibility that the plasmas have collisions (dissipation, absorption). Nevertheless, by heuristically using the results of Landau and Lifshitz and algebraically comparing them with the plasma equation of motion (Hora 1969) a solution was derived where there was only acceleration perpendicular to the plasma surface. The result was (Hora 1969) that the plasma equation of motion should be

$$f = -\nabla p + \frac{1}{c}\mathbf{j}\times\mathbf{H} + \frac{1}{4\pi}\mathbf{E}\nabla\cdot\mathbf{E}$$
$$-\frac{1}{4\pi}\frac{\omega_p^2}{\omega^2+v^2}\left(1+i\frac{v}{\omega}\right)\mathbf{E}\nabla\cdot\mathbf{E} - \frac{1}{4\pi}\frac{\omega_p^2}{\omega^2+v^2}\left(1+i\frac{v}{\omega}\right)\mathbf{E}\cdot\nabla\mathbf{E} \quad (4\text{-}17)$$
$$-\frac{1}{4\pi}\mathbf{E}\mathbf{E}\cdot\nabla\frac{\omega_p^2}{\omega^2+v^2}\left(1+i\frac{v}{\omega}\right),$$

where we see that following the well-known Schlüter term, Eq. (3–68) there appeared two additional nonlinear terms. Furthermore, it appeared to be necessary that the Coulomb force term [the third on the right-hand side of Eq. (4–17)] must be included. It can be seen that if any of these terms is dropped or any other term added, the force becomes a component along the plasma surface. Since this is forbidden because of momentum conservation for the collisionless case, we can conclude that the terms and only the terms in (4–17) completely describe the nonlinear force density in a plasma at non-transient conditions (i.e., time-independent laser irradiation).

Equation (4–17) contains the result for the more general case with collisions. From Eq. (4–11) for perpendicular incidence it was derived for the first time that the ponderomotive and non-ponderomotive terms appeared.

## 4.5 General Derivation of the Nonlinear Force and Hydrodynamic Foundation of the Maxwell Stress Tensor

After it was known on the basis of momentum conservation that Eq. (4–17) and only this equation is the non-transient solution of the equation of motion in a plasma, it was easier than (Schlüter 1950) to arrive at the complete derivation of the space charge quasi-neutral two fluid equations.

First for the non-transient case with the result of Eq. (4–17) (Hora 1991: Appendix C), the detailed derivation will be considered.

We begin with the Euler equations for ions and electrons which we write now in more detail than in Eqs. (3–63) and (3–64), especially with regard to the (last) viscosity terms determined by the electron–ion collision frequency $v$. For the ion fluid we have:

$$m_i n_i \left[ \frac{\partial}{\partial t} \mathbf{v}_i + \mathbf{v}_i \cdot \nabla \mathbf{v}i \right] = -\nabla \frac{3}{2} n_i K T_i + Z n_i e \mathbf{E} + \frac{Z n_i e}{c} \mathbf{v}_i \times \mathbf{H}$$
$$- \frac{m_i n_i m n_e}{m_i n_i + m n_e} v(\mathbf{v}_i - \mathbf{v}_e) + \mathbf{K}_i, \qquad (4\text{-}18)$$

and for the electron fluid of the plasma:

$$m n_e \left[ \frac{\partial}{\partial t} \mathbf{v}_e + \mathbf{v}_e \cdot \nabla \mathbf{v}_e \right] = -\nabla \frac{3}{2} n_e K T_e - n_e e \mathbf{E} - \frac{n_e e}{c} \mathbf{v}_e \times \mathbf{H}$$
$$+ \frac{m_i n_i m n_e}{m_i n_i + m n_e} v(\mathbf{v}_i - \mathbf{v}_e) + \mathbf{K}_e. \qquad (4\text{-}19)$$

In the following we will assume equilibrium state of equal electron and ion temperature, Eq. (2–16), since we are not studying the thermokinetic part of the total force density of the plasma but we are concentrating exclusively on the electrodynamic part, the nonlinear force.

The main assumption of Schlüter's two-fluid model is space charge quasi-neutrality such that (apart from microscopic fluctuations within a Debye length)

$$n_e \approx Z n_i . \qquad (4\text{-}20)$$

Under this assumption, the net velocity **v** of the plasma

$$\mathbf{v} = \frac{m_i n_i \mathbf{v}_i + m n_e \mathbf{v}_e}{m_i n_i + m n_e} \tag{4-21}$$

was approximated by Schlüter (1950) as

$$\mathbf{v} \approx \frac{m_i \mathbf{v}_i + Z m \mathbf{v}_e}{m_i + Z m}. \tag{4-22}$$

The difference of the electron and ion velocity can then, under the same assumptions, be used to determine the electric current **j** in the plasma

$$\mathbf{v}_e - \mathbf{v}_i \approx \frac{n_e \mathbf{v}_e - Z n_i \mathbf{v}_i}{n_e} \frac{e}{e} = -\frac{\mathbf{j}}{e n_e} \tag{4-23}$$

based on the definition

$$\mathbf{j} = e(Z n_i \mathbf{v}_i - n_e \mathbf{v}_e). \tag{4-24}$$

In order to achieve the equation of motion of the plasma, Eqs. (4–18) and (4–19) have to be added. The right-hand side results in

$$RHS = -\nabla p + (Z n_i - n_e) e \mathbf{E} + (Z n_i e \mathbf{v}_i - n_e e \mathbf{v}_e) \times \frac{\mathbf{H}}{c} \times \mathbf{K}_i + \mathbf{K}_e, \tag{4-25}$$

or, using (4–28) in

$$RHS = -\nabla p + \mathbf{E} e (Z n_i - n_e) + \mathbf{j} \times \frac{\mathbf{H}}{c} \times \mathbf{K}_i + \mathbf{K}_e, \tag{4-26}$$

where the viscosity terms cancelled.

Adding the left-hand sides of Eqs. (4–18) and (4–19) is more difficult. To obtain a recognizable result we have to add and subtract further terms to the first four terms stemming from Eqs. (4–18) and (4–19)

$$\begin{aligned}
LHS = &\, m_i n_i \frac{\partial}{\partial t} \mathbf{v}_i + m n_e \frac{\partial}{\partial t} \mathbf{v}_e + m_i n_i \mathbf{v}_i \cdot \nabla \mathbf{v}_i + m n_e \mathbf{v}_e \cdot \nabla \mathbf{v}_e \\
&+ m_i n_i \frac{Zm}{m_i} \mathbf{v}_e \cdot \nabla \mathbf{v}_i - m_i n_i \frac{Zm}{m_i} \mathbf{v}_e \cdot \nabla \mathbf{v}_i + m_i n_i \frac{Zm}{m_i} \mathbf{v}_e \cdot \nabla \mathbf{v}_e \\
&- m_i n_i \frac{Zm}{m_i} \mathbf{v}_e \cdot \nabla \mathbf{v}_e - m_i n_i \frac{Zm}{m_i} \mathbf{v}_i \cdot \nabla (\mathbf{v}_e - \mathbf{v}_i) + m_i n_i \frac{Zm}{m_i} \mathbf{v}_i \cdot \nabla \mathbf{v}_e \\
&- m_i n_i \frac{Zm}{m_i} \mathbf{v}_i \cdot \nabla \mathbf{v}_i + m_i n_i \left(\frac{Zm}{m_i}\right)^2 \mathbf{v}_e \cdot \nabla \mathbf{v}_e.
\end{aligned} \tag{4-27}$$

Here, the last term was added without any compensation by its negative value which is possible because it is negligibly small because of the square of the mass ratio. The realization that this and nothing else has to be added here is due to the ingenuity of Schlüter (1950), and produces the following reasonable result.

Sorting out all the terms of (4–27) and using the meaning of Eqs. (4–22) and (4–23), the left-hand side can be written as

$$LHS = m_i n_i \left[ \frac{\partial}{\partial t} \mathbf{v} + \mathbf{v} \cdot \nabla \mathbf{v} \right] + m_i n_i \frac{Zm}{m_i} (\mathbf{v}_e - \mathbf{v}_i) \cdot \nabla (\mathbf{v}_e - \mathbf{v}_i), \quad (4\text{-}28)$$

and the second term results from the terms No. 6, 7 and 9 in (4–27), and

$$mZn_i (\mathbf{v}_e - \mathbf{v}_i) \cdot \nabla (\mathbf{v}_e - \mathbf{v}_i) = \frac{m\mathbf{j}}{e} \cdot \nabla \frac{\mathbf{j}}{en_e} = \frac{1}{4\pi} \left( \frac{\omega_p}{\omega} \right)^2 \mathbf{E} \cdot \nabla \mathbf{E}. \quad (4\text{-}29)$$

Here, where the current densities $\mathbf{j}$ were expressed by the special case of fields oscillating with the frequency $\omega$, Eq. (3–68). We see that Schlüter's procedure for analyzing the left-hand side arrived immediately at the term which formally is identical with Kelvin's electrostatic ponderomotive force (1–1), remembering that $\mathbf{n}^2 - 1$ is equal to $-(\omega_p/\omega)^2$. Therefore, we arrive at the force density in the plasma of

$$f = m_i n_i \left[ \frac{\partial}{\partial t} \mathbf{v} + \mathbf{v} \cdot \nabla \mathbf{v} \right] = -\nabla p + \mathbf{E} e (Zn_i - n_e) + \mathbf{j} \times \frac{\mathbf{H}}{c} + \mathbf{K}_i + \mathbf{K}_e$$

$$- \frac{1}{4\pi} \left( \frac{\omega_p}{\omega} \right)^2 \mathbf{E} \cdot \nabla \mathbf{E}. \quad (4\text{-}30)$$

The essential problem remaining is how to handle the second term on the right-hand side of the equation. Schlüter had no option but to set this term to zero in agreement with his assumption (4–20) of space charge neutrality. The only reason not to do this now for laser plasma interaction at oblique incidence is the fact that we know from the WKB solution for p-polarization that there is a longitudinal electric field component appearing when the propagation vector of the wave is being bent in the stratified plasma (Hora 1969). We have therefore at least a high-frequency electric field to be taken into account. To do this at this point is justified by achieving the correct result proven by the momentum conservation of the interaction.

After accepting that the space charge term in Eq. (4–30) should not be neglected, the further question is, what formula for the space charge should be used for the inclusion of an effective high-frequency dielectric constant given by the refractive index according to the optical oscillation frequency $\omega$? Or should only be the vacuum dielectric constant be used as in Lorentz theory of metals (and plasmas)? The answer is that we have to use the dielectrically modified description, otherwise the momentum balance for oblique incidence cannot be achieved. Therefore, using

$$4\pi e(Zn_i - n_e) = \nabla \cdot \mathbf{n}^2 \mathbf{E} = \nabla \cdot \mathbf{D}, \qquad (4\text{-}31)$$

we can write the second and the last term on the right-hand side of Eq. (4–30) as

$$\frac{1}{4\pi}\mathbf{E}\nabla \cdot \mathbf{E} - \frac{1}{4\pi}(1-\mathbf{n}^2)\mathbf{E}\nabla \cdot \mathbf{E} - \frac{1}{4\pi}\mathbf{E} \cdot \nabla \mathbf{E}(1-\mathbf{n}^2). \qquad (4\text{-}32)$$

The last two terms can be joined by one differentiation to arrive at the general forced density in a plasma after deduction of the thermokinetic force at the formulation of the non-transient nonlinear force

$$f_{NL} = \frac{1}{c}\mathbf{j}\times\mathbf{H} + \frac{1}{4\pi}\mathbf{E}\nabla \cdot \mathbf{E} + \frac{1}{4\pi}\nabla \cdot \left(\mathbf{n}^2 - 1\right)\mathbf{E}\mathbf{E}. \qquad (4\text{-}33)$$

As mentioned before, that these and only these terms determine the non-transient nonlinear force is proven by the fact that when using the electromagnetic wave field for oblique incidence in Eq. (4–33), forces only perpendicular to the plasma surface are produced that agree with the condition of momentum conservation using WKB solutions for oblique incidence for collisionless plasma (Hora 1969).

The case for plasmas with collisions was treated later (Miller et al. 1979; Kentwell et al. 1980) and there was then a force density along the plasma surface which was the complete description of the usual radiation pressure in the plasma modified by dielectric swelling of the laser intensities.

Another reason that Eq. (4–33) is correct can be seen from the following algebraic transformation into the Maxwell stress tensor. This is the first derivation of the Maxwellian stress tensor from hydrodynamics whereas it was derived in the past only from elastomechanics. It may be suggested in this way that there is a formal link between eleastomechanics and hydromechanics. This is not trivial since it was outlined by Deng Ximing (1992) that a hydrodynamic description of optics is possible very close to the usual definition of optics but with a basic use of the photon flux.

For achieving the Maxwell stress tensor we consider the nonlinear force (4–33) in two parts

$$f_{NL} = \mathbf{A} + \mathbf{B}, \qquad (4\text{-}34)$$

where

$$\mathbf{A} = \frac{1}{c}\mathbf{j}\times\mathbf{H}; \qquad (4\text{-}35)$$

$$\mathbf{B} = \frac{1}{4\pi}\mathbf{E}\nabla \cdot \mathbf{E} + \frac{1}{4\pi}\nabla \cdot \left(\mathbf{n}^2 - 1\right)\mathbf{E}\mathbf{E}. \qquad (4\text{-}36)$$

The current density $\mathbf{j}$ can be eliminated by using the second Maxwell equation (3–72)

$$\mathbf{j} = \frac{c}{4\pi}(\nabla \times \mathbf{H}) - \frac{1}{4\pi}\frac{\partial}{\partial t}\mathbf{E}. \tag{4-37}$$

Inserting Eq. (4–37) into Eq. (4–35), we find

$$\begin{aligned}\mathbf{A} &= \frac{1}{4\pi}(\nabla \times \mathbf{H}) \times \mathbf{H} - \frac{1}{4\pi c}\left(\frac{\partial}{\partial t}\mathbf{E}\right) \times \mathbf{H} \\ &= -\frac{1}{4\pi}\mathbf{H} \times (\nabla \times \mathbf{H}) - \frac{1}{4\pi c}\left(\frac{\partial}{\partial t}\mathbf{E}\right) \times \mathbf{H}.\end{aligned} \tag{4-38}$$

Using the vector identity

$$\mathbf{H} \times (\nabla \times \mathbf{H}) = \frac{1}{2}\nabla \mathbf{H}^2 - \mathbf{H} \cdot \nabla \mathbf{H} \tag{4-39}$$

can be rewritten as

$$\mathbf{A} = -\frac{1}{4\pi}\left[\frac{1}{2}\nabla \mathbf{H}^2 - \mathbf{H} \cdot \overline{\nabla \mathbf{H}}\right] - \frac{1}{4\pi c}\left(\frac{\partial}{\partial t}\mathbf{E}\right) \times \mathbf{H}. \tag{4-40}$$

The expression $\mathbf{B}$ in Eq. (4–36) is

$$\mathbf{B} = \frac{1}{4\pi}\mathbf{E}\nabla \cdot \mathbf{E} + \frac{1}{4\pi}\nabla \cdot \mathbf{n}^2\mathbf{E}\mathbf{E} - \frac{1}{4\pi}\mathbf{E}\nabla \cdot \mathbf{E} - \frac{1}{4\pi}\mathbf{E} \cdot \nabla \mathbf{E} \tag{4-41}$$

which can be changed to

$$\mathbf{B} = \frac{1}{4\pi}\mathbf{n}^2\mathbf{E} \cdot \nabla \mathbf{E} + \frac{1}{4\pi}\mathbf{E}\nabla \cdot (\mathbf{n}^2\mathbf{E}) - \frac{1}{4\pi}\mathbf{E} \cdot \nabla \mathbf{E} \tag{4-42}$$

From Eq. (4–31),

$$\nabla \cdot (\mathbf{n}^2\mathbf{E}) = 4\pi\rho_e, \tag{4-43}$$

we note that the high-frequency field does not permit this oscillating charge density to be cancelled. This is how the transverse electromagnetic waves produce longitudinal (Langmuir) oscillations in a plasma even at frequencies which are not in tune with the plasma frequency. There is the experience from numerical studies that the electric field amplitude of the longitudinal waves is about one-tenth that of the driving transverse wave if other resonance mechanism [e.g. the Frösterling–Denisov resonance absorption or the perpendicular resonance (Hora et al. 1985; Goldsworthy et al. 1986)] do not increase the longitudinal wave amplitudes in a special way.

In Eq. (4–12) we can combine the first two terms for differentiation and add another term which is again subtracted in the following

$$\mathbf{B} = \frac{1}{4\pi}\left[\nabla\cdot\mathbf{n}^2\mathbf{EE} - \mathbf{E}\cdot\nabla\mathbf{E} + \frac{1}{2}\nabla\mathbf{E}^2 - \frac{1}{2}\nabla\mathbf{E}^2\right]. \tag{4-44}$$

The second and third term are combined to

$$\frac{1}{4\pi}\mathbf{E}\times(\nabla\times\mathbf{E}) = \frac{1}{8\pi}\nabla\mathbf{E}^2 - \frac{1}{4\pi}\mathbf{E}\cdot\nabla\mathbf{E}. \tag{4-45}$$

Using the first Maxwell equation (3–71) leads to

$$\mathbf{B} = \frac{1}{4\pi}\left[\nabla\cdot\mathbf{n}^2\mathbf{EE} - \frac{1}{2}\nabla\mathbf{E}^2\right] - \frac{1}{4\pi c}\mathbf{E}\times\frac{\partial}{\partial t}\mathbf{H}. \tag{4-46}$$

Combining the results of Eqs. (4–40) and (4–46) for Eq. (4–34) we arrive at

$$f_{\mathrm{NL}} = \frac{1}{4\pi}\nabla\cdot\left[\mathbf{EE} + \mathbf{H}\cdot\nabla\mathbf{H} - \frac{1}{2}\left(\mathbf{E}^2 + \mathbf{H}^2\right)\underline{\mathbf{1}} + (\mathbf{n}^2 - 1)\mathbf{EE}\right]$$
$$-\frac{1}{4\pi c}\frac{\partial}{\partial t}\mathbf{E}\times\mathbf{H}. \tag{4-47}$$

Using

$$\mathbf{H}\cdot\nabla\mathbf{H} = \nabla\cdot\mathbf{HH} - \mathbf{H}\nabla\cdot\mathbf{H}, \tag{4-48}$$

with a negligible last term because of Eq. (3–29) based on the fact that plasma has a permeability $\mu = 1$, and using the unity tensor **1** in Cartesian coordinates

$$\mathbf{1} = \mathbf{i}_x\mathbf{i}_x + \mathbf{i}_y\mathbf{i}_y + \mathbf{i}_z\mathbf{i}_z, \tag{4-49}$$

we arrive at the tensor formulation of the nonlinear force

$$f_{\mathrm{NL}} = \frac{1}{4\pi}\nabla\cdot\left[\mathbf{EE} + \mathbf{HH} - \frac{1}{2}(\mathbf{E}^2 + \mathbf{H}^2)\underline{\mathbf{1}} + (\mathbf{n}^2 - 1)\mathbf{EE}\right]$$
$$-\frac{1}{4\pi c}\frac{\partial}{\partial t}\mathbf{E}\times\mathbf{H}. \tag{4-50}$$

Using the Maxwell stress tensor

$$\mathbf{T} = \frac{1}{4\pi}\left[\mathbf{EE} + \mathbf{HH} - \frac{1}{2}(\mathbf{E}^2 + \mathbf{H}^2)\mathbf{1}\right] \tag{4-51}$$

with the components

$$4\pi \mathbf{T} = \begin{pmatrix} \frac{1}{2}(E_x^2 - E_y^2 - E_z^2 + H_x^2 - H_y^2 - H_z^2) & E_x E_y + H_x H_y & E_x E_z + H_x H_z \\ E_x E_y + H_x H_z & \frac{1}{2}(-E_x^2 + E_y^2 - E_z^2 - H_x^2 + H_y^2 - H_z^2) & E_y E_z + H_y H_z \\ E_x E_z + H_x H_z & E_y E_z + H_y H_z & \frac{1}{2}(-E_x^2 - E_y^2 + E_z^2 - H_x^2 - H_y^2 + H_z^2) \end{pmatrix} \quad (4\text{-}52)$$

we can write the nonlinear force using the Maxwell stress tensor as

$$f_{NL} = \nabla \cdot \left[ \mathbf{T} + \frac{\mathbf{n}^2 - 1}{4\pi} \overline{\mathbf{E}\mathbf{E}} \right] - \frac{1}{4\pi c} \frac{\partial}{\partial t} \mathbf{E} \times \mathbf{H}. \quad (4\text{-}53)$$

Formally this force density is identical with the result of Landau and Lifshitz (1966) derived from elastomechanics for a nondispersive and nondissipative fluid for low-frequency changes of the electromagnetic field. Here the expression (4–53)—derived from hydrodynamics—for the dispersive and dissipative plasma for high-frequency irradiation, however, is essentially different with respect to the presumed conditions. The factor of the **EE**-term needs to be interpreted as being proportional to the material density as in the case of Landau and Lifshitz (1966). The evaluation of the refractive index of plasmas, Eq. (3–78), fulfills just this relation.

From the stress tensor formulation (4–53) of the nonlinear force we can directly derive the force density for perpendicular incidence of infinitely spread plasma waves on a stratified plasma. We may use the propagation of the wave in the $x$-direction in which case all other derivatives than those by $x$ vanish and the linearly polarized laser field has the components $E_y$ and $H_z$ only. The nonlinear force from Eq. (4–53) using (4–52) is

$$f_{NL} = \frac{\mathbf{i}_x}{8\pi} \frac{\partial}{\partial x} \left( E_y^2 + H_z^2 \right). \quad (4\text{-}54)$$

This is then the exact derivation of the nonlinear force density for this case which previously, in Eq. (4–10), was derived more intuitively from the energy conservation law of electrodynamics (3–51).

Up to this point we have discussed the non-transient case only. The time dependence of the incident laser or microwave (or any other electromagnetic) radiation of sufficiently high intensity was considered to be stationary and unchanged in time. We discuss now the case that there is a switching-on or switching-off process of the radiation. This definitely is important if this switching is performed within 100 oscillations. Here, we again consider that time averaging over one oscillation cycle is performed. At the end of these lectures when considering the "pancake" of photons (Fig. 1–5) even shorter times of interaction will be considered and similar results as with the ponderomotive potential will be achieved.

The transient case was first discussed by Klima and Petrzilka (1972). Contrary to our case of (Hora 1969) the inhomogeneous plasma, a homogeneous plasma was used where a time dependent laser pulse was penetrating. The same nonlinear forces were derived as for our earlier case (Hora 1969) for inhomogeneous plasmas with time constant laser irradiation.

The work by Klima and Petrzilka (1972) covered the case of perpendicular incidence. Very complex problems occurred for the general case and it is the merit of Tskhakaya (1981), that he started to investigate these problems. An enormous controversy arose as to which the correct terms were for the transient case. Including Tskhakaya's (1981), six different elaborations were presented (Kono et al. 1981; Karpman et al. 1992; Stratham et al. 1983; Lee et al. 1983; Mulser et al. 1983) and there were—as Zeidler et al. (1985) were permitted to say even in such an exclusive journal as *Physics of Fluids*—six different schools and six different terms for the transient case. Even the work by Kentwell et al. (1984) was not included in these comparisons.

The most advanced result was the work by Zeidler et al. (1995) clarifying which terms are dominant. Their final transient formula covered most of their terms discussed earlier. The treatment, however, was still an approximate transient case for very slowly varying intensity. It turned out when we were going into more detail that a very small logarithmic term was still missing. Only with this term, the finally complete and general transient nonlinear force was derived (Hora 1985)

$$f_{NL} = \frac{1}{c}\mathbf{j}\times\mathbf{H} + \frac{1}{4\pi}\mathbf{E}\nabla\cdot\mathbf{E} + \frac{1}{4\pi}\left(1+\frac{1}{\omega}\frac{\partial}{\partial t}\right)\underline{\nabla}\cdot\mathbf{E}\mathbf{E}\left(\mathbf{n}^2-1\right), \quad (4\text{-}55)$$

or in the algebraically identical stress tensor formulation

$$f_{NL} = \frac{1}{4\pi}\nabla\cdot\left[\mathbf{EE}+\mathbf{HH}-\frac{1}{2}(\mathbf{E}^2+\mathbf{H}^2)\underline{\mathbf{1}}+\left(1+\frac{1}{\omega}\frac{\partial}{\partial t}\right)(\mathbf{n}^2-1)\mathbf{EE}\right]$$
$$-\frac{1}{4\pi c}\frac{\partial}{\partial t}\mathbf{E}\times\mathbf{H}. \tag{0.1}$$

We see that the difference in the non-transient case, Eqs. (4–33) and (4–50) is the time derivative in the bracket of such terms where the refractive index **n** appears. This is understandable: the switching on and off of the transverse electromagnetic radiation in the plasma corresponds simply to the transfer (or return) of electromagnetic field energy into the oscillating electrons.

Apart from this clear evidence of the completeness of the transient formulations, the final completeness was further proven by confirming that these and only these formulations (4–55) and (4–56) were Lorentz and gauge invariant. The completeness proof of these transient formulas could not be done as before in the non-transient case from the momentum balance, since there is a temporal change of the momenta involved. The proof of the completeness of the formulas (4–55) and (4–56) needed confirmation of Lorentz and gauge invariance (Rowlands 1990). Any term added or removed from the formulations (4–55) and (4–56) violates this invariance and would result in an incorrect formulation. Rowlands (1991, 1995) established that the general formulation of plasma theory in the invariant form results automatically in the fact that the magnetic susceptibility of the plasma has to be $\mu = 1$, in agreement with the specially derived result by H. Grad (1970) who said "Yes, Virginia, plasma is paramagnetic if you believe in Santa Claus."

## 4.6 GENERALIZED OHM'S LAW AND SOLITONS

The following steps show the complete derivation of the space charge quasi-neutral, see Eq. (4–20), two fluid equations with the generalized Ohm's law for plasmas. We generally point out the Schlüter (1950) derivation but add a minor extension with respect to solitons (Hora 1981, 1991). Subtracting the two Euler equations for the electron and ions after multiplying Eq. (4–18) by $Zm$ and Eq. (4–19) by $m_i$ the left-hand side is

$$LHS = m_i m \left[ Zm_i \left( \frac{d}{dt} \right)_i \mathbf{v}_i - n_e \left( \frac{d}{dt} \right)_e \mathbf{v}_e \right] \approx m_i m Z n_i \frac{d}{dt}(\mathbf{v}_i - \mathbf{v}_e), \quad (4\text{-}57)$$

where Eq. (4–20) was used. This implies also approximate equality of $(d/dt)_{e,i} = \partial/\partial t + \mathbf{v}_{e,i} \cdot \nabla$ and $d/dt = \partial/\partial t + \mathbf{v} \cdot \nabla$ using $\mathbf{v}_e \approx \mathbf{v}_i \approx \mathbf{v}$. The necessary neglect of coherent quiver motion in $\mathbf{v}_e$ does not affect the following conclusions. Another limiting case is where the spatial derivations are less compared with the $\partial/\partial t$ terms and thus no restriction is given to the amplitudes [if Eq. (4–20) is not violated]. Using Eq. (4-23), the left-hand side of the result of subtraction is then

$$LHS = \frac{m_i m}{e} \left[ \frac{d}{dt} \mathbf{j} - \frac{\mathbf{J}}{n_e} \frac{d}{dt} n_e \right]. \quad (4\text{-}58)$$

The second term in the bracket is usually neglected in order to achieve the well-known result (Schlüter 1950) that follows in Eq. (4–61).

The right-hand side from the subtraction of the Euler equations after adding and subtracting one more term is

$$\begin{aligned} RHS = &-m_e \nabla \frac{3}{2} n_i K T_i + m_i \nabla \frac{3}{2} n_e K T_e + \mathbf{E} e \left( m n_i Z^2 + n_e m_i \right) \\ &+ \left( Z^2 m n_i e \mathbf{v}_i + m_i n_e e \mathbf{v}_e \right) \times \frac{\mathbf{H}}{c} \\ &+ \left[ Z m^2 n_e \nu + m_i m n_e \nu \right] (\mathbf{v}_e - \mathbf{v}_i) \end{aligned} \quad (4\text{-}59)$$

Dividing both sides of $m_i m$ and neglecting $m n_i Z^2$ as compared to $n_e m_i$ results in

$$\frac{m}{e} \left[ \frac{d}{dt} \mathbf{j} + \nu \mathbf{j} \right] = \nabla n_e K T_e + \mathbf{E} e n_e - \mathbf{j} \times \frac{\mathbf{H}}{c} + Z n_i e \mathbf{v}_i \times \frac{\mathbf{H}}{c}. \quad (4\text{-}60)$$

Dividing by $en_e$, using $v_i \approx \mathbf{v}$, and using the pressure $p_e = p/(1 + 1 = 1/Z)$ at thermal equilibrium, we arrive at the generalized Ohm's law (diffusion equation)

$$\frac{4\pi}{\omega_p^2}\left[\frac{d}{dt}\mathbf{j} + \nu\mathbf{j}\right] = \mathbf{E} - \frac{1}{en_e c}\mathbf{j}\times\mathbf{H} + \mathbf{v}\times\frac{\mathbf{H}}{c}$$
$$+ \frac{1}{en_e}\nabla\frac{p}{1+1/Z}$$
(4-61)

in agreement with Eq. (3–69) (Schlüter 1950; Lüst 1959) which was stated earlier but now has been completely derived.

If we had not neglected the second term in the bracket of Eq. (4–58) the general Ohm's law is then

$$\frac{4\pi}{\omega^2}\left[\frac{d}{dt}\mathbf{j} + \mathbf{j}\left(\nu - \frac{1}{n_e}\frac{d}{dt}n_e\right)\right] = \mathbf{E} - \frac{1}{en_e c}\mathbf{j}\times\mathbf{H} + \mathbf{v}\times\frac{\mathbf{H}}{c} + \frac{1}{en_e}\nabla p_e. \quad (4\text{-}62)$$

This shows that an additional damping mechanism appeared, given by an effective collision frequency

$$\nu_{\text{eff}} = \nu + \nu_{\text{add}}; \quad \nu_{\text{add}} = -\frac{1}{n_e}\frac{d}{dt}n_e = -\frac{1}{n_e}\left[\frac{\partial}{\partial t}n_e + \mathbf{v}_e\cdot\nabla n_e\right]. \quad (4\text{-}63)$$

As we see, this expression can fully compensate the usual collision frequency $\nu$ (2–12) or (2–13)—or its quantum modification (2–13b)—arriving at a collisionless plasma or a conductor without resistivity or negative resistivity as in the Esaki-diodes. This property has been discussed (Hora 1995) in connection with high-temperature superconductors (Bednorz and Müller 1987).

The damping by the effective collision frequency $\nu_{\text{eff}}$ given in Eq. (4–63) (Hora 1981) has obviously not been recognized before. This can find a special interpretation for the case where one has a one-dimensional variation of $n_e$ only (using the $x$-direction) and especially

$$\frac{\partial}{\partial t}n_e + \mathbf{v}_e\frac{\partial}{\partial x}n_e = -\mu\frac{\partial^3}{\partial x^3}n_e. \quad (4\text{-}64)$$

The $x$-dependence of the electron density follows the special profile of the right-hand side. The damping is then due to Langmuir solitons where Eq. (4–64) is the Korteweg–de-Vries equation for the electron density $n_e$ (Langmuir waves). With respect to the necessary plasma inhomogeneity, we need to consider this case as Langmuir pseudo-waves, as has been clarified in all details as well as numerically (Eliezer et al. 1989). The dispersion relation $\mu$ is not necessarily the usual plasma dispersion but can be a more complex solution as we shall show with the following numerical example for numerically derived dynamic solitons (with $\mathbf{v}$ instead of the case with $n_e$ for the Langmuir solitons).

The additional collision frequency, now expressed with inclusion of the Langmuir solitons of Eq. (4–64)

$$\nu_{add} = -\frac{1}{n_e}\frac{d}{dt}n_e = \mu\frac{\partial^3}{\partial x^3}n_e \qquad (4\text{-}65)$$

indicates a dissipation (absorption) of transverse electromagnetic waves by conversion into longitudinal (Langmuir) waves. This is most important since this is a purely macroscopic-plasma hydrodynamic process. This is in complete contrast to the usual transfer of transversal electromagnetic waves into longitudinal waves. This can be described by the kinetic theory only as the stimulated Raman instability. Our relation (4–65) is therefore a macroscopic mechanism for damping of electromagnetic waves in a plasma without collisions and without Landau damping. We underline that Landau damping is basically a microscopic effect with an electron distribution different from the Maxwell distribution.

We should mention that there are ion acoustic solitons possible, if instead of $n_e$ in Eq. (4–64) the ion density $n_i$ is used (Bharatram and Shukla 1986).

When the plasma velocity $\mathbf{v}$ appears instead of $n_e$ in the Korteweg–de-Vries equation (4–64), we have the dynamic solitons. These correspond exactly to the nonlinear force in cases where the thermokinetic force is small and can be neglected at sufficiently high laser intensities interacting with the plasma (Fig. 4–2). In this case for perpendicular incidence of laser radiation we find from Eqs. (4–10) and (4–54)

$$-\mu'\frac{\partial^3}{\partial x^3}\mathbf{v} = \frac{1}{8\pi m_i n_i}\frac{\partial}{\partial x}\left(\mathbf{E}^2 + \mathbf{H}^2\right) = \frac{|f_{NL}|}{m_i n_i}. \qquad (4\text{-}66)$$

It happens that a relation for the dispersion function $\mu'$ can be derived numerically for the kind of computations described in Figs. 4–12 to 4–17. At the beginning of the laser interaction with arbitrary initial values (apart from selecting the low-reflectivity bi-Rayleigh case for the initial electron density), there is indeed not any correlation with the soliton properties as could be expected. After the laser has acted and produced the blocks of plasma moving mostly due to the nonlinear force, we have seen a remarkable agreement with soliton properties from the numerical results.

The numerical output is now described in more detail. We use the case of Fig. 4–14 and 4–15 for the neodymium glass laser intensity of $10^{16}$ W/cm$^2$. Figures 4–18 to 4–20 show the electromagnetic energy density profiles from the laser light in the deuterium plasma, the plasma velocities and the electron density, respectively, for the time 1.5 ps (straight curves) and 2.5 ps (dashed curves). At 2.5 ps the conditions of the solitons have just been reached as can be seen in Fig. 4–21 from the evaluation of Figs. 4–18 to 4–20 for the quantities interesting to the Korteweg–de-Vries equation. The evaluation of $(\mathbf{E}^2 + \mathbf{H}^2)/(8\pi)$ and of $\partial^3 \mathbf{v}/\partial x^3$ shows poles at identical depths $x$. The zero points and the location of the poles are very accurately given while the curves themselves contain some inaccuracies of the numerical output. It is remarkable how the poles and zeros of the nonlinear force correspond to the extrem and zeros of the electron density. From this we can derive the dispersion function

$$\mu' = \frac{\partial/\partial x(1-n_e/n_{ec})}{1-n_e/n_{ec}}.$$  (4-67)

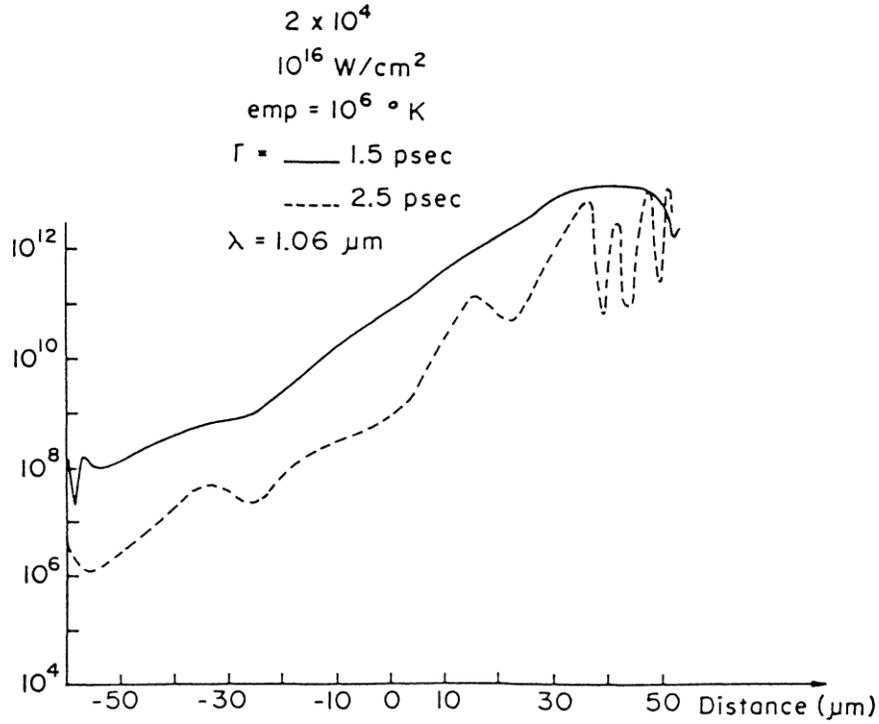

**Figure 4–18.** Electromagnetic energy density $(\mathbf{E}^2 + \mathbf{H}^2)/8\pi$ for the same conditions of a bi-Rayleigh profile with $\alpha = 2 \times 10^4$ cm$^{-1}$ as in Fig. 4–14 for the times 1.5 and 2.5 ps.

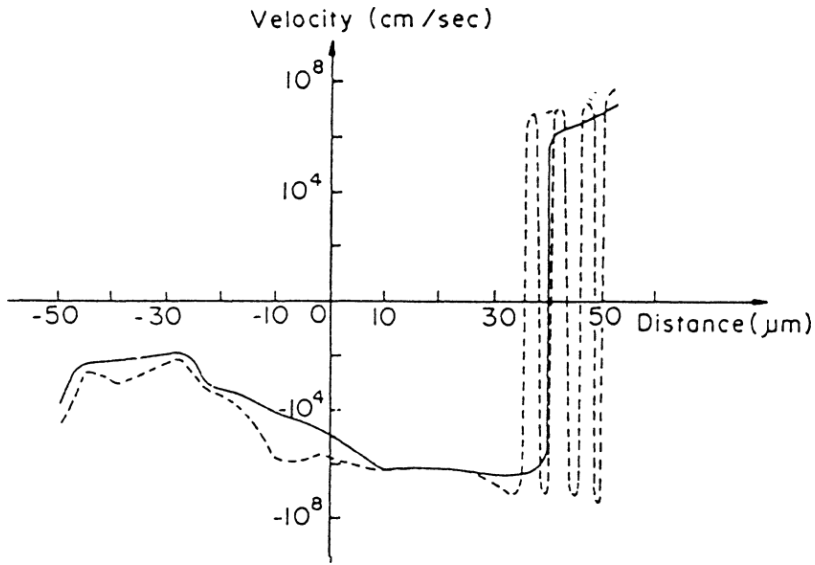

**Figure 4–19.** Velocity profiles for the case of Fig. 4–18.

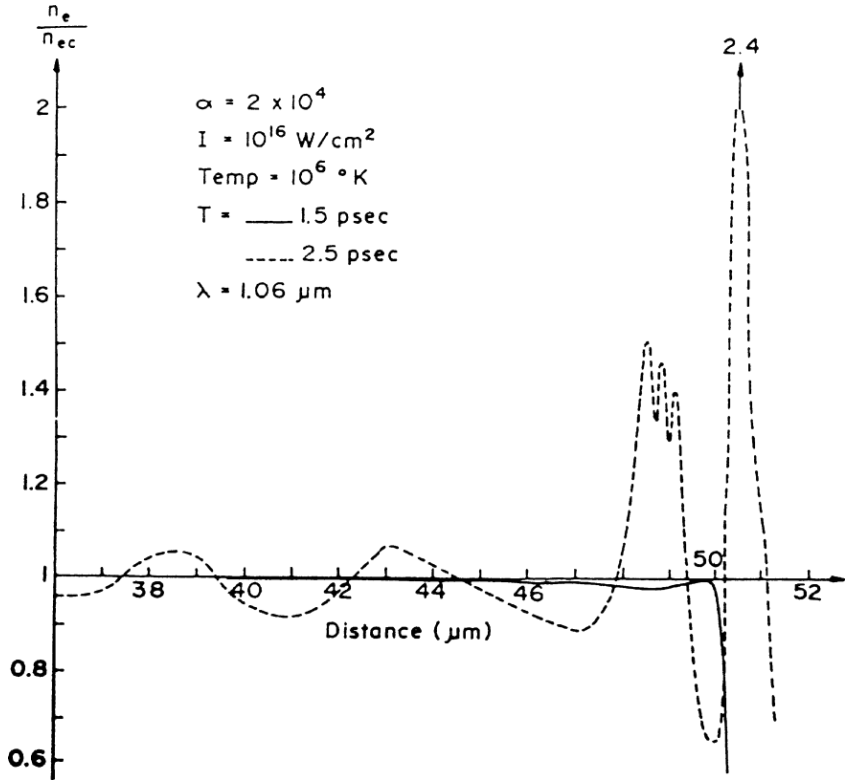

**Figure 4–20.** Electron density profiles for the cases of Fig. 4–18 showing the rippling at the time 2.5 ps.

With this we arrive (Hora 1981, 1991) at the complete identity of the nonlinear

force

$$\frac{\partial}{\partial t}v + v\frac{\partial}{\partial x}v = \frac{1}{8\pi n_i m_i}\frac{\partial}{\partial x}\left(\mathbf{E}^2 + \mathbf{H}^2\right) \tag{4-68}$$

with the Korteweg–de-Vries equation

$$\frac{\partial}{\partial t}v + v\frac{\partial}{\partial x}v = \frac{\partial \ln\varepsilon'}{\partial x}\frac{\partial^3}{\partial x^3}v, \tag{4-69}$$

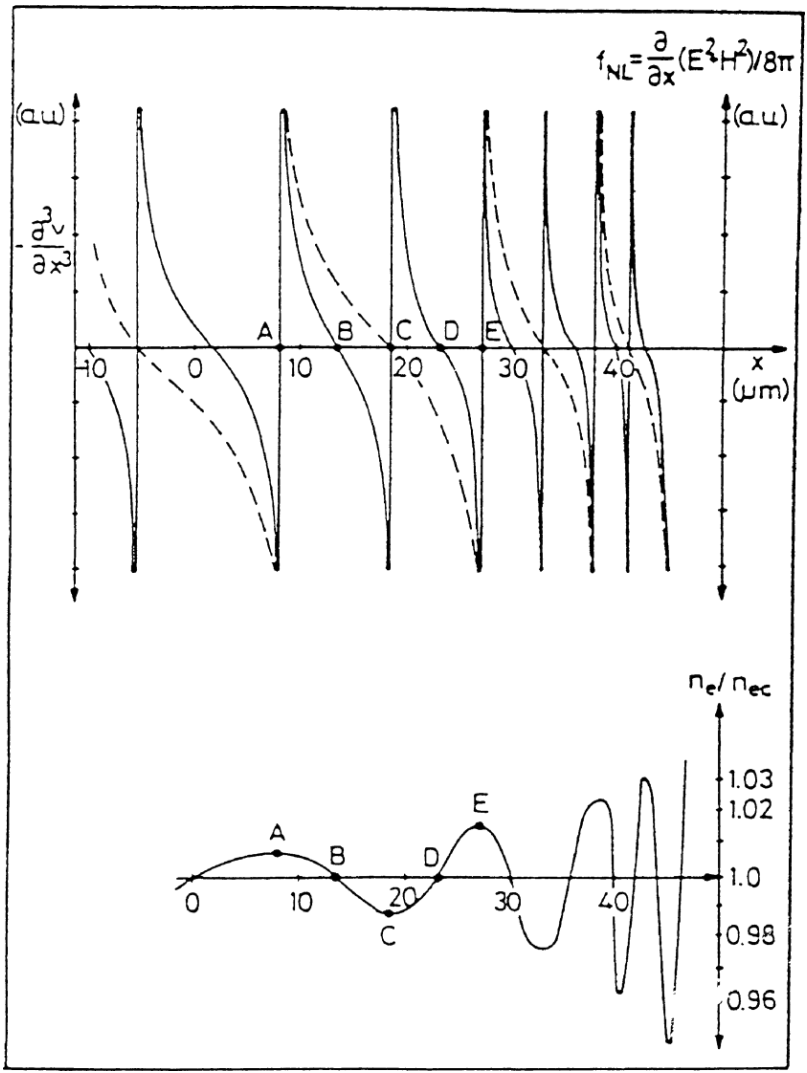

**Figure 4–21.** Following the requirements of the Korteweg–de-Vries equation, Eq. (4–68), the numerical results of Figs. 4–18 to 4–20 are used. $\partial^3 v/\partial x^3$ (———) and $\mathbf{f}_{NL}$ (– – –) are evaluated.

where the high-frequency (complex) dielectric constant $\varepsilon^J = \mathbf{n}^2$ (Eq. 3–78) is given by

$$\varepsilon = \varepsilon' + i\varepsilon''; \quad \varepsilon' = 1 - \omega_p^2/\omega^2 = 1 - n_e/n_{ec}. \tag{4-70}$$

Since the plasmas at the very high laser intensities in the case of Figs. 4–18 to 4–20 are practically collisionless we can compare the result with the spatial dispersion $\partial \ln \varepsilon / \partial x$ function which is well known from the theory of Försterling–Denisov resonance absorption (Hora 1991: Chapter 11.2) in connection with density rippling (Castillo et al. 1977; Peratt et al. 1977; Bocher et al. 1977; Blazenkov et al. 1980; Sauer et al. 1979).

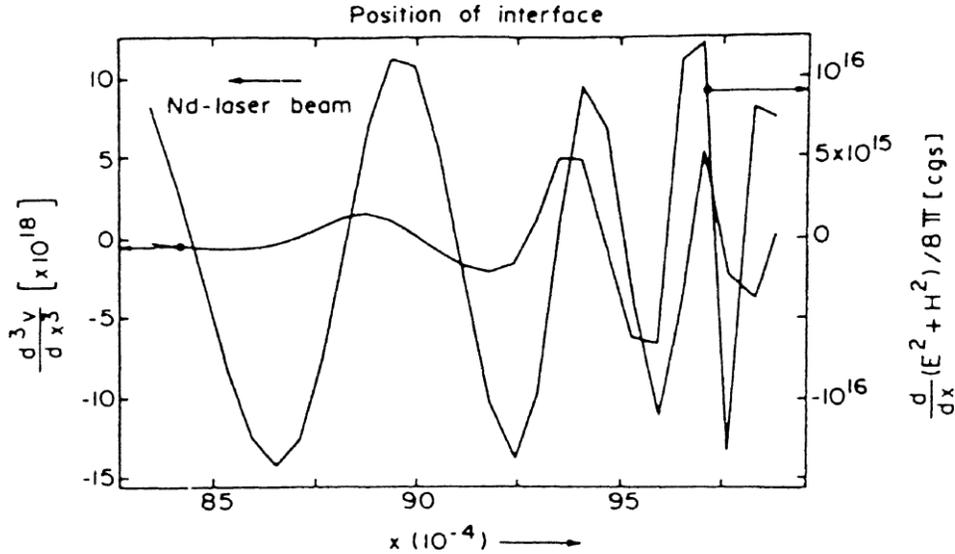

**Figure 4–22.** Evaluation of a numerical output similar to the case of Fig. 4–21 but for a weaker bi-Rayleigh profile with $\alpha = 103$ cm−1 . The deeper part of the plasma follows the Benjamin–Ono equation.

While the initial soliton work for plasmas (Kruskal and Zabuski 1970) used completely collisionless plasma and had then only oscillations of the Korteweg– de-Vries states, our computations with the realistic inclusion of (even weak) collisional dissipation shows how the nonlinear force converts an initially non-soliton state into that of an ideal soliton. Kruskal (1991), especially, acknowledged this first case (Hora 1979) for plasmas demonstrating the dissipation process and the building up of the solitons.

It should be mentioned that generation of solitons by the nonlinear force is not exclusively related to the Korteweg–de-Vries equation. Considering the Benjamin–Ono equation

$$\frac{\partial}{\partial t}v + v\frac{\partial}{\partial x}v = -H\frac{\partial^2}{\partial x^2}v = \frac{\partial}{\partial x}\frac{\mathbf{E}^2+\mathbf{H}^2}{8\pi} \tag{4-71}$$

with the Hilbert transform $H$, it was noted that another case of computation of the laser–plasma interaction results in numerical agreement with this equation (4–71) (Hora 1991: Fig. 10.23) in the plasma interior, while the periphery of the plasma corona shows agreement with the Korteweg–de-Vries equation.

# CHAPTER 5
# Hydrodynamic Plasma Properties with the Nonlinear Force

## 5.1 MOMENTUM TRANSFER

Based on the preceding results of the nonlinear force we now discuss the momentum transfer to the plasma (at perpendicular incidence for simplification). We shall see that plasma moving from vacuum into plasma increase their momentum. This can be seen from the reflection conditions at a discontinuous interface between plasma and a homogeneous plasma, resulting in the value given by the well-known Fresnel formulas.

A laser pulse between the times $t_1$ and $t_2$ of a cross-section $K$ moving into the $x$-direction has a *total energy* [from the definition of the energy density, Eq. (3–51) in vacuum]

$$\varepsilon_L = c \int_K dydz \int_{t_1}^{t_2} dt \frac{E_v^2(y,z,t)}{8\pi} , \qquad (5\text{-}1)$$

and a momentum, expressed by the number $N$ of all photons of energy $h\nu$ with Planck's constant $h$

$$P_0 = \varepsilon_L / c = Nh\nu / c . \qquad (5\text{-}2)$$

When the laser wave, having no reflection as in the WKB approximation, enters the inhomogeneous plasma as shown in Fig. 5–1, then the following momentum is transferred to the inhomogeneous plasma between a depth $x_1$ (in vacuum) and any depth $x_2$ in the plasma

$$P_{\text{inh}} = \int_K dydz \int_{x_1}^{x_2} dx \int_{t_1}^{t_2} f_{\text{NL}} dt . \qquad (5\text{-}3)$$

Using the nonlinear force $\mathbf{f}_{\text{NL}}$ from Eq. (4–10) (Hora 1969), the result for a plasma with collisions is given by using the absolute value of the refractive index and the electrical field amplitude $E_v$ of the laser

$$\begin{aligned}
P_{\text{inh}} &= -\int_K dydz \int_{t_1}^{t_2} dt \int_{x_1(t)}^{x_2(t)} dx \left[ \frac{\partial}{\partial x} \frac{E_v^2(y,z,t)}{16\pi} \left( \frac{1}{|\mathbf{n}|} - |\mathbf{n}| \right) \right] \\
&= -\frac{P_0}{2|\mathbf{n}_2|} \left( 1 - |\mathbf{n}_2| \right)^2 .
\end{aligned} \qquad (5\text{-}4)$$

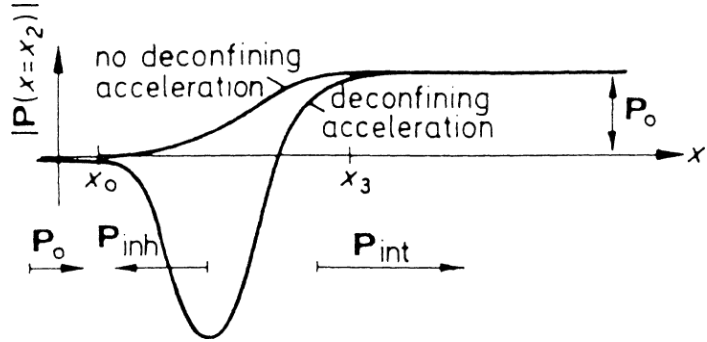

**Figure 5–1.** Momentum $P$ transferred to the plasma between the vacuum and the interior of the inhomogeneous plasma. At sufficient swelling, the momentum can be negative $P_{inh}$, showing an ablation of the corona up to the minimum value of the refractive index.

In Eq. (5–4) the $\mathbf{n}_2$ is the refractive index at the depth $x_2$ and the refractive index in vacuum (at $x_1$) is 1.

This plasma momentum is directed against the laser light [see the negative sign in Eq. (5–4)] reaching a minimum (highest absolute value) near the critical electron density $n_{ec}$ (see Fig. 5–1), causing the ablation of the plasma corona. The momentum transferred to the plasma interior, $P_{int}$, is then the value of the recoil of the ablation plus the photon momentum in the vacuum because the difference always has to be the photon momentum in the vacuum

$$P_0 = P_{int} - P_{inh} . \qquad (5\text{-}5)$$

The momentum of the photons in the plasma interior is, therefore,

$$P_{int} = P_0 + P_{inh} = \frac{P_0}{2}\left(\frac{1}{|\mathbf{n}|} + |\mathbf{n}|\right) \qquad (5\text{-}6)$$

as first realized by Hora (1969, Eq. 68a). Each photon in the plasma has the momentum

$$p\phi_{,pl} = \frac{\hbar\omega}{2|\mathbf{n}|c}\left(1 + |\mathbf{n}|^2\right), \qquad (5\text{-}7)$$

and the interpretation of the nonlinear force acting on the plasma ablation region in the case without reflection (WKB) is that the plasma has to be blown off in order to balance the growing momentum of the photons entering the plasma. What happens to the plasma below $n_{ec}$ when the photons are absorbed by the plasma, is that the radiation pressure increases due to the (dielectric) increase of the photon momentum (Peratt 1979; 1988; 1989).

The best proof of this result is given by comparing the case of a uniform (nearly collisionless) plasma with a real refractive index **n** with Fresnel's formulas. In order to squeeze a photon with this increased momentum into the bulk plasma, one needs a reflection $R$ of photons for momentum compensation in the vacuum where the light hits the interface. Remembering the photon momentum in vacuum

$$p_\phi \frac{\hbar \omega}{c}, \tag{5-8}$$

we have for momentum balance the relation

$$p_\phi + \tilde{R} p_\phi = \tilde{T} p_{\phi,\text{pe}} \tag{5-9}$$

where $\tilde{T}$ is the optical transmission coefficient given as

$$1 - \tilde{R} = \tilde{T} \tag{5-10}$$

Using our values for the photon momenta as determined by the nonlinear forces (4–7) and (4–8), we solve equations (5–9) and (5–10) for the two unknowns $R$ and $T$ and find using

$$\tilde{R} = \left(\frac{1-\mathbf{n}}{1+\mathbf{n}}\right)^2 ; \quad \tilde{T} = \frac{4\mathbf{n}}{(1+\mathbf{n})^2}, \tag{5-11}$$

the Fresnel formulas.

The nonlinear force ablation of the plasma corona (Figs. 4–1 and 5–1) is, therefore, just the necessary recoil to the inhomogeneous plasma when the photons are invading the plasma without reflection and increasing their momentum. A similar reflection of optical waves on a block of electrons in connection with Thomson scattering (Wu 2010) should be mentioned.

Comparing Eqs. (1–9) and (1–10) we see from Eq. (5–4), that the ponderomotive potential $\varphi$ drives the plasma corona as a nonlinear force towards the vacuum. Then we may conclude that the increase of the photon momentum when moving into an inhomogeneous plasma with increasing electron density is caused by the ponderomotive potential, by which the ablated plasma falls down like dielectric material by ponderomotion in electrostatics.

We come now to the important question of how the charge number $Z$ determines the ion energy of the plasma when blown off as nonlinear force ablated corona. The force driving the plasma corona when falling within the ponderomotive potential is

$$m_i n_i \frac{d}{dt} \mathbf{v}_i = f_{NL} = -\mathbf{i}_x \frac{E_v^2}{16\pi} \frac{n_e}{n_{ec}} \frac{\partial}{\partial x} \frac{1}{|\mathbf{n}|}. \tag{5-12}$$

The velocity $v_0$ gained by the acceleration $dv_i/dt$ arrives, as in the case of free fall

$$v_0 = \left(2 \frac{dv_i}{dt} x\right)^{1/2} \tag{5-13}$$

at a differential gain of

$$\delta v_0^2 = 2 \frac{dv_i}{dt} \delta x. \tag{5-14}$$

Substituting $n_e = Z n_i$ in Eq. (5–12) we find then for the gained kinetic energy in differential form

$$d\left(\frac{1}{2} m_i v_i^2\right) = m_i \frac{dv_i}{dt} dx = \left| \frac{E_v^2}{16\pi} \frac{Z}{n_{ec}} \frac{\partial}{\partial x} \frac{\exp(-\bar{k}x/2)}{|\mathbf{n}|} \right| dx. \tag{5-16}$$

The exact integral where the weak average absorption expressed by $\bar{k}$, Eq. (3–109), has been taken into account and is given by

$$\frac{m_i}{2} v_i^2 = \frac{E_v^2}{16\pi} \frac{Z}{n_{ec}} \int_{x_1}^{x_2} \frac{\partial}{\partial x} \left[ \frac{\exp(-\bar{k}x/2)}{|\mathbf{n}|} \right] dx$$

$$= \frac{E_v^2}{16\pi} \frac{Z}{n_{ec}} \left[ \frac{\exp(-\bar{k}x_2/2)}{|\mathbf{n}(x_2)|} - \frac{\exp(-\bar{k}x_1/2)}{|\mathbf{n}(x_1)|} \right]. \tag{5-16}$$

The ions of the plasma that fall within the plasma bulk driven by the ponderomotive potential then gain an energy of translation

$$\varepsilon_i^{trans} = \frac{E_v^2}{16\pi} \frac{Z}{n_{ec}} \left( \frac{1}{|\mathbf{n}|_{\min}} - 1 \right), \tag{5-17}$$

exactly given by $Z$ times the difference of the ponderomotive potential. If the dielectric swelling is very high (and the minimum absolute value of the refractive index is very low) the ion energy can be expressed by

$$\varepsilon_i^{compr} = \frac{E_v^2}{16\pi} \frac{Z}{n_{ec}} \frac{1}{|\mathbf{n}|_{\min}}. \tag{5-18}$$

This result confirms the observation that the nonlinear force interaction of lasers with plasmas when generating keV and more energetic ions results in a linear dependence of the ion energy on their charge $Z$.

The other question, however, is how laser produced plasma separates into a number of different plasmas, each with ions of one sort of $Z$ and with a spe- cial velocity of ion energy according to Eq. (5–18). It is evident that each of these plasma clouds, when first united after the laser produces the plasma, behaves separately as space charge neutral material (Fig. 1–1), being individually moved by the ponderomotive potential.

This tendency of the $Z$-separated plasmas to fall out from the ponderomotive potential as observed (Fig. 5–2), is dependent on the high plasma density, i.e., there must be much smaller Debye length than the size of these plasma "pieces of matter." We shall see later (Boreham et al. 1979) with large Debye lengths that only the electrons fall down the ponderomotive potential while the ions are practically untouched.

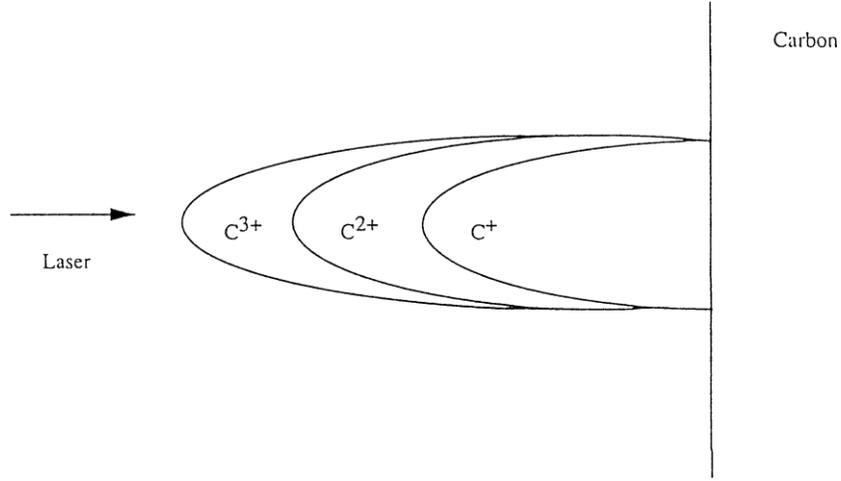

**Figure 5–2.** Separation of (space charge neutral) plasma with a different ion charge $Z$ due to the different ponderomotive force within the same ponderomotive potential structure (Boland et al. 1968).

The result (5–7) of the increased photon momentum in the plasma (Hora 1969) leads to a remarkable interpretation which was first expressed by Klima and Petrzilka (1972). With respect to the still unsettled Abraham–Minkowski controversy about the correct relativistic description of the electromagnetic energy in a dielectric medium, we find that the Abraham photon momentum

$$PA = \frac{h\omega}{\mathbf{n}c} \tag{5-19}$$

and the Minkowski photon momentum

$$PM = \frac{h\omega}{c}\mathbf{n} \tag{5-20}$$

each share 50 percent in Eq. (5–7)

$$P_{\phi,pl} = \frac{PA + PM}{2} \tag{5-21}$$

Since the results of the nonlinear force are fully proven experimentally in agreement with the theoretical values including Eq. (5–7) we can firmly say that for plasmas the Abraham–Minkowski controversy is solved: the photon momentum is half Abraham and half Minkowski.

The result that photons in plasmas have a higher momentum than in vacuum and the fact that photons leave plasmas like the sun or inertial confined plasmas with nearly no (Fresnel-like) reflection, means that there always has to be a mechanical recoil to the plasma corona. This is the case even when light emerges from higher plasma densities in the opposite direction of that of laser irradiation. This explains immediately the origin of the solar wind. Apart from one factor one must introduce "by hand," either the velocity or the ion density serves for an immediate explanation of the properties and the generation of the solar wind. At very strong radiation emission, the blowing off of the corona by the nonlinear force results in a self-confining force for the reacting plasma, which works against inertial expansion with adiabatic cooling, and whose magnitude is of such value that this self-confinement may increase the fusion reaction gain for such a spherical geometry. This aspect is of interest for energy generation by inertial confinement fusion.

## 5.2 Ponderomotion: the Electric Analogy to Alfvén Waves

We consider now the generation of magnetohydrodynamic waves, or Alfvén waves, produced by motion of plasma with a velocity $v_0$ ($y$-direction) perpendicular to a static magnetic field $H_0$ ($x$-direction). The Lorentz force will then cause a current density $j_z$ ($z$-direction) which results in an acceleration of the motion by

$$n_i m_i \frac{\partial v_y}{\partial t} = \frac{1}{c} j_z H_0. \tag{5-22}$$

Further differentiation by times gives

$$\frac{\partial j_z}{\partial t} = C \frac{n_i m_i}{H_0} \frac{\partial^2 v_y}{\partial t^2}. \tag{5-23}$$

In the diffusion equation (4–61), Ohm's law, we neglect nonlinear terms and fast oscillations of j ($\partial j / \partial t = 0$) and collisions ($v = 0$)

$$\frac{m}{e^2 n_e}\left(\frac{\partial j}{\partial t} + vj\right) = \mathbf{E} + \frac{1}{c}\mathbf{v} \times \mathbf{H} + ...;$$

$$0 = E_z - \frac{1}{c} v_y H_0. \tag{5-24}$$

The electric field $E_z$ is that generated by the current $j_z$ due to the motion of $v$ across $\mathbf{H}_0$. Differentiating Eq. (5–24) twice by time results in

$$\frac{\partial^2}{\partial t^2} v_y = \frac{c}{H_0} \frac{\partial^2}{\partial t^2} E_z. \tag{5-25}$$

Any fast motion of **E** will follow a wave equation (from the Maxwell's equations)

$$\nabla^2 \mathbf{E} = \frac{1}{c^2} \frac{\partial^2}{\partial t^2} \mathbf{E} + \frac{4\pi}{c^2} \frac{\partial \mathbf{j}}{\partial t}, \tag{5-26}$$

or from Eqs. (5–23) and (5–25)

$$\nabla^2 \mathbf{E} = \frac{1}{c^2} \frac{\partial^2}{\partial t^2} \mathbf{E} \left( 1 + 4\pi \frac{n_i m_i c^2}{H_0^2} \right). \tag{5-27}$$

This is a wave equation with a wave velocity

$$v_A = \frac{c}{\left(1 + 4\pi n_i m_i c^2 / H_0^2\right)^{1/2}} \tag{5-28}$$

called the Alfvén velocity. If $4\pi n_i m_i c^2 / H_0^2 \gg 1$, we find

$$v_A = \frac{H_0}{\left(4\pi n_i m_i\right)^{1/2}} = \frac{H_0}{\sqrt{4\pi \rho}}. \tag{5-29}$$

The Alfvén waves are of importance in plasmas with static magnetic fields. However, the high-frequency fields, and the velocity an ion gains by the nonlinear force, are similar. Following Eq. (5–17) for high swelling ($1/|\mathbf{n}| \gg 1$), the energy gained by the ion after being accelerated along the inhomogeneous plasma surface is

$$\frac{m_i}{2} v_i^2 = \frac{\overline{\mathbf{E}^2}}{8\pi} \frac{Z}{n_{ec}}. \tag{5-30}$$

Using $n_e = Z n_i$ we arrive at

$$v_i = \frac{|\mathbf{E}|}{\sqrt{4\pi n_i m_i}}. \tag{5-31}$$

Here, **E** and $n_i$ have to be taken from an area close to the cutoff density. The result (5–31) is similar to the Alfvén velocity, if **E** is considered instead of $\mathbf{H}_0$ for densities below cutoff, $|\mathbf{E}| \approx |\mathbf{H}|$ of the laser field. The direction of $v_i$ is that of $v_0$ in the initial derivation of the Alfvén wave. An interpretation of the connection of the nonlinear-force acceleration by the HF laser field with the Alfvén velocity is possible by considering the stepping through of the plasma along the HF wave maxima. The velocity $v_i$ given by the nonlinear force to the ions is the electric analogy to the Alfvén velocity.

## 5.3 PONDEROMOTIVE AND RELATIVISTIC SELF-FOCUSING

The shrinking of a laser beam in a dielectric medium (self-focusing) is due to the fact that the dielectric constant $\varepsilon$ in the relation (3–9a) has a nonlinear change (Chiao, Garmire and Townes 1963) for very high electric fields **E** [See Eq. (3–9a)]. This was observed in electrostatics for very high field strengths which were a little below that which caused electrical discharge (breakdown) through the material. The second-order constants measured in the last century have the same values as in the high-frequency fields with lasers.

In fully ionized plasmas, this dielectric effect cannot be used. But there is a mechanism for self-focusing based on the nonlinear force which was called "ponderomotive self-focusing" (Hora 1969a). Earlier, Askaryan (1962) discussed how the force due to the (radial) gradient of the electromagnetic energy density of a laser beam in a plasma [the nonlinear force, Eq. (4–10)] may be compensated by radial gas dynamic pressure. Using the expression (4–53), equilibrium between both forces appears if the laser beam drives plasma out from the beam axis by the nonlinear force that a density gradient of a pressure produces a compensating thermokinetic pressure [Eq. (4–1)]:

$$f_{th} = f_{NL}; \quad \nabla \cdot \left( \mathbf{T} - \frac{\mathbf{n}^2 - 1}{4} \mathbf{EE} \right) = \nabla n_e KT (1 + 1/Z) \frac{3}{2}. \tag{5-32}$$

Askaryan (1962) could do no more with this relation than discuss some general properties. The addition of the second physical relation describing the bending of the laser beam in the plasma by the optical constants arriving at total reflection, and of the third relation that these totally reflected beams had to be within an angle such that the diffraction condition for the first minima with respect to the beam diameter had to be fulfilled (Fig. 5–3), led to the evaluation of the laser power threshold $P$ for which the ponderomotive self-focusing occurs (Hora 1969a)

$$P \geq \frac{(1.22 \pi c)^2 n^3 m_e}{e^2 \left[ 2/\exp(1) \right]^{1/2} c_1^2 \left( 1 + \mathbf{n}^2 \right)}, \tag{5-33}$$

where the constant $c_1 = 1.63 \times 10^{-5}$ cgs units. With the plasma temperature in eV the threshold of power $P$ in watts is

$$P \geq \begin{cases} 1 \times 10^6 T^{-5/4} & \text{for } \omega_P \lesssim \omega; \\ 8 \times 10^3 T & \text{for } \omega_P \ll \omega. \end{cases} \tag{5-34}$$

The Rayleigh factor 1.22 in Eq. (5–33) takes care of the diffraction of the laser beam while this factor is unity for diffraction at a slit (Hora 1969a). The power threshold (5–34) depends on plasma density $n_e$ via the plasma frequency $\omega_p$, Eq. (2–4). Since $T$ is in the range of 10 eV, we see that ponderomotive self-focusing in a plasma has a threshold of about 100 kW to 1 MW. A slight modification of this result takes into account the radial variation of the plasma temperature due to the laser beam in the filament and was described by Sodha et al. (1973, 1974, 1976, 1978, 1979) and Tewari et al. (1975) reproducing the result Eq. (5–34), and by Yu et al. (1974), Siegrist (1977), Mavaddat et al. (1980) as well as others.

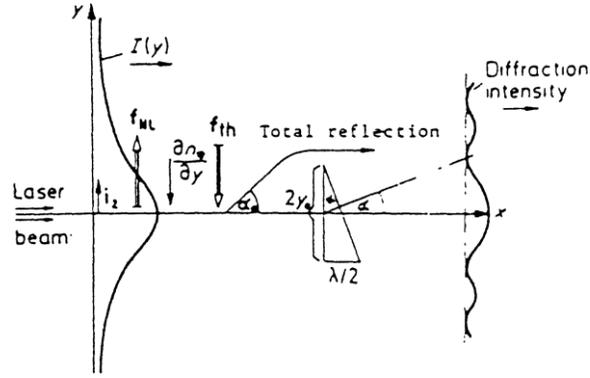

**Figure 5–3.** Scheme of a laser beam of a lateral intensity decrease $I(v)$ in plasma, producing nonlinear forces $\mathbf{f}_{NL}$ in the plasma, rarifying the axial regions until being compensated by the thermokinetic force $\mathbf{f}_{th}$ of the gas dynamic pressure. The density gradient causes a total reflection of partial beams. The diffraction condition for permitting partial beams of an angle of propagation less than total reflection for achieving the first diffraction minimum is the final condition for deriving the self-focusing threshold.

The result of Eq. (5–34) for low-density plasma arriving at the same value of $8 \times 10^3$ $T$ was reproduced in different ways in derivations by Palmer (1971), Shearer et al. (1973) and Chen (1974), where agreement with (Hora 1969a) was correctly mentioned. The same agreement was achieved without any reference to the preceding work by Kaw, Schmidt and Wilcox (1973), which was produced between submission and revision of the paper of Shearer and Eddleman (1973). Even the co-author Kaw was not permitted to give reference to his own basic work [Lindl et al. (1971)]. It is strange that nowadays many references to ponderomotive self-focusing are given only to the paper by Kaw et al. (1973).

Ponderomotive self-focusing stops when the whole plasma is depleted from the laser beam center. In this case, the electromagnetic energy density is equal to the thermokinetic pressure

$$\frac{\mathbf{E}^2 + \mathbf{H}^2}{8\pi} = n_e K T_e \left(1 + \frac{1}{Z}\right) \quad (5\text{-}35)$$

corresponding to Eq. (4–13) and Fig. 4–2. The corresponding laser intensity is just at the level of $I^*$ which determines the diameter of the self-focusing filament. The agreement with the experiments (Korobkin et al. 1968) is evident: a 10 MW ruby laser pulse shows a diameter of 3 µm corresponding to a laser intensity in the $10^{14}$ W/cm$^2$ range. This depletion mechanism was measured in detail (Richardson et al. 1971). We see that self-focusing in the plasma above MW laser power gener- ates ions up to keV energies and to the separation of plasma groups with linearly depending energy on ion charge $Z$, Eq. (5–17).

Relativistic self-focusing is basically an optical effect and refers to the nonlinear force only to the extent that the focusing down to beam diameters of 0.6 wavelengths produces such high laser intensities (and corresponding ponderomotive potentials) that extremely high ion energies ($Z$-separated) are produced.

Relativistic self-focusing is based on the fact that the absolute value of the refractive index

$$|\mathbf{n}| = \left[\left(1 - \frac{\omega_p^2}{\omega^2 + \nu^2}\right)^2 + \left(\frac{\nu}{\omega}\frac{\omega_p^2}{\omega^2 + \nu^2}\right)^2\right]^{1/4} \tag{5-36}$$

[from Eqs. (3–86) and (3–87)] has a dependence on the laser intensity $I$ when the plasma frequency has a relativistic change due to relativistic mass increase of the electron for the quiver motion, Eq. (3–97). The same relativistic change appears for the collision frequency (Hora 1981, 1991) though this is a weak effect only. The result is that the laser wavelength in vacuum $\lambda_0$ becomes longer in the plasma

$$\lambda = \frac{\lambda_0}{|\mathbf{n}(I)|}. \tag{5-37}$$

As can be seen from the relativistic relations (3–97), the wavelength in the plasma is shorter at higher laser intensities (e.g., at the intensity maximum $I_{max}$ of the laser beam) than at lower intensities, e.g., at $I_{max}/2$

$$|\mathbf{n}(I_{max})| > |\mathbf{n}(I_{max}/2)|. \tag{5-38}$$

This leads to the fact that a Gaussian laser beam of a diameter $d_0$ in vacuum with a plane wavefront in the plasma gets its wavefronts bent as shown in Fig. 5–4. This bending lets the beam shrink down to a diameter given by the diffraction limit. This is rougly one wavelength. Measurements by Rode (1984) and Basov et al. (1987) clarified experimentally on the basis of relativistic self-focusing and nonlinear force producing MeV ions that the relativistically self-focused beam has a diameter of 0.6 wavelengths.

From the geometry of Fig. 5–4 the self-focusing length $l_{SF}$ is

$$l_{SF} = \left[d_0\left(\rho_0 + \frac{d_0}{4}\right)\right]^{1/2}, \tag{5-39}$$

and the geometric ratio

$$\frac{|\mathbf{n}(I_{max}/2)|^{-1}}{(d_0/2 + \rho_0)} = \frac{|\mathbf{n}(I_{max})|^{-1}}{\rho_0}. \tag{5-40}$$

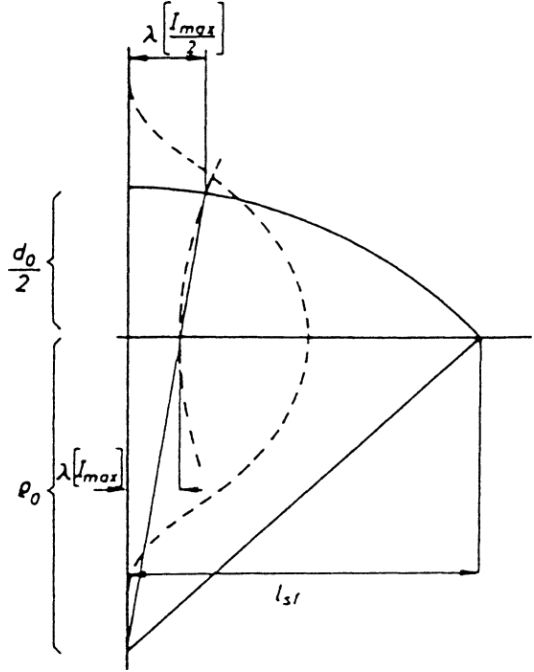

**Figure 5–4.** Evaluation of the relativistic self-focusing length from the initial beam diameter $d_0$ and from the effective wavelengths. The relativistic effects cause a shorter wavelength at the maximum laser intensity $I_{max}$ than at the half maximum intensity.

We find the ratio of the self-focusing length $l_{SF}$ to the initial beam diameter in vacuum $d_0$ as

$$\frac{l_{SF}}{d_0} = 0.5 \left[ \frac{|\mathbf{n}(I_{max})| + |\mathbf{n}(I_{max}/2)|}{|\mathbf{n}(I_{max})| - |\mathbf{n}(I_{max}/2)|} \right]^{1/2}. \tag{5-41}$$

The evaluation for various neodymium glass laser intensities and (uniformly assumed) plasma densities is given in Fig. 5–5. Obviously the relativistic effect works even at laser intensities 1000 times lower than the relativistic threshold (3–96), a phenomenon well known from the relativistic instability (Tsintsatse 1976). For electron density close to the critical value $n_{ec}$, the beam shrinks very fast within a depth of the initial laser beam diameter. These values were numerically evaluated including the influence of the relativistic change of the collisions while observing the nonlinear deviation of the absorption process (Fig. 5–6). The results of Fig. 5–5, which are valid for all intensities below and above the relativistic threshold, agree with the values derived for laser intensities much lower than the critical density (Spatschek 1978).

    Subsequent acceleration of ions of charge number $Z$ no longer depends on the laser wavelength but only on the laser power (Fig. 5–7). These results were fully

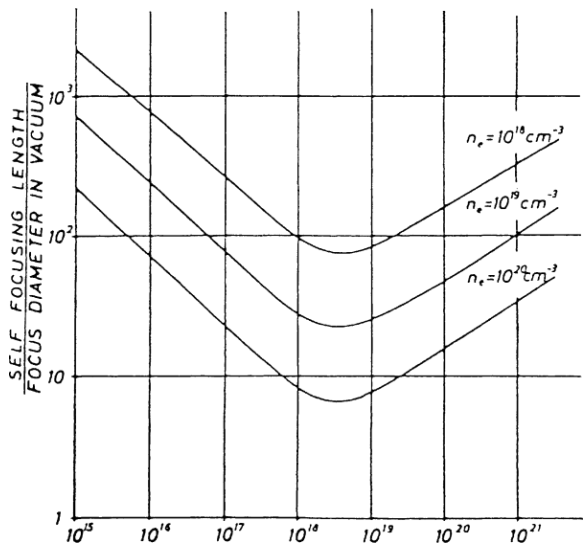

**Figure 5–5.** Calculated self-focusing lengths divided by the laser beam diameter for neodymium glass laser radiation for various plasma densities depending on the laser intensity (Hora 1975).

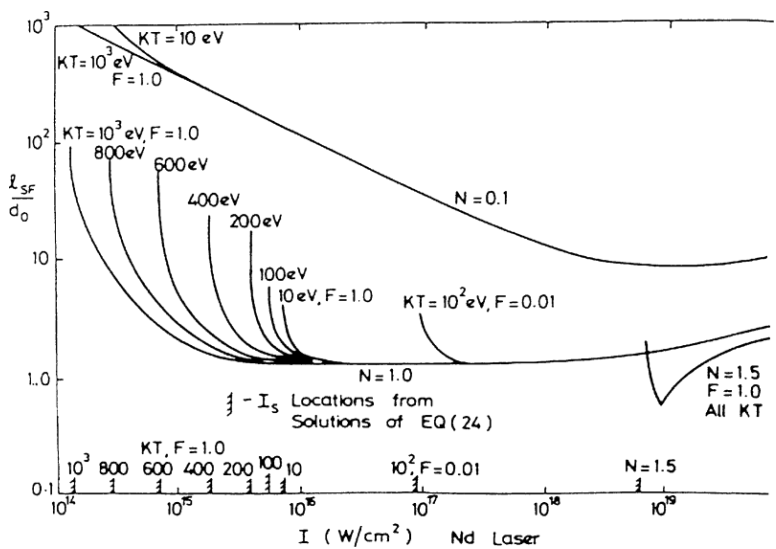

**Figure 5–6**. Ratio of the self-focusing length $\ell_{SF}$ over the initial laser diameter $d_o$ for laser beam intensities near the relativistic threshold of $3 \times 10^{18}$ W/cm$^2$ for neodymium glasslaser radiation for varying plasma temperatures. The plasma density is equal to the nonrelativistic cutoff value ($N = n_e/n_{ec} = 1$) and 10% of this value ($N = 0.1$), respectively. The factor F is given by an effective collision frequency $\nu_{eff} = \nu/F\ 2/3$ to demonstrate possible anomalous effects (Kane et al. 1978).

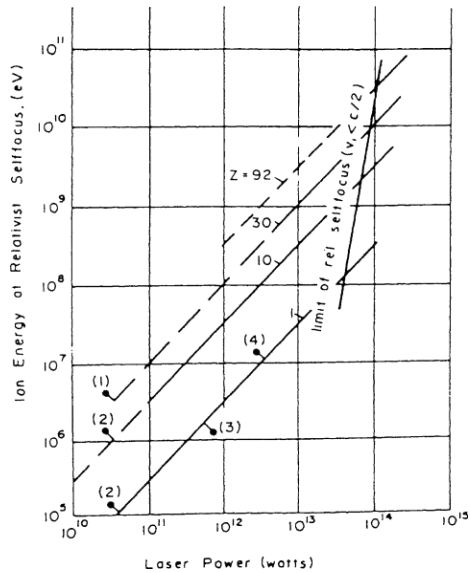

**Figure 5–7.** Energy of $Z$-times charged ions at relativistic self-focusing for (sufficiently short) laser pulses of various laser powers following the theory (Hora et al. 1978). Measured values are from (1) Luther-Davies et al. (1976) and Siegrist et al. (1976), (2) Ehler (1975), (3) Haas et al. (1976), Manes et al. (1977), and (4) Godwin et al. (1978).

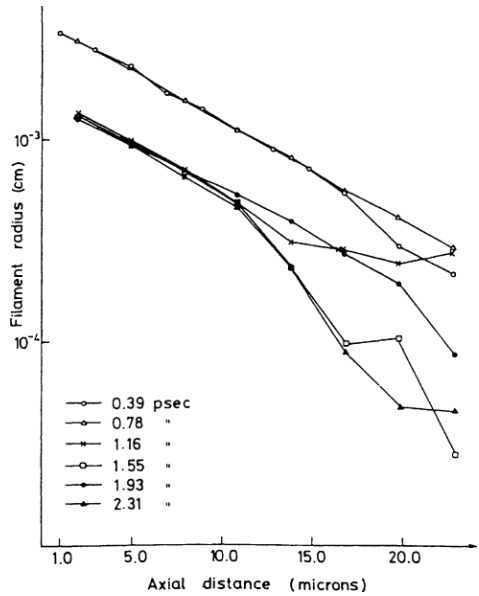

**Figure 5–8.** A $10^{13}$ Watt Nd glass laser pulse from vacuum incident on $Sn^{38+}$ plasma of $10^{21}$ cm$^{-3}$ density. Filament radius as a function of depth into the plasma at various times (Jones et al. 1982).

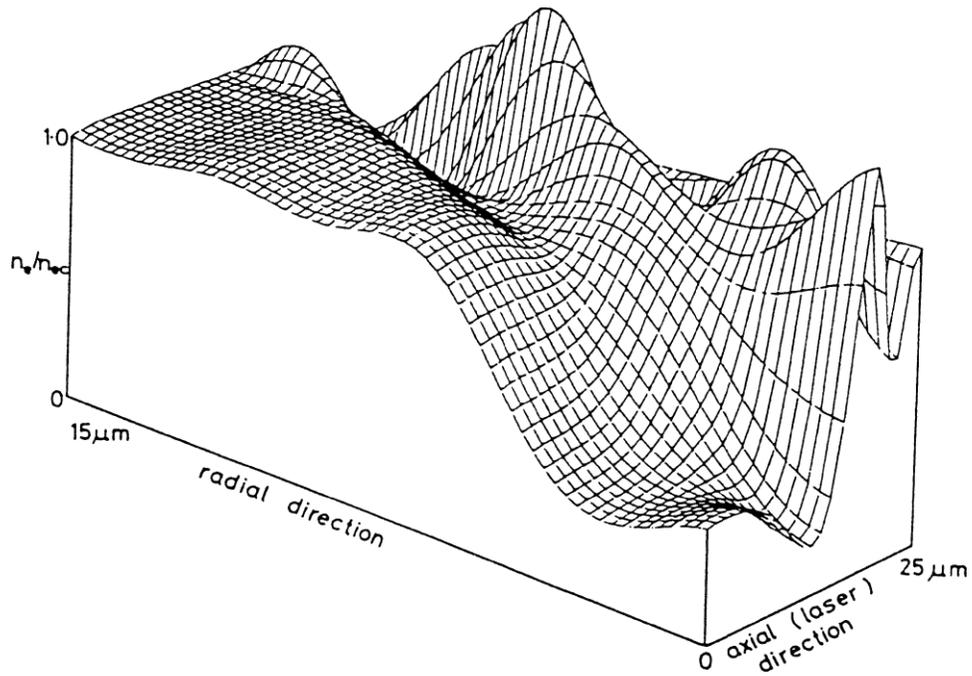

**Figure 5–9a.** Electron density profile at 1.16 ps after beginning irradiation, same conditions as for Fig. 5–8.

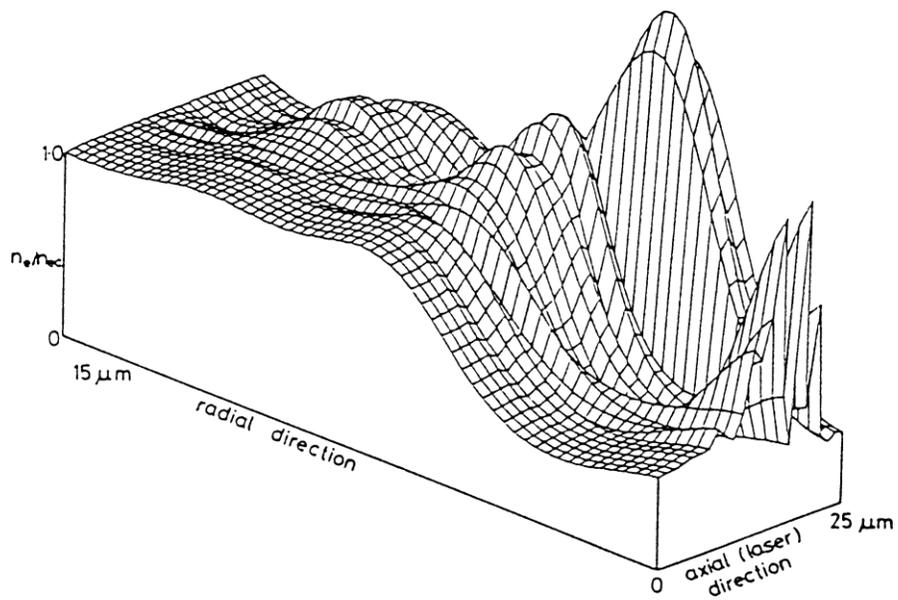

**Figure 5–9b.** Electron density profile at 1.93 ps after beginning irradiation, same as Fig. 5–8.

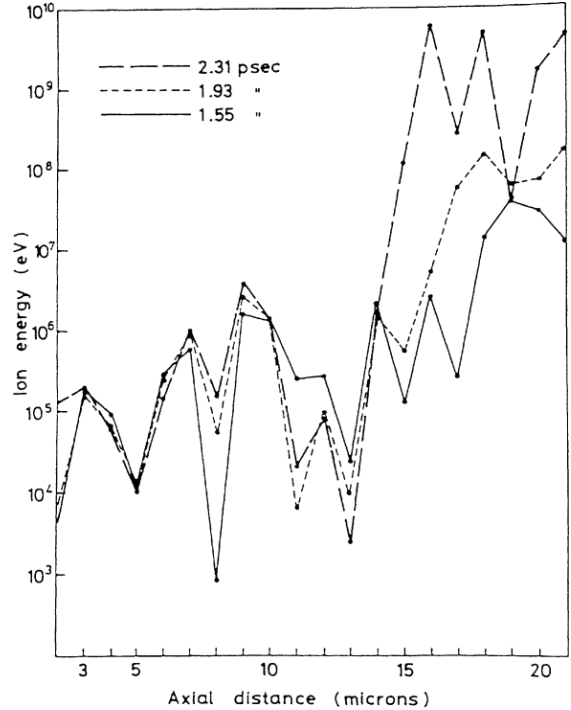

**Figure 5–10.** Center line ion energies as a function of axial depth into the plasma at various times (ps) for the conditions of Fig. 5–8.

reproduced by the measurements of MeV ions from laser produced plasmas (Ehler 1975; Luther-Davies et al. 1976) up to 500 MeV ions (Begay et al. 1983). The three-dimensional cylindrically symmetric simulation of relativistic self-focusing including all relevant nonlinear forces, the thermokinetic effects and the collisions, resulted in a fully dynamic following up of the process. If the beam intensity grew too slowly (20 ps for TW pulses of neodymium glass lasers) the beam drilled a hole in the plasma and no relativistic self-focusing occurred; the plasma was driven out of the beam by the nonlinear forces during this rather long time. Only during laser pulses of a few ps rise time, relativistic self-focusing was seen with rapid shrinking of the Nd:g laser beam of initially 20 μm diameter (Fig. 5–8) showing all details of the dynamics (Figs. 5–9a and 5–9b). The ion energies of 38-times ionized tin reached 6 GeV at 10 TW laser power (Fig. 5–10). Later measurements with the ANTARES $CO_2$ laser for the same laser powers arrived at 6 GeV ions (Gitomer 1984). About enhanced relativistic self-focusing see Nanda et al. (2014).

# CHAPTER 6
# Single Particle Derivation of the Nonlinear Force

## 6.1 QUIVER DRIFT AND ELECTRIC DOUBLE LAYERS

Up to this point we have described the laser plasma interaction and the generation of the nonlinear force exclusively on the basis of plasma hydrodynamics and Maxwell's theory. The hydrodynamics was limited to space charge quasi-neutrality according to Schlüter's assumption for deriving the two fluid equations (3–67) and (3–69) which with the requirement of momentum conservation of obliquely incident laser radiation on plasma had to be extended to Eqs. (4–62) and (4–33) or (4–55) where some high-frequency electrical charge effects had to be realized. The essential point for the appearance of the nonlinear force was that there was a phase shift between **E** and **H** of the waves in an inhomogeneous medium according to a WKB approximation (Hora et al. 1967), which could be understood from the Rayleigh case of inhomogeneous media (Hora 1957) (see Fig. 3–7).

We now look for another derivation of the nonlinear force based on the single particle motion of an electron in the laser field in an inhomogeneous plasma and then we contrast the case of any low-density plasma. We are considering the geometry of Fig. 6–1 where laser radiation from vacuum penetrates into a plasma with continuously growing electron density and therefore decreasing (real part) of the optical refractive index until the critical density. The motion of a single electron in vacuum is the quiver motion where the electron follows the electric field **E** with the vacuum amplitude $E_V$ of the linearly polarized laser, oscillating in the $y$-direction and results in an electron velocity $v_y$. Taking the cross product of this velocity with the magnetic field results in a Lorentz force such that the electron will perform a longitudinal oscillation with twice the optical frequency resulting in figure-8-like motion (Fig. 6–1, vacuum range). The longitudinal motion is a $v_y/c$ effect, therefore a somewhat relativistic process. We shall see that this is essential even at laser intensities many orders of magnitudes less than the relativistic threshold, in fact for the whole subrelativistic range.

Inside the plasma, the figure-8-like quiver motion is modified first by a larger amplitude of the electric field, expressed here by the WKB approximation, Eq. (3–108),

$$\mathbf{E} = \mathbf{i}_2 \frac{E_v}{n^{1/2}} \cos F .\qquad(6\text{-}1)$$

We can neglect the very weak absorption in the underdense range of the plasma corona (shown in Fig. 3–4) and can use the real part only of the refractive index $n$,

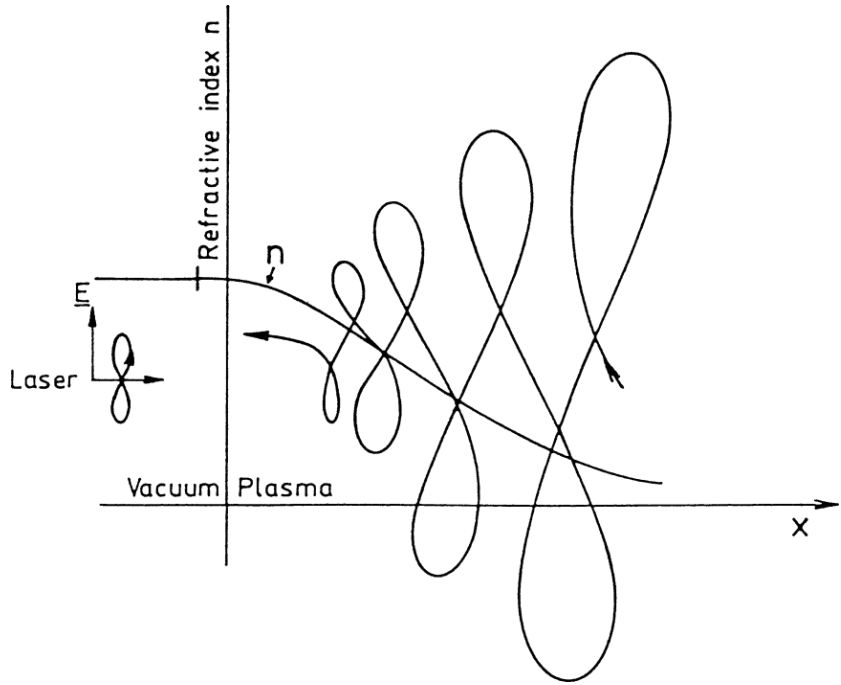

**Figure 6–1.** Refractive index **n** in vacuum and in an inhomogeneous plasma corona depending on the depth $x$ with monotonously increasing electron density. If laser light is perpendicularly incident with **E** polarized in the plane of drawing, an electron performs a closed figure-8-like motion in vacuum. Inside the plasma the deformed *eight* is swelled up [see Eq. (6–1)] due to the decreasing denominator $n$ to values 0.1 or much less, and the phase shift causes the electron to not stand at *eight* but drift toward vacuum (towards lower electron density).

Eq. (3–91). The decreasing **n** to one tenth and much smaller values. This results in the increasing (swelling) of the laser field amplitude or of its intensity or for its electromagnetic field density inside the plasma. The magnetic field of the laser wave in the WKB approximation (3–111) is

$$\mathbf{H} = \mathbf{i}_3 \frac{c}{2\omega} \frac{E_v}{n^{3/2}} \frac{dn}{dx} \sin F - \mathbf{i}_3 E_v n^{1/2} \cos F, \qquad (6\text{-}2)$$

using the representation

$$F = -\int^x \frac{\omega n(x)}{c} dx + \omega t. \qquad (6\text{-}3)$$

The quiver motion is determined by the velocity of the electron in the $y$- direction driven by the Coulomb force of the **E**-field of the laser to produce an

Acceleration

$$\dot{v}_{ey} = \frac{e}{m} E_y . \tag{6-4}$$

In the $x$-direction we have the acceleration

$$\dot{v}_{ex} = \frac{e}{mc} v_{ey} H_z \tag{6-5}$$

due to the Lorentz force caused by the magnetic field (6–2) and the velocity given by Eq. (6–1) from integrating Eq. (6–4)

$$v_{ey} = \frac{e}{m\omega} \frac{E_v}{n^{1/2}} \sin F . \tag{6-6}$$

In the $x$-direction, we find a net acceleration when substituting Eqs. (6–6) and (6–2) into Eq. (6–5) or a net force acting on the electron by multiplying the electron mass $m$. The force density to the $n_e$ electrons per volume is given as the nonlinear force to the electron gas

$$\begin{aligned} f_{e,NL} &= n_e m v_{ey} H_z \mathbf{i}_x \\ &= \mathbf{i}_x n_e \frac{e}{c} \frac{e}{m\omega} \frac{E_v}{n^{1/2}} \sin F \left( \frac{c}{2\omega} \frac{E_v}{n^{3/2}} \frac{dn}{dx} \sin F - E_v n^{1/2} \cos F \right). \end{aligned} \tag{6-7}$$

Time averaging results in

$$f_{e,NL} = \mathbf{i}_x \frac{1}{4} \frac{e^2}{m\omega^2} \frac{E_v^2}{n^2} \frac{\partial n}{\partial x} = -\mathbf{i}_x \frac{\omega_p^2}{16\pi\omega^2} \frac{\partial}{\partial x} \frac{E_v^2}{n} . \tag{6-8}$$

We see that the nonlinear force of the electron cloud in the plasma is the same as what we derived for the net plasma fluid determined by the ion mass $m_i$ and the net plasma velocity $v_x$ in the $x$-direction, Eq. (4–9),

$$f_{NL} = \mathbf{i}_x n_i m_i v_x = -\mathbf{i}_x \frac{(1-n^2)}{16\pi} \frac{\partial}{\partial x} \mathbf{E}^2 . \tag{6-9}$$

The result is that the electrons quiver in the plasma interior not simply by performing a continuous 8-figure as in the vacuum but (Fig. 6–1) an open eight drifting towards lower plasma density. In this way the electron achieves a net translation velocity where the excess of the quiver amplitude corresponding to an excess quiver energy is converted into translative energy of motion. As seen from the time averaging in Eq. (6–7), this *quiver drift* is the result of the phase shift between **E** and **H** of the laser wave field in the inhomogeneous plasma.

     A further result is that the force density on the electrons is equal to the force density acting on the whole plasma as confirmed by all the preceding chapters. This

means that in a plasma, the nonlinear force is acting on the electrons but the inertia of the acceleration is determined by the ions. This can be understood only if we assume that between the laser pulled electron gas and the ions a strong electric force is produced which drives the ions towards the vacuum. This electric field was not seen in the preceding theory because the internal fields in a plasma were assumed to be zero according to the assumption of space charge quasi-neutrality. Nevertheless, the nonlinear force works as is well known from the computed and measured cavitons.

This result modifies our conclusion with regard to the "ponderomotive potential". From the result of the $Z$-separation, Fig. 5–2, we had the view that each plasma consisting of the cloud of $Z$-charged ions and the added neutralizing electrons behaved like the dielectric material in Fig. 1–2 when it is driven by the ponderomotive potential of electrostatics. This is formally correct, but we learn from the single electron quiver drift that each of the $Z$-separated plasma clouds are not simply acting like space charge neutral plasmas but are pulled by the laser field acting on the electrons, and the ions follow by electric field attrac- tion.

This fact was the starting point of the genuine two fluid model described in Section 6.4 where the fields between the electron and ion fluid, the consequent electric double layers and the internal electric field in any inhomogeneous plasma can be seen, modified by the high-frequency property of the driving laser field. There result several new effects of resonances and second harmonic generation.

A final remark is necessary concerning other "single electron" derivations. Starting from the fact of vector calculus relationship

$$\nabla \mathbf{E}^2 = 2\left[\mathbf{E}\cdot\nabla\mathbf{E} + \mathbf{E}\times(\nabla\times\mathbf{E})\right], \tag{6-10}$$

Chen (1974) claimed a derivation of the nonlinear force $f_{NL}$ from the quiver motion of an electron according the right-hand side of Eq. (6–10). This derivation does not at all provide the nonlinear force in a plasma based on the phase difference between E and H nor the quiver drift process as explained by Fig. 6–1. Chen even claimed that our (Hora 1969) derivations only refer to the last term in Eq. (6–10), while everyone can see the result of the left-hand side of Eq. (6–10).

What is intriguing is that the expression on the right-hand side of Eq. (6–10) seems with the first term to be formally identical to the expression of Kelvin's electrostatic ponderomotive force (1–1). Chen claims that the second term refers to our nonlinear force (Hora 1969) only, just the term which formally is not Kelvin's ponderomotive force. This is a remarkable confusion with our result (Hora 1969) since the whole left-hand side of Eq. (6–10) is the nonlinear force and not only the part on the right-hand side of Eq. (6–10) which is not identifiable with the ponderomotive force.

In no way does this view alter the exceptional merit of the work of Chen (1974) (Hora 1992: Section 9.5) where he, along the lines of the two terms of Eq. (6–10) fully summarizes and quantitatively evaluates all the parametric plasma instabilities. This classification of the instabilities by the nonlinear force is indeed unique, since the preceding derivations of the instabilities, based on Landau's derivation of the parametric effects merging in Mathieu's differential equation (Spatschek 1978) with the characteristic small stable and broad unstable ranges as used for understanding the Paul trap [(Paul et al. 1955); Oraevski and Sagedeev (1963); Silin (1965); DuBois and Goldman (1967); Nishikawa (1968)], did not directly show this generality by one scheme of derivation.

## 6.2 THE BOREHAM EXPERIMENT CONCLUDING LONGITUDINAL OPTICAL FIELDS AND PREDICTION OF THE MEYERHOFER FORWARD DRIFT

In order to see the nonlinear force mechanism acting on the electron fluid, the experiment of Boreham (Boreham et al. 1979, 1979a) is essential. A neodymium glass laser beam (Fig. 6–2) was focused in a chamber filled with helium pressures between $10^{-6}$ and $10^{-3}$ Torr. From the focus with a laser beam intensity $I$ of $10^{16}$ W/cm$^2$, electrons were emitted which after a rather free flight could be collected through a grid and the energy spectrum measured. The maximum energy was $\varepsilon_e = 1$ keV (Boreham et al. 1979, 1979a).

The laser ionizes the helium atoms by a tunneling process (Keldysh 1965) as could be seen in the Boreham experiment by measuring the temporal sequence of the emission of the first and the second electron (K.G.H. Baldwin and B.W. Boreham 1981; Fittinghoff et al. 1992). While it is still beeing discussed whether or not the electrons have some initial energy of eV from the ionization process, what happens is that the electron experiences a nonlinear force driving it out of the laser beam. Since the gradient of the laser electric field $\mathbf{E}^2$ is in the radial direction only we have a nonlinear force on the single electron taken from Eq. (6–8) and (naïvely!!) taking the nabla operator in cylindrical coordinates instead of $(\partial/\partial x)$ in Eq. (4–10)

$$f_{e,\mathrm{NL}} = \frac{e^2}{2m\omega^2} \frac{\partial}{\partial \mathbf{r}} \mathbf{E}^2 . \qquad (6\text{-}11)$$

Remembering that the time average of $\mathbf{E}^2 = E_v^2/2$ the radial integration of (5–11) from the beam center to a very large radius results in an energy of the electron of translation

$$\varepsilon_e^{\mathrm{transl}} = \frac{e^2 E_v^2}{4m\omega^2} . \qquad (6\text{-}12)$$

This energy is equal to half of the quiver energy of the electron (3–93), i.e., the average kinetic energy the electron has when oscillating in the laser field which is equal to the ponderomotive potential $\varphi$, Eq. (1–9), when taking the time-averaged value of $\mathbf{E}^2$,

$$-\phi = \varepsilon_e^{\mathrm{transl}} = \varepsilon_{\mathrm{osc}}^{\mathrm{kin}} = \frac{\varepsilon_{\mathrm{osc}}}{2} = \frac{I}{2cn_{\mathrm{ec}}} . \qquad (6\text{-}13)$$

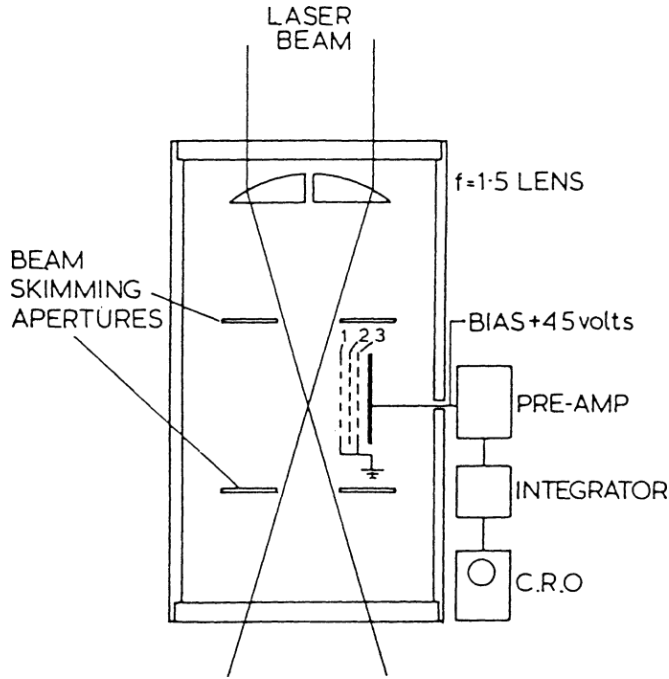

**Figure 6–2.** The Boreham experiment (Boreham et al. 1979, 1979a) focusing a neodymium glass laser beam to $10^{16}$ W=cm$^2$ in $10^{-6}$ to $10^{-3}$ Torr helium with radial emission of electrons of up to keV energy.

Taking the neodymium glass laser intensity of $I = 10^{16}$ W/cm$^2$ and the critical electron density for this wavelength of $n_{ec}$, we arrive at the maximum electron energy of 1.04 keV as measured.

Here we have a case where the electron gas is accelerated by the laser beam and the ions are nearly untouched, contrary to the earlier result of the plasma acceleration where the whole inertia is given by the ions, and the neutral plasma behaves like a neutral dielectric material in a field with ponderomotive forces. In the case of the Boreham experiment the ions did not follow, because the density was too low. A measurement of the emitted electrons on the background gas density (Fig. 6–3) was fully linear at low densities. Only for such densities where the laser focus diameter was close to the Debye length, was a deviation from the linear dependence observed (Boreham et al. 1979), Fig. 6–3. This result was one of the points clarified by these measurements.

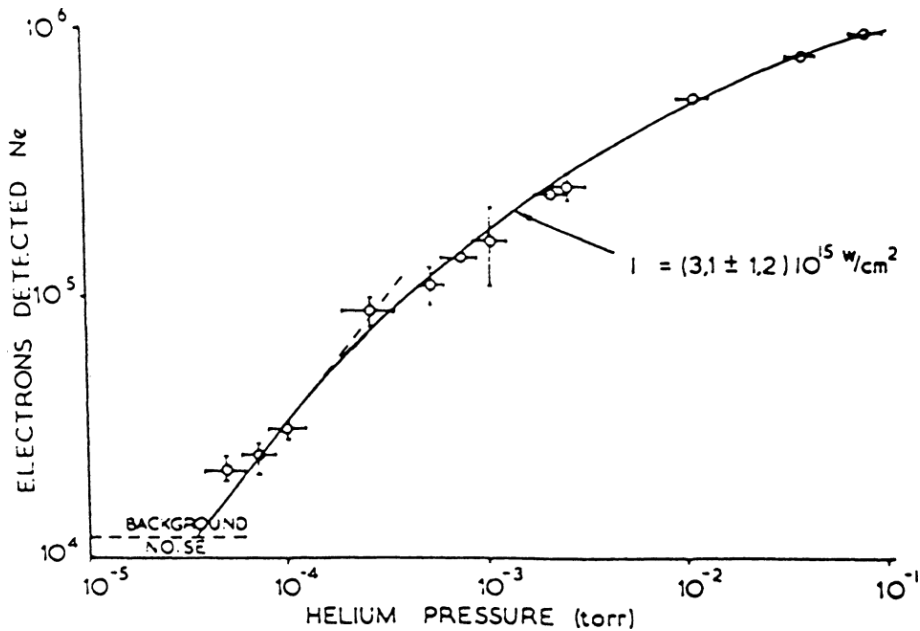

**Figure 6–3.** Number of electrons emitted from the focus of a laser beam in helium gas of varying pressure if the laser intensity is about $10^{15}$ W/cm$^2$. The linear increase on the number of electrons is saturated at about $2 \times 10^{-4}$ Torr, due to the Debye length becoming equal to the focus diameter.

The easiest interpretation is that the electron has been falling down the ponderomotive potential.

The alternative is, in view of Fig. 6–1, that the electron has its quiver motion and due to the radial decay of the electric field strength, the figure-8-like motion changed into a quiver drift showing how (like in Fig. 6–1 for the plasma) the electron quiver kinetic energy is converted into translative energy of radial motion. This seems to be in full agreement with the energy relations (5–11) to (5–13). When examining the details, however, problems arise which could have been over- looked in view of the just-mentioned picture of the ponderomotive force using the time averaged value of the electric laser field. The solution of these difficulties, however. led to very interesting new insights of the properties of the laser beam.

As can be seen with rather elementary mathematics, the quiver drift in such a radius, which has the direction of the **E**-field of the laser, really does arrive at the translative energy given by half of the maximum quiver energy or by the ponderomotive potential (6–13) as expected (Hora 1991: Section 12.3). However, if one follows up the quiver drift in the direction of the **H**-vector, the energy gain is zero, contrary to the observation that with linearly polarized laser light, there is no polarization dependence of the electron emission.

What happened? We must remember that the transverse field in a laser beam is not the only field component of the Maxwellian field. We must remember that this is only the case for infinite plane waves, not for beams. For a finite beam one can construct longitudinal fields from the Maxwell equations. These longitudinal fields indeed were shocking since it is the pride of Maxwell's theory to explain the earlier unbearable observation that light is only transversal without any longitudinal component. Now one has to have longitudinal light components. As has been shown first with such a laser beam where the transverse *and* longitudinal field components could be expressed with elementary functions (Hora 1981), it was shown that the tiny additional part of the longitudinal components changed the quiver motion along the magnetic field direction from zero to the measured value equal to that in the **E**-direction.

This was the first Maxwell exact discovery of longitudinal optical components which were zero at the beam axis and had a shape exactly given by the self diffraction of the beam. The longitudinal components were phase shifted by 90º such that no Poynting energy flow was in radial direction as it had to be (Hora 1981). A preceeding observation of the longitudinal components was the result of a microwave study with the expressed surprise of the authors (Lax et al. 1978), whose result, however was not the Maxwellian exact solution but was based only on the ray optical paraxial approximation, see also other approximations (Carter et al.1972; Agraval et al. 1979; Deng 1992; Barton et al. 1989). Here (Hora 1981) we needed the Maxwellian exact solution.

When using a laser beam with a Gaussian radial intensity decay, the solutions cannot be expressed by elementary functions but by a series of Bessel functions in a stepwise iteration process. This was done in a thesis by Cicchitelli (1989) (Cicchitelli et al. 1990) and the result was tested in the same way as before by the polarization independence of the Boreham experiment. The numerical result of the iteration with vanishing numerical error arrived at the values shown in Figs. 6–4 and 6–5. Including the Maxwellian exact longitudinal field for a Gaussian beam, the force density was completely independent of the polarization. The same calculation with the transversal laser field components (all in vacuum) resulted in a strong polarization dependence of the radial force density.

The forces were calculated in this case (Cicchitelli et al. 1990) using the tensor formulation of the nonlinear force with all the relevant components including the Maxwell stress tesnor, Eq. (4–50). This example also shows that the nonlinear force has to be in the fully exact formulation, and not naïvely taking the radial gradient as described before in Eq. (6–11). This is an example of *how dangerous it can be in nonlinear physics* if such a naïve assumption leads to a correct theoretical confirmation. These difficulties could be seen from the beginning when looking at the Maxwell stress tensor (4–52). The components $ij\ ij$ and similar show, that the dominating vector components of the laser field $E_y^2 - H_z^2$ are connected with a minus and therefore vanish. This is contrary to our earlier case of perpendicular laser incidence, Eq. (4–54), where a plus is between these vectors. It would be very unscientific to ignore this problem, simply saying, "The (naïvely assumed and wrongly applied) ponderomotive potential (incidentally arriving at the correct answer) is the final solution."

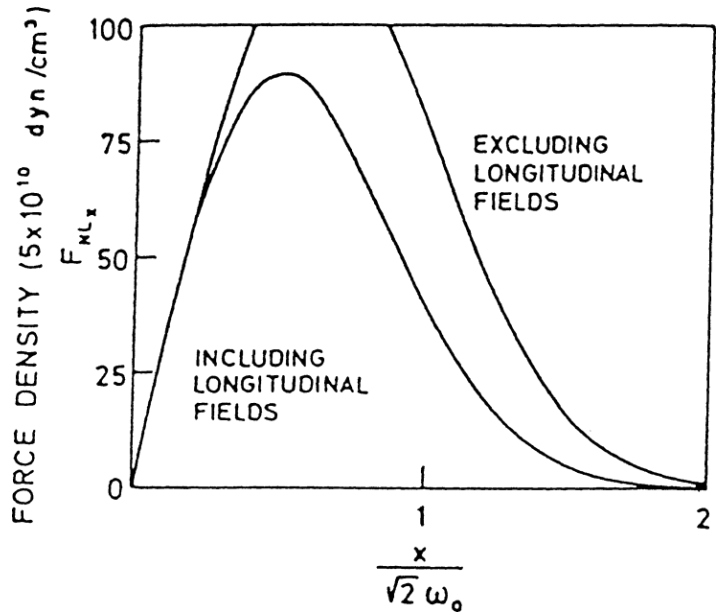

**Figure 6–4.** Radial force density (dyn=cm3) in a linearly polarized Gaussian neodymium glass laser beam of 13 μm diameter (half maximum intensity width) of $1\times10^{15}$ W/cm$^2$ intensity in the direction of the **E** vector for the two cases including and excluding longitudinal laser field components plotted against the distance, relative to the spot-size parameter from the center of the beam. [Note that (0; 0; z) is the axis of symmetry of the beam.]

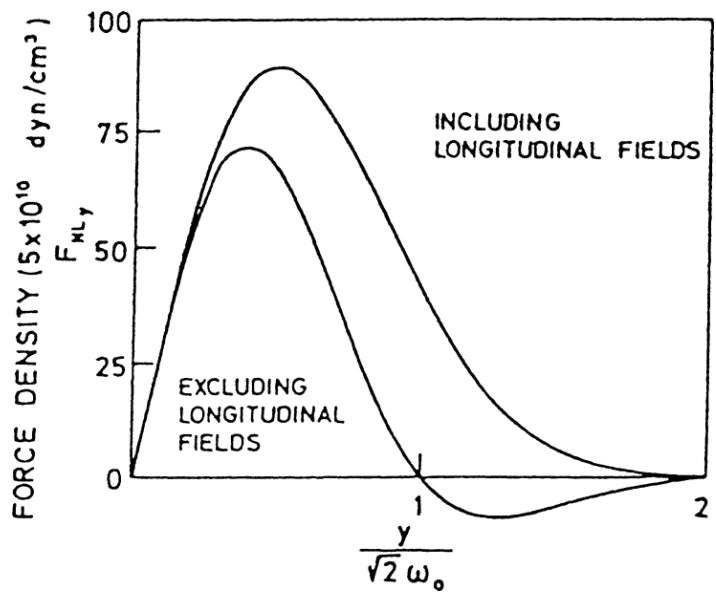

**Figure 6–5.** Same as Fig. 6–4, for the direction of the **H** vector.

There is a further aspect of the Boreham experiment with respect to the momentum transfer of optical energy into the electron motion for radial emission from the beam. This is a collisionless (nonthermal) nonlinear transfer of the optical energy by the nonlinear forces into mechanical energy of electron motion. The optical energy taken from the laser beam for the radial emission is $m_r v^2/2$. This corresponds to a momentum of the optical energy which has to be transferred to the electron as an axial momentum

$$mv_e = \frac{mv_r^2}{2c} \quad \text{or} \quad \frac{v_e}{v_r} = \frac{v_r}{2c}. \tag{6-14}$$

The momentum of the optical energy going into the radial electron emis-sion causes therefore a forward direction given by an angle $\Theta = 90^0 - u$, Fig. 6–6,

$$\tan u = \frac{v_e}{v_r} = \frac{v_r}{2c}, \tag{6-15}$$

using Eq. (6–14). Remembering that the electron energy of translation afte re-emission from the laser beam is $mv_r^2/2 = \varepsilon_e^{trans}$, Eq. (6–12), we arrive after substituting $v_r$ into Eq. (6–15) at

$$\tan u = \left(\frac{\pi I}{c}\right)^{1/2} \frac{e}{m\omega c} \tag{6-16}$$

[see Eq. (17) of Hora et al. (1984)].
When Meyerhofer was measuring this angle using laser intensities about 10 times below the relativistic value, full agreement with this prediction was reported (Meyerhofer et al. 1996). Equation (6–16) is included in the following relativistic approximation used by Meyerhofer et al. (1996), Meyerhofer (1997)

$$\tan \Theta = \cotan u = \left[\frac{2}{\gamma - 1}\right]^{1/2}. \tag{6-17}$$

Approximating the Lorentz factor for the subrelativistic case

$$\gamma - 1 = \frac{1}{\left(1 - v^2/c^2\right)^{1/2}} - 1 \approx \frac{v_r^2}{2c^2} \tag{6-18}$$

we find from Eq. (6–17)

$$\tan u = \left[(\gamma - 1)/2\right]^{1/2} = \frac{v_r}{2c}, \tag{6-19}$$

in agreement with our earlier result (Hora et al. 1984), Eqs. (6–15) and (6–16).

At the time when the Meyerhofer measurement was predicted we had discussed the case for the inverse of emission of the electron from a laser beam. The use of a stationary laser beam would result in the following: if the energy of radial motion of the electron is less than the ponderomotive potential value, the electron will be reflected. If the energy is higher it will be transmitted without any net exchange of energy (neglecting Thomson scattering and quantum effects). But if there is a nonstationary (transient) laser pulse and the electron is injected halfway into the laser beam, with an energy equal to the ponderomotive potential, it will then have no transitive energy at the beam axis and the quiver energy into which the transla- tive energy was converted will then be converted into optical pulse energy when the laser pulse is switched off. The first calculations showed that such an optical amplification can be done only for a very long wavelength in the millimeter range since the electron beam density for the injection is limited by space charge effects.

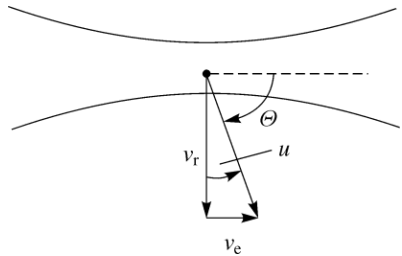

**Figure 6–6.** Forward (axial) momentum $mv_e$ of radially emitted electrons from the laser beam for momentum conservation (transfer of momentum of optical energy converted into electron motion).

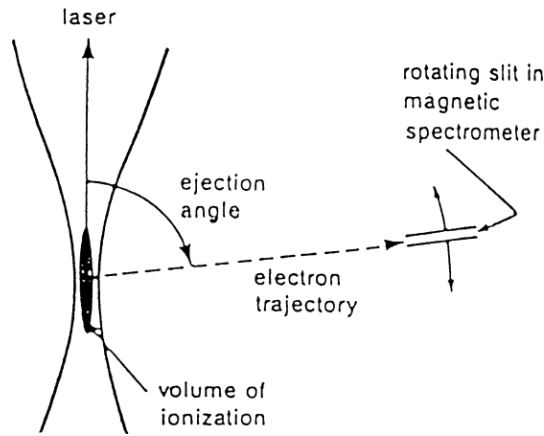

**Figure 6–7.** Experimental proof of the forward direction of the emitted electrons when radially emitted from the laser beam, Eq. (6–16), as measured by Meyerhofer et al. (1996).

To overcome this problem for electrons, the injection of solid state clusters was considered, see Figs. 6–8 to 6–10 (Kentwell et al. 1986, 1986a) with some forward momentum as calculated in Eq. (6–16) and experimentally confirmed (Meyerhofer et al. 1996). It has been calculated that laser amplifiers up to the X-ray range are possible with such clusters. The necessary cluster density was not available for a long time, but the recent laser welding experiments arrive at the interesting cluster densities (Matsunawa et al. 1996).

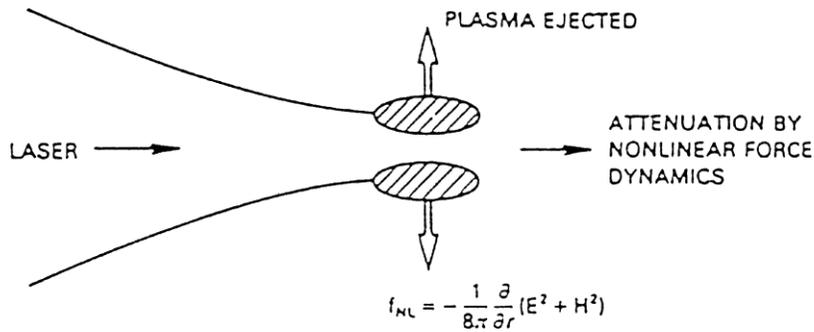

**Figure 6–8.** Scheme of the ponderomotive (nonlinear force) self-focusing with the ejection of plasma clouds from the center of the laser beam causing a collisionless transformation (absorption) of optical energy into kinetic energy of plasma.

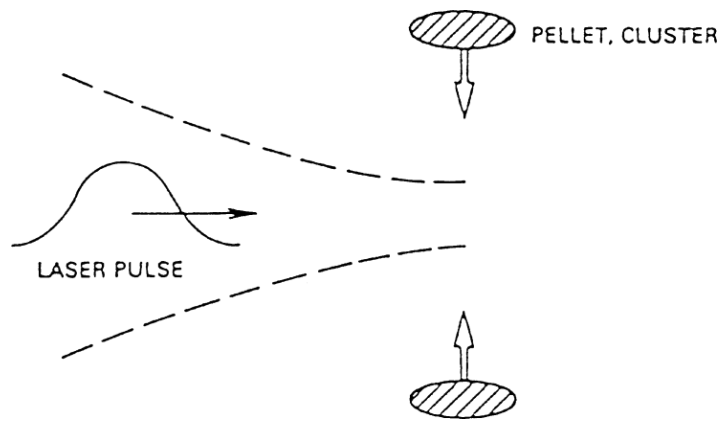

**Figure 6–9.** Inversion of the self-focusing process by injection of plasmas into the center of a laser beam.

There is the further question of the momentum transfer in the radial direction. This is considered as the change of the angular momentum of the laser beam causing a spiralling and not a fully radial emission of the electrons from the laser beam (Boreham et al. 1993).

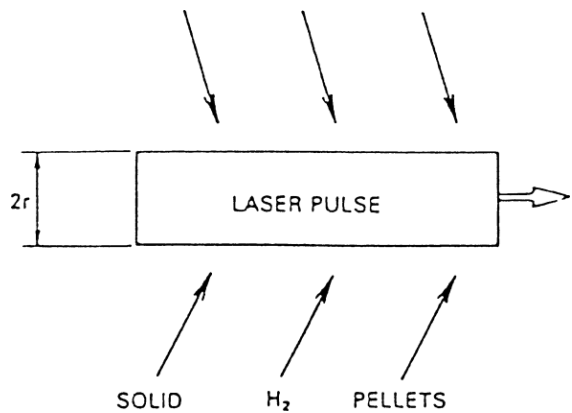

**Figure 6–10.** Box-like scheme of a laser pulse for the calculation of the amplification by lateral injection of clusters.

# 6.3 A REMARK TO STEPHEN HAWKING AND THE NONLINEARITY PRINCIPLE

What has the theory of the ugly plasmas with their swamp of mostly classical plasma physics and hopeless confusions with uncontrollable parameters to do with such a brilliant work as quantum gravity, black holes, string theory, "theory of everything" or the $N = 8$ supergravity which Stephen Hawking is following up? I did indeed alienate the reader from some general questions as to whether or not one should go into the depth of theory in general. I asked whether we should have stayed with a trivial explanation of the Boreham experiment (1979, 1979a) by using a partially misunderstood ponderomotive potential or whether we should have gone into the more complex problems and contradictions to clarify the observations which led to the discovery of the Maxwellian exact longitudinal components of light beams (Hora 1981; Cicchitelli et al. 1990).

This all led us to discuss and evaluate the term *nonlinear physics* in a more general way.

Let us talk about Stephen Hawking's inaugural lecture, delivered on April 29, 1980, when as the new Lucasian Professor of Mathematics at Cambridge University, his address "Is the End in Sight for Theoretical Physics?" startled "his listeners by announcing that he thought it was" (Ferguson 1992). This statement has power since Newton, the founder of systematic physics, and Dirac, who predicted antimatter and discovered the quantum electrodynamics, were Lucasian Professors before. Hawking noted in a celebration in Westminster Abbey on November 13, 1995, when a plaque for Dirac was unveiled and Hawking referred to him as the greatest English theoretical physicist after Newton.

Hawking envisages that in the near future one may discover an equation of all forces, a kind of "theory of everything," and then theoretical physics will no longer have to exist; all will then be solved. This ambitious goal is indeed much more sophisticated than the answer which was given to the 18-year-old Max Planck at the University of Munich when he asked the theoretical physicist Prof. Jolly whether he would recommend him to study this field. Jolly answered, "Don't do this, all is known in physics and there is nothing left to be discovered" (Planck 1948). This was in 1876.

Carl Friedrich von Weizsäcker presented the main lecture at the meeting of the German Physical Society in 1970 in Munich in the presence of more than 2000 physicists, speaking on the topic "Will Physics Knowledge Have a Saturation?" He stated at the beginning that he was not going to defend the opinion of Prof. Jolly, mentioned in the preceding paragraph. This is a trivial case. But von Weizsäcker explained his more pessimistic view that he was entitled to then as a professor of philosophy and prominent physicist, to state that there may be a saturation of knowledge. He considered quantum theory and the theory of relativity. It may be assumed that three discoveries of this caliber per hundred years could be possible, but can this go on for a thousand years and more? Von Weizsäcker was pessimistic to expect this and this may appear to be convincing.

Hawking assumes that if the dream of "the theory of everything" comes thrue, no more work will have to be done. The unification in a formula of the nuclear force, of the electric force, of the weak force and of the gravitational force (grand unified theory) has indeed gone one step forward by unifying the weak force and the electric force by the electroweak theory resulting in Nobel prizes and, as sub-stantial result, in Rubbia's consequent discovery of the W and Z particles at CERN in Geneva. But one should realize what is not known: the electroweak theory cannot predict the difference of the forces or produce the ratio between the electric and the weak force. This value has to be put into the theory from observations "by hand." Would it really be the end if one would arrive at the grand unified theory? I explained (Section 3.1) from bitter experience how difficult it is teaching at a faculty of electrical engineering how Maxwell discovered the unification of the theory of electricity and of magnetism and how this was not an end but a beginning. Maxwell's revolutionary predictions needed many years to be confirmed, elaborated and made understandable apart from the applications which changed the society fundamentally. And still there are power engineers who have not accepted Maxwell and go around his theory like a cat around a hot stew. These people are the consultants of publishers and did not hesitate in written statements in 1994, naturally, to suppress the publication of books (Hora 1994) which try to cover the period after 1864. So, even if the grand unified theory will have been written down, a lot of theory will have to be done to exhaust its new views and to spread its ideas at least for many dozens of years.

But there may be another reason why theory will not end even if the grand unified formula is found. I expect that it is nonlinear physics in a general sense which now should be described. This may be considered as opening new dimensions of physics after following the clear path of Newton. This nonlinear physics is a reason why studying physics will be enormously attractive in the future. When I spoke with Richard Feynman about this topic he simply said that this is not new because all is nonlinear in physics, as we well know. Here, we are not referring to this kind of nonlinearity. We are not referring to the well-known numerous relations in physics which are basically nonlinear. The best example is the attempt before Maxwell to explain electromagnetism by mechanics (a "unified theory") by comparing the magnetic field lines with the stream lines of a fluid, or by the stress fields in elastomechanics. This failed, although there were correct and fruitful results derived—for example, Kelvin's ponderomotive force, Eq. (1–1). The reason for the failure is that there is no linear relation between mechanics and electrodynamics in general. The relation is nonlinear (quadratic) as can immediately be seen in Eq. (4–51) with (4–52) showing that the mechanical force density in a material is given by quadratic expressions of the force quantities of the electromagnetic field (binary products of **E** and **H** component.

Feynman's well-known nonlinearity which we are not referring to is, for example, the relation between **E** and **D**, Eq. (3–8), at very high values which shows the nonlinear expansion of the dielectric constant $\varepsilon$ of the following kind

$$\mathbf{D} = \varepsilon_0 \varepsilon \mathbf{E}(1 + \varepsilon_1 \mathbf{E}^2 + ...). \tag{6-20}$$

This relation was mentioned for Townes' self-focusing theory (Chiao et al. 1963) where $\varepsilon_1$ for lasers was the same as from last century's measurements of electrostatics. This type of connection between an enormous number of physical quantities is well known. This type of nonlinear extension is always fruitful for gaining new knowledge. This can be seen by beginning with Dirac's (spinor) equation of the particles which led to the prediction of antimatter which people could not have dreamed about for thousands of years. A nonlinearization of this spinor equation was treated by Heisenberg (1960) and Vigier (1958) and the masses of elementary particles were derived. Also, the fine structure constant could be calculated (Heisenberg 1960) to a value of 120 (instead of 137.036..), which produced astonishing results.

In contrast, our problem of nonlinearities originates from the experience of our solution to explain the Boreham experiment (Section 6.2). When we worked with the transverse components of the laser beam only, the resulting radial force on the electrons in the laser field was zero, contrary to the experiment, but after using the very tiny addition of the longitudinal components, derived as exact Maxwell's solutions, the result changed from no forces at all to the forces in agreement with the observations.

This experience is most important. The little additions for using an exact solution compared with the earlier (unprecisely assumed) rather good approximations with the transverse wave components changed the result from completely wrong (zero) to right. Usually, when treating nonlinear physics, one begins with an expansion of the related functions and uses the second order approximation. Our experience teaches us that such a procedure can lead to totally incorrect predictions, not only to marginally differing results within an error bar. Since theoretical physics aims to analyze the laws of nature and to (mathematically) derive new phenomena which can then be proven experimentally, our experience of nonlin- ear physics is fundamental. It means that we *must not use approximations only but must use the best possible and complete solutions of linear physics as presumptions for nonlinear predictions*.

This is a difficult condition. It requires more effort to have the best linear physics phenomena available, which therefore it is an important task for future generations of physicists. Further it requires an enormous amount of additional effort in mathematics and computations than before. This is true, especially, in eliminating numerical instabilities and inaccuracies, which sometimes ran into numerically caused chaos which has nothing to do with physics, at least in well known cases. Indeed one cannot predict everything one may think of, but it is true that this theoretical physics type of nonlinear prediction will produce properties (later experimentally proved) which are so numerous and so surprisingly new that no one could ever have dreamed of them. Einstein's prediction of the laser (Einstein1917) is an example of a seemingly undreamable fact which needed a genius to be revealed. This type of *nonlinear physics will open a basically new dimension of knowledge which can be earned by hard theoretical physics and mathematical work.* Apart from this basically new dimension of knowledge, even more diversi- fied applications in technology and society will emerge if one is optimistic about the cleverness of mankind.

This way of nonlinear thinking is also of general importance in logistics and not only in mathematical physics. There is an example (Hora 1987, 1992) from the history of fusion energy. After the very last experiment in which Lord Rutherford was involved with Harteck and Oliphant (Oliphant et al. 1934) with the discovery of the main nuclear fusion reactions showing an enormous amount of energy available from heavy water, Lord Rutherford forbade Oliphant in 1937 from performing experiments to gain nuclear energy (Oliphant 1972). The uncontrolled explosive reaction of this kind was achieved in 1952 by Teller (1987) repeated shortly by Sakharov (1989).

The less explosive, controlled fusion energy production for future power stations has been the aim mainly since 1951 (Spitzer 1951). It was intended to fire a beam of 100 kV heavy hydrogen nuclei (deuterium) to a target with super heavy hydrogen (tritium) as was done initially (Oliphant et al. 1934) but with higher currents to obtain an exothermic controlled reaction. Then Spitzer explained that the fusion reaction cross sections are at least 300 times smaller than the cross section for the fast deuterons to hit an electron in the cold target, with the tritium losing energy as heat. Therefore some reactions will occur but there should never be an energy gain with this ion beam experiment using a cold target.

This argument of Spitzer was difficult to accept and very important persons like Nobel Laureate E.O. Lawrence, Mark Oliphant and many others simply said that one has to use higher ion beam currents and this will become exothermic. Spitzer could only smile: a factor of 300! It was decided then that instead of the cold target, the whole reacting medium had to have electrons and ions at the high temperatures (100 million degrees and more) so that electron collisions would not cause an energy loss and one could have a fusion power reactor. This line was then followed up by confining the high-temperature plasmas with magnetic fields and more than 30 billion dollars has been invested, bridging the gains by many orders of magnitude, but still has not reached break-even. Apart from the unsolved basic physics, the technological problem of the generated strong wall erosion in the reactor is now well known and practically mortal to this concept. It has been estimated that one needs an additional 150 billion dollars and more than 50 years (Maisonier 1994) to build a magnetic confinement fusion reactor.

But even more surprising is that Spitzer's argument was wrong even though the argument is logically, mathematically and physically correct. The mistake is that the argument is correct only in a linear way. Nonlinear physics does provide exothermic beam fusion.

The important persons in 1951 did not have the key word *nonlinear physics*, to counter Spitzer when using very high intensity ion beams and when the use of a cold target is possible. The very complex (nonlinear) hydrodynamics resulting from the intense particle beam (or later laser beam) irradiation results in a physics process including ablation and compression mechanisms of the target with the fusion material. We know today that 2000 times solid state densities of compressed material (Azechi et al. 1991) have been achieved. The necessary drivers with ion or width laser pulses of megajoules energy in a few nanoseconds are present-day technology (Rubbia 1993) for a power station.

The tragedy is that the incorrect linear argument of Spitzer led to expensive research in the wrong direction. The use of beam-driven fusion pellets may possibly provide a fusion reactor within less than 20 years if a crash program for about 15 billion dollars would be initiated, e.g., using laser beams (Hora et al. 1994), especially if the volume ignition is used (Hora et al. 1978) and evaluated sub- sequently by J.A. Wheeler, R. Kirkpatrick, S. Colgate and others at Los Alamos (Lackner et al. 1994). There are also good prospects for heavy ion beam fusion (Rubbia 1993a; Eliezer et al. 1994) if one ignores the useless space charge calculations for heavy ion beams (Möhl 1993) because better methods of space charge neutralization are available.

The wall problem does not exist for the type of inertial confinement fusion reactor because the necessary 50 cm thick lithium blanket for converting fusion neutrons into heat and new tritium fuel is adjacent to the reaction. The lithium salt or ceramic pebbles within a viscous fluid or fixed by centrifugal forces to the reactor wall do not have any of the problems mentioned with the walls of magnetic confinement fusion. The shock from the controlled laser triggered explosion hitting the fluid is softened by the thermal spread of the reacting plasma and is reduced compared with that of a chemical explosion by the square root of the ratio of

$$\left(\frac{\text{nuclear energy}}{\text{chemical energy}}\right)^{1/2} \geq 1000.$$

A 100 MJ fusion energy gain causes less shock momentum than the explosion of

6 g TNT.

This is the error resulting from linear thinking instead of the nonlinear reality. For magnetic confinement fusion fortunes are still wasted instead of solving the energy problem by inertial confinement fusion (Hora 1997). The philosophy of nonlinear physics and of the new dimension of theoretical physics activity beginning now—contrary to the statement of Stephen Hawking—is a product of the research on the nonlinear force and the ponderomotive potential explained here. Fortunately we were not discouraged and frus- trated after the Boreham experiment and have not rested on the argument that all is explained by the ponderomotive potential.

But even the hard work of finding the correct physics theory with the complete tensor description of the nonlinear force and including the first derivation of the exact solution for the longitudinal optical field component was nearly suppressed. One referee from *Physical Review* (Cicchitelli et al. 1990) was very positive; it was possibly Prof. Melvin Lax from Columbia University who was happy that his result of longitudinal optical components was confirmed so convincingly by our exact solutions. The other referee strongly opposed the publication of the paper saying that all is known about the pondermotive potential and the presented theory is not needed. Fortunately, a third referee helped the journal editors accept the paper. It should be added as a general remark that Kibble (1966) never used the expression "ponderomotive potential" for the interaction of low-density electrons with a laser beam. This was before the plasma case of this interaction with the first derivation of the dielectric effects was discovered (Hora et al. 1967). Kibble used the expression "$\mathbf{E}^2$ field-gradient force" and had the correct and intriguing view that electrons hitting a laser beam experience "an electron optical refraction," a view which is fully justified. But what is then the difference to the refraction of the electrons in the potential produced by the Coulomb field of an electrostatic lens? This difference between the Coulomb force and the (stationary) potential produced by the nonlinear force was definitely not clarified in 1966 (Dewar 1977). The view of electron refraction may be useful for a further examination of the elaborations presented here.

## 6.4 Double Layers and the Genuine Two-Fluid Model. Stuttering Interaction and Laser Beam Smoothing

The double layer result derived from the knowledge of the single particle motion of the electron fluid being pulled forward by the nonlinear force with the ions following due to the adhesion by strong electric fields between the electron and ion fluid was the starting point to look into this mechanism with the electric fields in more detail. After we had learned from the final derivation of the nonlinear force that plasmas are not free from internal macroscopic electric fields at least in the high-frequency range, it was one step further to question these electric fields. This question was heretical since it had been sacrosanct that plasma must not have internal electric fields. The results on these fields from cosmic plasmas or from geophysical plasma at the earth poles elaborated by Alfvén (1981), Fälthammar (1988) and some other enthusiasts was described very skeptically (Kulsrud 1983) as "fields which are not intuitively clear."

We knew about internal fields and established a detailed computation with a real time and realistic description of the plasma including collisions, indeed, non-linearly generalized for high-intensity laser fields but of the classical value for the normal plasma state. The code included the equipartition time for energy transfer from electrons to ions and vice versa, e.g., for those cases where the nonlinear force plasma dynamics were producing heat in the ions by adiabatic compression while the electron fluid would be heated at a different level. The thermal conductivity was included in the classical way as well as the collision produced viscosity.

This assumption for the thermal conduction may need a correction. Indirectly comparing computations for the laser compression and heating of spheri- cal pellets with experiments resulted in an inhibition of thermal conductivity. The classical Spitzer value of the electron thermal conductivity has to be corrected by an empirically determined factor of between 50 and 100 when deuterium pellets were irradiated. This inhibition factor corresponded very well with the fact that the electric double layer between the hot and the cold plasma prevented the electrons from transporting their energy such that the energy transport was no longer by the electron thermal conductivity but only by ion thermal conductivity. This could be explained very easily by the double layer, Fig. 2–11, contrary to numerous very obscure and extremely complicated theoretical models for understanding the inhi- bition factor. The ratio between both conductivities is given by the square root of the mass ratio, i.e., the thermal conductivity is reduced by a factor of 71 for the case of deuterium plasma. In all our calculations we ignored this inhibition mechanism since the thermal conduction processes do not have much influence on our results for picosecond laser–plasma interactions.

The genuine two fluid equations were the basis, for the geometry of perpendicular incidence of laser radiation, on a plasma in the following conservation equations:

(i) Continuity for electrons and ions:

$$\frac{\partial}{\partial t}(m_e n_e) + \frac{\partial}{\partial x}(m_e n_e v_e) = 0;$$

$$\frac{\partial}{\partial t}(m_i n_i) + \frac{\partial}{\partial x}(m_i n_i v_i) = 0.$$

(ii) Equations of motion for electrons and ions

$$\frac{\partial}{\partial t}(n_e m_e v_e) = -\frac{\partial}{\partial x}(n_e m_e v_e^2) - \frac{\partial}{\partial x}(n_e k T_e) - \frac{1}{8\pi}\frac{\partial}{\partial x}(E_L^2 + H_L^2)$$
$$-\frac{1}{4\pi} E \frac{\partial}{\partial x} E - m_e n_e (v_e - v_i) \nu;$$

$$\frac{\partial}{\partial t}\left(n_i m_i v_i\right) = -\frac{\partial}{\partial x}\left(n_i m_i v_i^2\right) - \frac{\partial}{\partial x}\left(n_i k T_i\right) + \frac{Z}{4\pi} E \frac{\partial}{\partial x} E + m_i n_i \left(v_e - v_i\right) \nu \, ;$$

(iii)   Energy conservation for electrons and ions

$$\frac{\partial}{\partial t}\left(n_e \frac{3}{2} k T_e\right) = -\frac{\partial}{\partial x}\left(n_e v_e \frac{3}{2} k T_e\right) - \frac{3}{2}\frac{k}{m_e} T_e n_e \frac{\partial}{\partial x} v_e$$

$$+ \frac{1}{m_e}\frac{\partial}{\partial x}\left(\kappa_e \frac{\partial T_e}{\partial x}\right) - \frac{3}{2}\frac{k}{m_e} n_e \frac{T_e - T_i}{\tau} + \frac{1}{m_e} W \, ;$$

$$\frac{\partial}{\partial t}\left(n_i \frac{3}{2} k T_i\right) = -\frac{\partial}{\partial x}\left(n_i v_i \frac{3}{2} k T_i\right) - \frac{3}{2}\frac{k}{m_i} T_i n_i \frac{\partial}{\partial x} v_i$$

$$+ \frac{1}{m_i}\frac{\partial}{\partial x}\left(\kappa_i \frac{\partial T_i}{\partial x}\right) + \frac{3}{2}\frac{k}{m_i} n_e \frac{T_e - T_i}{\tau} \, ;$$

and (iv) the Poisson equation (now considered with time dependence)

$$\frac{\partial E}{\partial x} = -4\pi e \left(n_e - Z n_i\right)$$

for the seven unknowns $n_e$, $n_i$, $T_e$, $T_i$, $v_e$, $v_i$ and the electric field $E$.

The seven equations determine the $x$- and $t$-dependence of the seven quantities together with initial conditions and with boundary conditions given by the time dependence of the laser radiation. This is evaluated for each time step by the spatial dependence of $W$ as a solution of Maxwell's equations of the electromagnetic wave including all reflections and absorption properties within the varying inhomogeneous plasma.

The propagation direction of the laser light is $x$ and the (linearly polarized) electric laser field amplitude is called $E_L$ (previously $E_y$) and the magnetic field $H_L$ while the electric (Langmuir) field in the plasma determining the plasma oscillations in the $x$-direction is $E$.

These equations can first be used to see how an inhomogeneous plasma behaves without laser irradiation. Having the initial conditions that the 1 keV deuterium plasma has an electron and ion distribution which is a linear ramp at time $t = 0$ increasing from half the critical neodymium glass electron density within ten wavelengths (10 µm) to the critical density (Fig. 6–11), the resulting longitudinal electric field is zero at the beginning. But then the faster electron fluid expands faster down the ramp until it is stopped electrostatically and returned, causing oscillations at the plasma frequency, depending on the actual electron density in Fig. 6–11. Since we have included the whole plasma physics including the collisional damping, we see in Fig. 6–12 a damping of the oscillations after 12 periods. After 40 periods the oscillations have nearly gone and a uniform internal electric field of about $10^6$ V/cm maximum has been created in agreement with the fact that the 1000 eV plasma decayed within $10^{-3}$ cm. These kinds of internal electric fields in a plasma are those which were realized by Alfvén (1981), Fälthammar (1987) and others.

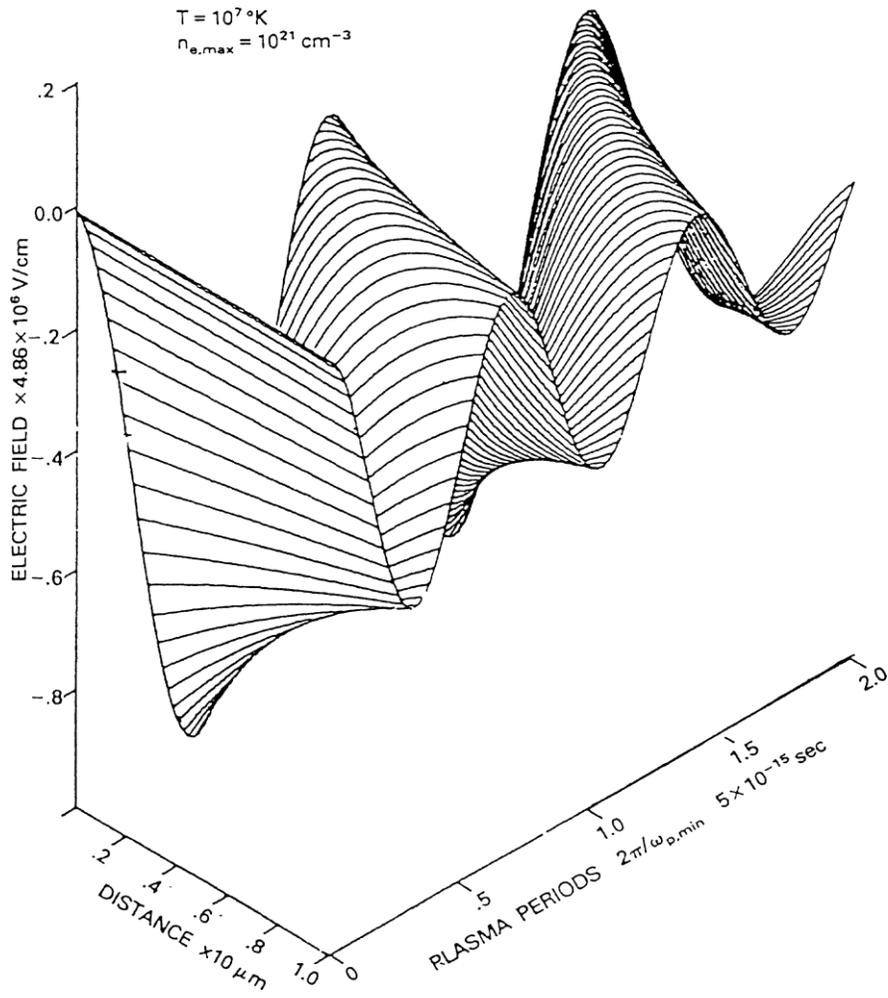

**Figure 6–11.** Time dependent development of the longitudinal dynamic electric field $E_s$ along the density with an initial resting ramp of linear plasma of initial temperature $10^7$ K and $5 \times 10^{20}$ cm$^{-3}$ at $x = 0$ and $10^{21}$ cm$^{-3}$ at $x = 10$ μm (Lalousis et al. 1983).

When switching on the laser field, the motion of the electron fluid can be computed including the generated longitudinal Langmuir oscillations whose electric field amplitude reached about one tenth (!) of the transverse laser field amplitude. The longitudinal Langmuir oscillations in the plasma can be seen, e.g., in Fig. 6–13, when a carbon dioxide laser radiation of 1/30 of the relativistic inten- sity is hitting a 15 wavelength long (initially parabolic) plasma ramp. These kinds of internal electric fields of very strong oscillations due to laser radiation, indeed, no longer permit conservative conditions, and it is then impossible to speak about "ponderomotive potentials."

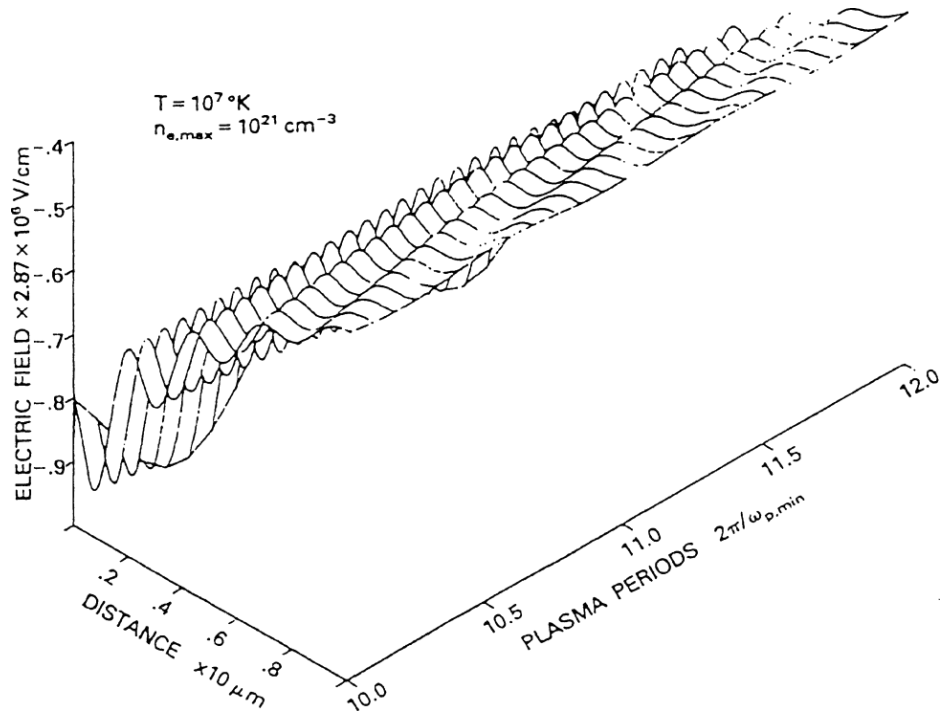

**Figure 6–12.** Same as in Fig. 6–11 for times 10 to 12 plasma oscillation periods (Lalousis et al.1983).

    The nonconservative character can be seen in Fig. 6–14 when a similar interaction (as in Fig. 6–13) is produced and when a single electron is followed up moving through such fields. The electron can receive a net loss or a net gain of energy. A keV electron gains approximately 100 keV energy (Eliezer et al. 1995). This type of laser plasma interaction may play a role in electron acceleration by lasers including plasmas (Joshi et al. 1994) where the laser automatically drives strong Langmuir waves without needing any beat wave configuration (Tajima 1985). It has been noted that 25 TW laser pulses have produced 30 MeV electrons in large numbers (Umstadter 1996), and the plasma acceleration mechanism may be similar to that in the experiments by Kitagawa et al. (1992) or by Joshi et al. (1994). However, there the number of the resulting MeV ions was about 4 orders of magnitude less than in the case of Umstadter (1996).
    Apart from the numerical results when solving the conservation equations at the beginning of this subsection (Lalousis et al. 1983) analytic result can be derived from the oscillation equation

$$\frac{\partial^2 E}{\partial t^2} + \nu \frac{\partial E}{\partial t} + \omega_{p0}^2 E$$
$$= E_{s0}\omega_{p0}^2 + \frac{4\pi e}{m_e}\frac{\partial}{\partial x}\frac{E_L^2 + H_L^2}{8\pi} + 4\pi e\nu\left(n_i v_i - Z n_e v_e\right), \quad (6\text{-}21)$$

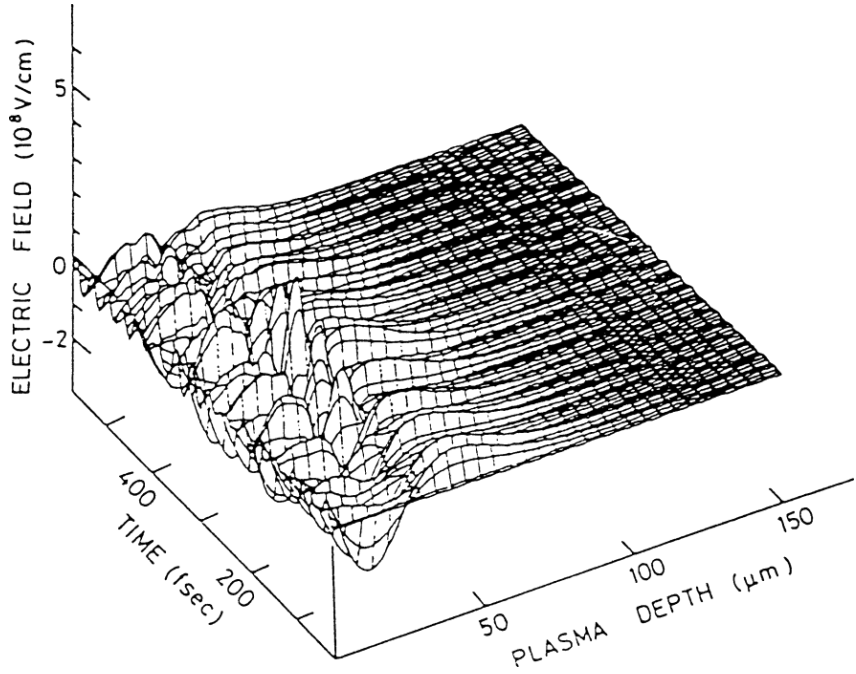

**Figure 6–13.** Longitudinal electric fields in a plasma of 150-μm depth developing in time for an initially parabolic density profile irradiated by $10^{15}$ W/cm$^2$ CO$_2$ laser light.

where

$$E_{s0} = \frac{4\pi e}{\omega_{p0}^2} \left[ \frac{\partial}{\partial x}\left( \frac{3n_i kT_i}{m_i} + Zn_i v_i^2 \right) - \frac{\partial}{\partial x}\left( \frac{3n_e kT_e}{m_e} + n_e v_e^2 \right) \right]$$

and

$$\omega_{p0}^2 = 4\pi e^2 \left( \frac{n_e}{m_e} + \frac{Z^2 n_i}{m_i} \right).$$

The solution arrives at the longitudinal electric ("electrostatic" Langmuir) fields, now called $E_s$ in an inhomogeneous (genuine two-fluid) plasma

$$\begin{aligned}
E_s = & \frac{4\pi e}{\omega_P^2} \left[ \frac{\partial}{\partial x}\left( \frac{3n_i kT_i}{m_i} + Zn_i v_i^2 \right) - \frac{\partial}{\partial x}\left( \frac{3n_e kT_e}{m_e} + n_e v_e^2 \right) + \frac{1}{m_e}\frac{\partial}{\partial x}\frac{E_L^2 + H_L^2}{8\pi} \right] \\
& (1 - \exp(-vt/2)\cos\omega_c t) \\
& + \frac{\omega_p^2 - 4\omega^2}{\left(\omega_p^2 - 4\omega^2\right)^2 + v^2\omega^2} \frac{4\pi e}{m_e} \frac{\partial}{\partial x}\overline{\left(E_L^2 + H_L^2\right)}\cos 2\omega t \\
& + \frac{2v\omega}{\left(\omega_p^2 - 4\omega^2\right)^2} \frac{4\pi e}{m_e} \frac{\partial}{\partial x}\overline{\left(E_L^2 + H_L^2\right)}\sin 2\omega t,
\end{aligned} \quad (6\text{-}22)$$

where the first line contains a term with the electron temperature $T_e$ which term is identical with Langmuir's ambipolar field as we saw first in the Debye sheath of the plasma surface (Fig. 2–2). All the other terms were new (Hora et al. 1984, 1985; Goldsworthy et al. 1986). All pressures in the plasma generate these fields, as well as the ion pressure or the kinetic pressures at varying magnitudes. We see, however, from the oscillating term in the bracket of the second line in Eq. (6–22) that we have an oscillation at the plasma frequency $\omega_p$ as seen in Figs. 6–11 and 6–12, damped by the collsion frequency $v$, even if there is no laser field. But with the laser field the electromagnetic field energy density [first part of second line in Eq. (6–22)] does contribute, in general, in a nonconservative way.

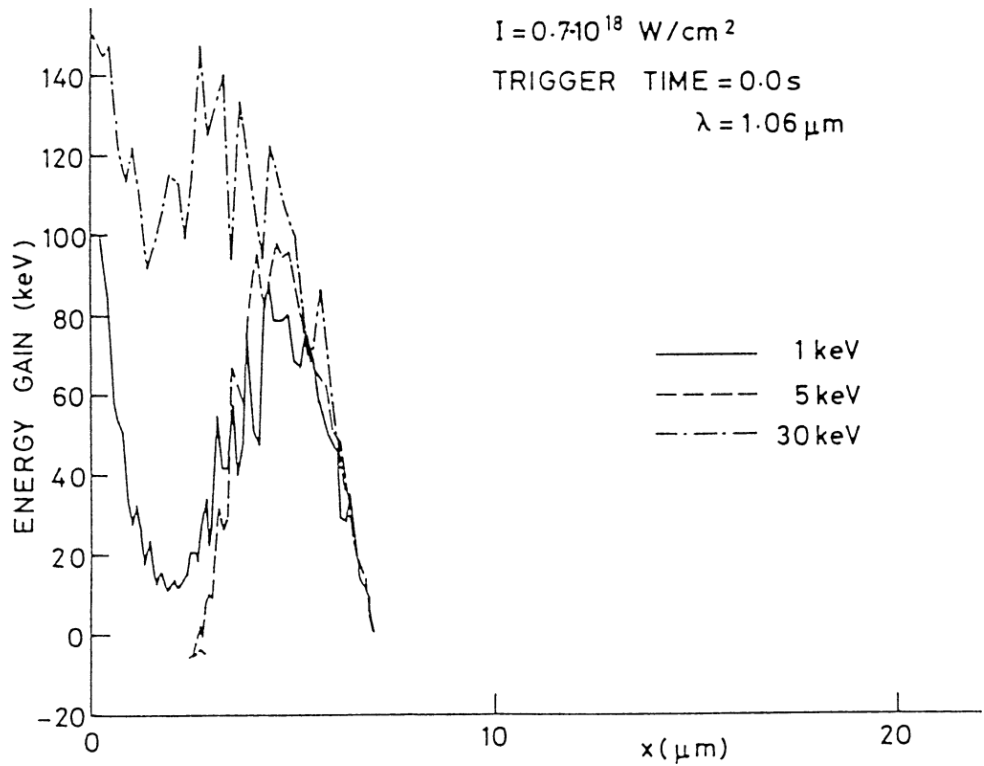

**Figure 6–14**. Computation of the energy gained by electrons starting with initial energies of 1, 5, or 30 keV at times zero from x = 7 μm towards negative x in a 20 wavelength thick plasma slab of initially parabolic density profile (Hora et al. 1984, 1989) at irradiation (from the left-hand side) by a neodymium glass laser of $7\times10^{17}$ W/cm$^2$ intensity.

    What is most interesting is that the last term in Eq. (6–22) results in a resonance driven by the laser radiation at four times the critical density. This can be especially strong when the caviton process results in a very strong steepening of the plasma density near the critical density. This new type of resonance (Hora et al. 1985) acts at perpendicular incidence (as was searched for earlier) of the laser radiation and has a second harmonic oscillation (Fig. 6–15).

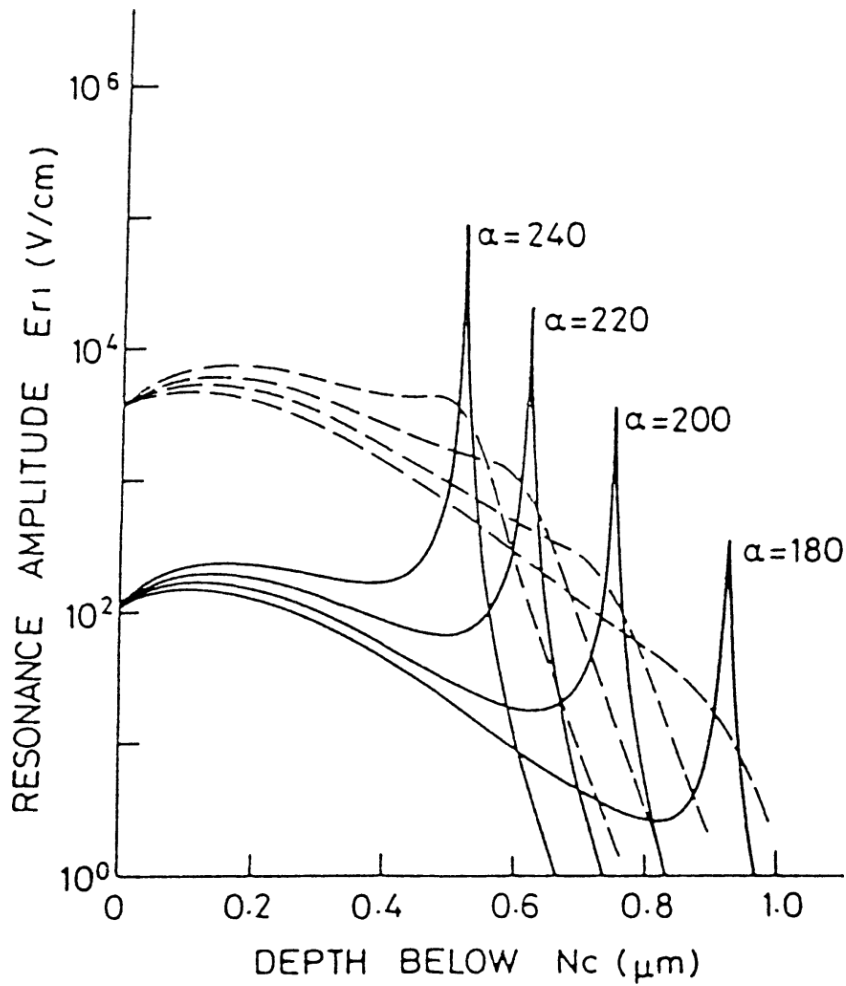

**Figure 6–15.** Amplitude of the longitudinal electric field second harmonic resonance oscillation according to Eq. (6–22) calculated for neodymium glass laser fields of $10^{14}$ W/cm$^2$ for linear density profile of varying steepness (parameter $\alpha$) depending on the depths below the critical density. Dashed curves are for the optical constants of a plasma of temperature 100 eV. Fully drawn curves are for the same plasma temperature, however, with the nonlin- ear optical constant including the quiver motion, Eq. (3–94), of the electrons by the laser field (Goldsworthy et al. 1988).

The second to last term in (6–22) has another important meaning: it refers to second harmonic generation in the plasma corona with nearly the same ampli- tude, even for very low electron density. This was ideally demonstrated experi- mentally (Alexandrova et al. 1985), Fig. 6–17, and could in no way be explained by parametric instabilities which only result from special electron densities, never from the whole plasma corona, especially not from the very low density at the periphery. The results of Fig. 6–16 explained this immediately, as well as the 50 μm long spatial oscillation as seen in Fig. 6–16. There is also a short-range (few-wavelength) modulation of the second harmonic emission which just at the time of the calculations of Fig. 6–16 was measured (Tan et al. 1987; Zhunqi et al. 1986). Broad band laser irradiation prevented the second harmonic generation in the whole plasma corona, and the modulation disappeared.

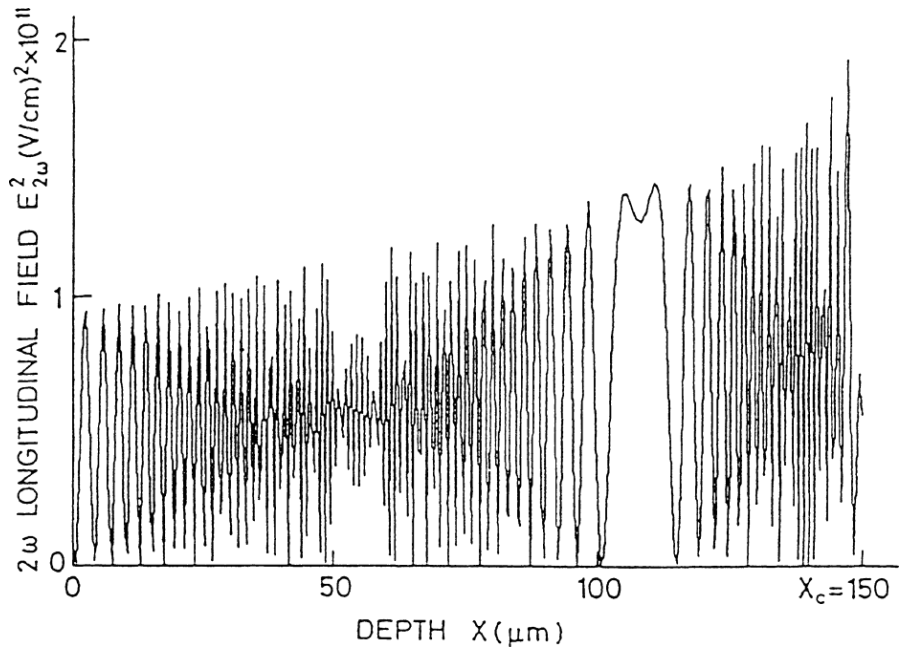

**Figure 6–16**. Amplitude of the second harmonic reactive response term (second to last term in Eq. (6–22) for a linear plasma corona from x = 0 to a critical density at 150 µm for $10^{14}$ W/cm$^2$ incident Nd glass laser irradiation [Goldsworthy 1988; Goldsworthy et al. 1988, 1990)]. The fine oscillations correspond to the observed striations (Tan et al. 1987) and the long periods of about 50 µm correspond to the observations by Aleksandrova et al. (1985), see Fig. 6–17.

We see that these interesting new results of the genuine two-fluid models were derived after starting the genuine two-fluid models to which we were motivated by the strong electric fields between the nonlinear force driven electron fluid, Fig. 6–1, and the following ions. The double layer between electrons and ions was another important new result treated by intense laser irradiation. It resulted in surface tension and in the stabilization of surface waves against the Rayleigh–Taylor instability (Eliezer et al. 1989). It even led to the first theory of surface tension of metals (for the degenerate electron gas) in agreement with measurement (Hora et al. 1989). The other considered model of surface tension using the jellium model arrived at incorrect negative surface tensions which is never possible with the plasma double layer surface tension model (Hora et al. 1989). This model was applied to the surface of nuclei (Hora 1991a, 1992), resulting in an explanation of Hofstadter's decay of the charge at the nuclear surface in the equilibrium leading to the well-known density of nuclei, and at (about six times) higher density in the mass independence of the particles involved due to the relativistic change of the Fermi energy. This explained the possibility of the quark–gluon plasma without any possible surface structure. Further, it clarified that nuclei with more than about 350 nucleons will never be stable. Connetions sould be mentioned to surface waves Kaw et al. 1970) and 1-D solutions (Kaw et al. 1992),

The genuine two-fluid computation led to the explanation of the observed inverted double layers as soon as the cavitons were generated (Eliezer et al. 1983). The calculation is seen in Fig. 6–18 with the corresponding electromagnetic energy density "trapped" in the caviton, Fig. 6–19, and the longitudinal fields now available from the theory, Fig. 6–20 (Hora et al. 1984a).

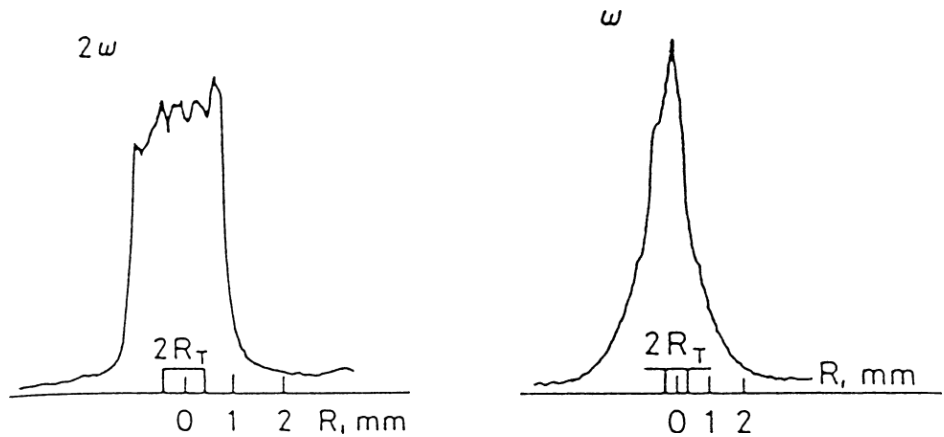

**Figure 6–17.** Side-on emission intensity distribution from a pellet of a diameter 2*R* irradiated by a rectangular 2 ns neodymium glass laser pulse in Delfin (upper part). The second harmonic emission is in the lower part and is nearly rectangular apart from oscillations. The narrow profile in the upper part corresponds to the strongly decreasing density in the outer plasma which, however, emits the second harmonic with unchanged strength even at very low density (Aleksandrova et al. 1985)

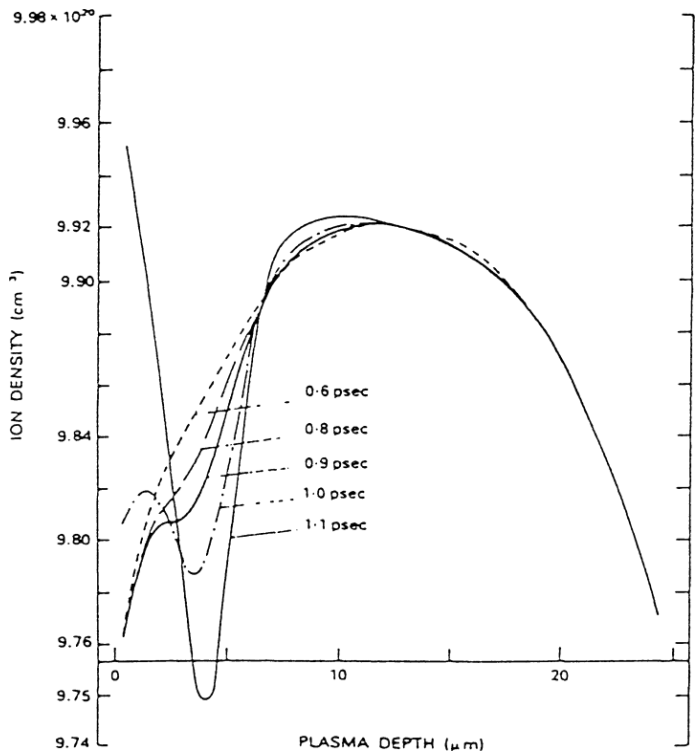

**Figure 6–18.** Ion density of a 25-μm thick hydrogen plasma slab initially at rest and 1- keV temperature irradiated from the left side by a $10^{16}$ W/cm² Nd:glass laser. At t = 0.6 ps the density is very similar to its initial value. The energy maximum near *x* = 4 produced a cavitation by nonlinear forces (Hora et al. 1984a).

The genuine two-fluid model also explained another very important observation: the stuttering or stochastic pulsation of the laser plasma interaction of about 20 ps duration. It was observed (Maddever et al. 1990) that a constantly incident laser intensity is changing the reflectivity of the irradiated plasma between a few percent and more than 95% in a pulsating way going up and down stochastically within a duration of about 10 to 40 ps. The plasma is accelerated to a few times $10^7$ cm/s velocity at each low reflectivity phase and then the interaction stops until the next low reflectivity occurs.

This could be understood from the earlier computation of Fig. 4–5: the laser light is first reflected like a mirror at the critical density, accelerating the whole corona to high velocity but causing also the self-generated density ripple (from nonlinear force pushing the plasma into the nodes of the partially standing wave). Then the nearly 100% reflectivity from a phase reflection at the ripples appears until, within a few picoseconds, the density ripple is abolished by hydromotion, and the process repeats. The pulsating acceleration of plasma blocks and the change of the localization of the reflectivity have been observed (Maddever et al. 1990). This stuttering is a basic mechanism in laser-produced plasma discovered only few years ago after so many other parametric mechanisms were incorrectly made responsible for the observed anomalies.

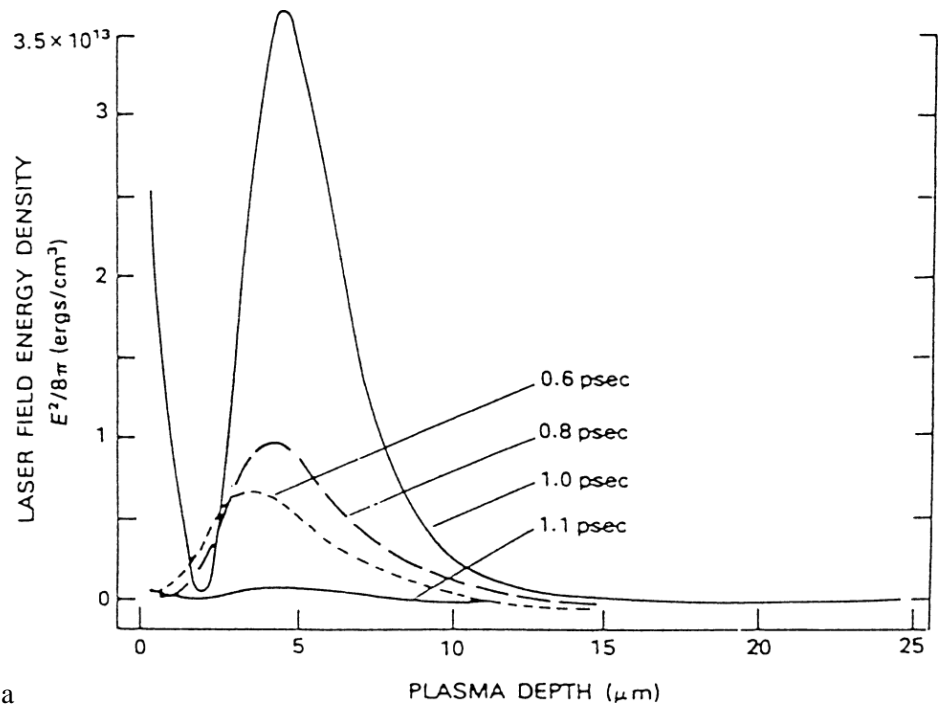

**Figure 6–19.** Density of the electric field energy of the laser (without the electrostatic fields generated within the plasma) for the case of Fig. 6–18.

Applying the genuine two-fluid model we see the following (Hora et al. 1992), Fig. 6–21. The initially linear density profile at laser irradiation receives a laser driven hit toward lower density until the density ripple appears and the accel-eration stops. A few ps later the ripple disappears hydrodynamically and the laser interaction works again until the next ripple appears. The same stochastic pulsa- tion can be seen from the electromagnetic energy density (Fig. 6–22) or from the ion velocity (Hora et al. 1992). If, however, a broad-band laser radiation is inci- dent, Fig. 6–23, the pulsation, for example, of the velocity field, disappears. This is the mechanism for smoothing of laser beams for a better interaction with plasma which— independently from the discovery of the stuttering—was intuitively discovered, the random phase plate smoothing (Kato, Mima et al. 1984), the spatial incoherence (Lehmberg and Obenschain 1983), and the broad-band irradiation (Deng et al. 1984).

The initial aim of this smoothing was to suppress the filamentation of the laser beams by self-focusing (Hora 1969a, 1975), and measurements by Labaune et al. (1992) showed how filamentation was reduced if a sufficiently fine structure random phase plate was inserted into the laser beam. Against these expectations however, there were structures seen perpendicular to the filamentation fully under- standable as a picosecond stochastic stuttering process if there was not sufficient smoothing. At sufficient smoothing, this stuttering was suppressed in a way simi- lar to our numerical results with the genuine two-fluid model when stepping from Figs. 6–21 and 6–22 to 6–26.

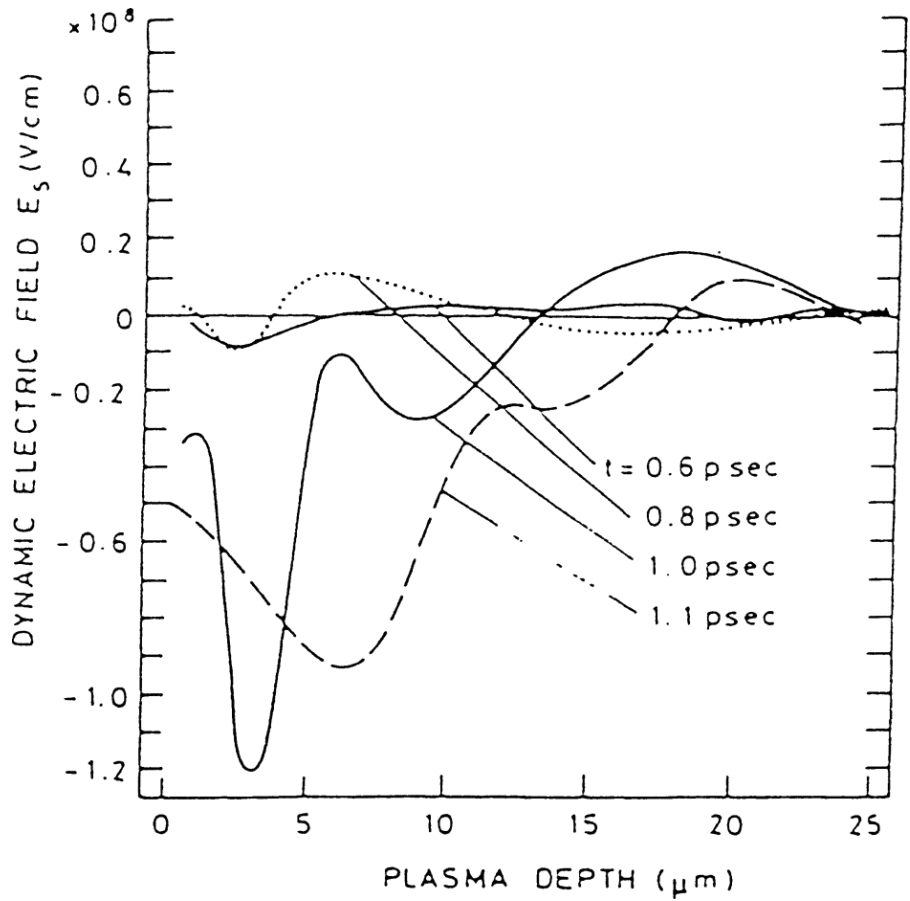

**Figure 6–20.** Electric field $E_s$ inside the plasma of Fig. 6–3, dynamically evolving with absolute values beyond $10^8$ V/cm near the caviton produced by the nonlinear laser forces.

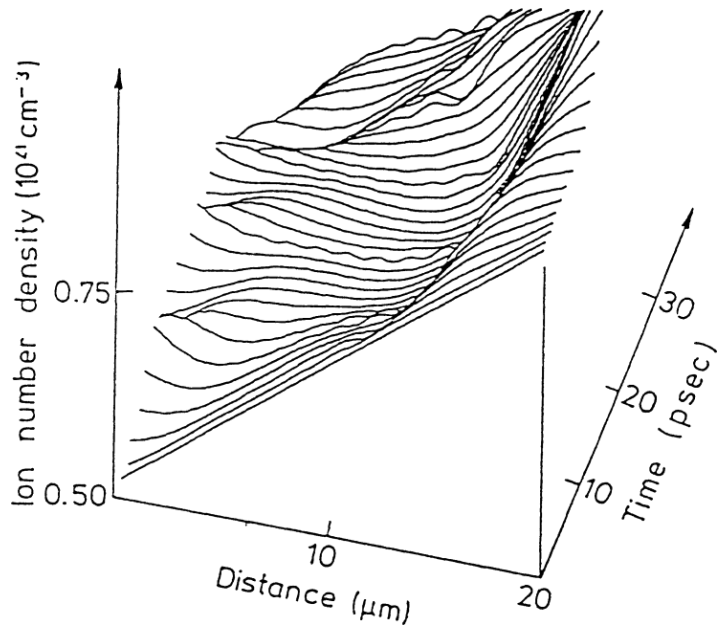

**Figure 6–21.** Time dependence of the ion density profile of an initial ramp when irradiated by neodymium glass laser radiation of $10^{15}$ W/cm$^2$ from the left-hand side with a pulsating generation and relaxation of density ripples at about 5, then at 25 and 40 ps.

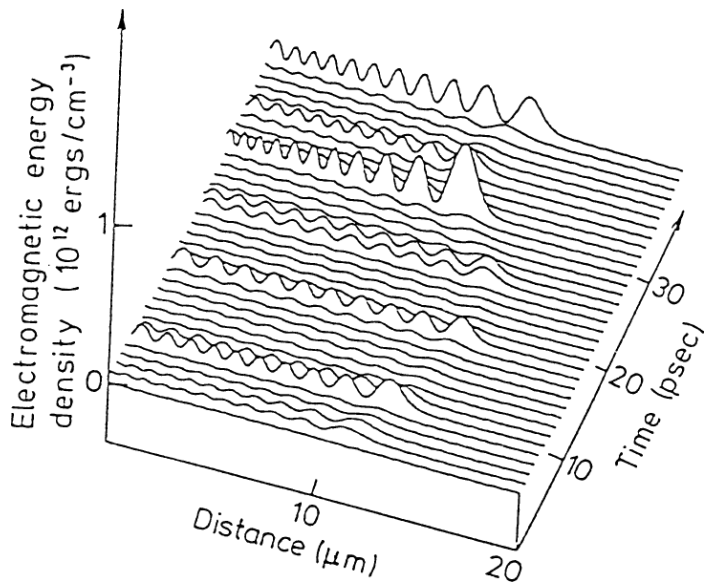

**Figure 6–22.** Time dependence of the electromagnetic energy density of the laser field as for the case of Fig. 6–21 showing times of strong penetration of the light to the critical density (strong maxima) and those with low penetration.

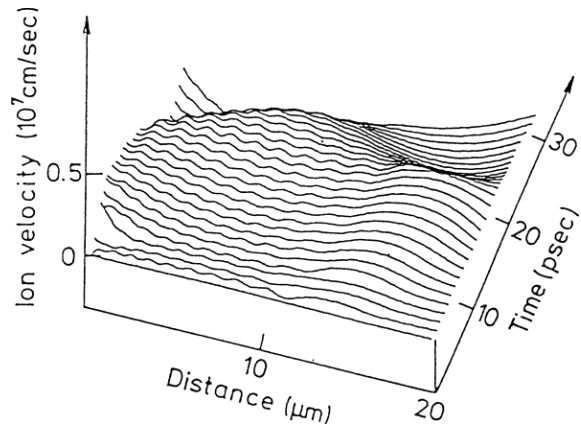

**Figure 6–23.** Ion velocity profiles depending on time of the plasma motion at the same laser irradiation as the cases in Figs. 6–21 and 6–22, however for a broad-band laser spectrum with three laser frequencies each 0.5% apart (Hora et al. 1992).

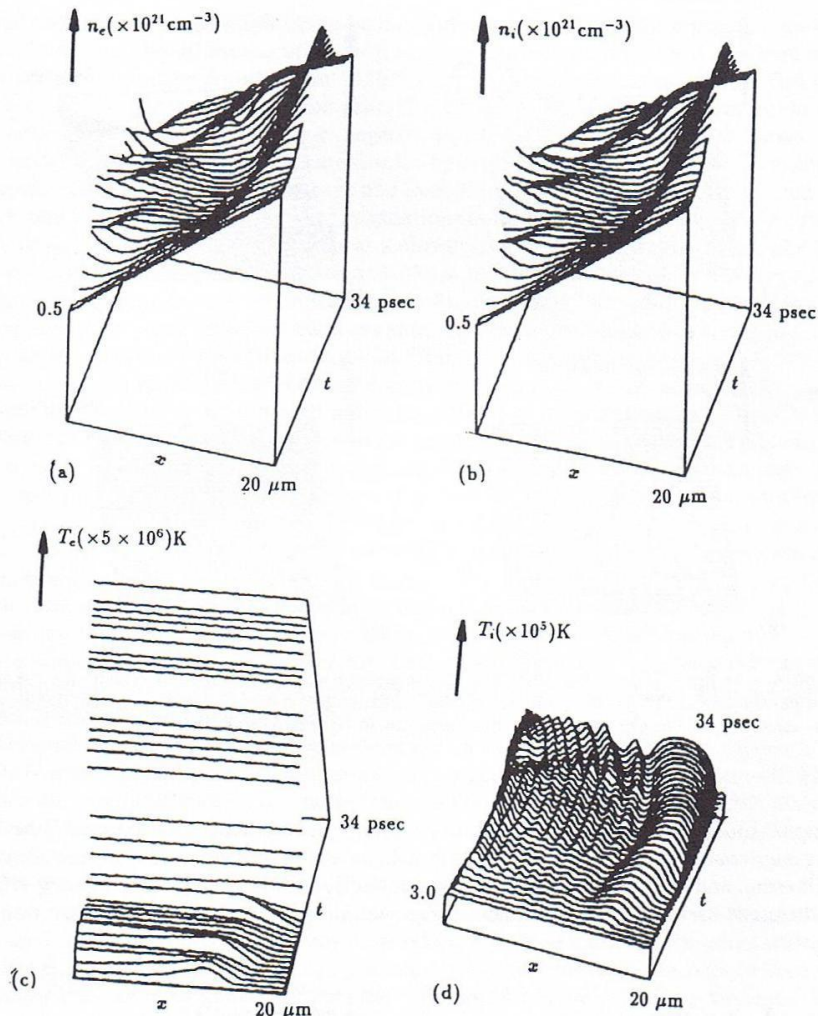

**Fig. 6-24.** Further details from the computations without smoothing of the cases of Figures 6-21 and 6-22: (a) electron density, (b) ion density, (c) electron temperature, (d) ion temperature (Boreham et al 1997).

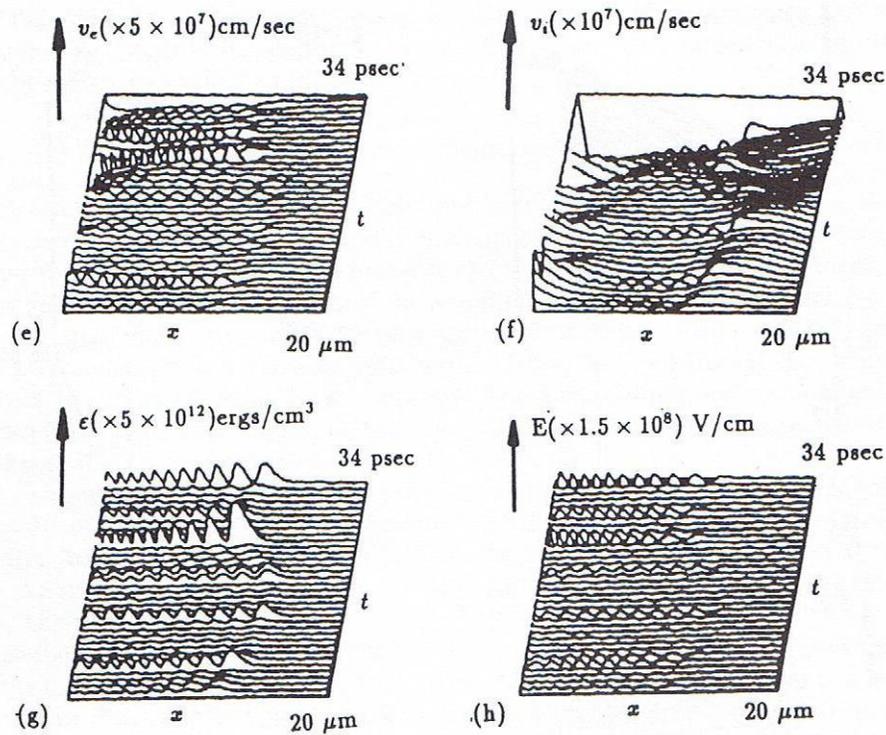

**Fig. 6-25.** Continuing Fig. 6-24: (e) electron velocity, (f) ion velocity, (g) energy density, and (h) electric field stgrength of the laser in the plasma.

It is interesting to mention the following results. The electron temperature (c) is slowly growing by direct laser-plasma interaction by thermal collisions of the quiver motion and later increasing by the equipartition mechansim with the ions. The ion temperature (d) is determined by adiabatic dynamics from the nonlinear force pushing off the plasma within the nodes of the partial standing waves which were not eliminated by smoothing as in Fig. 6-23. The velocities of the electrons (e) and of the ions (f) are all negative for the depths x of the interaction showing the motion against the laser direction at these values of x. The electron velocities within the standing wave ripples are rather high because the genuine two-fluid computation (Lalousis et al. 1983; Hora et al. 1984) shows how the electron fluid is driven along x by the nonlinear forces of the partial standing waves of the laser field and the ion fluid follows the generated field in this directed way. Figures (e) to (g) also show how the length of the ripples is increasing on the depth x according to the decreas of the refractive index on x resulting in an increase of the dielectric corrected laser wave lenth in the plasma up to the very long maximum in the same way as the effective laser intensity (Eq. 4-14) is swelling by the decreasing optical refraction **n** as seen in (d) from the nonlinear force adiabatically generated ion temperature close to the critical density.

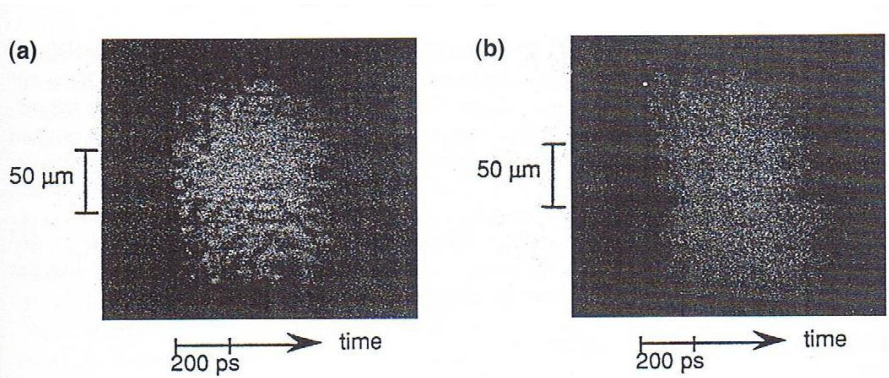

**Fig. 6-26.** Measurements of the side-on pictured of a laser irradiated plasma with insufficient (a) and sufficient (b) smoothing (Labaune et al. 1992).

Comming back to the result of smoothing to eliminate the pulsating interaction, Fig. 6-23 – with the directly measured stuttering interaction by Maddever et al. (1991) to be suppressed – this can be seen directly form the measurements of Labaune et al. (1992). It is shown that not all smoothing is sufficient and only higher resulution of the random phase plates (Kato et al. 1982; 1984) is necessary. Fig. 6-26 demonstrates in (a) the side-on picture of the emitted radiation from a laser produced plasma with lower resulution. There is no elimination of the self-focusing. Self-focusing can still be seen with the horizontal chanels. In addition also with vertical structures can be seen generated by the pulsating partial standing wave field produced nonlinear force generated structures. Only with higher resolution of the random phase plates (b), the smoothing is perfect without stuttering interaction and without filaments. The success with the smoothing (Hora 2006) led to the highest fusion gains at direct drive (Azechi et al. 1991). Some modified smoothing technique combining broad band (Obenschain et al. 1989) and related techniques (Skupski et al 1983) was used to achieve the highest direct drive volume burn/igntition of fusion (Soures et al. 1996), see Section 9.

An important consequemce of this result is that the appropriate smoothing will eliminate the most undesired anomalous effects in laser–plasma interaction which obviously are not due to the parametric instabilities but due to the stuttering. It is very probable then that the appropriately smoothed laser irradiation for inertial fusion energy can use the longer wavelength. In this case the now reached 1.9 MJ megajoule lasers (Hurricane et al. 2014) with the third harmonic of neodymium glass lasers (Campbell, Teller et al. 1997, Campbell 2006) may drive laser fusion much better at the fundamental wavelength with pulses of 5 MJ without much change or increase of costs (Hora et al. 1999; Hora 2006; Roth 2014). This suppression of the stuttering interaction by appropriate smoothing—to be developed by sophisticated experiments parallel to computations (Hora et al. 1992, 1999) —may solve the very serious problems of the design of the National Ignition Facility NIF (Gwynn 1999): this could have led also to a possible cost reduction of the laser beams in the future with respect to damage problems of optical components in the third harmonics and in expensive crystals for harmonics generation (see Section 9) and may still be of interest (Roth 2014). The techniques for about 1% laser broadband smoothing are encouraging (Bibeau et al. 2000) and the suppression of various instabilities was measured (Bodner et al. 1998), but the fact might not sufficiently have been realized that this is related to the stuttering (ps-pulsation) in a fundamental way (Hora et al. 1992, 1999).

# CHAPTER 7
# ADVANED LASER ACCELERATION OF ELECTRONS

The groundbreaking work about the beat-wave acceleration of electrons by Tajima and Dawson (1979) within the field of two laser beams with different frequency needed from the beginning some clarification. This was achieved by Joshi (Joshi et al. 1984, Joshi 2006). The merit to move this into the focus of interest was last not least reached by Sessler (1985) when he initiated conferences by the American Institute of Physics AIP led by Channel (1982) and Joshi and Katsouleas (1985). The fact that an optical wave packet of an infinite symmetric plane wave cannot transfere optical energy into free electrons (Sessler 1985) was to be understood while laser acceleration of free electrons was know e.g. from the Boreham experiment (Section 6.2) The advantage of the very high electric fields of laser pulses for accelerator application was very well interesting but the fields were directed into the wrong direction.

     The topic of this book is describing basic ingrediences about nonlinearities, the extreme high gradients of laser energy densities, the longitudinal field components of laser beams, and the extreme longitudinal "static" fields in double layers. These were well recognized in the earlier stages and were continued and filled up with the beginning of the AIP conference (Joshi et al. 1985) but the entire breakthrough was possible only by Mourou's discovery of the chirped pulse amplification CPA (5et al. 1986) where the numerically predicted (Hora 1981) and finally measured (Sauerbrey 1996) ultrahigh acceleration of plasma blocks led to the large scale step of physics developments, see the following Section 8.

     It may be of some advantage to recognize first several achievements of the laser driven electron acceleration in this Section 7 with reference to the basic property of nonlinearity (Hora 1988; Evans 1988) and the clarification of asymmetries in wave packets (Scheid et al 1989) before realizing and appreciating the next steps in Section 8.

## 7.1 FREE WAVE ACCELERATOR

There has been a demand for a long time to find alternatives to the classical accelerators for very high energies. The biggest existing electron accelerator was LEP (large electron positron) whose 27 km circular tunnel was changed into the Large Hadron Accelerator LHC at CERN in Geneva, now accelerating nuclei up to conditions producing Higgs particles. Since high power lasers can provide electric field amplitudes which are six or more orders of magnitude higher than that of other accelerators with electrostatic or microwave fields it is interesting to consider lasers for accelerators. This is despite the fact that the field is in the wrong (transverse) direction while one needs longitudinal acceleration. We shall see that the Lorentz force, the nonlinear force and/or the ponderomotive force can find a way out. This is therefore an example of the single electron motion by the nonlinear (ponderomotive) force.

We have seen from theory and numerical solutions at an early stage that the nonlinear force in plasmas after relativistic self-focusing should accelerate ions to 6 GeV energy (Jones et al. 1982) in agreement with later related measurements (Gitomer 1984). This was well acknowledged for the discussion of the laser accelerator (Sessler 1982). It should also be mentioned that double layer and internal electric field conditions in laser produced plasmas due to nonlinear forces can accelerate electrons to rather high energies (Fig. 6–14) in a very spontaneous and uncontrollable way. The high laser intensities in the plasma generate nonlinear forces for which ponderomotive potentials are no longer possible because of strong temporal variations of the laser and plasma conditions. The result is that neodymium glass laser intensities of $7 \times 10^{17}$ W/cm$^2$ speed up keV electrons to 100 keV simply by moving through the laser irradiated plasma (Eliezer et al. 1995). We estimated that in the case of the experiments with up to $10^{20}$ W/cm$^2$ laser intensities after relativistic self-focusing (Umstadter 1996), acceleration of electrons in the plasma to 20 MeV energy were possible, but again for sporadic electrons only and not for the bulk of all electrons in the interaction area observed. The other acceleration process of beat-wave acceleration (Tajima et al. 1979) uses the longitudinal field of the Langmuir oscillations or the Langmuir waves if one has a fully homogeneous (uniform) plasma since any inhomogeneous plasma permits only pseudo-Langmuir waves (Eliezer et al. 1989). This led to the developments of wake-field accelerator (Katsoules 2004).

Contrary to the laser accelerator, use of plasma effects is the scheme to apply the interaction with electrons by electromagnetic wave fields in vacuum. This *free wave accelerator* is basically free from plasma conditions. We need to watch the plasma processes secondarily only in some cases with respect to relativistic self-focusing (Hora 1975, 1991; Häuser et al. 1988, 1992). The scheme of the laser free wave acceleration was intended at looking into straightforward mechanisms where plasma and its highly complex properties are not involved. The first step was to provide conditions where the electrons were trapped within the intensity minima of a laser interference field by the nonlinear force and then accelerated when these minima are moved 7-1.

Since there was some concern whether this would be possible at all, a detailed numerical study with the necessary high-order solution of the motion of the electrons in the intensity minima and subsequent acceleration of the wave field clearly proved this mechanism (Cicchitelli et al. 1990). Figure 7–2 shows the trapping of electrons within two maxima of $\mathbf{E}^2$ of a standing laser wave field. It is very important to realize that the trapping is not a simple movement of the electrons to the nodes of the standing waves by the nonlinear force, as initially predicted by Weibel (1957) and others, but that the electrons are quivering and bouncing between two equipotential points in the standing wave field. This cannot be seen in the first and second approximations. For deriving an acceleration, at least the third order approximation is necessary. In the computations, iterations up to the eleventh order were necessary in Fig. 7.2 to arrive at converging results (Cicchitelli et al. 1990). The earlier picture of Weibel (1957) has to be modified in the sense that the electrons are in time average and have a higher concentration at the nodes of the standing wave.

When moving such a standing wave field with a constant acceleration we can then follow up the bouncing mechanism mentioned earlier and see that the trapped electrons are accelerated in the same way as one could have calculated trivially (Fig. 7-3). This very sophisticated exercise was necessary since there were some influential authorities who bluntly rejected the possibility of such a trapping and acceleration mechanism; they dictated that there is a slip of the electrons through the $\mathbf{E}^2$ maxima. This slip-through is indeed possible at very high acceleration (contrary to the normal conditions in Fig. 7–3), but even then a remaining net acceleration is achieved. Some historical steps of these developments are summarized (Hora et al. 2000).

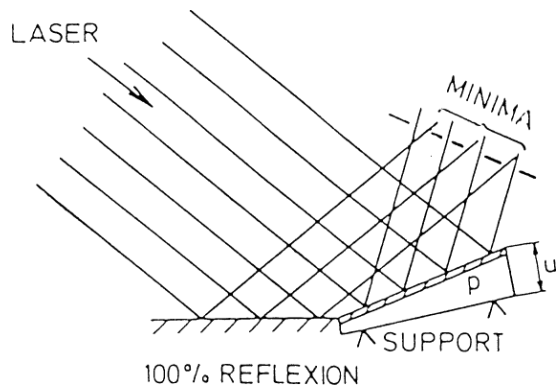

**Figure 7-1.** Acceleration of electrons by moving the minima of an interference field of one laser beam by a Fresnel mirror whose one mirror part is being turned in a controlled way piezoelectrically by a voltage $u$.

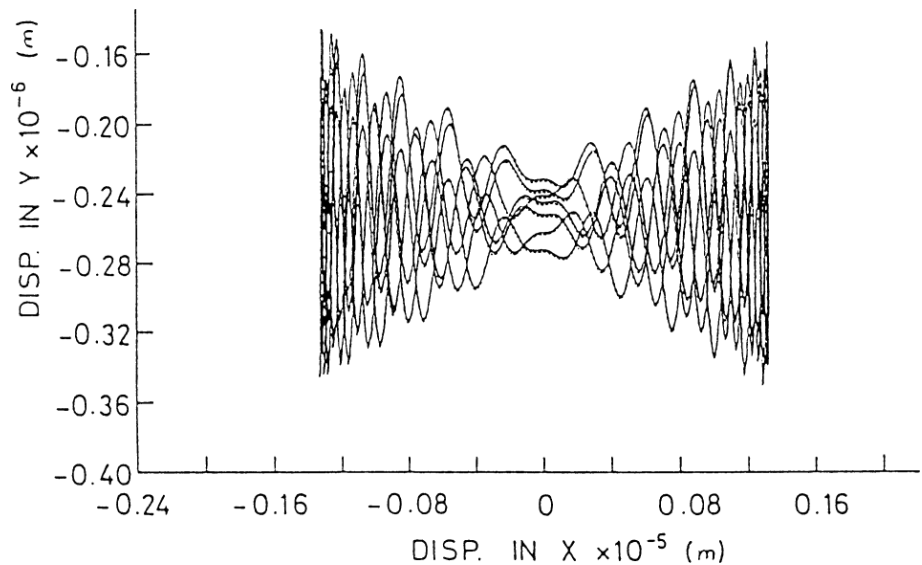

**Figure 7–2**. Trace of an electron in a standing wave field of a carbon dioxide laser of intensity $1 \times 10^{13}$ W/cm$^2$ starting from the half maximum intensity of the standing wave field. The mirror is stationary and the electron is trapped between two adjacent antinodes (Cicchitelli et al. 1990a).

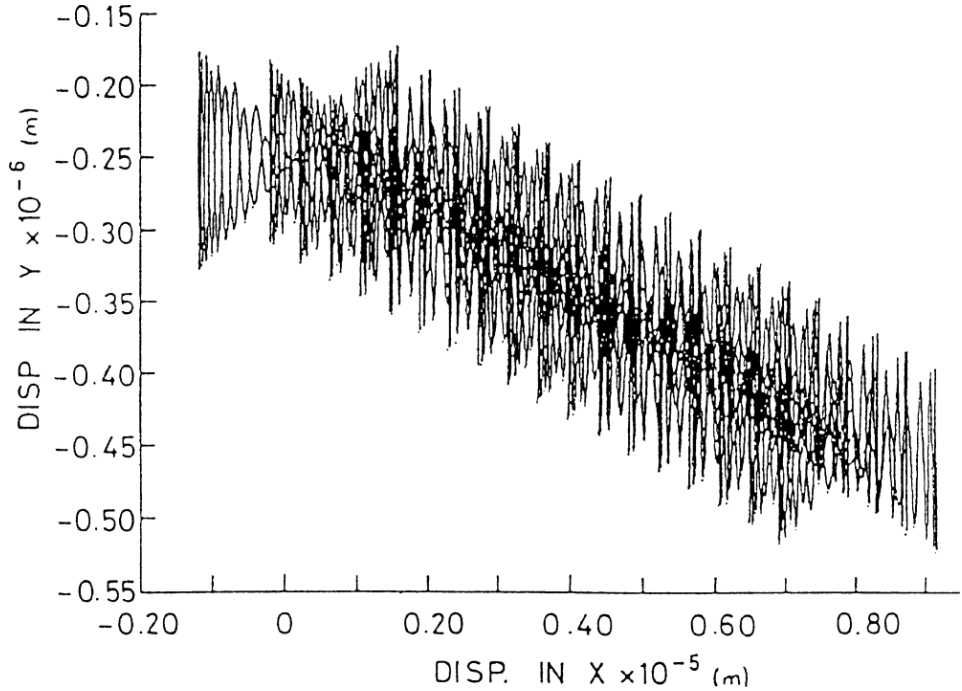

**Figure 7–3.** Trace of an electron in a standing wavefield of a carbon dioxide laser of intensity $1\times10^8$ W/cm$^2$ starting from the half-maximum intensity of the standing wave field. The mirror is accelerated with a parabolic velocity profile from zero to $1\times10^8$ cm/s in 7.6 ps then held at constant velocity for the remaining time (Ciccitelli et al. 1990a).

The second step was to evaluate the energy an electron receives within half a cycle of the laser wave. The relativistic motion could then be solved analytically in closed form (Scheid et al. 1989; Häuser et al. 1994). If the plasma wave prop- agates in the $x$-direction with the electric field oscillating in the $y$-direction, then the Lorentz $\gamma$-factor and the distances $x$ and $y$ that the electron is moved are

$$\gamma = 1 + \frac{2\delta^2}{k^2}; \quad x = \frac{3\delta^2\lambda^3}{32\pi^2}; \quad y = \frac{\delta\lambda^2}{4\pi}; \quad z = 0, \qquad (7\text{-}1)$$

where $\lambda$ is the wavelength and $\delta = (eE_0)/(mc^2)$ using the mass $m$ and charge $e$ of the electron and the amplitude $E_0$ of the laser electric field. Expressed by the laser power $P$, the Lorentz factor is

$$\gamma = 1 + \frac{2^{5/2}eP^{1/2}}{mc^{5/2}\pi} = 1 + \sqrt{\frac{P}{2.7\times10^9\,\text{W}}} \qquad (7\text{-}2)$$

After this gain the electron (within the phase symmetric plane wave) loses this energy within the following half cycle (Scheid et al. 1989), returning this energy to the initial value in the $y$-direction and being shifted by $2x$, showing no net energy gain at all (Sessler 1982, 1988). If the electron were to keep the gained energy, the second half-cycle would have to be removed by rectifying the laser beam (Scheid et al. 1989). Optical rectification has been achieved experimentally (Bonvalet et al. 1995 Rau et al. 1997). Alternatively, the electron can be moved into a sidewise cut wave (beam), and leaves the beam after half a cycle before being slowed down.

The same mechanism can be followed up numerically for the more realis- tic case (Häuser et al. 1994). If a laser wave packet (sometimes called a pancake of photons, Fig. 1–5) of a femtosecond duration is hitting a resting electron, its exact motion can be followed up. The fully phase-symmetric wave without any rectification and a Gaussian lateral intensity decay moves the electrons first side-wise by the transversal electric field component (Fig. 1–5) and then axially due to the Lorentz force. Because of the laterally varying intensity, the electron receives a net energy gain (Häuser et al. 1994) after the (phase symmetric) laser pulse has passed. This depends on the coordinates which the electron has at the beginning with respect to the beam center. The laser beam is described exactly by the complete Maxwellian solution including the previously unknown longitudinal field components (Cicchitelli et al. 1990), necessary to be added to the usual transverse components. The maximum net energy $\varepsilon_{max}$ of such an accelerated electron can gain—derived from higher order numerical results including the longitudinal field—is 57.5% of the energy $\varepsilon_0$ given for the plane wave solution. This is the free wave accelerator known since 1991 (Häuser et al. 1992a) in full agreement with the independently derived results of Woodworth, Kreisler and Kerman (1996). An example of the acceleration of electrons to TeV energy is given in Fig. 7–4.

## 7.2 UMSTADTER EXPERIMENT OF MeV ELECTRONS

We are able to explain, in detail, the experiment of Umstadter (1996) with accelerating $10^8$ electrons to about 30 MeV energy by a 400 fs 25 TW neodymium glass laser pulse. Basically the conditions are comparable with the Boreham experiment (Section 6.2). The nonlinear force acceleration happened in both cases to a free ensemble of electrons undisturbed by ions. This can be seen that the Debye length was essentially larger than the ensemble diameter for the electron cloude accelerated radially from an electron beam (Boreham et al. 1979). The same conditions are given at the very much higher laser intensities in the case of the laser beam crossing a gas jet. This condition of the Debye length was crucial (Boreham et al. 1979) to avoid an interaction with ions from the interaction volme. For the case of the experiment (Umstadter 1996a) with a peak power of 25 TW (only possible by Mourou's chirped pulse amplification CPA, see later section 8.2) we arrived from Eq. 7-2) with

$$\gamma = 97.225 \quad \text{or} \quad \varepsilon_0 = 49.7 \text{MeV} \qquad (7\text{-}3)$$

and therefore with inclusion of the longituidinal field components for a reduction to 57.5%, at

$$\varepsilon_{max} = 28.6 \text{MeV} . \qquad (7\text{-}4)$$

This value is in rather good agreement with the measured 30 MeV taking into account the measuring error in this femtosecond laser pulse technique. It is interesting to note that our computer result was achieved only if the longitudinal laser field (Hora 1981:Chapter 12.3; Cicchitelli et al. 1990) was included, otherwise the calculated energy $\varepsilon_0$ of about 50 MeV would have resulted.

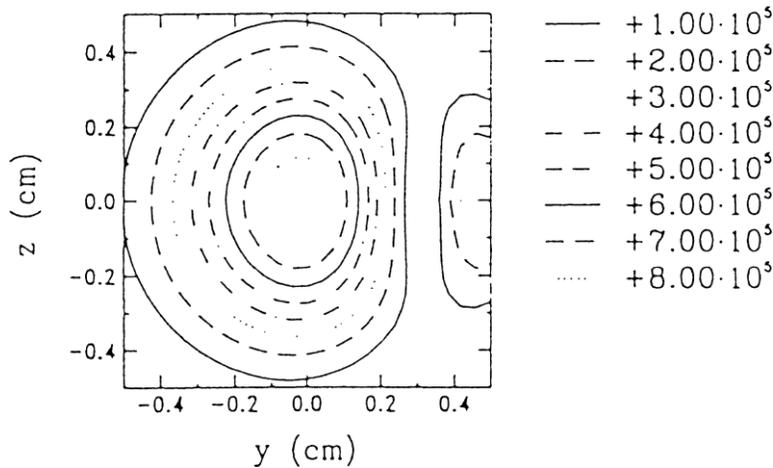

**Figure 7–4.** The Lorentz factor $\gamma$ of electrons initially at rest at $y$ and $z$ with respect to a one full-wavelength Gaussian $CO_2$-laser pulse of 0.5 cm radius and maximum electric field strength $3 \times 10^{12}$ V/cm using the exact laser field, linearly polarized with $E$ in the $z$-direction, and including the longitudinal components (Häuser et al. 1994).

A further interesting agreement of the calculations with the measurements is the angle of the electron emission. As was shown (Umstadter 1996), the diameter of the pulse of the 100 million 30 MeV electrons is about 2 cm at 8 cm distance from the interaction area corresponding to our calculated beam aperture of $u = 14.2^0$. Our calculation based on the half-wave acceleration model taking $\tan(u/2) = y/x$ from Eq. (7-1) for higher order approximation arrives at an angle of $15.9^0$, which value may be reduced by a small percentage for the realistic photon pancake case.

If this free wave acceleration is the main mechanism in the experiments (Umstadter 1996) one can conclude that the experiment with the 2 petawatt laser pulse (Perry et al. 1994; Perry 1995; Cowan et al. 1999), has produced again a similarly large (or even considerably higher) number as before in the case of Umstadter (1996) where the electron energy will be about 300 MeV electrons (Key et al. 2000) based on the square root increase of the electron energy on the laser power, Eq. (7-1). The calculation without the longitudinal laser field would have arrived at 500 MeV electrons. These numbers were expected also (Woodworth 1995) from the independent calculations (Woodworth et al. 1996). We have evaluated before in much detail (Häuser et al. 1992a; 1994) how a static magnetic field of 10 Tesla parallel to the laser $E$-field increases the electron energy, as first suggested from general considerations by Apollonov et al. (1988). The electron energy for the 2 petawatt pulse can then increase by the factor 2 in agreement with the calculations of Woodworth et al. (1995).

It is important to note that the electrons next to the optimized initial resting position with respect to the pancake center will achieve lower energies. Energy of this order of magnitude (similar to the cases as shown before, see Fig. 7-4) will be given to the major part of all electrons within the interaction region, thus explaining why such an astonishingly high number of electrons ($10^8$) was observed in the 20 to 30 MeV range, while the plasma beat wave acceleration (Kitagawa et al. 1992; Joshi et al. 1994) produced *about four orders of magnitude fewer electrons in the 30 MeV range.*

It should be mentioned that, surprisingly, the optimum energy also corresponds formally to the energy the electron would gain by the "ponderomotive potential," though this analogy is really only formal since there are complex switching on and off processes within the photon pancake. This is all covered by the nonlinear force (Hora 1969, 1991), or, identically, by the direct single particle motion described by the relativistic equation of motion including the Lorentz force (Häuser et al. 1992a, 1994).

Further, we have established that the relativistic self-focusing process (Hora 1975; Häuser et al. 1992; Kalashnikov et al. 1992; Pukhov et al. 1996; Nickles et al. 1996; Shkolnikov et al. 1997; Salamin et al. 1998; Hartemann et al. 1998; Wang et al. 1998) is possible under the conditions of the experiment of Umstadter (1996) where, however, the relativistic change of the electron mass results in a self-focusing length according to the initial theory (Hora 1975) of 57 wavelengths. This may relate realistically to the experiment. The number of electrons after tunnel ionization (Baldwin and Boreham 1981; Fittinghoff et al. 1992) with the smallest possible volume still exceeds $5 \times 10^8$ electrons, well in agreement with the experiment. We are aware of the necessary sidewise motion of the acceleration which is the range of a beam diameter of 7.4 μm (Häuser et al. 1992a, 1994) and which may correspond to the beam even before the final smallest diameter of relativistic self-focusing has been reached (Hora 1975). But this is sufficient for the acceleration, and the number of electrons within the corresponding volume has been estimated to be about $10^9$ electrons.

It should be mentioned that the intensity of blackbody radiation of temperature $mc^2$ intensity is close to this order of magnitude (Eliezer et al. 1986; 2002: see there chapter 1.5). It was mentioned in this connection that the energy distribution of electrons in this blackbody radiation is not longer that of the Fermi–Dirac statistics but modified to the strong coupling to the radiation, possibly following an intermediary statistics (Gentile 1940). It was calculated that the interaction of the radiation with the electrons is inelastic (pair production, etc.) to a part (Hora 1978b) of $15/\pi^5 = 4.9\%$ and the rest is elastic interaction (quiver motion), see p.13 of Ref. Eliezer et al, (2002). Gravitational effects at these extremely high optical intensities were also discussed (Hora and Novak 1977).

These results can be treated also for the question whether the Unruh radiation (Unruh 1976) has an acceleration $a$ of the electrons proportional to the temperature $T$ of the thermal bath or whether $a$ is proportional to $T^2$. The latter case can be related to the subrelativistic conditions while proportional to $T$ is a case characteristic to the relativistic electron energy at temperatures above $mc^2$, as can be seen in the similar derivation of the electron recoil similar to Einstein's proof of the needle radiation (Einstein 1917). This permitted a physics derivation of the fine structure constant (Hora et al. 1977) in order to express one of the three fundamental constants $e$, $c$, and $h$ by two of the others. Against precedings expectations, it was clarified that there is a difference between the Unruh radiation and the Hawking radiation (Stait-Gardner et al. 2006; Hora et al. 2011).

## 7.3 NONLINEARITY SOLVES LINEAR SUPERPOSITION QUESTION OF LAWSON-WOODWORTH PROBLEM

Considerable attention is given to the aim how to accelerate electrons by lasers— preferably in vacuum—to very high energies. The very high electric fields $E$ in a laser beam are by orders of magnitudes higher than in the classical particle accelerators, e.g. the initially reached (Perry et al. 1994; Cowan et al. 1999) petawatt laser pulses are focused in vacuum to intensities $I = 10^{20}$ W/cm$^2$ corresponding to an amplitude of the high frequency electric field of 2.7 × $10^{11}$ V/cm, about hundred times higher than the electric field in a hydrogen atom at the Bohr radius. One of the first conferences (Channel 1982) to use these fields for particle acceleration by lasers well appreciated (Sessler 1982) that ions can be accelerated to 0.5 GeV energy immediately (Begay et al. 1983, Haseroth et al. 1996) by irradiation of solid targets. The dielectric plasma effects cause a shrinking of the laser beam by relativistic self-focusing (Hora 1975, 1991; Esarey et al. 1997) to diameters of about half the wave length (Häuser et al. 1992) as measured (Basov et al. 1987) and understood from wave optics (Castillo et al. 1984) including a soliton mechanism (Häuser et al. 1992).

For the acceleration of electrons it was underlined (Sessler 1982) that the (transversal) *E*-field of the laser light goes into the wrong direction. Though we are focusing here only on the laser acceleration of electrons in vacuum, we should mention marginally that there was an early extensive discussion how the addition of plasma effects with its longitudinal electric fields (in the interesting direction!) can be used including plasma wave effects as the beat wave (Tajima 1985), wake field (Katsouleas et al. 1989) or the laser driven large amplitude longitudinal pseudo-wave (Eliezer et al. 1995) acceleration. These mechanisms, without deciding which of them, later produced small numbers of accelerated electrons in the range of few MeV (Kitagawa et al. 1992).

The mechanism of laser acceleration of electrons in vacuum was excluded in the earlier discussions by the well known result that a plane wave packet of electromagnetic radiation with symmetric phase properties can never transfer energy to a free electron if Thomson or Compton scattering or the Kapitza–Dirac effect is ignored. This result (Sessler 1982, 1988) is well known in the literature as an exact solution of the Maxwellian equations since the fiftieths, as reproduced later (Scheid et al. 1989). This fact from an exact solution of the equation of motion of electrons in a wave packet of infinitely spread plane waves is known as the Lawson–Woodward theorem. Since any electromagnetic field can be produced by linear superposition of plane waves, the impossibility of electron acceleration by lasers in was concluded (Sessler 1982).

Discussions with Lawson (1989) were about the trapping of electrons in vacuum within the intensity minima of standing wave or interference fields (Hora 1988a) and how the electrons are accelerated by moving the intensity minima, whether the electrons are really moved and do not slip through the intensity maxima. It was then shown by an extensive numerical work (Cicchitelli et al. 1990) with convergences only after up to the ninth iteration and realizing that the motion was only of a third order effect, that the electrons trapped in the intensity minima are not statically pushed to the minima as Weibel (1957) suggested but that there is a dynamic bouncing of the electron motion between equivalent field potentials. It was shown fully convincingly that the motion of the interference field carries on the electrons and result in an acceleration in agreement with the trivial calculation.

With respect to the argument of Sessler (1982) in the sense also of the Lawson–Woodward theorem there was the nontrivial question that if a phase symmetric plane wave packet cannot accelerate electrons, and since any electromagnetic field can be produced by linear superposition of infinite plane waves that there can never be an acceleration of electrons by laser fields in vacuum. This argument overlooked the fact that the superposition was linear while the electromagnetic forces to the electrons are basically nonlinear (Hora 1969, 1985) as clarified (Hora 1996) by distinguishing between the Lorentz forces from ponderomotive processes and how this is generalized in the nonlinear force (Hora 1969, 1985,1991, 1996). The complexity of the physics in connection with the classical ponderomotive force and the nonlinear force has been elaborated (Hora 1996, 1999) and apart from clarifying numerous points, several open questions were underlined especially with very short time interaction (Hora et al. 1996a).

One key question for the laser acceleration of electrons in vacuum was the breaking of the symmetry of "plane waves" and/or "phase symmetry in a wave packet" such that laser acceleration of electrons in vacuum does happen contrary to the before mentioned arguments. An active modulation of the phase by electro-optical crystals for the superposition of two laser wave fields was elaborated (Hora 1988), see Fig. 8-5, as a first step especially highlighted (Evans 1988) as further evaluated by Du and Xu (1999).

The next step was to study plane waves with completely phase asymmetric pulses by using "rectified" laser fields where only half waves were used while the following half waves were eliminated as in ac-rectifiers (Scheid et al. 1989). The advantage was that exact solutions for the single electron motion using the relativistic Lorentz force were derived. These results were used to calculate the *maximum energy* an electron achieves during such half-wave interaction in order to find the laser pulse parameters for TeV electron generation, e.g., by injection of electrons into box-like laser beam profiles (Häuser et al. 1992a). But even when using phase symmetric Gaussian beams, an enormous acceleration was seen if the beam width had a minimum value and a single wave length pulse was running over the initially resting electron. The final electron energy was very close to the before mentioned maximum energy (Häuser et al. 1994). It was essential that the earlier discovered (Hora 1981; Cicchitelli et al. 1990a) Maxwellian exact laser fields including the necessary longitudinal component had to be used.

The longitudinal components for a single beam acceleration was decreasing the electron energy (Häuser et al. 1994) against the initial expectation and contrary to the scheme of a two beam crossing acceleration scheme (Caspers et al. 1991; Scully 1990; Takeuchi et al. 1998). All these results were considered with hesitation in view of the linear superposition argument of Sessler (1988) or the alternatively formulated Lawson–Woodward theorem. A breakthrough of about the laser acceleration of electrons in vacuum appeared with the work of Woodward et al. (1996), where essentially the same results and about the same gained maximum electron energies (modified since the longitudinal laser field was not included) were derived independently and in a different way than before (Häuser et al. 1994).

The final persuation – after inclusion of the longitudinal components arriving the at the measured values - for the acceleration of electrons by lasers in vacuum was given by the experiment at Limeil–Valenton (Malka et al. 1997; Lefebvre et al. 1989) where the gain of MeV energy by electrons interacting with lasers in vacuum was measured. Very extensive computations followed (Hartemann et al. 1998; Wang et al. 1998) such that there was no more any doubt that the mechanism does work. This acceleration was described (Sprangle et al. 1996) as "vacuum laser acceleration" or as "violent acceleration" (Wang et al. 1998), in the scheme of "free wave acceleration" (Woodward et al. 1996), as "ponderomotive scattering" (Hartemann et al. 1998) or as the "vacuum beat wave acceleration" (Sprangle et al.1996).

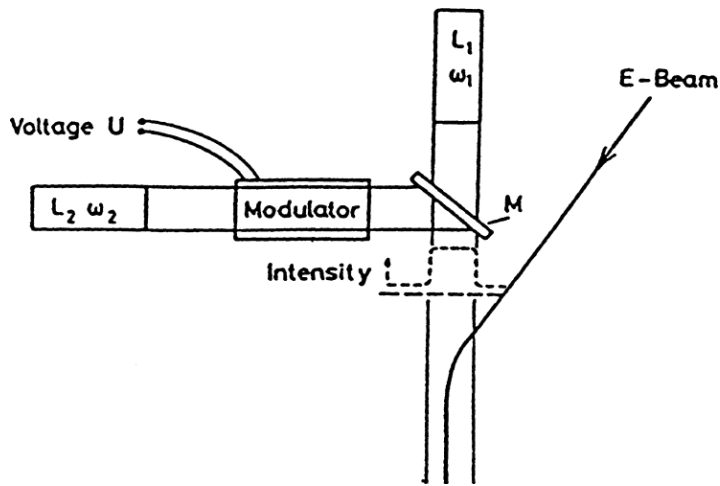

**Figure 7-5.** Superposition of laser beams $L_1$ and $L_2$ with frequencies $\omega_1$ and $\omega_2$ by a mirror M ans using an active phase modulation by applying a voltage $U$ to an electro-optical modulator causing a controlled motion of the minima of the interference field into which electrons from a beam are injected for acceleration by the acceleration of the intensity minimum (Hora 1988).

It is curious to understnad why the experimental results of the electron acceleration by laser in vacuum (Malka et al. 1997, Lefebvre 1998) were not accepted (McDonald 1998) as facts. The difficulties to understand this acceleration were based on similar confusions (McDonald 1989) as known from the beginning (Hora 1988; Evans 1988) of these consideration, while the fully clear theory had been published before (Häuser et al. 1994) or some other convincing theoretical facts had been given, e.g., by Mora et al. (1998). The earlier work (Scheid et al. 1989; Häuser et al. 1982a) using half wave interaction are now supported (Rau et al. 1997) in view of the experimentally verified ingenious realization of rectified laser beams(Bonvalet, Migus et al. 1995) of subcycle laser pulses (Domier, Luhmann Jr. et al. 1995; Raman, Bucksbaum et al. 1996).

While the basic mechanism of the laser acceleration of electrons in vacuum is now settled (Hartemann et al. 1998; Sprangle et al. 1996, Wang et al. 1998) since the first breakthrough in 1988 (Hora, Evans) there is an interesting problem to be considered when comparisons between experiments and theory and computations are to be done now in the details for the next measurements. It has been experienced before (Hora 1981; Cicchitelli et al. 1990a) that a wide range of discrepancies can appear if nonlinear processes are to be analyzed theoretically.

Inacuracy problems are well known in physics, especially when treating nonlinear problems, see the remarks by Feynman (Hora 1996, 1999 Chapter 6.3). We have to point out here that these inacuracies are not gradually only by adding or neglecting higher order terms or by differences of the results by few percentages. We have to realize that a basically new phenomenon has appeared in nonlinear physics where the addition or neglection of very tiny quantities can change the results from completely right into wrong or effects from "yes" into "no" contrary to observations. This will be more detailed in the following section. This paper describes then a similar example in the theory of nonlinear physics with the case of the relativistic electron acceleration by lasers in vacuum (Hoelss 1998, Hoelss et al. 1998a, 1999) as will be reported here in more detail.

This experience should teach how cautiously one has to proceed with the theory of nonlinear processes. On the other hand it opens a door of systematically discovering basically new effects of nonlinearity as a very new dimension of exploration in physics in contrast to the view of the saturation of physics knowledge articulated by Stephen Hawkings and Carl Friedrich von Weizsäcker, see Section 6.3 of Hora (1996). This again is the reason that the next section is first discussing the "principles of nonlinearity" before the mentioned results of our studies on the accuracy of computations for the laser acceleration of electrons in vacuum will be presented in the then following Sections.

## 7.4. A SCALE FOR MAXIMUM ELECTRON ENERGY

We summarize some earlier results from exact calculations about the max- imum energy an electron can obtain in a laser field. The calculation (Scheid et al. 1989) for a fully rectified laser field of infinite plane waves shows exact solu- tions for the motion of the electron inside the first half wavelength. The electron is moved first to the side, i.e., perpendicular to the direction of the wave along the $E$-field (linearly polarized in the $x$ direction) and then in a bent motion in forward propagation direction ($z$-direction) due to the Lorentz force by the $H$-field (in the $y$-direction). At very high laser intensities the relativistic motion of the electron is driven nearly completely into the direction of the laser radiation.

The exact analytical solutions (Scheid et al. 1989) for this half wave acceleration of the initially resting electron result in a relativistic $\gamma$-value or a translative energy $E$ for neodymium glass lasers of 1053 nm wave length in

$$\gamma = 1 + 1.62087 \times 10^{-18} I \quad (I \text{ in W/cm}^2), \tag{7-5}$$

$$E = 8.283 \times 10^{13} I \text{eV} . \tag{7-6}$$

During this acceleration the electron performs a sidewise motion in the direction of the electric field (for $\gamma \gg 1$)

$$x = 4.74 \times 10^{-14} I^{1/2} \text{cm} \tag{7-7}$$

and a motion along the direction of the propagation of the laser field

$$z = 3.20 \times 10^{-23} I \text{ cm} \tag{7-8}$$

Taking a box-like cross section of the plane wave with a square side of $x$, we express the $g$-value and the translative energy $E$ of the electron in terms of the laser power $P = Ix^2$ :

$$\gamma = 1 + (P/8.551 \times 10^8)^{1/2} \quad (P \text{ in W}) \tag{7.9}$$

$$E = P/(3.275 \times 10^3)^{1/2} \text{eV} \tag{7.10}$$

These results implied the important condition that the cross section of interaction is exactly determined by the value of the sidewise motion $x$. Formulas (7-9) and (7-10) are independent of the wave length.

These values are the highest possible energies $E$ an electron can achieve in the laser field by an exact plane wave and half wavelength interaction. Figure 2 shows the result where the intensity $I$ is independent of the wave length. The forward motion of the electron in the $z$-direction (Eq. (7-8) is given by $X$ and the sidewise motion along the electric field ($x$-direction, Eq. (7-7)) by $Y$ for neodymium glass lasers. One example is the question of the values for reaching TeV electron energy without speculating how to rectify the laser wave and how to produce a half-wave pulse. The result is that one needs a laser intensity of $I = 1.21 \times 10^{24}$ W/cm$^2$; the sidewise extension is $x = 0.521$ mm. The length of interaction, i.e., the path along the electron is carried by this pancake-like laser pulse (see Hora 1996, Fig. 1.5) at nearly the speed of light and receiving its energy by shifting from the front edge to the end edge of the laser half-wave, is $z = 38.6$ cm. Due to the sidewise motion, the laser power $P = Ix^2$ is then $3.275 \times 10^{21}$W $= 3275$ exawatts. Such an accelerator with a length of 39 cm only instead of the dozens of kilometers of classical linacs indeed needs enormous laser capacities which finally may result in lower costs than the costs of a linac. For the TeV electrons it was estimated that the energy loss by bremsstrahlung is sufficiently low and that a luminosity of up to $10^{33}$ cm$^{-2}$s$^{-1}$ may be achieved (Häuser et al. 1992).

The just mentioned values are the absolute maximum energies one can reach with the mentioned intensities and powers for a half wave acceleration. In practical cases one cannot have the rectified pulses and not the box-like cross sections of plane waves. Going to a cylindrical laser beam with the radius $x$, the maximum $g$-value or the energy $E$ of the electron after a half-wave is (Häuser et al. 1994) (independent of wave length)

$$\gamma = 1 + P/(2.69 \times 10^9)^{1/2} \quad (P \text{ in W}), \tag{7-11}$$

$$E = P/(1.03 \times 10^{-2})^{1/2} \text{ eV}. \tag{7-12}$$

But even this rectified wave field is not easily possible. Using a Gaussian beam of one full wave length with symmetric phase, nevertheless the sidewise motion of the electron into areas of lower intensity does result in an energy loss of a few or several percent below the mentioned maximum values. If the calculation includes the longitudinal components of the laser field, again a reduction by about 40% of the electron energy occurs (Häuser et al. 1992, 1994).

The result is that acceleration of the electrons in the laser field up to the range of the maximum energies can be expected. For the 2 petawatt laser at Livermore (Perry et al. 1994; Cowan et al. 1999) the maximum electron energy to be gained per interaction is given by

$E_{max} = 441$ MeV (for 2 petawatt, wave length independent) (7-13)

which again may be reduced to 60% or a similar value when the exact laser field with the longitudinal components will be applied. The minimum radius of the beam for this interaction is $x = 0.011$ mm for a wave length of 1053 nm. The use of this minimum radius, given by Eq. (7-7), is essential for the design of experiments. If the radius is too small, the maximum electron energy, Eq. (6) or (7) or a little below these values can never be reached. Larger focus values may provide an electron motion within several wave length of the laser beam and may result even in higher values than the here given single-half wave exact maximum values.

There is a simple understanding for the acceleration of the electrons ex-pressed by Mori (1999). Since the magnetic field cannot transfer energy to the electron but can only bend the sidewise motion due to the electric field, it is interesting that all the before mentioned results can be explained in an energy gain of the electron by integrating $E$ along the sidewise motion. The result is exactly the energy which the electron receives by the complicated final motion into the forward direction caused by the Lorentz-force mechanism. The only question is how far does the integration go into the $x$-direction. The answer is indeed the very complicated motion described (Häuser et al. 1994; Hora 1988; Sprangle et al. 1996; Hartemann et al. 1998; Wang et al. 1998), resulting in the energies in the range of Eqs. (2), (6) or (7). Therefore the energy input by the electrical laser field goes indeed in the wrong direction but thanks to the Lorentz force the electron trajectory is bent in the direction of the laser beam.

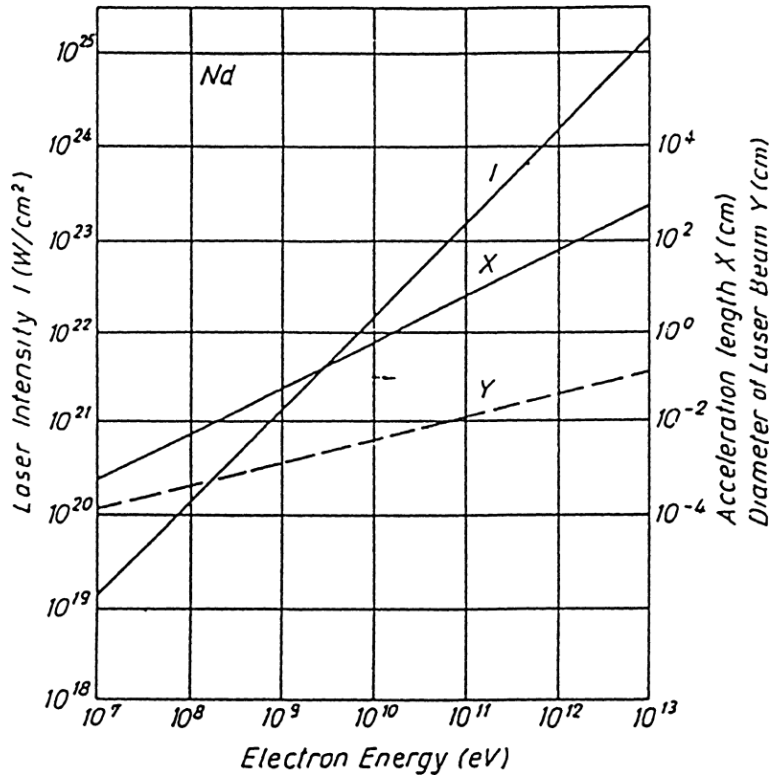

**Figure 7-6.** Maximum electron energies gained by electrons by sidewise injection in a nearly box-like laser pulse of given intensity $I$ with the necessary minimum width $Y$ (corresponding to $x$ in the preceding formulas) resulting in an acceleration length $X$ (corresponding to $z$) before the electron is ejected from the beam after a half wave length interaction (Häuser et al. 1992).

As an example for the agreement of the calculated values of Fig. 2 with measurements we mention the following detailed analysis of the Umstadter (1996, 1996a) experiment. The experiment was producing more than $10^8$ 30-MeV electrons when a 30 TW laser pulse hits an atmospheric pressure gas puff. This was the right order of the number of electrons expected from the here discussed electron acceleration by lasers in vacuum contrary to the much smaller numbers of MeV electrons measured before (Kitagawa et al. 1992). The measurement of Um- stadter et al. (1966, 1996a) could immediately be explained (Hora et al. 1997) by the acceleration of the electrons in the "vacuum" including relativistic self focus- ing (Basov et al. 1987, Castillo et al. 1984; Esarey et al. 1997; Häuser et al. 1992; Hora 1975, 1991; Hora et al. 1978a; Jones et al. 1982; Lezius et al. 1998; Hain et al. 1997; Kumar 1998; Lezius et al. 1989). The number of all electrons in the volume of the focus with a radius fulfilling the optimum condition of Eq. (3) is $10^9$ what is close to the measured number of Umstadter et al. (1996, 1996a). The energy of the electrons following Eq. (7) is near 50 MeV which has to be reduced by 40% due to the longitudinal laser field arriving just at the measured 30 MeV.

Exactly the same mechanism is the basis of the explanation of similar experiments using self-focusing by a numerical particle in cell PIC description (Kalashnikov et al. 1994, Meyer-ter-Vehn et al. 1999).

## 7.5. PARAXIAL APPROXIMATION AND EXACT PRESENTATION OF THE LASER FIELD

Taking the complexity of the accuracy principle of nonlinearity into account, it is no surprise that some attempts for the numerical computation of the energy gained by an electron in a laser beam may be much lower than the experimental value though the initial conditions of position and energy of the electron before the interaction and the phase of the interaction are further parameters to be taken into account. The energy gained by an initially resting electron when put into a neodymium glass laser beam of $1.2 \times 10^{24}$ W/cm$^2$ and 0.168 mm half width ra- dius with lateral coordinates $x$ (along the electrical field) and $y$ (along the magnetic field) is shown in Fig. 7-6 (Hoelss 1998). The parameters were chosen that a max- imum energy towards TeV were to reached. The longitudinal laser fields were all Maxwellian exact. It should be mentioned that these results are similar to that cal- culated before for the corresponding conditions of a carbon dioxide laser pulse (Häuser et al. 1994, see Fig. 7-6) though these calculations had only an approxima- tion for the longitudinal laser field. We note here that the little deviations of the longitudinal field for the carbon dioxide case from the exact case are not affect the result remarkably. We also underline that the maximum energy of the electrons shown in Fig. 7-6 well reach the order of magnitude of the absolute maximum value of Fig. 7-5 since the energies reached for the case of the realistic acceleration is about half of the absolute maximum.

We demonstrate now examples of the computations with different repre- sentations of the laser field. Figure 7-7 shows (Hoelss 1998) the result for a laser field from the exact calculation using the superposition of plane waves according to Cicchitelli et al. (1990a). The electric field component $E_x$ depending on the ra- dius $r$ of the beam is shown in (a) with its maximum value at the beam propagation length $z = 0$ of the focal center and how this changed following the propagation of the beam along higher values of $z$. The Gaussian decay of $E_x$ at $z = 0$ along the coordinates $x$ and $y$ is shown in (b). The longitudinal component of the electric field, $E_z$ at $z = 0$ depending on $x$ and $y$ is shown in (c) and the then necessary longitudinal component $H_z$ of the magnetic field in (d).

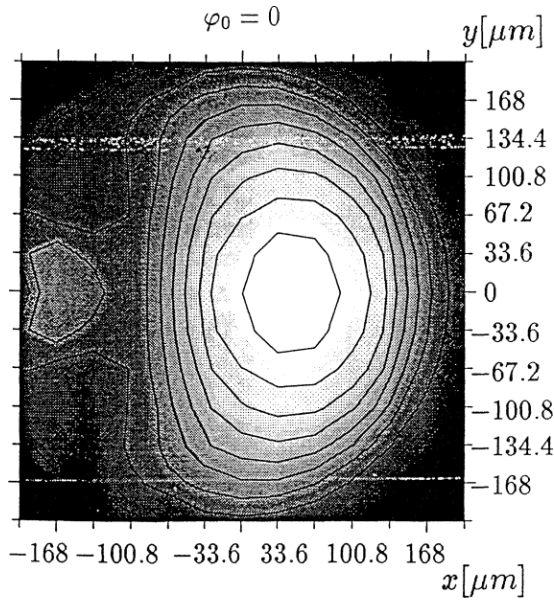

$\varphi_0 = 0$

Nd-Glass-Laser ($\lambda = 1.054 \mu m$)
with $I = 1.234 \cdot 10^{24} W/cm^2$

**Figure 7-6.** A Gaussian neodymium glass laser pulse of intensity $I = 1,234 \times 10^{24}$ W/cm$^2$ and 0.168 mm half width radius hits an electron at a phase $\varphi_0 = 0$ initially located in a position given by the coordinates $x$ and $y$ with respect to the beam cross section gaining an energy expressed by $g$ of 450 000 at the innermost closed curve, the next of 400 000 etc. ($g = 1.96 \times 10^6$ corresponds to an electron energy of 1 TeV). All field components are Maxwellian exact.

When calculating the same field from the paraxial approximation based on an angular spectrum method, the fields appear visibly at the very same diagrams as shown in Fig. 7-7. However, when looking to the values of div $E$ and rot $H$, see Fig. 7-8, the paraxial approximation results in values different from zero while the Maxwellian exact values are necessarily zero.

It is then no surprise, that the trajectories of the electron motion are very different between the calculation with the exact field and that using the approximation. Figure 6 shows the trajectories for the motion of an electrons in a stationary (time independent) neodymium glass laser beam (1053 nm wave length) of intensity $I = 1.23 \times 10^{22}$ W/cm$^2$ and a beam width of 0.016 mm. The trajectory for the exact calculation differs very strongly from the angular spectrum paraxial (Osman et al. 1999) approximation (Hoelss 1999).

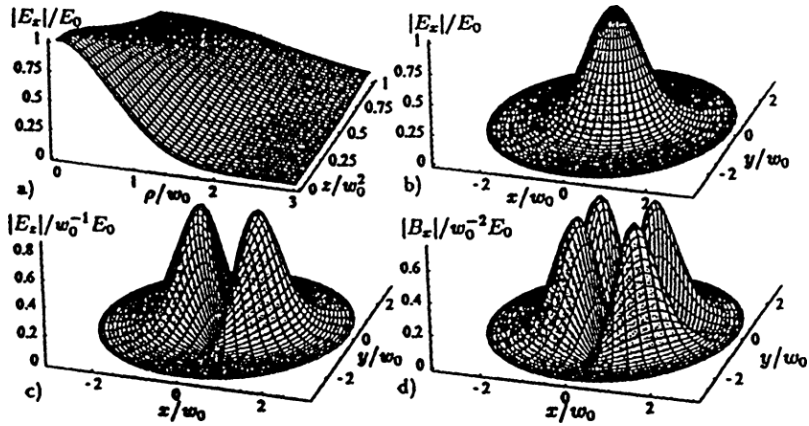

**Figure 7-8.** Beam profiles (Hoelss 1999) for the exact calculation of a laser field with a Gaussian transversal component $E_x$ depending on the radius $r$ developing along the propagation direction $z$ (a) having a dependence on the coordinates $x$ and $y$ of the cross section in the focal areas (b). The Maxwellian exact following longitudinal component of the elec- tric field, $E_z$ has a cross section shown in (c) and the then necessary magnetic field needs a component $B_x$ additional to the Gaussian transversal component $B_y$ (d).

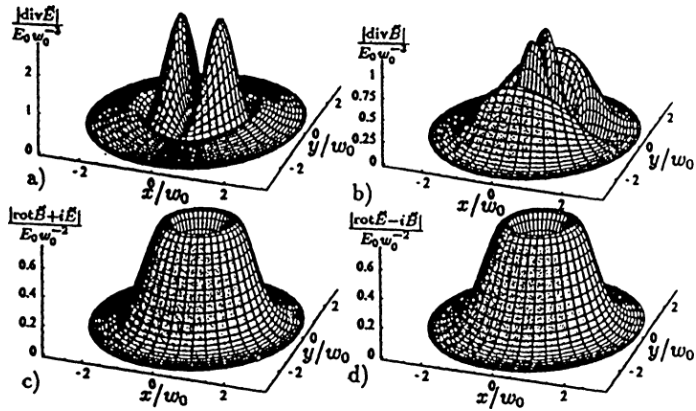

**Figure 7-9.** The fields from the paraxial angular spectrum approximation (Hoelss 1999) look very similar to the exact solutions of Fig. 4, but when evaluating $\text{div}\,E$ and $\text{rot}\,H$, the values shown in the diagrams are different from zero while the exact solutions are zero.

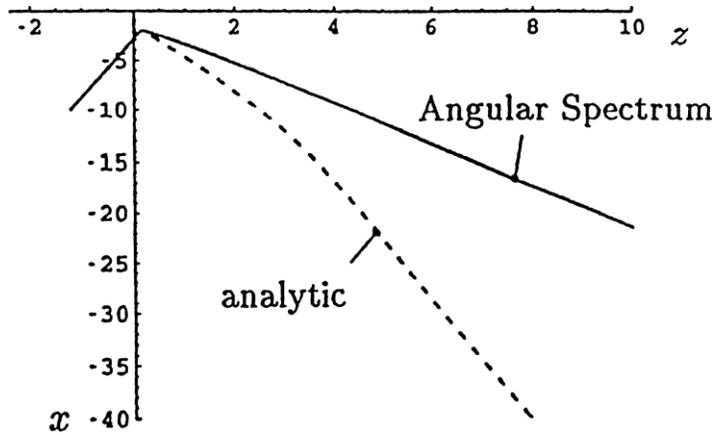

**Figure 7-10.** Demonstration of the very different trajectories of electrons moving through an exact laser fields (analytic) and that of the angular spectrum paraxial approximation (Hoelss 1998) for a neodymium glass laser beam of $I = 1.23 \times 10^{22}$ W/cm$^2$ intensity and beam width of 0.016 mm.

## 7.6. PHASE DEPENDENCE OF THE RELATIVISTIC ACCELERATION OF ELECTRONS IN THE LASER FIELDS IN VACUUM

Apart from the use of an exact or approximating description of the laser field for the electron acceleration in vacuum, there is a strong dependence on the phase of the laser field when injecting the electron in a stationary beam. This can be seen from the case described in the following (Wang et al. 1998). Figure 7-11 describe the geometry of the injection of an electron in extra-axial position into the laser beam. There can be an inelastic interaction as shown by the dotted trajectories and an elastic interaction shown by the fully drawn line where a deviation of the propagation of the electron by an angle $f$ occurs. The time dependence of the position $x$ of the moving electron and of its $\gamma$-value is shown in Fig. . The injected electron has an initial energy of 25 MeV which is nearly unchanged (apart from a very little increase) after the interaction in the case of the elastic interaction while the inelastic interaction results in an electron energy of 1.5 GeV.

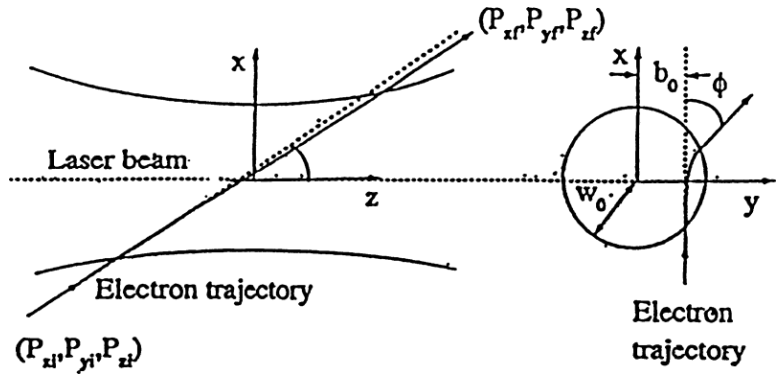

**Figure 7-11.** Dependence of the electron motion at injection into a stationary laser beam at an extra-axial position. The dotted trajectory of the electron corresponds to an inelastic interaction and the fully drawn trajectory with a bending of the electron trajectory by an angle $f$ shows the case of an elastic interaction (Wang et al. 1998).

The criterion as a necessary condition for the correctness of the result can be seen by comparison with the box-like calculation including a reduction by the longitudinal field of a beam instead of a box. The highest possible energy is then 1.86 GeV which value is not too much higher than the achieved value of 1.5 GeV which resulted from a series of computations until the optimized conditions for the selection of the parameters of phase, direction and energy for the injection of the electron into the stationary laser beams were found.

As an example how the resulting electron energies depend on the phase of the laser field, the calculation of the final electron energies (expressed by $\gamma$-values) at the same conditions as Fig. 7-6 but with a phase of 0.51 radian instead of zero are shown in Fig. 7-8.

The experience was that cases had been found where whatever initial conditions for the electrons were chosen, there was nearly no acceleration in the laser field. The essential result of these treatment is that as long as not nearly the mentioned maximum electron energies are achieved, either the used approximation of the laser field is not sufficiently accurate or the initial conditions for the computations are not optimized or both insufficiencies are determining the results in rather discrepancy to experiments.

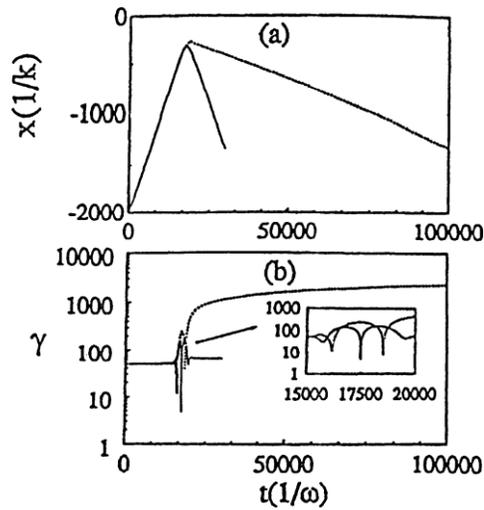

**Figure 7-12.** Example (Wang et al. 1998) of an inelastic interaction (dotted line) of an elec- tron at an extra-axial injection into an electron beam showing the position $x$ (measured in reciprocal wave numbers $k$) and the actual $g$-value depending on the time $t$, compared with the elastic interaction (fully drawn line). The calculation is for a neodymium glass laser beam of intensity $I = 1.23 \times 10^{22}$ W/cm$^2$ of 0.033 mm diameter. The electron with inelastic interaction gains an energy $E = 1.5$ GeV.

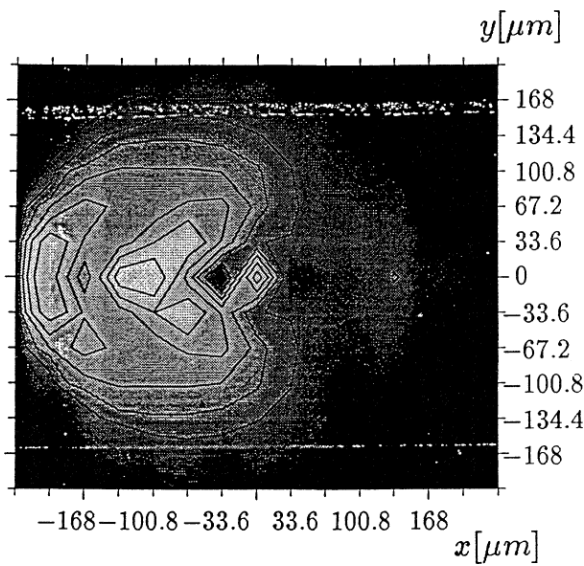

**Figure 7-13.** Same conditions as in Fig. 7-6 but with an initial phase $\varphi_0 = 0.52$ for the interac- tion of the laser beam with the electron (Hoelss 1989).

## 7.7. CONCLUDING REMARKS

In view of the rather fundamental aspects about the *accuracy principle of nonlinearity* applied in this paper (Section 6.3), it seems to be indicated to underline the following. There will be no change in the settled parts of physics. Newton's mechanics for calculating the motion of planets including Einstein's modifications will be always correct, or the fact that the time dilatation in a space craft flying nearly with the speed of light will result in much less aging of the astronaut than for his twin in the earth, or the quantum mechanical calculation of spectral lines. The problems came when Spitzer's logically, mathematically and physically correct *prediction* in 1951 of the impossibility of beam fusion and the need of magnetically confined fusion energy production *turned out* to be nevertheless *completely wrong* (Hora 1987, 1992) because he was using linear physics and missed nonlinearity.

The same was with the prediction (Hora 1988) of the strong acceleration of electrons by laser fields in vacuum. Its impossibility was claimed from exact solutions of Maxwell's theory while the contrary of the prediction was shown experimentally (Malka et al. 1997). The theoretical prediction, however, could be based on fully exact theory and computations only. This was experienced by evaluation of the polarization independence of the experiment of Boreham (1979) where the mathematical transparent calculations with an exotic (triangular) laser beam showed that neglecting of a very tiny nonlinear component could change the whole result from completely wrong into correct (Hora 1981, Section 12.3).

It is then no surprise that relativistic computations of electron acceleration by lasers in vacuum can result in totally different predictions if the mathematical ingredients are not fully exact as shown again in the examples of this setion. This result, on the other hand demonstrated that the new nonlinear physics—if done with sufficient accuracy—will permit predictions of phenomena of which nobody could have dreamed before.

It may be admitted that it could have been somehow boring for the readers to direct their attention to several trvial facts of free electron interactions with laser fields, but it was not trivial to sort out various knowledge as the impossibility that a plane symmetric laser-wavepacket cannot convert energy to free electrons while other geometries can do this due to nonlinearity. Why it is essentially this nonlinearity of forces (beginning with Kelvin 1845) that linear superposition of electromagnetic fields leads to overcome exact results in linear physics. The crucial role of sufficiently large Debye lengths in laser accelerated electron clouds was needed in the Boreham experiment (Boreham et al.1979) in the same way as in the Umstadter experiment (Umstadter et al. 1996) to arrive at the exact relativistic electron motion. Only with the correct longitudinal fields in laser beams it was arrived theoretically without at the measured 30 MeV intense electron beams and their angular dependence. This was a direct result without the need of PIC computations (Wilks 1990; Wilks et al. 1992).

Acceleration of electrons by half wavelength longitudinal electrostatic Langmuir beat waves as postultated initilally from superposition of two laser waves wth different wave lenghts (Tajima et al. 1979) with using ns laser pulses, did not succeed (Kitagawa 1992; Joshi et al. 1994) showing only many orders of magnitude smaller numbers of MeV electrons. The beat-wave acceleration assumed a homogeneous plasma background for the electron acceleration. During the ns interaction, this condition was destroyed by the plasma dynamics. This could be avoided by using ps or fs laser pulses. Because plasma ions were involved – a property excluded in the considerations of this section – the then discovered PIC computations (Wilks 1990; Wilks et al 1992) could be ideally used to numerically describe this electron acceleration (Pukhov et al. 1996; Esirkepov et al. Phys. Rev.Lett. 92, 175003) resulting in further very detailed properties including bubble generation. This led to the measured acceleration of electron bunches or wakes up to GeV energy on a cm scale (Leemans et al 2006, 2009; Esarey et al. 2009) leading to 3GeV (Corner et al. 2012) by using the sub-picosecond laser pulses of more than PW power. This arrives at the goal of Sessler (1982) for a laser-alternative for electron acceleration with very short lengths. The electron acceleration by not fully symmetric laser plane wave packets was repeatedly treated including pulse durations in the atto-second range (Wu et al. 2010; 2012) being experimentally confirmed (Wu 2014). It is another rather exotic result, that the interaction leads to generation of the 57th higher harmonics (Gane'ev et al. 2005).

Other topics are mentioned on very high intensity laser interaction with electrons or by the vacuum polarization. The recent years' advancements in laser technology justify the hopes that extremely high intensity laser interaction may be available within a reasonable time (Perry et al. 1994). Laser pulses of petawatt power (Cowan et al. 1999; 2000) have been produced and the possibility of exawatt ($10^{18}$W) lasers has been considered (Mourou 1994; Bulanov et al 2004; Mourou et al. 2013). The long expected electron–positron pair production by lasers (Hora 1973;1973a) has been verified with the petawatt laser pulses in very large numbers (Key 2000, Wilks et al. 2009) even raching the record positron density of $10^{16}$ positrons/cm$^3$ (Chen Hui et al. 2009), or at laser interaction with 46 GeV electrons including photon–photon scattering (Burke et al. 1997).

This opens up the number of expectations for the extremely high intensity laser interaction as outlined by Kirk McDonald (1985) in which the nonlinear Thomson scattering was one point (as a source of bremsstrahlung which later turned out to be rather low, see Häuser et al. 1992a, 1994) and where the laser fields cause violent accelerations to the electrons. McDonald (1998) hesitated to accept a net acceleration of electrons by laser fields in vacuum in view of controversial discussions despite clear and transparent measurements (Malka et al. 1997; Lefebvre et al. 1997) and despite the theoretical expectations (Häuser et al. 1992a, 1994).

In a more heuristic way, McDonald (1985) compared the violent acceleration of the electrons with the gravitational acceleration $g$ in the Hawking radiation of black holes. According to Hawking (1974, 1975) there is a relation with a black hole temperature $T$

$$kT = \frac{hg}{4\pi^2 c} . \qquad (7\text{-}14)$$

This is interesting in connection with the earlier noted (Hora 1973, 1973a; Seely 1974) pair production due to the vacuum polarization (Heisenberg 1934; Heisenberg et al. 1936; Schwinger 1951). This appears when an electric field $E$ in vacuum would accelerate an electron within a Compton wavelength $\lambda_C = h/(mc)$ from rest to an energy $mc^2$. The necessary field strength $E$ is

$$E = \frac{m^2 c^3}{he} . \qquad (7\text{-}15)$$

This field is 4.4143 × $10^{13}$ cgs = 1.3243 × $10^{16}$ V/cm. If this is the amplitude of a high-frequency (laser) field, the laser intensity is

$$I = 5.891 \times 10^{27} \, \text{W/cm}^2. \qquad (7\text{-}16)$$

100 exawatt laser pulses (Mourou 1994) of about 1000 nm wavelength may well be focused to such intensities.

This is to be compared with the Hawking radiation for a temperature T = $mc^2$. The gravitational acceleration is then

$$g = 4\pi^2 mc^3 / h = 1.46 \times 10^{32} \, \text{cm/s}^2. \qquad (7\text{-}17)$$

The acceleration $a$ of an electron in a static field $E$ for pair production by vacuum Polarization

$$a = mc^3 / h \qquad (7\text{-}18)$$

is of comparable magnitude where a modification for the laser case has to be taken into account for the modification by the relativistic quiver motion. It has been clarified that there is a physical difference between Hawking and Unruh radiation (Stait-Gardner et al. 2006; Hora et al. 2011). These applications are related to extremely high laser intensities (Christopoulos et al. 1988; Deutsch et al. 2006)

# CHAPTER 8
# ULTRAFAST ACCELERATION OF PLASMA BLOCKS BY THE NONLINEAR FORCE

____________________________________

A radical novelty was experienced by Sauerbrey (1996) when measureing the Dopplershift of spectral lines in the reflected light from a laser produced plasma irradiated by a laser pulse resulting in an about hundredthousand times higher acceleration than ever measured in a laboratory. The generated macroscopic plasma plasma blocks were the result of laser interaction. Attention was given that this was theoretically predicted an elaborated in numerical details see Figures 4-12 to 4-15 in 1978 (Hora et al 1978; Hora 1981: p. 179) as nonlinear discovery tracing back to Kelvin's ponderomotion (William Thomson 1845). The long way to clarify these facts is paved with a lot of obstacles last not least that the initial theory (Hora 1969) used plane wave geometries while the self-focusing of the laser beams in the irradiated plasma (Hora 1969a) and the relativistic self-focusing (Hora 1975) destroyed these plane geometry assumptions.

Furthermore, these developments needed an exclusion or essential reduction of thermodynamic and gasdynamic pressure effects which usually are dominating at interaction with laser pulses of ns duration causing delays of electron and ion thermalization and equipartition processes, losses by radiation emission and complications by instabilities. Against these complications, the simplification of the physics is possible by very short duration of laser pulses in the ps range and shorter, when microscopic quantum processes are dominating even for the macroscopic scale. This simplification by the physics of the atoms was envisaged by Edward Teller, when he proposed that physics should first be taught from this side and then going to the macroscopic phenomena with the theromostaticstic chaos (Teller 2001) and complex phenomena as analysed by Lord May (May 1972). When Teller was asking Bohr about his opinion, the answer was better to stay with the usual way to teach first the macrospic physics.

This all needed sub-picosecond laser pulses of extremely high powers above the terawatt and petwatt range which was reached by Mourou's Chirped Pulse Amplification CPA (Stickland et al. 1985; Mourou 1994; Miley 1994; Perry et al 1994; Mourou et al. 1998; Mourou et al. 2004; Mourou et al 2008) where his ICAN-scheme beyond exawatt—Figures 8-2 and 8-3 (Mourou et al 2013, Hora et al 2014) —is going to be developed. A further problem was the difficulty to repeat Sauerbrey's ultrahigh acceleration experiments (Földes et al 2000) where a number of reasons are due to the need of extremely high quality of laser pulses with respect of pulse contrast and having single mode pulses. This was solved by the most pioneering work of Jie Zhang (Zhang et al. 1998) and thanks to the extreme high quality of his lasers and and thanks to his incommensurable merit of attention to notice basic differences of results in contrast to the broad stream of usual observations. The similar merit to notice most exceptional experimental phenomena is due to Badziak et al (1999).

Chapter 8 focuses on the case of plane geometry interaction of laser irradiation on targets. The question of limited radius of laser beams – especially with respect to energy losses for nuclear fusion application – will be discussed in Chapter 10.

## 8.1 EXTREME LIGHT: CHIRPED PULSE AMPLIFICATION

The following developments were possible only by  very high amplification of laser pulses of picoseconds (ps) an shorter duration. This has been discovered with CPA (Chirped Pulse Amplification) about which Gerard Mourou with his PhD projects finally succeeded in stretching and compressing of weak ps laser pulses and to perform the amplification in the stretched stage (Strickland et al 1985). As Mourou (2010) explained, the most difficult problem was the stretcher-compressor element which discoverey came to him when skiing as documented when he returned to the laboratory and re-directed the work of the student Maurice Pessot who became co-author of the patent application. The aspect of Zettawatt lasers is for fundamental studies to reach the Heisenberg-Schwinger pair production by vacuum polarization (Dunne, G.V. 2009; 2011; Hora et al 2011) for laser intensities above $10^{28}$ W/cm$^2$ or generation of charged particles of PeV energy as step beyond the present best accelerator technology, Fig. 8-1.

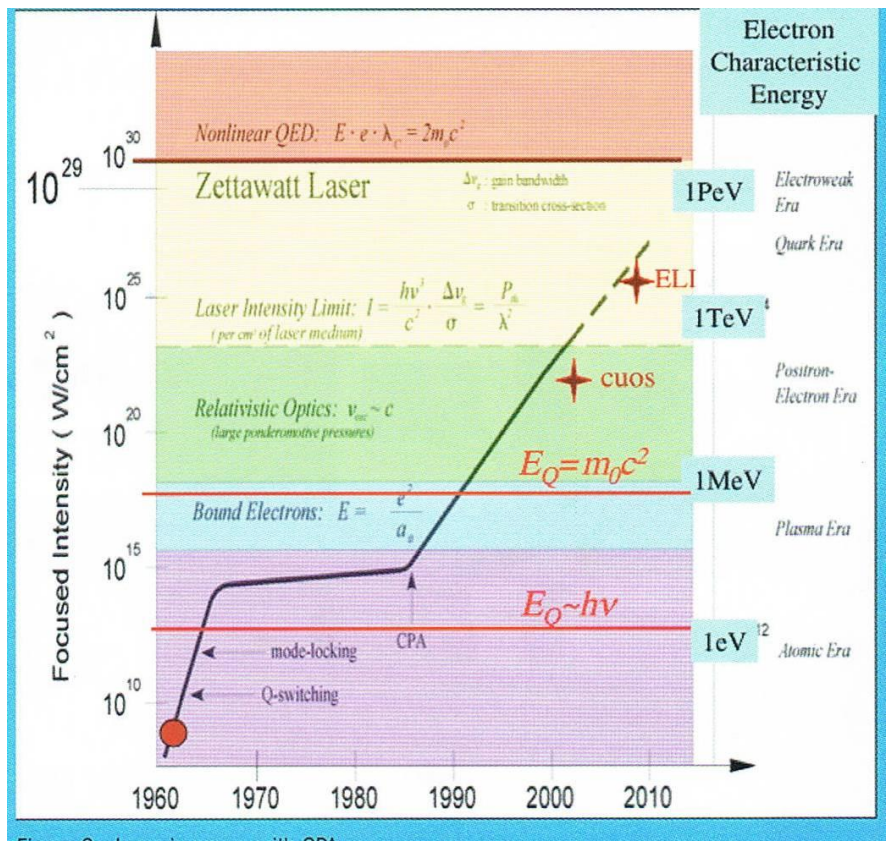

**Figure 8-1.** Diagram about laser intensities (Mourou 2011) of lasers on time with the significant turn at 1996 of the discovery of CPA.

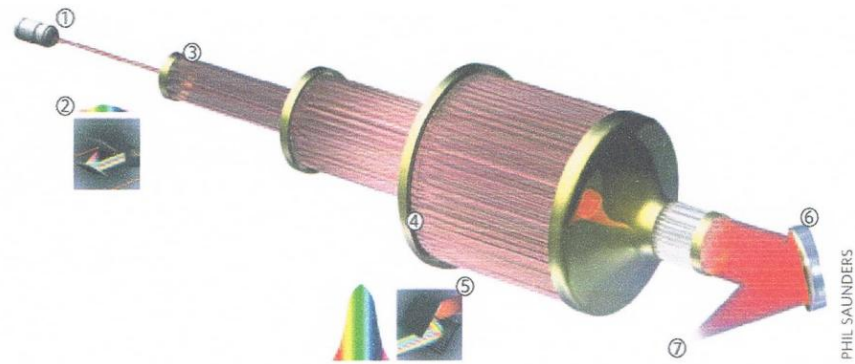

**Figure 8-2.** ICAN laser fiber amplifyier project for producing single mode highest quality sub-picosecond laser pulses (Mourou et al 2013).

For the repetition of the following described ultrahigh acceleration process, the very high quality of the laser pulses is essential with special attention to the single-mode output. This is automatically fulfilled with the fiber laser amplifiers, Fig. 8-2 (Mourou et al. 2013). In order to reach higher laser powers, a scheme in Fig. 8-3 was described, how exawatt (EW) pulses may be reached (Hora et al 2014a). The advantage is that a spherical geometry of with radially directed axes of the fiber output avoids further optical focusing as it was still necessary in Fig. 8-2.

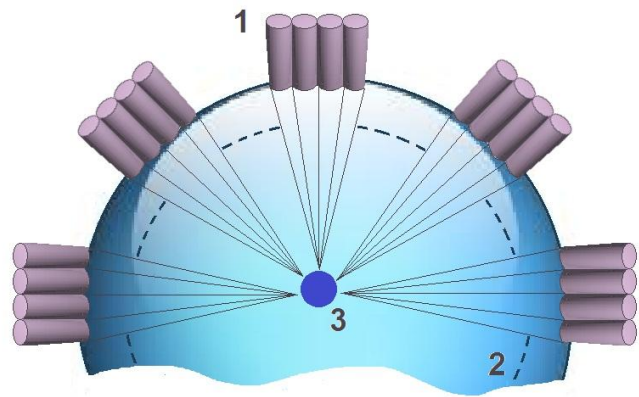

**Figure 8-3.** Generation of a spherical shrinking laser pulse from radially directed fiber ends at a radius 1 to irradiate a spherical solid state fusion fuel 3 with the concentric grid 2 of positive electric charge for slowing down the generated ions of the reaction to gain their energy, especially alpha particles from HB11 reactions (Hora et al. 2014).

It was estimated that a fiber laser bundle of one decimeter cross section may produce ps laser pulses of kJ energy. Producing this on the sphere 1 in Fig. 8.3 with one meter radius results in a pulses of 1.25 EW power. With a radius of 10m and single mode focusing to wave length diameter results in the mentioned higher intensity than $10^{28}$ W/cm$^2$ underlining that the mentioned projections in Fig. 8-1 cannot be excluded.

## 8.2 THEORETICAL PREDICTION AND MEASUREMENT OF ULTRAHIGH ACCELERATION

The results of 1978 (Hora et al. 1979; Hora 1981) for laser intensities of $10^{18}$ W/cm$^2$ from Figs. 4-14 and 4-15 are selected and drawn together in Fig. 8-4 showing the plane geometry computations at perpendicular incidence on a plane deuterium target target of initially bi-Rayleigh density profile of Fig. 4-12. The evaluation of the electron temperature due to the time delay by the collision frequency at this general hydrodynamic calculations wascomparably low compared with the ion energy whose energy was resulting form the nonlinear (ponderomotive) force acceleration by the high laser intensity. This result was understandable because it was a mostly non-thermal conversion of the electromagnetic laser energy into the the macroscopic energy of motion of the two generated plasma blocks, one moving against the laser light (positive velocity v) and another moving parallel with the laser beam into the target.

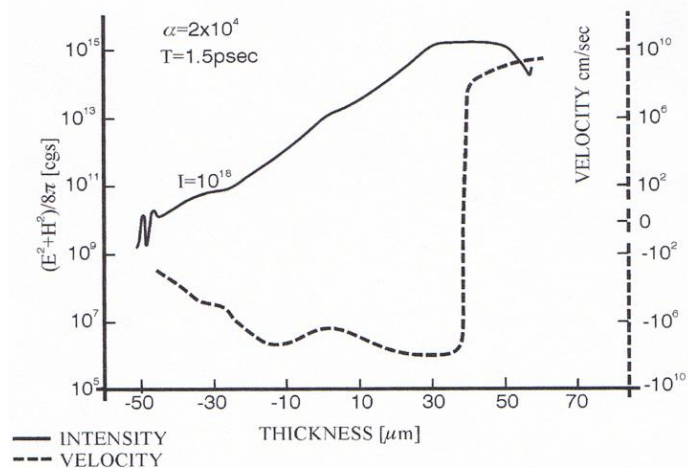

**Figure 8-4.** $10^{18}$ W/cm$^2$ neodymium glass laser incident from the right hand side on an initially 100 eV hot very low reflecting bi-Raleigh deuterium plasma profile (4-12) at initial time t=1, results at time t=1.5 ps of interaction in a velocity distribution v(x) on the depth x and in an energy density of the laser field $(E^2 + H^2)/8\pi$. The dynamic development had accelerated the plasma block of about 20 vacuum wave length thickness of the dielectric enlarged skin layer moving against the laser and another block into the plasma showing ultrahigh $>10^{20}$ cm/s$^2$ acceleration.

A schematic picture of the result of this ultrahigh plasma acceleration for a plane target and for the interaction of the laser pulse focused at the usual 30 wave length diameter may be shown in Fig. 4-5 with the two accelerated plasma blocks where is was estimated that the Debye length for te separation between the laser driven electron cloud was sufficiently small compared with the thickness of the blocks.

The motion of the plasma can be seen also in Fig. 8-4 for the thickness range between 50 wave lengths, the initial position of the bi-Rayleigh plasma block at the beginning of the laser interaction, and the drawn position of the acceleration v at 1.5 ps up to about 55 wave lengths. The velocity is about $10^9$ cm/s in agreement with the printout of the velocity, and the acceleration is above $10^{20}$ cm/s$^2$.

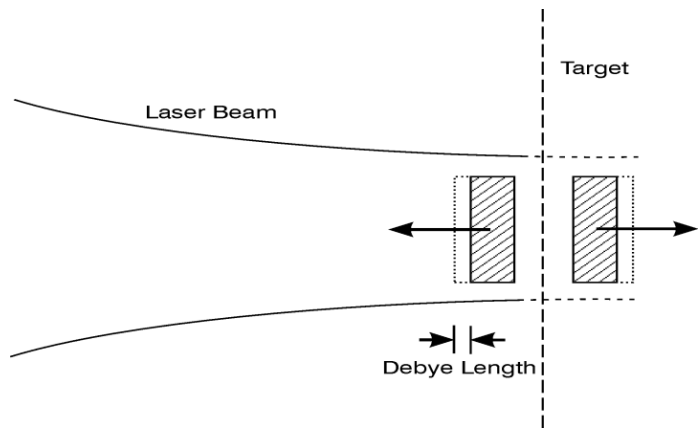

**Figure 8-5.** Scheme of skin depth laser interaction where the non-linear force accelerates a plasma block against the laser light and another block towards the target interior. In front of the blocks are electron clouds of the thickness of the effective Debye lengths.

Sauerbrey's (1996) measurements were evaluated in details (Hora et al 2007) where the swelling factor, Eq. (4-14), about the actual increase of the elecgtromagnetic energy density of the laser field above the vasuum value had to be detemined. In the cases of cavitons, double layers and profile steepening at interaction of intense microwaves comparable to high intensity lasers, very high swelling factors could immediately be seen from the 25 times increased electric field of microwave, Fig. 4-8, (Wong & Stenzel 1975). For the comparisons of the experiments of Sauerbrey (1996) the concluded swelling was in the range of several comparable laser and plasma conditions (Hora et al. 2007).

Another more serious problem was the repetition of the ultrahigh acceleration experiment. The complexity with providing and the control of the onditions of the Doppler measurements was well acknowledged. This condition was to work with an extrmel high contrast ratio as this was needed for other experimental reasons of suppressing superradiance. It was then the success with similar KrF laser pulses (Földes et al 2000) that the ultrahigh acceleration was directly measured again by the Doppler line shift, Fig. 8-6. The acceleration was then about ten times lower than in the case of Sauerbrey according to the lower laser intensity in agreement with the theory. The KrF laser was of the same design as that used by Sauerbrey following the development of Szatmari and Schäfer (1988;1994) based on their very extensive studies of this kind of gas lasers.

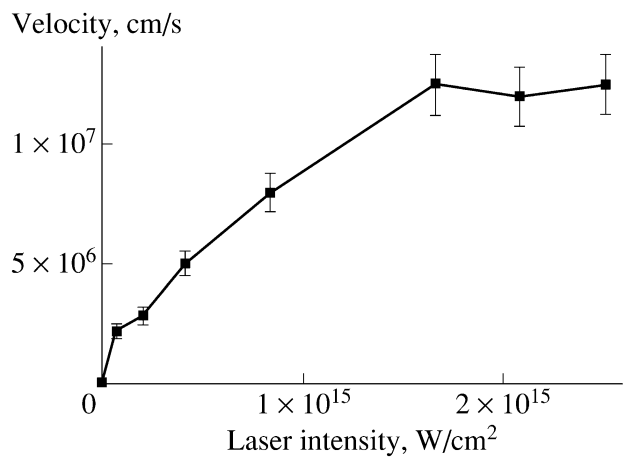

**Figure 8-6.** Intensity dependence of the velocity of the plasma front from the Doppler line shift of the reflected from the irradiation of 700fs KrF laser pulses on an aluminium target (Földes et al 2000).

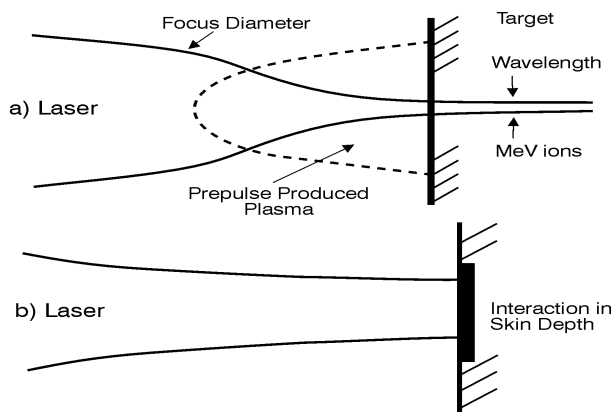

**Figure 8-7.** Scheme for demonstration of the essential different geometry of the laser-plasma interaction volumes for subsequent volume-force nonlinear electron acceleration with separation by the ion charge Z. In case (a), the pre-pule generated plasma before the target causes instantaneous relativistic self-focusing of the laser beam to shrink to less than a wave length diameter with very high acceleration due to the strong gradient of the laser field density. In case (b), the nearly not present or too thin plasma in front of the target permits only interaction in the skin depth with much lower ion energies but nearly ideal plane plasma geometry.

The crucial importance for the successful measurement of the ultrahigh acceleration of plasma blocks was – next to the necessary extremely high contrast ratio – that the relativistic self-focusing was eliminated. Following Fig. 5-4 in Section 5.3, the relativistic change of the electron mass in an intense laser field above the relativistic threshold with the quiver energy equal to the electron's $mc^2$, casus the relativistic change of the refractive index and causes a focal bending of the pane wave front of a laser beam such that the beam reaches a diameter of half of the wave length (Hora 1975). Nanosecond laser pulses with a modest energy of few eV produced than such high intensities, that the generated ions received energies far above the MeV by nonlinear (ponderomotive) force acceleration.

Following this theoretical prediction it was convincingly measured first by Luther-Davies et al (1976) after indications could be concluded from the measurements by Ehler (1975). It was understandable that celebrated physicists could not accept how 4 Joule laser beams could produce more than 10 MeV ions whose properties were determined in a most sophisticated way by Rode using techniques of Sklizkov et al (Basov et al. 1977) in full agreement with the relativistic self-focusing theory (Basov et al 1987). In the mean time the generation of very highly charged MeV to GeV ions from relativistic self-focusing is common knowledge while the generation of a similar group of charge-separated ions was observed (Rohlena et al 1996; 2006; Jungwirth et al. 2005) as Gitomer's (1986) hot-electron process. The necessary thermalization if the quivering electrons by collisions led to a splendid confirmation of the quantum modification (Haseroth et al 1996; Hora 2003) for explaining the measured ion energies by Clark et al (2001).

The acceleration of a respectably high number (still orders of magnitudes lower than from block acceleration) of MeV to GeV ions due to relativistic self-focusing is now common knowledge. But the following very rare and exceptional cases had to be taken into account leading to the discovery by very sophisticated attention of experimentalists. This was drastically different to the common knowledge. This is the main and exceptional reason why Sauerbrey (1996) could measure the ultrahigh acceleration. The theory for the ultrahigh acceleration, Fig. 8-4, was based on plane plasma wave fronts irradiating a plane target. This is in contrast to the usual process with relativistic self-focusing, part a) of Fig. 8-7, showing the usual observations, while the plane geometry, part b) of Fig. 8-7, was needed for the ultrahigh acceleration of plasma blocks.

After several studies of x-ray emission from laser produced plasmas, Jie Zhang (Zhang et al 1998) decided to clarify one of his rare observations. After it became common knowledge that high intensity laser pulsed relativistic self focussing in targets led to emission of highly charged ions above 100 MeV energy it was understandable that the high intensities in the thin filaments of Fig. 8-7, case (a), emitted always high intensity hard (energetic short wave length) x-rays. However, there were few cases where the x-rays were soft and of low intensity. This happened when for other reasons, more than terawatt laser pulses of ps duration were hitting targets where the contrast ratio was very high, i.e. where the prepulse was cut off by an intensity ratio of more than $10^7$ was suppressed before of the ps pulse. This was then the case (b) of Fig. 8-7 where prepulses had not generated a plasma plume for relativistic self-focusing as in Fig. 8-7a. In order to prove this, Zhang separated a small part of less than 1% of the man pulse and pre-iradiated the target by a time t* before the main pulse. When t* was 10ps or 40 ps, the x-ray emission was unchanged soft and of low intensity. This changed dramatically at t* of 70 ps and more when all was usual with very hard and intense x-rays. The laser beam according to Fig. 8-7 (case b) had a diameter of 30 wave lengths at the target and it was easy to estimate from the theory of relativistic self-focusing (section 5.8) that this time was sufficient to generate the plasma plume neede for the relativistic self-focusing filament.

This was the essential proof that that the Sauerbrey (1996) experiment fortunately and exceptionally had avoided the relativistic self-fousing!

The other results drastically differing from that of the broad stream of observations - and this was not rejected from publication by people usually and notoriously bound to conformistic knowledge! – were the measurements by Badziak et al (1999). Firing laser pulses in the range of ps-TW of high contrast ratio did not

result - as usually - in the emission of the MeV ions going into all directions. Instead, the fast ions were highly directed against the laser beam. The fast ions did not have the 22 MeV energy as theoretically expected form relativistic self-focusing but were of 0.5MeV (!). And against all the usual observations, their number did not change when varying the laser energy E (Fig. 8-8) while all other parameters were unchanged. Their energy was increasing exactly linearly on E. How could this be explained? The answer was easy: there were no complications with plasma plumes from pre-pulses as in Fig. 8-7a, therefore no relativistic-focusing with the filament, and only the constant plasma block of the (dielectrically increased) thickness of the skin-layer given by the swelling factor (Eq. 4-4) participated in the directed acceleration. This resulted in an explanation as a skin-layer acceleration (Hora et al 2002a, Hora 2003) in full agreement with Fig. 8-5 and with the ultrahigh accelerated plasma blocks measured by Sauerbrey (1996).

The publication of the results (Hora et al 2002a) was possible only because the Elsevier editor (Wolfgang Schleich) had the wisdom to realize and the courage to ignore the very destructive reports of referees. This remak should not critisize all the usually positive and helpful normal refereeing procedures.

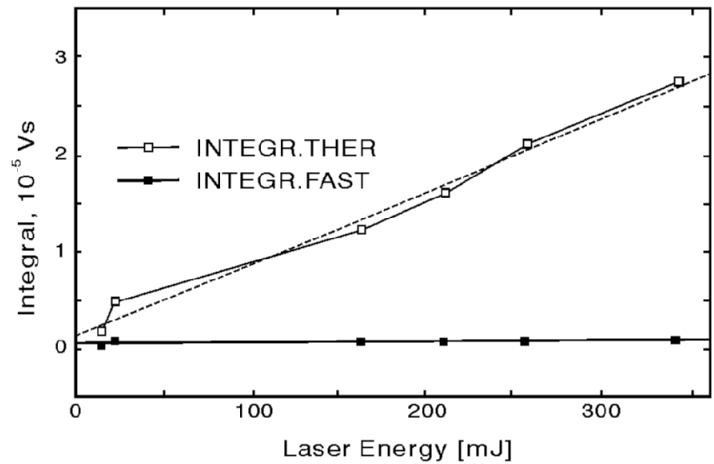

**Figure 8-8.** Badziak et al (1999) effect of anomalous ion emission: Number of (integrated signal) for the number of emitted fast and thermal ions from a perpendicular irradiated copper target at neodymium glass laser irradiation of 1.2 ps depending on the laser pulse energy focussed to 30 wave length diameter with suppression of a prepulse by $10^8$.

When Sauerbrey was trying to repeat the ultrahigh acceleration with a stronger KrF laser (Teubner et al. 1996) with a specifically experienced team using very specialized diagnostic facilities, this did not succeed with the acceleration. It finally turned out that this and and other lasers were not of the mentioned high quality level while producing interesting and important results for other phenomena. The main problem was that single mode quality laser pulses were necessary for the ultrahigh accleleration. This was evident for the measurements with the KrF laser of Sauerbrey (1996) and the repetition by Földes et al 2000. In retrospect this confirmed the extreme high quality of the Ti:Sapphir laser used by Zhang et al (1998) and of the neodymium glass laser used by Badziak et al (1999). It sould be reminded that the fiberlaser ICAN, Figs. 8-2 and 8-3 produces single mode emission (Mourou et al 2013, Hora et al 2014).

This experience is an interesting example about the progress in physics in the case of a rather complicated situation with an experimental task, how to proceed for reaching the truth. The importance of measuring the ultrahigh acceleration by Sauerbrey (1996) needed the exceptional and very especial support by the work of Jie Zhang (Zhang et al. 1998) and of Baziak et al (1999) who followed up very strange observations within the usual broad stream of experiments. And even then, the importance was not appreciated in a way it deserved. Is it too uncomfortable to work with the needed extremely high class experimental acuracy with single mode PW-ps laser pulses to study more details of the ultrahigh acceleration? And even if these facts were theoretically clarified since 1978 (Hora 1981) to be a proof for very fast processes by excluding the thermodynamic complications of laser interaction with nanosecond pulses?

This is another example how a very sophisticated experiment should not be ignored just because it is complicated and requires special attention for leading to a higher quality of knowledge. There is a similar situation with the repetition of the Kapitza-Dirac effect about crossing of an electron beam using the standing laser wave to measure diffraction at the laser field nodes (Freimund et al. 2001), and with the quantum modulation of electron waves (Schwarz et al. 1969) where a medium is placed into the crossing area. The very sophsitcated experiment could not quickly be repeated in the rush but this wide spread criticisms. Only by careful measurements (Andrick et al; Weingartshofer et al 1983) needing a several years lasting, highly experienced preparation arrived at a less noticed repetition against irresponsible cricisms. The quantum modulation measured by Schwarz (Schwarz et al. 1969) with second order electron beating is in agreement with the later achieved theory of photon propagation in transparent solids by Peierls (1976) which fact was a splendid mutual confirmation (Hora et al. 2013a) both for the the theory of Peierls and the measurements of Schwarz. The same is with the theoretical predicted quantum property of the electron wave modulation (Schwarz et al 1969) which was proved by the not intended measurement of Weingartshofer et al (1983) to confirm the correspondence principle of electromagnetic electron interaction predicted (Schwarz et al. 1969; Hora 1991).

## 8.3 PICOSEOND PLASMA BLOCK INITIATION FOR FUSION

The following considerations are restricted to plane geometry interaction of very intense laser pulses of less than about one picoseconds duration as far as these can experimentally be verified as in the examples just described for ultrafast acceleration of plasma bocks. Very many other cases deviating from this plane geometry description are well known and fill a large number of publications with highly varying interaction results which are well very important to evaluate other specific phenomena. With these mentioned restrictions we are now following the results first published by Chu (1972) and Bobin (1974) about laser initiation of fusion flames in solid density fusion fuel, go then to an updating by later results, and refer the cases of other geometries than the plane wave interaction to Section 10.

Similarities and overlap with these phenomena have to be taken into account as well as the different processes seen from the optical properties of dielectric highly enlarged plasma skin layers and to what extend these optical phenomena are related to the plasma phenomena of Debye sheaths known from internal electric fields in inhomogeneous plasmas, well covered by the computations of Target Normal Sheath Acceleration (TNSA) discovered by Scott Willks (Wilks 1990; Wilks et al 1992). Several properties of these widely used techniques are known from macroscopic plasma dynamics, as the acceleration normal to the target normal (Section 4.4) which resulted automatically from the theory for collisionless plasmas (Hora 1974) with some minor deviation from the normal direction if electron collisions (usual or quantum corrected see Hora 1981; Hora 1991: section 2.6) are included if laser radiation is obliquely incident on a plane target. TNSA confirms also the importance of the earlier ignored internal electric fields in plasmas which were first recognized from genuie two-fluid hydrodynamics with inclusion of the Poisson equation, Section 6.4 (Lalousis et al 1983) resulting in double layers (Hora et al 1984), surface tension, and stabilization of surface waves up to wave lengths of about about 80 Debye lengths (Hora et al. 1989).

This all began with the extensive earlier PhD work of Chu (1972) to find out what lasers are necessary for irradiation of soid density deuterium-tritium (DT) fusion fuel (see Eq. 9-1a) for an ignition. The result was that pulses in the range of picoseconds (ps) are necessary to initiate the process to develop from the surface into the volume of the fuel. Bobin preferred the expression "fusion flame" for this kind of initiation process for the subsequent reaction. As will be shown below, a fusion reaction is produced within the whole fuel volume over which the generated two dimensional flame front is spreading into the fuel as a shock wave with extremely high velocities far exceeding Mach 3000. This is the reason why the expression "flame" is used with restriction with reference to the short time picoseconds initiation process.

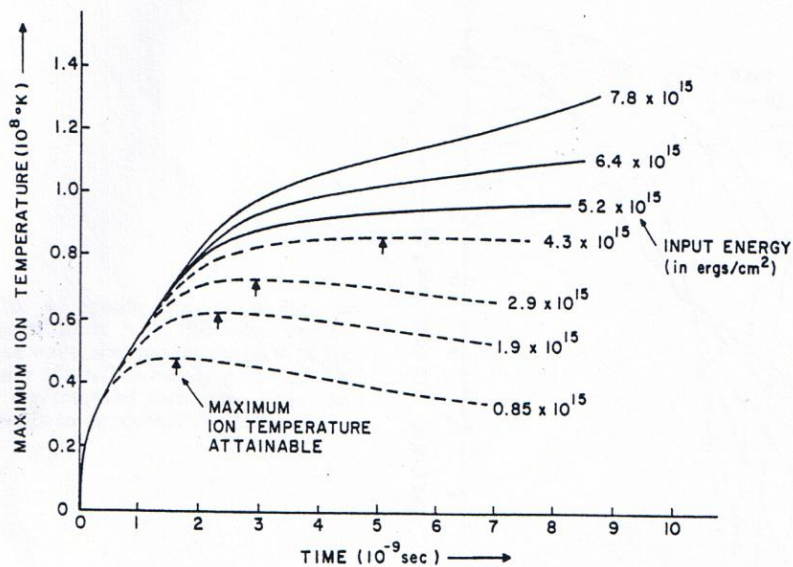

**Figure 8-9.** Maximum ion temperature in solid density DT depending on time after the deposition of an input energy flux density E* during one picosecond at the surface of the fuel depending on time. Decaying dashed curves show no ignition and fully drawn curve above a threshols of $E^*=5 \times 10^8$ J/cm$^2$ are in the range of ignition.

Before using plasma hydrodynamic computations at laser interactions using the genuine two-fluid model with the conservation equations for continuity, momentum and energy given on the beginning of Chapter 6.4, we use the initially reduced hydrodynamics with a plasma one fluid formulation in the treatment of Chu (1972) up to the cases repeating his computations (Hora et al 1988) for comparison of the results. Fig. 8.9 shows the essential results of Chu (1972) for demonstration of the ignition threshold for the ps fluxdensity of the input energy E*. In order to explain the updating of later discovered plasma processes, first an evaluation is shown by using Chu's single plasma fluid hydrodynamic model for comparisons in subsection 8.4.1 while the use of the more general genuine two-fluid model with the generated internal electric fields (Section 6.4) is used to the generation of the flame propagation as result of ultrafast propagation of the shock front.

## 8.4 BLOCK IGNITION WITH UPDATING OF CHU-BOBIN FUSION FLAMES

The updating of the initial conditions used by Chu (1972) was to elaborate first better understanding
(a) how the dielectric plasma properties are increasing the skin layer of the plasma following the knowledge from using the bi-Rayleigh density profile of Fig. 4-2 and reduction of thresholds is possible,
(b) the influence of the inhibition factor of reduced thermal conduction, and
(c) the collective effect of the stopping power works, for
(d) updating the hydrodynamics of the Chu-Bobin results (Hora et al 2008)
(e) comparing with results of Chu
(f) ignition at low compression.
At this stage it should be underlined that the application of the ultrahigh acceleration of plasma blocks for nuclear fusion goes back to 2002 (Hora 2002; Hora et al 2002a; Hora & Miley 2004; Miley et al 2005) and support received by the IAEA/Vienna Coordinated Research Program 13508 under the Directorship of Dr. G. Mank following activities at the Abdus Salam International Centre of Theoretical Physics in Trieste/Italy.

## 8.4.1 OPTIMUM THICKNESS OF ACCELERATED PLASMA BLOCKS AND REDUCED THRESHOLDS

It is the aim to arrive at comparbly deep plasma layers in the ps generated blocks. A first choise is to arrange this by geometric motion of the plasma block, Fig. 8-10. Another way for reaching a deep skin depth for the block acceleration is the increase of the depth as given by dielectric properties of the plasma defined by the selected initial density profile of the plasma. A special case was shown in Fig. 8-4 where the initial density of a deuterium plasma density before interaction with the laser irradiation was a bi-Rayleigh profile in order to have a very low reflectivity of the advantage of j or aiming block ignition for laser fusion following Fig. 8-4, it is important that the initially laser accelerated block in the area A should receive the highest possible thickness by the nonlinear force acceleration.

It was well known from one dimensional hydrodynamic computations before 1980, Fig. 8-11 (Lawrence 1978; Hora et al. 1979; Hora 1981) and selected for laser irradiation with $10^{18}$ W/cm$^2$ laser intensities that a deuterium plasma received up to 20 vacuum wavelengths thick blocks accelerated to velocities of about $10^9$ cm/s within 1.5 ps irradiation. An example of these results is shown in Fig. 8-11 where a compressing plasma block with a thickness of nearly 60 vacuum wave lengths was generated after 450 fs irradiation by a $10^{16}$ W/cm$^2$ laser intensity on a deuterium plasma with very specifically prepared initial density (Lawrence 1978:

p.104). The following new computations (Cang et al 2005) resulted in many details of this thick block generation (Sadighi et al 2008; Yazdani et al 2009).

The problem is related to the propagation of electromagnetic waves in media with varying refractive index $n^2 = 1 - (n_e/n_{ec})/(1 + i\nu/\omega)$, where $n_e$ is the electron density, $n_{ec}$ is the critical electron density where the plasma frequency is equal to the laser frequency $\omega$, and $\nu$ is the electron collision frequency depending on the locally varying electron density and the temperature of the plasma including nonlinear generalizations by the electron quiver motion in the laser field and on rela-

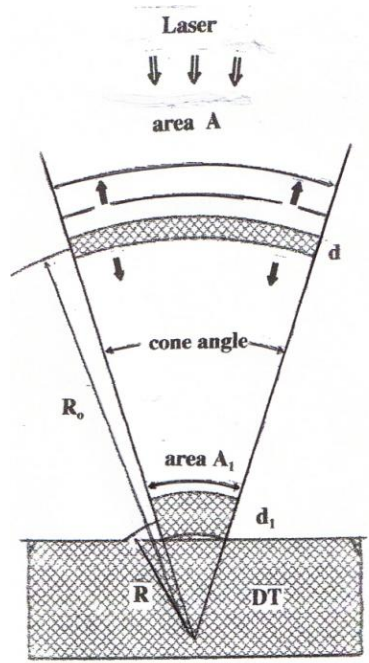

**Figure 8-10.** Schematic description of a spherical laser irradiation on a solid DT layer (area A) producing a block layer moving against the laser and another one of thickness d into the cone with space-charge neutral and radially directed ions of energies of about 80 keV. The modestly heated block expands to a higher thickness $d_1$ but smaller area $A_1$ to hit solid DT at a radius R for igniting fusion (Hora et al 2007).

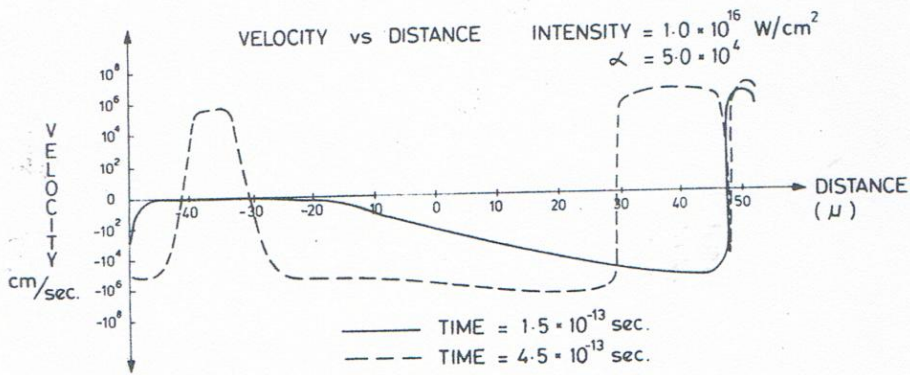

**Figure 8-11.** Velocity profile for laser irradiation on a deuterium plasma with initially 100 μm deep bi-Rayleigh profile (Fig. 4-12) of 100 eV temperature resulting in a nearly

undistorted compressing plasma block of 60 μm depth after 0.45 ps irradiation by a
neodymium glass laser of $10^{16}$ W/cm$^2$ intensity (Lawrence 1978).

tivistic effects (Hora 1981: Sect. 6;) and the quantum correction following the suggeston by Bethe that wave-mechanical diffraction of electrons has to be taken into account (Hora 1981a; 2003; Hora et al. 2012). The plasma frequency is $\omega_p = (4\pi e^2 n_e/m)^{1/2}$. Due to the local (x-dependent) variation of **n**, the wave equation cannot be solved by elementary functions (as sin or cos) but by higher (Bessel-, Legendre- etc) functions. Approximative solutions are familiar from quantum mechanics, with the WKBJ (Wentzel-Krames-Brillouin-Jordan) method (Eq. 3-104).

One exception with solutions by elementary functions was possible for the very special case where the spatially variation on the x-coordinate for collisionless plasma was given by, discovered by Rayleigh (1880)

$$n = 1/(1 + \alpha x) \tag{8-1}$$

where the solutions for the wave equation of the electric field E of the laser were exactly expressed by elementary functions with an amplitude $E_o$

$$E(x) = (1+\alpha x)^{1/2} E_o \exp\{\pm(i/2)[(2\omega/c\alpha)^2-1]^{1/2} \ln(1 + \alpha x)\} \tag{8-2}$$

With these exact solutions, the reflection between a homogeneous and an inhomogeneous Rayleigh medium could be solved precisely (Schlick 1904) and it was clarified by Hora (1957) that there are only two exact solutions in the inhomogeneous optical medium for a wave moving to +x and one to –x and no internal reflection. *"Internal reflection"* was wrongly suggested from the many-layer approximation. This result of no internal reflection was then shown generally with any medium (not only for the Rayleigh case) as a rather surprise by Osterberg (1958).

The Rayleigh medium has another special importance when studying the nonlinear (ponderomotive) forces generated by a laser field in plasmas. It was known from electrostatics that electrons can be moved by a ponderomotive force if there are gradients in the electric field E given by $-\nabla E^2$. It was the merit of Weibel (1958) to demonstrate that the same forces on electrons in vacuum appear also time averaged in the high frequency fields of microwaves, see Fig. 4-8 (Wong et al. 1975). The evaluation of these forces for laser propagation in plasmas including the inhomogeneous dielectric properties (Hora 1969) resulted in a the nonlinear force density (Eq. 4-56) after subtracting gas dynamic, thermo-kinetic forces

$$\mathbf{f}_{NL} = \nabla\bullet[\mathbf{EE} + \mathbf{HH} - 0.5(\mathbf{E}^2 + \mathbf{H}^2)\mathbf{1} + (1+(\partial/\partial t)/\omega)(\mathbf{n}^2-1)\mathbf{EE}]/(4\pi)$$
$$- (\partial/\partial t)\mathbf{E} \times \mathbf{H}/(4\pi c) \tag{8-3}$$

where **H** is the laser field vector, 1 is the unity tensor, ω the laser radian frequency, c the vacuum speed of light, and **n** is the (complex) refractive index. The prove that these and only these terms of the forces are correct, was derived from momentum conservation for the non-transient case (Hora 1969) and by symmetry reasons for the transient case (Hora 1985). For simplified geometry, the force (8-3) can be reduced to the time averaged value of Eq. (8-1)

$$\mathbf{f}_{NL} = -(\partial/\partial x)(\mathbf{E}^2+\mathbf{H}^2)/(8\pi) = -(\omega_p/\omega)^2(\partial/\partial x)(E_v^2/\mathbf{n})/(16\pi) \tag{8-4}$$

where $E_v$ is the amplitude of the electric field of the laser. Within the plasma, the square of the electric field is increased by a swelling factor

$$S = 1/n \tag{8-5}$$

With respect to the result of the Rayleigh profiles, Eq. (8-2), the main limitation is that propagating waves are to be considered requiring an oscillating exponential function. This is fulfilled as long as – see Section 3.4

$$4\omega^2/(c^2\alpha^2) - 1 > 0 ; \quad \alpha < \alpha_o = 1.1 \times 10^5 \text{ cm}^{-1} \tag{8-6}$$

where the limit $\alpha$ is given for neodymium glass lasers.

The very unique property of the Rayleigh profile consists in the fact that the the interaction of the laser field in such a medium causes a (nearly) constant force producing a uniform acceleration and motion of the whole block to a (nearly) undistorted DT plasma block, corresponding to monochromatic ions. Considering mostly cases where $(\mathbf{n}^2 -1) = - n_e/n_{ec}$ to be close to unity where $n_{ec}$ is the critical density with $\omega = \omega_p$ the Rayleigh profile (Eq. 8-1) results in a locally constant force because of

$$\nabla \mathbf{E}^2 = E_o(d/dx)(1/n) = \alpha. \tag{8-7}$$

confirming that the whole plasma is then accelerated as an undistorted block. This property of the Rayleigh profile with respect to the nonlinear (ponderomotive) force is very significant and important to generate uniformly fast moving plasma blocks for the applications.

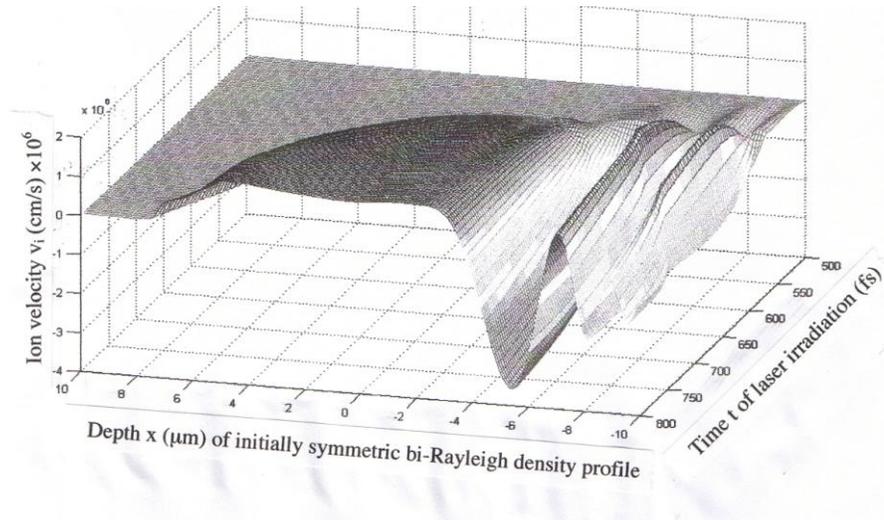

**Figure 8-12.** Genuine two fluid calculation of ion velocity for an initially bi-Rayleigh density profile of 20 μm depth with $\alpha=1.02\times10^4$ cm$^{-1}$ (Eq. 8-6) and 10 eV temperature by neodymium glass laser irradiation of intensity $10^{15}$ W/cm$^2$ of 300 fs duration.

The results of the undistorted plasma blocks of thickness up to 20 times the vacuum wave length by laser irradiation of Rayleigh density profiles was seen in the numerical hydrodynamic one fluid studies, (Hora 2004; Hora et al 2007) of non linear force acceleration in plane geometry long years preceding the confirmation in the first exact measurement by Sauerbrey (1996; Hora et al 2007) thanks to his first use of TW-ps laser pulses with a contrast ratio above $10^8$. The thickness of the skin layer in the Rayleigh profile with the appropriately selected α−value was increased by the swelling factor S = 1/n which value could well be higher than 20 even with inclusion of the collision frequency (Hora 1982, see Eq. (6.48)). The example of a block of even 60 vacuum wave length thickness from these computations was shown in Fig. 8-11.

This all was essential in the clarification of the anomaly of TW-ps laser pulse interaction with targets for driving the plasma in area A of Fig. 8-10 as a skin layer aceleration process under avoiding relativistic self-focusing. The reasonable result for the one-fluid computations can be understood from the following estimations. A neodymium glass laser pulse of $10^{18}$ W/cm$^2$ irradiated a deuterium plasma of initially 100 eV temperature and a Rayleigh profile with α = $2\times10^4$ cm$^{-1}$. At the interaction time of 1.5 ps, the electric field **E** of the laser was so stongly swelled that the laser field energy density was up to 20 times higher than in vacuum (Fig. 8-4). In the same way the thickness of the skin layer was increased by a similar factor and a plasma block of more than 15 vacuum wavelength depth was moving against the laser light nearly undistorted with velocities up to few $10^9$ cm/s. A similar block was moving into the plasma below the critical density, Fig. 8-12 showing the inverted double layers as known form Fig. 6-18 and 6-19 (Hora et al 1984).

It was evident that the conditions had to be selected in some very specific way. On the one hand, the laser intensity had to be of such a value that heating was not much influencing the profiles in the plasma to avoid optical reflection, partially standing waves and subsequent density rippling as seen before (Figures 10.20a and 10.20b of Ref. Hora 1981) at times after 2.5 ps while the block conditions were well preserved at the time 1.5 ps. On the other hand the laser intensity had to be rather high, close to $10^{18}$ W/cm$^2$ to adjust for the DT fusion conditions.

Fig. 8-4 shows an example with generation of a compressing plasma block of nearly 20 wavelength depth appearing at early times of 0.40 ps after the irradiation of a laser pulse of $10^{16}$ W/cm$^2$ using the numerical code (Cang et al 2005) for comparison with Fig. 8-11. The compressing block is indeed of a lower depth than in Fig. 8-12. From the acceleration of the plasma against the laser light, one finds an acceleration of $2\times10^{19}$ cm/s$^2$ in some similaritiy to the result of Sauerbrey (1996). Based on the vacuum electric field of the laser for the intensity of $10^{16}$ W/cm$^2$, a swelling factor, Eq. (8-5) of S = 3.75 could be derived. This is similar to the evaluation (Hora 2003) of the swelling factor at the initial skin layer acceleration experiments (Badziak et al 2004). In these estimations, the plasma density was approximated by the value of the critical density of $3.3\times10^{-3}$ g/cm$^3$ for the deuterium plasma.

For cases closer to the conditions of the geometry in area A of Fig. 8-10, calculations were performed with bi-Rayleigh deuterium plasma targets of initial thickness of 20 μm. Fig. 8-12 shows the result for laser intensity of $10^{15}$ W/cm$^2$ of a 300 fs pulse on a plasma with initially temperature 10 eV where the compressing block has a depth of 8 vacuum wave lengths. Fig. 5 represents the result for $10^{15}$ W/cm$^2$ where the compressing block of 10 μm is not with a fully homogenous velocity (monochromatic ions) to show an example how the initial conditions for the computations have to be fit for the aim of achieving thick blocks for the laser fusion scheme according to Fig. 8-10. Details of the computations are reposted in related papers (Yazdani et al 2009; Sadighi et al 2010).

Studies about the mechanism how the directed plasma blocks interact at the area $A_1$ of Fig. 1 with a DT target were based on the work of Chu (1972). This is a hydrodynamic model and one has to be aware, that the mechanisms of the interpenetra tion of the hot plasma hitting the cold DT fuel may need another more detailed model. An earlier attempt (Hora 2008 et al, Ghoranneviss et al 2008; Malekynia et al 2010) lead to a reduction of the hydrodynamic ignition threshold by a factor 20. A more detailed study could be based on a treatment with PIC techniques (Esirkepov et al 2004) which fact has to be taken into account when this chapter is treating only the hydrodynamic side of the process.

The question is about the exorbitantly high energy flux density E*, Eq. (2) needed for ignition of uncompressed DT. When Chu (1972) derived this value, several later discovered processes in plasmas were not known. This refers manly to two phenomena (a) the reduction of the thermal conductivity between hot and cold plasma given by the inhibition factor F* and (b) the reduction of the stopping length of the generated alphas from the fusion reaction in the plasma due to collective effects.

## 8.4.2 INHIBITION FACTOR

The reduction of the thermal conductivity of the electrons by the factor F* was discovered in an empirical way from the evaluation of experiments for laser fusion. Experiments were performed with targets of different layers and the diagnostics by x-rays etc. resulted in a reduction of the thermal conduction by factor F* = 33 (Young et al 1977). Other experiments resulted in a reduction by a factor 100 (Deng et al 1982). Several theories tried to explain these results were assuming magnetic fields, ion-acoustic turbulence or Weibel instabilities, the closest to arrive at the clear facts was that of Tan Weihan (1985) Gu Min (1985) based on the Krook equation leading to pressure effects since these are causing ambipolar fields and therefore internal electric fields.

The final solution was the *theory of electric double layers* with their strong internal electric fields within the plasma (Lalousis et al 1983; Hora et al 1984; Hora 1991). To illustrate this, the problems with these internal electric fields in inhomogeneous plasmas have to be explained. How difficult it was to understand these fields may be seen that these fields were fully known to the Stockholm school working about the ionospheric plasmas, see the review by Fälthammar (1991), but in contrast, nearly all physicists believed that there are no electric field inside of plasmas. Kulsrud (1983) reviewed Alfven's book (1981) just after Alfven had received the Nobel Prize with the statement "Alfven's electric fields which are intuitively not clear". Indeed there is some relation between the Alfven waves and the electric fields as these appeared in the laser interaction with plasmas as the nonlinear ponderomotive forces (Hora 1991, section 12.4) based on the same mathematical formulation. The knowledge of these fields was fully familiar in the studies of plasmas above the atmosphere since nearly 100 years from the studies of the polar light of the Stockholm pioneering plasma school beginning with Birkeland (see Fälthammar 1988) who qualitatively suggested some particle emission from the sun. This phenomenon was then discovered as the solar wind whose velocity and ion current density of the involved protons was calculated quantitatively first by Biermann (1951) from evaluating the photographs of a comet motion in agreement with later measurements with space crafts.

Mentioning – as before - Kulsrud's (1983) book review should not be understood as a criticism. This remark was most helpful to overcome an insufficiency within the then existing usual knowledge of the plasma state. It had been tacitly assumed that all plasmas cannot have internal electric fields due to the fact that the electric conductivity of plasmas is of similar orders of magnitudes as in metals. Undergraduate students learn how in a homogeneous metal, any generated electric field is decaying on time within atto- or femtoseconds. If a piece of metal is located within an external electric field, this decay of any internal field leads to the generation of electric double layers at the surface of the metal and then the discussion of electrostatics without any time dependence is beginning. The fact that there is a most complicated time dependent mechanism involved for this generation of the electron layers at the metal surface could always be neglected because of the short times. However, since the recently discovered mechanisms due to atto- or femtosecond laser pulses are known, these dynamics of the electric fields in plasmas as in metals cannot be ignored. It should be underlined that the situation in a metal at times longer than femotseconds is correct only within a uniform metal. What is significant is that under inhomogeneous spatially and/or temporally conditions as in plasmas, the mentioned conclusions even for much longer times are highly complicate.

The merit of Kulsrud is the shake up against the usually assumed prejudice in plasma theory. He formulated it while most of all other authorities tacitly and without any doubt went ahead "intuitively" with the wrong assumption. The very detailed knowledge of the Stockholm school about the internal electric fields in plasmas was ignored or marginalized as a kind of heresy though most of the plasma experiments for magnetic confinement fusion or at laser-plasma interaction are always inhomogeneous plasmas even with inclusion to complicate temporal dependences which otherwise even lead to further complications. The excuse for the situation in extraterrestrial plasmas is just in the fact that there is a long time dependence at these very low plasma densities and there is the very large spatial geometry that the internal electric fields in plasma could not be ignored. It also should respectfully be admitted that the action of electric fields in the equation of motion of a plasma, in the generalized Ohm's law as an expression of diffusion (see Eq. 6.7 of Ref. Hora 1991), the ambipolar term, was well known as pressure gradient defining an expression of electric fields.

The elimination of any electric field was the principle of Schlüter's (1950) plasma hydro-dynamic equations which was valid for spatial dimensions larger than the Debye length

$$\lambda_D = \{kT/(4\pi n_e e^2)\}^{1/2} \tag{8-8}$$

using the plasma temperature T with the Boltzmann constant k, and the density $n_e$ and charge e of the electrons. This limit (8-8) to larger lengths permits the condition of space charge neutrality when adding the Euler equations of motion for electrons with that of the ions to arrive at a force density in the plasma

$$\mathbf{f} = \mathbf{f}_{th} + \mathbf{f}_{NL} \tag{8-9}$$

consisting in the thermokinetic force

$$\mathbf{f}_{th} = -\nabla p \tag{8-10}$$

given by the gasdynamic pressure p and the general nonlinear force as total force desity **f**. This equation is algebraically identical Eq. (4-55) with

$$\begin{aligned}\mathbf{f}_{NL} = &\ \mathbf{j}\times\mathbf{H}/c + \mathbf{E}\rho + \mathbf{P}\bullet\nabla\mathbf{E}/4\pi + (1/\omega)(\partial/\partial t)\mathbf{E}\nabla\bullet(\mathbf{n}^2 - 1)\mathbf{E}/4\pi \\ &+ [1 + (1/\omega)\partial/\partial t](\mathbf{n}^2 - 1)\mathbf{E}\bullet\nabla\mathbf{E}/4\pi\end{aligned} \tag{8-11}$$

It was shown that these identical formulations (8-11) and (4-56) are the final and general expressions of the time dependent (transient) equation of motion derived by solving of a long controversial discussion proving that these and only these terms for the nonlinear force Hora (1985) are gauge and Lorentz invariant (Rowlands 2006).

The formulation (4-56) is that of the Maxwellian stress tensor including the dielectric response and transient (time dependent) behavior of the fields. The formulation (8-11) explains the parts acting in the nonlinear force. Here one recognizes on the right hand side first the Lorentz term $\mathbf{f}_{Lorentz} = \mathbf{j} \times \mathbf{H}/c$ with the plasma current density $\mathbf{j}$ and the vacuum velocity of light c, then the Coulomb term $\mathbf{E}\rho$ with the electric charge density $\rho$ and as the third term the Kelvin ponderomotive term

$$\mathbf{f}_{Kelvin} = \mathbf{P} \bullet \nabla \mathbf{E}/4\pi = (\mathbf{n}^2 - 1)\nabla \mathbf{E}^2/8\pi - (\mathbf{n}^2 - 1)\mathbf{E} \times (\nabla \times \mathbf{E})/4\pi \qquad (8\text{-}12)$$

The remaining terms in Eq. (8-11) are new nonlinear terms which were derived for the general equation of motion in plasmas from the studies of laser interaction. The proof for the final generality of Eq. (8-11) was given by momentum conservation for the non-transient case ($\partial/\partial t = 0$) and for the transient case by symmetry (8-11). The inclusion of the term $\mathbf{E}\rho$ in (8-11) was enforced by momentum conservation (Hora 1969) for electric charges $\rho$ due to oscillations with the laser radian frequency $\omega$.

For the correct interpretation it is necessary to mention that Kelvin's ponderomotive force is identical with the nonlinear Schlüter term

$$\mathbf{j} \bullet \nabla(1/n_e)\mathbf{j}m/e^2 = (\omega_p^2/\omega^2)\mathbf{E} \bullet \nabla \mathbf{E}/4\pi \qquad (8\text{-}13)$$

remembering the definition of the electric polarization $\mathbf{P}$ and the refractive index without collisions

$$\mathbf{P} = (\mathbf{n}^2 - 1)\mathbf{E}/4\pi \ . \qquad (8\text{-}14)$$

This term (8-13) was the only nonlinear term in (8-11) which was derived in a very sophisticated way by Schlüter (1950) which did not appear in the derivation from the kinetic Boltzmann equations (Spitzer 1956). All other and the transient terms were the result of studies on laser-plasma interaction (Hora 1969; 1985).

From Kelvin's ponderomotive force (8-12) follows formally an expression of the "field gradient force" [as a more general expression than Eq. (8-4)], or the "electrostriction" for collisionless plasma ($\mathbf{n}$ without imaginary part)

$$\mathbf{f}_{NL} = (\mathbf{n}^2 - 1)\nabla \mathbf{E}^2/(8\pi) \qquad (8\text{-}15)$$

This can be used for the case of perpendicular incidence of plane laser waves on an inhomogeneous plasma of one dimensional geometry e.g. along the coordinate x. For the same conditions, the stress tensor description produces a force density into the x-direction as it was used in Eq. (8-4).

Formulation (8-15) led to the common expression of "ponderomotive force". As is known for (plane wave) perpendicular incidence of laser radiation on a plasma, the Schlüter term is then zero. Nevertheless there is a force of the form of Eq. (8-15). In this case, however, the nonlinear force $\mathbf{f}_{NL}$ is the result of the Lorentz term in Eq. (8-11). This confusion of the definitions is avoided if one uses the general expression of the nonlinear force (8-11) for the electrodynamic part of the force density in plasma.

This is valid for any incidence, for plasma with collisions and including time dependence of the fields.

These results of the nonlinear force with clear proofs by experiments (see section 6.1 and Fig. 6-1) were derived from the space neutral plasma. Nevertheless this was the access to see the internal electric fields within high density plasma similar to "Alfven's electric fields" (Kulsrud 1983) leading to a direct understanding of the inhibition factor.

The derivation of the force (8-11) from single particle motion (section 6.1) demonstrated that the forces are mostly acting to the electron cloud within the (space charge neutrally assumed) plasma and push or pull the electrons clouds generating electric double layers such that ions in the cloud are following by electrostatic attraction. These are exactly the electric fields of the space plasma following Alfven (1981) as seen also in experiments (Hershkovitz 1985) between two homogeneous plasmas each having different density or temperature to produce the ambipolar field as a double layer in a transition region. The whole dynamic mechanisms of these electric fields including plasma collisions could be studied by the genuine two fluid hydrodynamics (Lalousis et al 1983, Hora et al 1984, Hora 1991) leading to an established and detailed knowledge about the double layers with Alfven's (1981) electric fields.

As an example how the electric field in plasmas were marginalized, it should be mentioned (Eliezer et al 1989) how the radial electric field in a magnetically confined discharge plasma causes a high speed rotation by the $\mathbf{E}\times\mathbf{B}$ forces. This happens also in mirror machines and in tokamaks and can be used for isotope separation (Hora et al 1973). This rotation was measured from the side on observed Doppler shift of $H_\alpha$-lines exactly arriving at the calculated velocities from the $\mathbf{E}\times\mathbf{B}$ forces, while the explanation (Sigmar et al 1974) ignored this and related it to a banana-plateau regime consistent with neoclassical theory. The clear rotation in tokamaks was then measured by Bell (1979) and Razumova (1983). The realization of electric fields in plasmas and double layers led to the surface tension in plasmas (Hora et al 1989) and to the first quantum theory of surface tension in metals. A further generalization of this Debye layer model led to nuclear forces with consequences for quark-gluon plasmas (Hora 2006a).

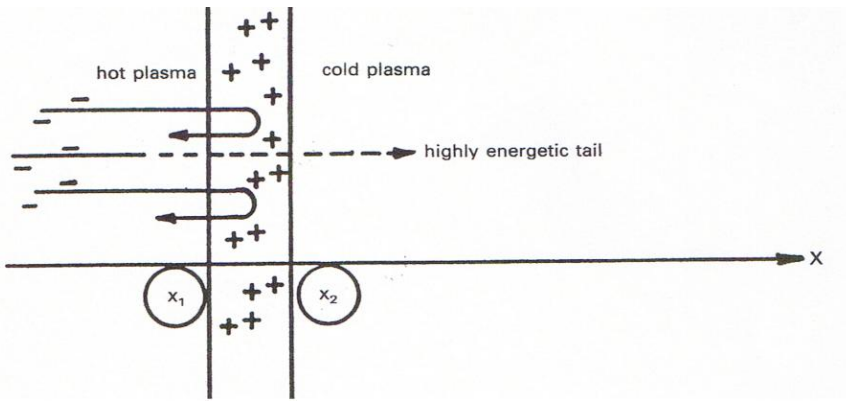

**Figure 8-13.** Double layer between hot and cold plasma with depletion of the high velocity electrons until the ions produce such a potential that the electrons in the hot part are reflected. Thermal transport is determined by the ion thermal conductivity.

Based on this knowledge, it was then straight forward to understand the inhibition factor F* as a double layer effect (Cicchitelli et al 1984; Hora et al 1985). For simplified conditions of a plasma surface expanding into vacuum (Hora 1991, see Figure 2-2), or at the interface between a hot and a cold plasma as in the following conditions of the hydrodynamic computations of Chu (1972), or at a wall confining a plasma, the Debye layer is generated showing a depletion of electrons. The electrons from the plasma interior are electrically reflected at the ions which remain in the double layer whose positive charge results in an electron return current of the electrons back into the plasma.. The potential step (Fig. 2-2) given by kT/2 (one dimension!) corresponds to the work function of the plasma similar to that of a metal surface following the generalization of the Richardson equation for the transmission of exceptionally energetic electrons to produce the thermionic emission.

This detailed elaboration was given to understand the cloud layer process for the initially only empirically noticed strong reduction of the thermal condiuctivity at a plasma surface or at an interface between tow plasmas of different temperatures T, Fig. 8-13. The thermal conduction is performed only by the ions because the electrons in the double layer have been removed.

The thermal conduction is then performed by the ions only and in the equation of energy conservation for the electrons one has to take the ionic thermal conductivity

$$K_i = K_e(m_e/m_i)^{1/2} \qquad (8\text{-}16)$$

instead of the electron conductivity $K_e$, determined by the mass $m_e$ of the electrons and that $m_i$ of the ions. Using the average ion mass of a 50:50 DT plasma, the square root in (8-16) defines the inhibition factor of F*=67.5 in agreement with the semi-empirical evaluation with values between 33 (Young et al 1977) and 100 (Deng et al 1982). For a wide spread double layer of an inhomogeneous plasma the hydrodynamic evaluation results in summary at the same potential step (Alfven 1981; Lalousis et al 1983; Hora et al 1984) to justify the same inhibition in general.

## 8.4.3 COLLECTIVE EFFECT FOR ALPHA PARTICLE STOPPING

After the just discussed problem of thermal transport, the question of the transport properties for the stopping of the DT fusion produced alpha particles in plasma are important for the ignition. Chu [see Eq. (7) of Ref. Chu 1972] used the Winterberg approximation for the binary collisions combining roughly all the numerical models mostly following the Bethe-Bloch theory. An comprehensive summary of these models was given by Stepanek, especially for the alphas of the DT reaction (see Figure 6 of Ref. Stepanek 1981) where the Bethe-Bloch stopping length R for binary collision of the ions with electrons increases as

$$R \propto T^{3/2} \qquad (8\text{-}17)$$

depending on the plasma temperature T.

A diredtly visible discrepancy appeared with the measurements by Kerns et al (1972) at the Air force Weapons Laboratory of the Kirtland Air Force Base where an electron beam with 2 MeV energy and 0.5 MA current of 2mm diameter was hitting deuterated polyethylene $CD_2$. The penetration depth of the electrons was measured by changing the thickness d of the $CD_2$ and the saturation of the emission of fusion neutrons at d = 3mm was a proof of the much shorter stopping than in the Bethe-Bloch theory predicted. An explanation of the value d was immediately possible when Bagge's (1974) theory of the stopping of cosmic rays was applied where the interaction of the charged energetic particles was to be taken by the whole electron cloud in a Debye sphere for the electrons and not by binary electron collisions. The discovery of this collective interaction was by Denis Gabor (1953) following the work of S.R. Milner who derived the Debye screening before Debye. Detailed results were reported (Ray et al 1977; 1977a) based on an analysis using the Fokker-Planck equation and quantum electrodynamics. Another drastic difference of the stopping length of the Bethe-Bloch theory was measured in a direct way (Hoffmann et al 1990).

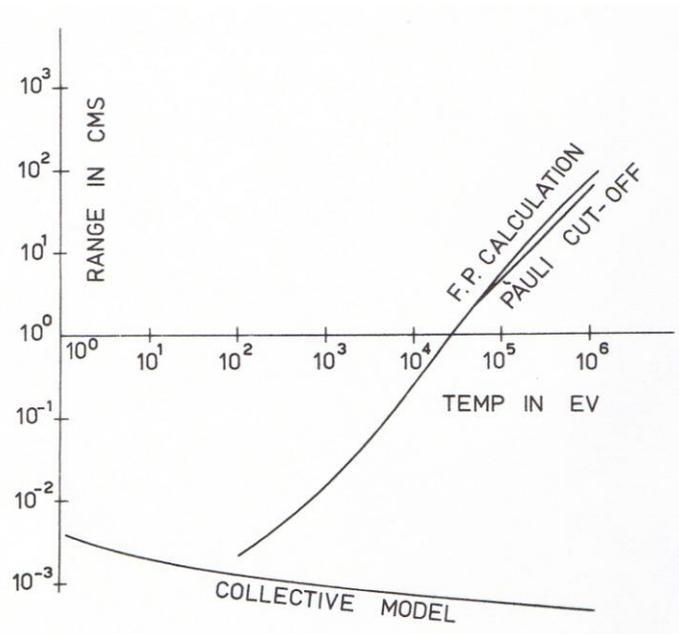

**Figure 8-14.** Temperature dependence of the stopping length R for alphas of 2.89 MeV in a hydrogen-boron(11) plasma with binary electron collision [Fockker-Planck F.P. collisions and quantum electrodynamic (Pauli) cutoff] and collisions with the electron collective in a Debye sphere (Ray et al 1977b).

In strong contrast to the $T^{3/2}$ dependence, Eq. (8-17), the stopping length was nearly temperature independent. The results are for the 2.89MeV alphas in a hydrogen-boron(11) plasma in Fig. 7 are nearly identical with those from the DT reaction (Stepanek 1981, see there Fig. 6). It can immediately be expected that such a discrepancy will change the fusion ignition significantly. This was the reason that a strong reheat occurs in an inertially confined DT fusion pellet leading to the discovery of the volume ignition for Inertial Fusion Energy IFE (Hora et al 1978) later confirmed by Kirkpartick and Wheeler (1981) where John Wheeler's close knowledge of the related physics was helpful. This was further confirmed by numerous other authors (Basko 1990; He et al 1994; Martinez-Val et al 1994; Atzeni 1995) where the robustness of volume ignition against spark ignition (Lindl 1994) with nearly the same fusion gains was underlined by Lackner et al (1994). The ideal and natural adiabatic hydrodynamics of the reacting DT plasma was shown that only this arrived at the highest measured fusion gains (Hora et al 1998) on the way to ignition (Miley et al 2005).

The more precise expression with the very slight decrease of the stopping length R on the temperature T for DT as shown (Stepanek 1991) can be approximated by

$$R = 0.01 - 1.7002 \times 10^{-4} T \quad \text{cm} \tag{8-18}$$

where the temperature T is in keV.

## 8.4.4 HYDRODYNAMIC CALCULATIONS

In order to see the importance of the collective effect of the stopping power in the hydrodynamic equations, first the results of Chu (1972) are going to be reproduced with a minium of changes in initial the conditions he had used before, but with adding now the collective stopping length R and the inhibition factor F*. It is to be underlined from the preceding section, that the collective effect and the inhibition were not at all known at the time of Chu's treatment. The hydrodynamic equations are used as close as possible on the same assumptions of Chu (1972). The equations of continuity and reactions ($D + T \rightarrow \alpha + n$) may be combined to yield as equations of mass conservation

$$\frac{\partial \rho}{\partial t} + \frac{\partial}{\partial x}(\rho u) = 0 \tag{8-19}$$

and

$$\frac{\partial Y}{\partial t} + u \frac{\partial Y}{\partial x} = W \tag{8-20}$$

where $\rho$ is the mass density, u is the plasma velocity and Y is the fraction of material burned, defined by

$$Y = (n_\alpha + n_n)/(n_D + n_T + n_\alpha + n_n)..$$

W is the reaction rate function, given by

$$W = \frac{1}{2} n(1-Y)^2 \langle \sigma v \rangle.$$

It is obvious that (8-21) is the same as the mass conservation equation, due to the small percentage (~0.35%) of mass transformed into energy. In the equation for Y, the n's are the particle densities, and the subscripts are for the different particle species. In the equation for W, the n stands for the total number density of the ions.

The equation of motion expressing the conservation of momentum is

$$\frac{\partial u}{\partial t} + u\frac{\partial u}{\partial x} = -\rho^{-1}\frac{k}{m_i}\frac{\partial}{\partial x}\left[\rho(T_i + T_e)\right] + \rho^{-1}\frac{\partial}{\partial x}\left[(\mu_i + \mu_e)\frac{\partial u}{\partial x}\right], \qquad (8\text{-}22)$$

in which pressure and viscosity terms are included. $\mu_{i,e}$ are the viscosity coefficients whose values are taken to be

$$\mu_{i,e} = \frac{0.406 m_{i,e}^{1/2}(kT_{i,e})^{5/2}}{e^4 \ln \Lambda},$$

where $\ln \Lambda$ is the usual Spitzer logarithm.

The ion and electron temperature equations are expressing the conservation of energy

$$\frac{\partial T_i}{\partial t} + u\frac{\partial T_i}{\partial x} = -\frac{2}{3}T_i\frac{\partial u}{\partial x} + \frac{2m_i}{3k\rho}\mu_i(\frac{\partial u}{\partial x})^2 + \frac{2m_i}{3k\rho}\frac{\partial}{\partial x}(K_i\frac{\partial T_i}{\partial x}) + W_i + \frac{T_e - T_i}{\tau_{ei}} \qquad (8\text{-}23)$$

and

$$\frac{\partial T_e}{\partial x} + u\frac{\partial T_e}{\partial x} = -\frac{2}{3}T_e\frac{\partial u}{\partial x} + \frac{2m_i}{3k\rho}\mu_e(\frac{\partial u}{\partial x})^2 + \frac{2m_i}{3k\rho}\frac{\partial}{\partial x}(K_e\frac{\partial T_e}{\partial x}) + W_e + \frac{T_i - T_e}{\tau_{ei}} - A\rho T_e^{1/2}$$

were included on the right-hand side are the pressure, viscosity, conductivity, thermonuclear energy generation, equilibration terms, and energy transfer terms $W_1$ and $W_2$ following Chu (1972). The last term on the right-hand side of (8-14) is the bremsstrahlung term. The thermal conductivity of the electrons $K_i$ for the case of inhibition has to follow from Eq. (8-16).

For the following reported computations the bremsstrahlung is based on the electron temperature $T_e$ working with Eq. (8-13) of Chu (1972) with the maximum at $x = 0$, thus,

$$W_i + W_e = A\rho T_e^{1/2} + \frac{8}{9}(k/m_i)(1/aT_e^{1/2}) + \frac{2}{9}(T_e/t) \qquad (8\text{-}24)$$

Eq. (17) is a little different from Eq.(20) of Chu (1972) where $T_i = T_e$ is assumed while the following computations with the collective stopping has to be for general temperatures.

The $\alpha$ particles are assumed to deposit their energy in the plasma. They have a mean free path for plasma of solid state density DT is in the case of Chu (1972, Eq. (7)) given by the Winterberg approximation of the binary collisions from the Bethe-Bloch theory and in the following computation according to the stopping length of collective effect given by Eq. (8-18). The action of the stopping with the collective effect is expressed by the temperature T from elimination of Eq. (8-18). For the calculation of the collective effect we added a term to right hand of Eq. (8-24) Thus

$$W_i + W_e = A\rho T_e^{1/2} + \frac{8}{9}(k/m_i)(1/aT_e^{1/2}) + \frac{2}{9}(T_e/t) + P \qquad (8\text{-}25)$$

Where P is the thermonuclear heating rate per unit time obtained from the burn rate and the fractional alpha particle deposition:

$$P = \rho\phi E_\alpha f \qquad (8\text{-}26)$$

$$\phi = \frac{dW}{dt} = \frac{d}{dt}(\frac{1}{2}n(1-Y)^2\langle\sigma v\rangle) \qquad (8\text{-}27)$$

$E_\alpha = 3.5 Mev$ and f is fraction of alpha particle energy absorbed by electrons or ions, has been given by

$$f_i = (1 + \frac{32}{T_e})^{-1} \quad \text{and} \quad f_e = 1 - f_i \qquad (8\text{-}28)$$

In the equations after (8-26) the temperatures of the electrons and of the ions were used to be equal T as used in Eq. (8-27) for the following numerical evaluations.

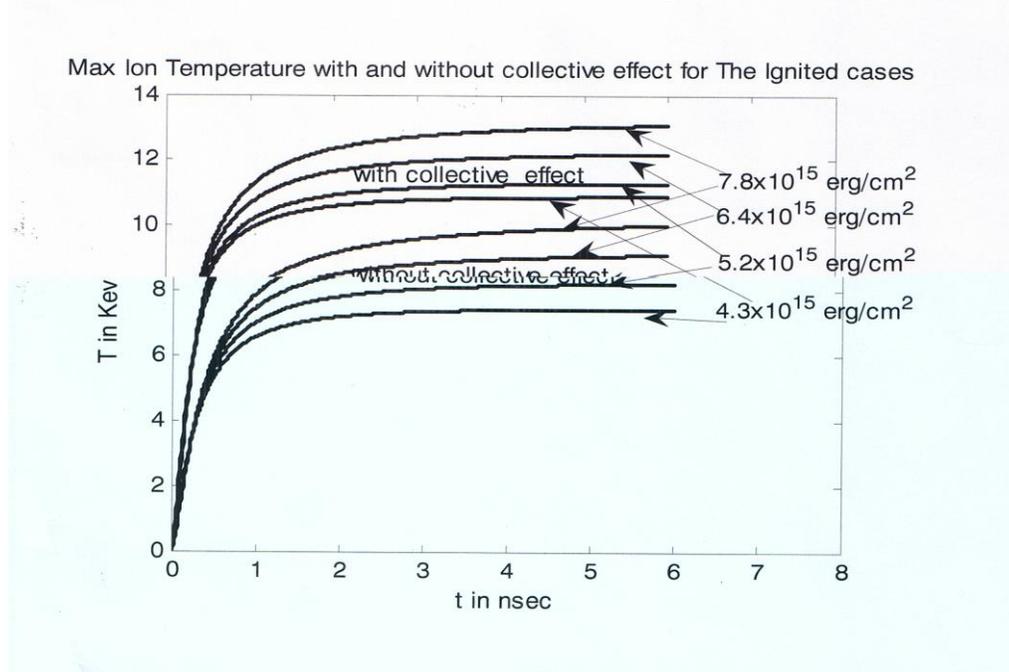

**Figure 8-15.** Results for re-calculation of the characteristic of Chu (1972) without inhibition as in Figure 8-9 and without and with collective effect. Without collective effect, ignition is reproduced at energy flux density E* = $4.3\times10^8$ J/cm$^2$ as achieved by Chu (1972), Figure 8-9, with the characteristic merging in constant temperature at times above 4 ns. Collective effect results in higher temperatures above ignition.

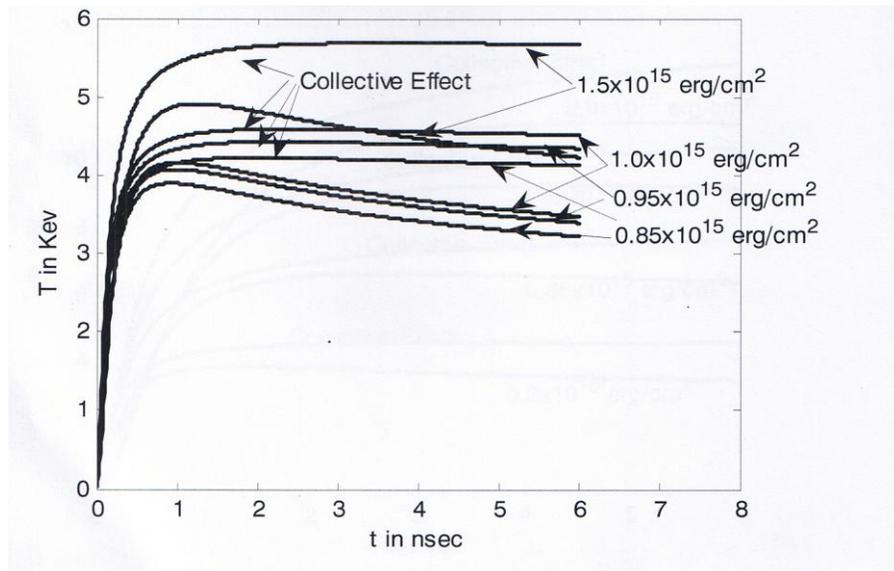

**Figure 8-16.** Same as Figure 8-15 with lower E* showing the ignition threshold at E* = 0,95×10⁸ J/cm².

Figure 8-9 summarized the results of Chu (1972) for the temperature T on an irradiated solid state DT target depending on time where the most characteristic case is for the ignition energy flux density $E^* = 4.3 \times 10^{15}$ erg/cm² = $4.3 \times 10^8$ J/cm² where the curve merges into a constant temperature T on time. This E* is then the ignition threshold $E_t^*$ as explained in more details by Chu (1972) in full agreement with Bobin (1974).

When using the changes by the inhibiation factor and the collective stoping power were similar to the valus of Chu, but with lower thresholds (Hora et al. 2008, Ghorannviss et al. 2008, Malkynia et al. 2010). Figure 8-15 reproduces the temperatures reported by Chu (1978) very well at times above about 1 ns without collective effects. The discrepancies at lower times t are not essentially different and may be due to some differences in the computation codes. Some details about these discrepancies were discussed before for cases without collective effect but only with the inhibition factor where specific numerical evaluations were shown and an effect of a slightly retrograde dependence was elaborated (Ghoranneviss 2008). As expected, the results with the collective effects arrive at higher temperatures T. In order to find the threshold temperature at these conditions, results at lower parameter E* are shown in Fig. 8-16 where the characteristics show ignition at E* at about $10^8$ J/cm².

Inclusion of the inhibiton factor (Ghoranneviss et al 2008) of F* = 67.5 on to of the collective effect results in characteristics in Figure 8-17. The ignition threshold is then

$$E_t^* = 2 \times 10^7 \text{ J/cm}^2 \tag{8-29}$$

This result shows a decrease of the ignition threshold due to the inhibition mechanism and due to the collective effect for the stopping of the alpha particles of the DT reaction by a factor 21.5 (Hora et al. 2008).

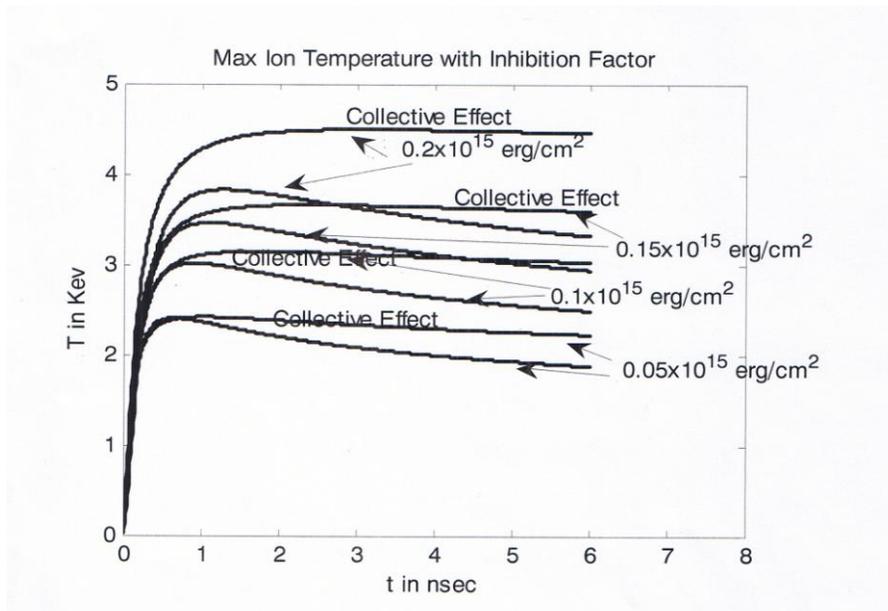

**Figure 8-17.** Characteristics with inhibition factor and with collective effect resulting in ignition at $E_t^* = 0.2 \times 10^8$ J/cm$^2$

This again – as mentioned before – appears as a high value which may not simplify the conditions for block ignition (Hora 2002; 2003; Hora et al 2007) though this hydrodynamic analysis is only a part for the problem. The interpenetration problem cannot be covered by hydrodynamics and there are good arguments that Wilks et al (1992) code techniques (Esirkepov et al 2004, Chen Hui et al 2005; Klimo et al 2006) may lead to further clarification of the ignition problem though the transport problems with respect to heat conduction and stopping power may be on a stronger basis from hydrodynamics used for the presented results up to the moment. An encouraging preliminary result about the interpenetration was achieved before (Hora 1983) with another eventually possible reduction be a factor 20. Adding up the estimated reduction of the threshold arrive then at values close to or less than

$$E_{It}^* = 10^7 \text{ J/cm}^2 \tag{8-30}$$

as an optimistic limit.

Summarizing, the results of Chu (1972) for the side-on ignition of uncompressed, solid density DT by using the irradiation of ps laser pulses for initiating a fusion flame was reproduced and the later discovered inhibition factor F* for a reduction of thermal conductivty as a double layer process was included as well as Gabor's collective stopping power of ions at the very high plasma densities differing from the Bethe-Bloch stopping by binary collisions. In view of the ps initiation process, the necessary energy flux densities of the laser irradiation of few $10^8$ J/cm$^2$, corresponding to intensities of $10^{20}$ W/cm$^2$ was reduced to a flux near or below $10^7$ J/cm$^2$. These values are in the range achieved of being aimed with high priority by the developed Chirped Pulse Amplification CPA considered in the first subsection. The computations for DT reactions show ignition all at an electron temperature above T > 4 keV = T* in agreement with this threshold for more fusion energy generation than bremsstrahlung emission. (Hora et al 2008, Ghoranneviss et al. 2008, Malekynia et al 2010). This corresponds to the nearly two-dimensional area of energy deposition for the initiation of the fusion flame during the ps interaction process by the laser produced plasma block. Any lower temperature T* is possible only with volume processes of alpha re-heat and/or re-absorption of bremsstrahlung, see following section for the volume igntion (Hora et al. 1978).

## 8.4.5 FUSION REACTIONS AND HIGH SPEED FLAME FRONTS

The following results are based on computations using the genuine two-fluid model based on the conservation equations, see first equations of Section 6.4 (Lalousis et al. 1983). Limitations for the block ignition are given by the just reported minimum thresholds of the energy flux density E* of the energy irradiated on the DT fuel how this is compatible with the need of not too high laser intensities I. These have to be e.g. for neodymium glass laser intensities between $10^{19}$ and (closer to) $10^{20}$ W/cm$^2$. The limit for I is given by the condition that the energy of the accelerated ions in the ps duration ignition ares has to be close to 80 keV corresponding to the resonance maximum of the DT reaction cross section. This intensity is to be modified by the swelling factor S which depends on the chosen parameters of the nonlinear (ponderomotive) force interaction of the laser beam with the plasma layer in the area $A_1$ of Fig. 8-10.

As a very preliminary estimation for an example, irradiation of a laser pulse of 10 kJ energy during 1 ps on a cross section of $10^{-4}$ cm$^2$ corresponds to an intensity of $10^{20}$ W/cm$^{-2}$. Up to 0.5 times of the irradiated laser energy can be converted into the kinetic energy of the DT ion block, equivalent to an energy flux density of $5\times10^7$ J/cm$^2$. The thickness of the compressing block moving parallel to the direction of the laser beam is assumed to be 5 μm by choosing the conditions as explained in Section 3. If a conical motion of this block as shown in Fig. 8.10 is performed up to a beam cross section of $10^{-4}$ cm$^2$ a block of plasma with the directed energy of the DT ions of 80 keV will be achieved of about 0.1 mm cylindrical diameter and about 1mm length. The energy flux density of $5\times10^7$ J/cm$^2$ should just meet the requirements for ignition of solid DT as elaborated in Section 8.4.3.

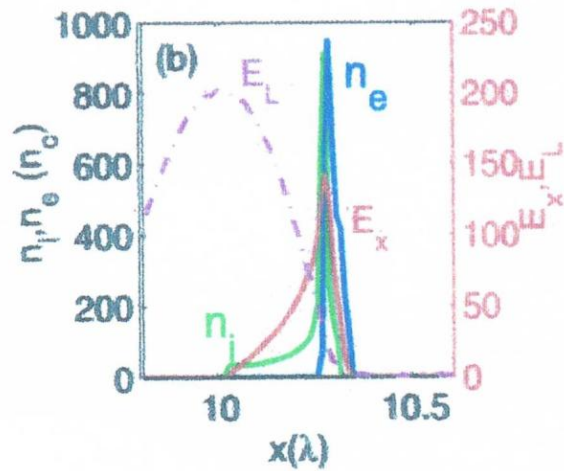

**Figure 8-18.** PIC computation of the laser field amplitude $E_L$ and the longituidinal electric field $E_x$ in the double layer with generated electron $n_e$ and ion $n_i$ densities at initial depths x of the energy deposition at the initiation of the fusion flame at laser intnsty $10^{20}$ W/cm$^2$ (Zhang, He et al. 2011).

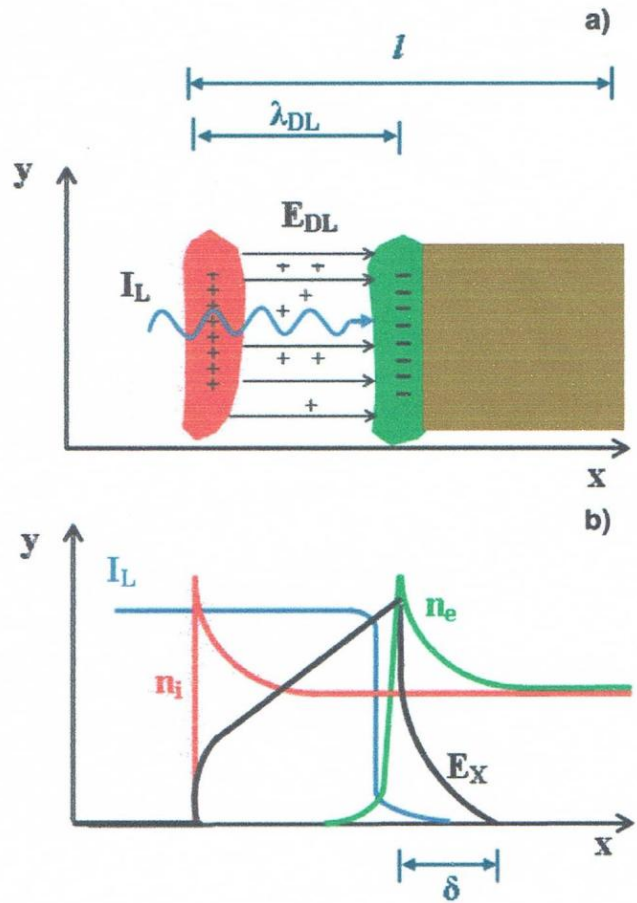

**Figure 8-19.** Double layer model for describing the ultrahigh acceleration process for the ps-laser-initiation of the fusion flame process (Eliezer et al. 2014; 2014a.)

This is an example only for demonstration that the discussed conditions for ps initiation of ignition may be fulfilled. A number of questions for this ignition by the laser driven ion beam are still open similar to the consideration about driving with the 5 MeV electron beam (Nuckolls et al 2002). These parameters refer to the role of interpenetration of the energetic plasma block within the DT fuel, whether the length of the block is optimised, what the details will be for preparing the DT layer in the area $A_1$ of Fig. 8-10 for generating the block as considered with respect to a block with a minium of distortion and optimised swelling, the optimised temperature in the range around 100 eV of the generated block due to thermalising mechanisms during the interaction at $A_1$ to fit with the lengthening of the block before reaching the area $A_2$, and others.

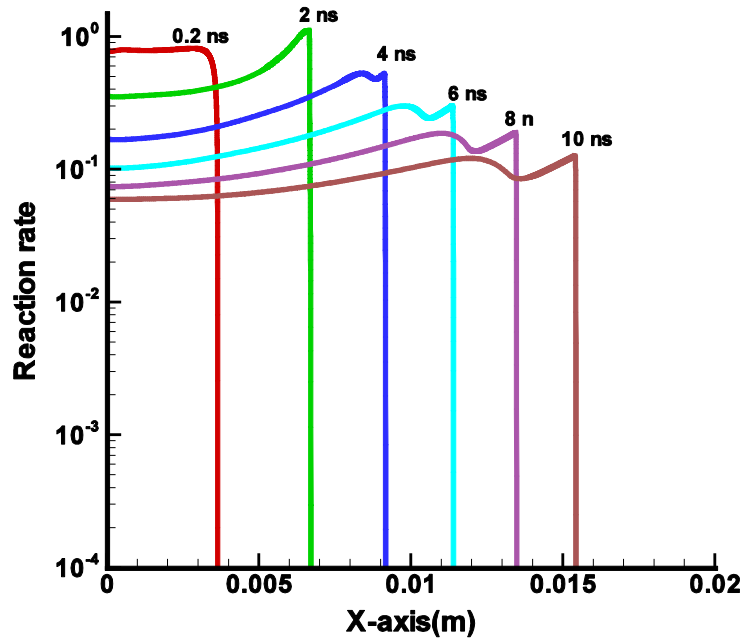

**Figure 8-20.** Reaction rates (to be multiplied by $10^{36}$ m$^{-3}$s$^{-1}$) in solid DT at 1 ps pulse of energy flux E*= $3\times10^8$ J/cm$^2$ KrF laser irradiation depending on the fuel depth x at different times up to 10 ns (Lalousis et al. 2013).

The following multi-fluid computation with adding the fluid of alpha particles to that of the electrons and ions, are then all for solid DT fuel with using laser intensities of $10^{20}$W/cm$^2$, laser pulse duration 1ps and deposition of the laser energy mainly in the directed DT ions in a depth of 5 μm of the fuel target if not otherwise noticed at the figures Lalousis et al (2013). The deposition depth is estimated from calculations of the stopping length including Fokker-Planck collisions and quantum electrodynamics (Pauli) cutoff (Ray et al 1977), estimations of the plasma interpenetration for the Chu-Bobin side-on ignition (Hora 1983) where indeed more evaluations are needed. The result with using PIC computations is shown in Fig. 8-18 (Zhang, He et al. 2014). There is some similarity about the generated electric fields by describing the acceleration of the plasma block (Naumova, Schlegel et al. 2009) acceleration as a double layer process, (Fig. 8-19 (Eliezer et al 2014).

A justification for the assumptions about the ps-laser-deposition process may be derived also from an agreement of results for relativistc acceleration of ions of 70 MeV by the used hydrodynamic model (Moustaizis et al. 2013: Figs. 9&10) which was achieved be extensive PIC computations (Gaillard et al. 2011) and analytical studies (Mourou et al. 2006; Eliezer et al 2014b). For further work related on PIC, see references Schwoerer et al. (2006), Hegelich et al. (2006), Borghesi et al. 2002, Naumova et al. (2009), Robinson et al. (2008), Esirkepov et al. (2002), Bulanov et al.(2003) Bulanov et al. (2010) and Xu et al. 2005. The very ntense ion beams are important also for and alternative option of fast ignition (Roth et al. 2005)

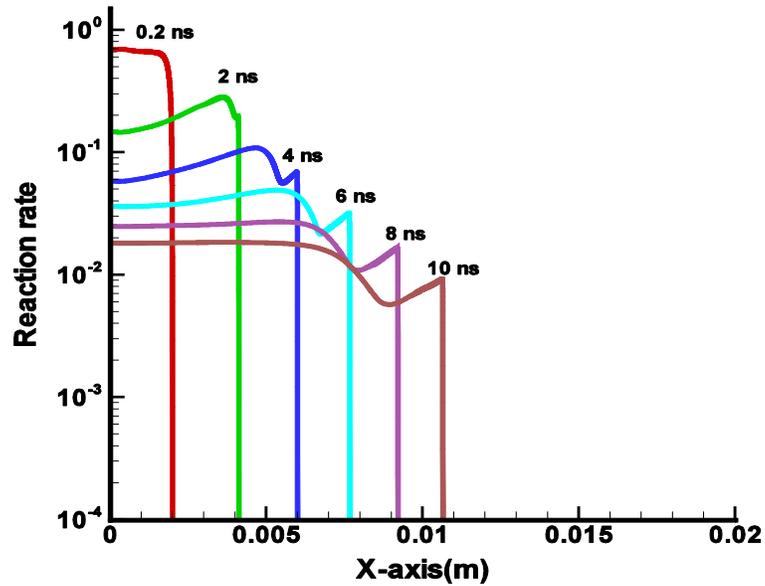

**Figure 8-21.** Same as Fig. 8-20 for energy flux density $E^*=10^8$ J/cm$^2$ of the igniting ps laser pulse.

One question is remaining about the fusion reactions in the plasma behind the fusion flame. This was printed out in Fig. 8-20 for a laser intensity of $3\times10^{20}$ W/cm$^2$ in Fig. 8-20 and for three time less intensity in Fig. 8-21. This shows that the reaction is persisting in the while volume passed by the flame with some minor decay in the case of Fig. 8-20 and a stronge decay in Fig. 8-21. Nevertheless the reaction gain for both cases is comparable, see Fig. 8-22. It is remarkable that for the lower laser initiation energy, the gain is not very much lower. The reason is that the initiating laser energy is lower in the last case for building up a good gain.

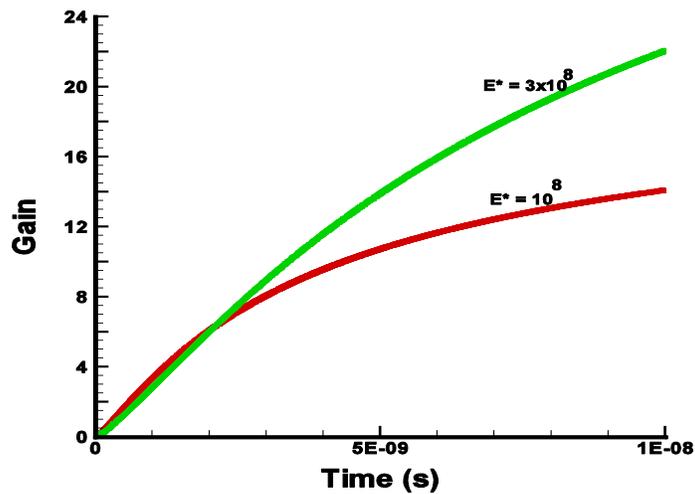

**Figure 8-22**. DT fusion reaction gains for the cases of Figs. 8-20 and 8-21 (Lalousis et al 2013).

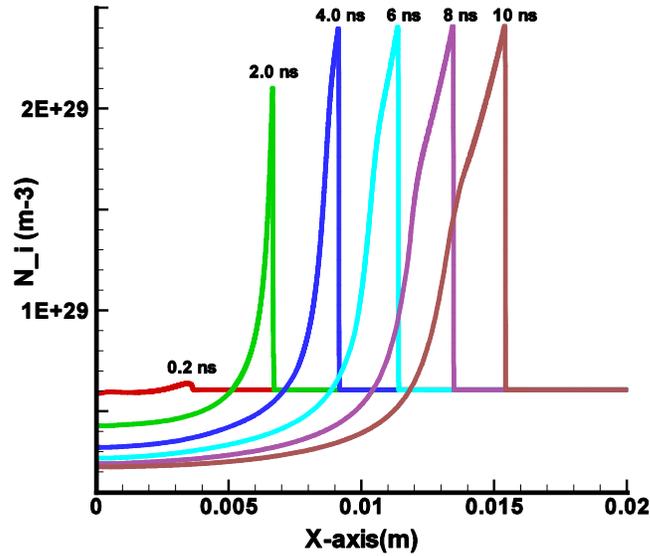

**Figure 8-23.** Ion density depending on the depth x of the propagating fusion flame at same times as in Fig. 8-20 .

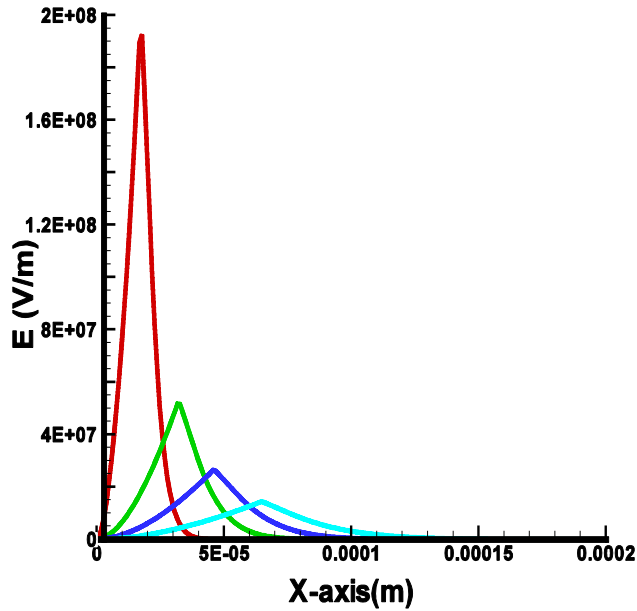

**Figure 8-24.** Longitudinal electric field E depending on the depth X in the DT fusion fuel close to the interaction range of the ps laser pulse with an energy flux $E^* = 10^8$ J/cm$^2$ for the cases of Fig. 8-23, for the times with decresing maxima: 40ps; 400ps; 1 ns; 2 ns

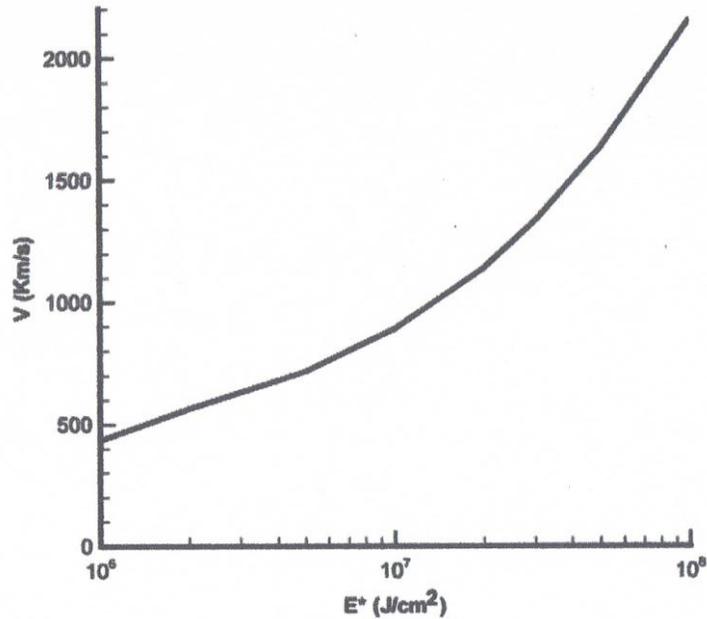

**Figure 8-25.** Velocity of the fusion flame at 2 ns at varying energy flux density E* of the laser irradiation (Lalousis et al. 2012).

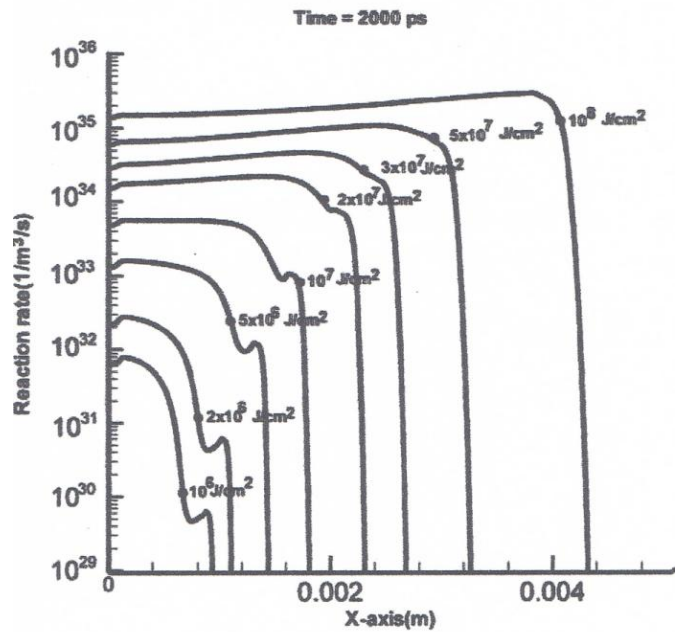

**Figure 8-26.** Reaction rate depending on the depth x at 2 ps for various energy flux E* of the ps laser initiation.

Generation of increase of shock densties at the top of the fusion flame could be derived from early computations (Cang et al. 2005; Kant et al. 2012; V. Nanda et al. 2014a). The more details derived from application of the multifluid computations showed an increase of the ion density as compression at the fusion flame whicih value was four times of the density of the untouched DT fuel inot which the flame was propagating (Fig. 8-23). The increase by a factor 4 is just the value of the Renkine-Hugoniot analytical approximation. The increase of the the thnickness of the shocked plasma is an effect of the while thermal collision processes whith were all included in the computations. What was unexpected was that the compression at the flame front was not instantly happening but only more than 2 ns after the flame was initiated and even at 0.2 ns, nearly nothing could be seen from the compression. The longitudinal electrc field depending on the depth x at gven times is shown in Fig. 8-24.

The properties of the shocks as known for shock ignition for nanosecond laser pulse irradiation for fusion (Betti et al 2007; Eliezer et al 2011) are detailed in Section 10, but it may well be interesting how the initial stage of the flame generation during the first hundred picoseconds of our evaluations may relate to these studies. From a series of plots as in Fig. 8-23, the velocity of the flame propagation can be derived. It is always slowly decaying on time. The velocity in the range below 50 ps has values exceeding 10,000 km/s and at few nanoseconds, the velicies are above Mach 3000. Fig. 8-25 shows the velocity at 2 ns at varying laser energy flux E* for initiation of the flame. The ignition is perfect above $E^* = 10^8$ $J/cm^2$ at the velocity of 2300 km/s.

The fact of ignition for $10^8 J/cm^2$ with the laser intensity of $10^{20} W/cm^2$ can be seen from the reaction rate, Fig. 8-26 where the the rate along the depth x is not decreasing. For E* less than $10^7$ $J/cm^2$, no ignition is happening as seen from the decreasing rate. It has to be realized, that evauations by PIC computations, Fig. 8-18, have to clarify the steps of the interaction for the initiation peocess of the fusion flames. Experimental results of emission of protons from thin film targets show a nonuniform behaviour with different groups of ions (Zepf et al. 1996; 2003). Irradiating diamond layers of nanometer thickness (Steinke et al. 2011) result in total absorption of layers of 2.3% thickness of the laser wavelength. There is not any tunnelling, and PIC results are declared as not sufficient. The maximum transfer of radiation to 6 pm diamond layers with the then sufficiently low Debye layer may be explained as an ideal nonlinear absorption process (Hora 2012).

It remains to be mentioned that Sauerbrey's (1996) experimental discovery of the ultrahigh acceleration was close to be realized under similar conditions (Kalashnikov et al. 1994). But for the clear result of the line shift at the Doppler effect, extensive elaborations and clarification were needed by Sauerbrey (1996). A similar prelude of the acceleration of 400 MeV ions without relativistic self-focusing (Clark et al 2001) by inclusion of the quantum modified collision frequency (Haseroth et al 1996, Hora 2003) was given by the measurement of fusion by 1.3 ps laser pulses at $10^{19}$ $W/cm^2$ neodymium glass laser intensity. The gains – converted for DT – of nearly 0.1% (Norreys et al. 1998) directed well towards the here considered plasma block ignition for fusion as a modification of fast ignition (Tabak et al. 1994).

# CHAPTER 9
# Laser Driven Fusion with Nanosecond Pulses

With the discovery of the laser (Maiman 1960) and following the work of Schawlow and Townes (1958) by Collins et al. (1960) it was evident that this extremely high energy concentration to short time and small volumes may offer a solution of the energy problems by fusion energy as immediately realized by the center of Edward Teller (2001, 2005) as explained by Nuckolls (2007). Sakharov (1982) underlined in an unusually expressed way how he from the beginning was fascinated by this concept. Nearly the whole generation of energy in the Universe is produced as in the sun from the nuclear fusion of hydrogen into helium converting the released mass difference $\Delta m$ of the nuclei into energy E according to Einstein's relation $E = \Delta mc^2$ where c is the vacuum speed of light.

## 9.1 General Approach

In the following attention is given to the reaction of heavy with very heavy hydrogen, deuterium D and Tritium T respectively (DT) given by the relation

$$D + T = {}^4He + n \quad + 17.2 \text{ MeV} \qquad (9\text{-}1a)$$

producing helium He and a neutron n. This is the easiest fusion reaction and was used in the very first manmade exothermic explosive fusion reaction on 1 November 1952 (Teller 2001). Another interesting reaction is the fusion of the isotope 11 with a proton (Miley 1972; Hora 1975a)

$$^{11}B + p = 3\,{}^4He + 8.999 \text{ MeV} \qquad (9\text{-}1b)$$

This HB11 reaction primarily does not produce neutrons but it is very difficult to achieve instead of the explosive DT reaction. It is the aim since many years studies to produce this in a controlled way in a power station by confining the plasma by magnetic fields (ITER or Wendelstein W-7 stelarator project (Grieger et al. 1981) etc.) while the use of lasers for achieving break even may be close to be realized by the largest laser on earth with the NIF (National Ignition Facility) at the LLNL (Lawrence Livermore National Laboratory) near San Francisco/CA.
     This Chapter 9 summarizes the studies using laser pulses of about nanosecond (ns) duration where the interaction with the plasma for fusion is mainly determined by heating and gas-dynamic pressures for high compression of the fuel and subsequent thermally driven fusion reactions. This is basically different from the results reported in the following Chapter 10 using laser pulses of picoseconds (ps) duration where thermal interaction is essentially avoided for initiating the reaction in favour of direct transfer of laser energy into macroscopic plasma motion avoiding thermal losses initially and reducing later losses by thermal and radiation processes, instabilities and delays. These new developments are in an early stage based on the discovery of Mourou's Chirped Pulse Amplification (CPA) (Section 8.1) for producing extremely high laser powers for pulses shorter than ps duration (Strickland

et al. 1985; Mourou (1994); Perry et al. (1994) Mourou et al. (1998); Mourou et al. (2013)) down to attosecond (as) duration (Krausz et al. 2008).

The ns-laser pulse was the only possible option before Mourou discovered and developed the CPA method for the extreme laser powers. The classical development was based on direct drive where a spherically irradiating laser pulse is hitting the fusion fuel contained in a pellet and the ablative expansion of the heated surface leads to compression and thermally ignites the fusion reaction. We have here to go back to the initial studies with comparable low fusion gains but where a turning point was discovered numerically (Hora et al 1978) when the reheat of the generated alpha particles in DT and the varying partial re-absorption of bremsstrahlung was producing a volume ignition. We have to explain how this was crucial in achieving high neutron gains.

The reheat process has been shown to be essential (Hurricane et al. 2014) to increase the reactions gained when using an alternative laser fusion scheme, the indirect drive with (central core) spark ignition where the initial low fusion gains are now reaching record values growing to initial expectations. This is going to be explained following the scheme of indirect drive in Fig. 9-1a. The success of this is expected from detailed computations similar to measurements where the hohlraumstrahlung from nuclear explosions produced high gain fusion energy form irradiated DT targets.

Nuclear fusion under the irradiation of hohlraum radiation is a crucial motivation for the indirect laser irradiation concept (Nuckolls 2007, Lindl 1994) of which deeper elaborated details are not intended to be presented in this book. The studies of this kind of pellet compression may go back to the key result for demonstrating the convincing success of radiation driven laser fusion (Broad 1986; Phipps 1989) referring to underground nuclear explosion experiments where the generated Planck radiation was irradiated on a fusion pellet and a non-disclosed very high DT reaction gain was measured. This is a substantial result for controlled laser driven fusion to arrive at very high efficient energy production differing and compared with the otherwise acknowledgeable result by the JET (Joint European Torus) magnetic confinement fusion experiment (Keilhacker 1999) with nearly break-even results, where 16 MW fusion energy was produced from irradiation by 20 MW neutral beam injection and 4 MW microwave irradiation. This JET result does not include the energy losses for driving the neutral beam, the microwaves and the "target" which in this case is the running of the tokamak for the raction. This experiment was considered also as a kind of "neutral beam injection fusion" (Hora 2004) as a case of nonlinearity, see Sction 6.3.

The experiment at LLNL is one of the most advanced achievements in laser technology on earth (Haan et al. 2011; Glenzer et al 2011, 2012) and is the only large scale fully established experiment closest for generation of controlled nuclear fusion energy for power stations. But it resulted initially in lower than expected neutron gains (Clery 2012). The higher gains may have been expected (Fig. 9-1b) derived from direct drive volume ignition experiments (Hora 2013) where the reheat in the reaction was essential (Hora et al 1978, 1998). A simplified and not sophisticated reaction gain for comparison may be based on the number of DT neutrons per energy $E_L$ of the incident laser pulse. Table 9-1 is a list of measured results with exception of the extrapolated values expected from direct drive volume ignition confirmed up to the irradiation of 35 kJ laser energy.

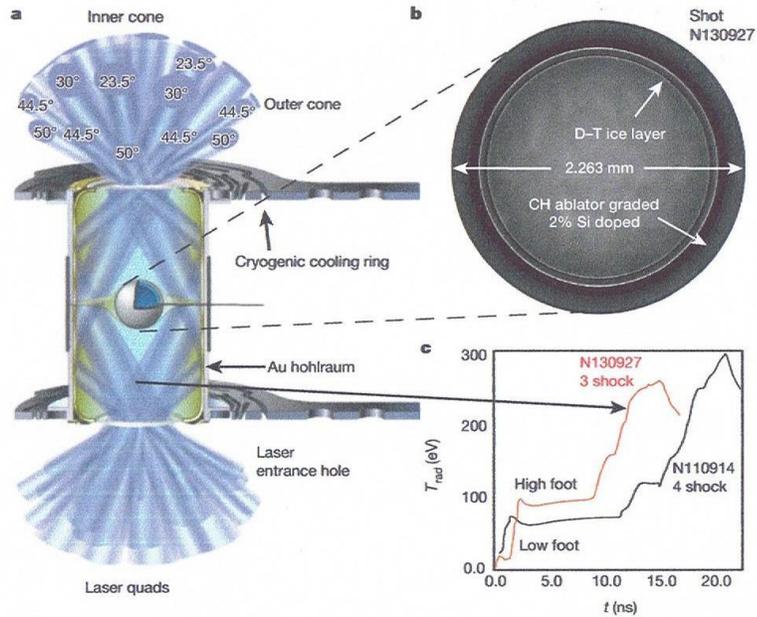

**Fig. 9-1a.** Indirect drive laser fusion with spark ignition at LLNL (Hurricane et al. 2014). a) 192 laser beams enter to upper and lower opening of the cylindrical volume to irradiate its inner surface for conversion of the laser radiation into x-ray Planck hohlraum radiation of temperature $T_{rad}$. This radiation hits the pellet b) to compress it and ignites in an inner core a high temperature reaction which is developing as a fusion detonation wave into the surrounding high density modest temperature plasmas for producing the computed high gain net fusion energy. c shows the measured time dependence of $T_{rad}$.

**Table 9-1.** Comparing neutron numbers N and the energy of the incident laser energy $E_L$ measured at experiments for comparison

| Neutrons N | $E_L$ | energy gain | Year | Laboratory |
|---|---|---|---|---|
| $10^{12}$ | 1 kJ | 0.0027 | 1985 | ILE Osaka (Yamanaka et al. 1986a) |
| $10^{13}$ | 10 kJ | 0,0027 | 1990 | ILE Osaka (Azechi et al. 1991) |
| $10^{14}$ | 35 kJ | 0.0076 | 1995 | LLE Rochester NY (Soures et al. 1996) |
| $2\times10^{14}$ | 1.8 MJ | 0.00054 | 2012 | LLNL Livermore CA (Glenzer et al, 2012) |
| $5\times10^{15}$ | 1.8 MJ | 0.007 | 2013 | LLNL Livermore CA (Hurrricane et al. 2013) |
| $6\times10^{15}$ | 1.9 | 0.008 | 2013 | LLNL Livermore CA (Hurricane et al. 2013) |
| $1.5\times10^{18}$ | 1.5 MJ | 2.72 | | numerical result by direct drive volume ignition |

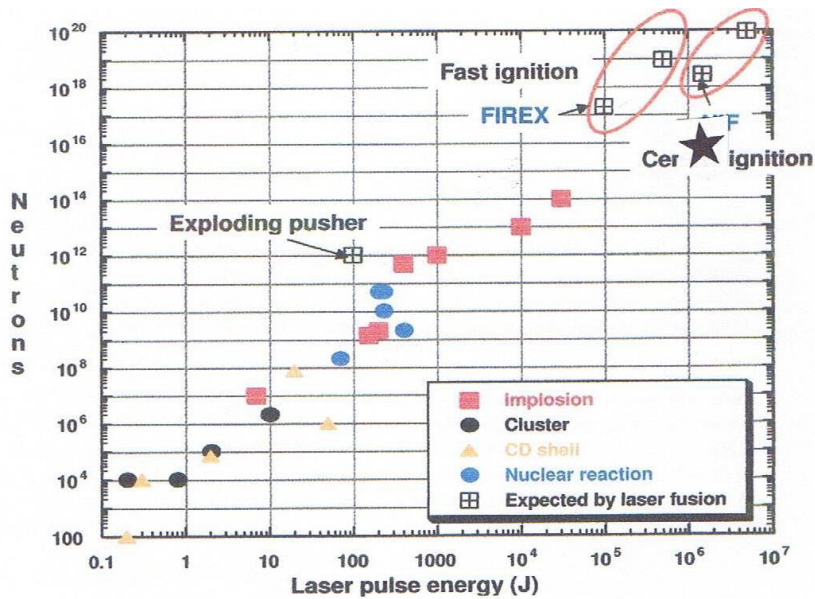

**Fig. 9-1b.** Measured highest neutron gains N per incident laser energy (full signs) with the highest value reported (Hurricane et al 2014) given by an asterisk (see, Nakai 2008; Hora et al. 2011a)

The extrapolation is taken from Fig. 9-1b where the asterisk is from the results at Livermore (Hurricane et al 2014) showing the achievement with indirect drive spark ignition which is different from all the measured results with direct drive and volume burn as a pre-stage of volume ignition in Fig. 9-1b. The experiments at Livermore in 2013 are more than 10 times higher than in 2012 (Hurricane et al. 2014). The very detailed evaluation showed that there was an increase of the reheat by a factor of about 10. This reheat effect for indirect drive spark ignition may not directly be compared with the drastic change of the fusion at laser direct drive (Hora et al 1998), but with some approximation there is a reasonable similarity as it is shown within this Section.

Why not to use the dirct drive with the hightest gains measured up to 35kJ input laser pulse energy and extrapolated to 2 MJ based basically on alpha reheat in Fig. 9-1b, while indirect drive (Fig. 9-1a) resulted in lower gains where only a partial increase could be shown to be by reheat (Hurricane et al 2014). The important reason for using the irradiation by Planck's hohlraum radiation of an equilibrium black body temperature above 300eV is given by measurements of the first completed historical success of controlled inertial confinement fusion for the convincing realization of fusion power stations in contrast to magnetic confinemet fusion. The details of this results are not published and only the results of "very high gains" became known through the New York Times (Broad 1988) reporting on otherwise classified experiments where the the Planck radiation came form experiments of underground nuclear explosions. It is crucial that these experiments possibley may have to be repeated for power generation. Many details for the power station have to be explored for the indirect drive (Lindl 1994) after the performance of underground explosions have been banned internationally.

The other question is whether it was possible by the underground nuclear reaction to measure whether this was a uniform adiabatic volume ignition (Hora et al 1978; Lackner, Colgate et al 1994 – as those with the highest measured gains, Fig. 9-1b, to be discussed in many more details – compared with the reaction of spark or cetral core ignition, on which the experiments of Fig. 9-1a are based (Haan et al 2011; Hurricane et al 2014). These different reaction schemes will be considered at least in the basic properties in this Chapter 9. The question may be open at this stage, whether the experiments with the explosions (Broad 1988) could be performed in a way that a distinguishing between volume or spark ignition had been followed up by experiments. These comments are rather individual remarks and must not be confused with the extremely higher dimension of technological and computational knowledge achieved with the excepoally highly developed experiment (Haan et al. 2011, Hurricane et al. 2014).

## 9.2. A PRELIMINARY VIEW

Among the different options, one has to consider different schemes for compressing the DT fuel plasma. There is the spark ignition model with isochoric (Kidder 1974;1976; Bodner 1981) or isobaric (Meyer-ter-Vehn 1982) conditions, as well as the ideal isentropic model of volume ignition with self-heating and re-absorption of bremsstrahlung giving a very high increase in temperature, as first shown by Hora and Ray (1978), see also Hora et al (1979). Further extensive work has been well recognized (Kirkpatrick and Wheeler 1981; Hora and Miley 1984; Cicchitelli et al. 1988; Kasotakis et al. 1989a,b,1997; Basko 1990, 1993; Pieruschka et al. 1992; Martinez-Val et al. 1993a,b, 1994; Eliezer and Hora 1993; Khoda-Bakhsh 1993; He and Li 1994; Tahir and Hoffmann 1994; Oparin and Anisomov 1994; Anisimov et al. 1994; Atzeni 1995; Hora et al. 1995;1996). The fact that a collapsing shell merges nearly into an isentropic compression of the self-similarity model (Heckmann 1942, Hora 1991: Section 5) was shown by Velarde et al. (1986).

We shall give a preliminary view of current results in this Subsection. Some early developments of volume ignition are summarized in Subection 9.3. Volume ignition provides gains (Subection 9.4) comparable to those of spark ignition. These comparisons will indeed show that there are very different parameters of geometry and burn efficiencies in the different cases that lead finally to similar high gains. The advantage of volume ignition has already been explained, namely, the compression scheme is much simpler and more "robust", and has fewer problems with respect to instabilities (Lackner et al. 1994) compared with the difficulties of spark ignition (Meyer-ter-Vehn 1996; Teller et al. 1997). In Subsection 9.5, we compare several experimental results giving high laser fusion gains with self-similar volume compression computations, and in Subsection 9.6 we draw some conclusions as to how experiments with the present volume-compression results may lead to volume ignition.

The unexpected result is that the *method* of isentropic (self-similar volume) compression leads to agreement with current experiments. This is very surprising in view of the very sophisticated analytical and numerical methods used in studying non-isentropic isobaric spark generation.

Figure 9-2 is a compilation of the best laser fusion gains of the earlier experiments . The line for DT neutron yields larger than $10^{11}$ is from Hora (1991: Fig. 13.15). Against the trend of decreasing fuel density with increasing neutrons the line increases according to the ILE results (Takabe et al. 1988), which may roughly follow an increasing line corresponding simply to an isothermal condition with the yield given by the square of the density. The results of LLE Rochester (Soures et al. 1996; McCrory et al 1997; Seka et al 1997) with $10^{14}$ neutrons from 30 kJ direct-drive laser pulses with DT densities of $4n_S$ (1 g/cm$^3$), exactly fit the calculated line in Fig. 9-2. This agreement of the Rochester result with the estimated isothermal line

encouraged us to compare these optimized fusion gains with the volume compression (self-similarity) model. A more detailed discussion of this figure with respect to the parameters such as the mass, temperature, etc., for each point is too complex in view of the different sources from which the data have been taken. Here we have given a continuation of similar diagrams published earlier, where the recent results from Soures et al. (1996) show a remarkable fit with those of Hora (1991). This is quite surprising, since the goal in laser fusion has been to achieve conditions for spark ignition rather than for volume ignition, although the present gains are far below the conditions for ignition, and what has been observed so far is a simple burn (on the way for reaching reaching ignition). It also underlines the later further studied double shell evaluations (Amendt et al. 2005).

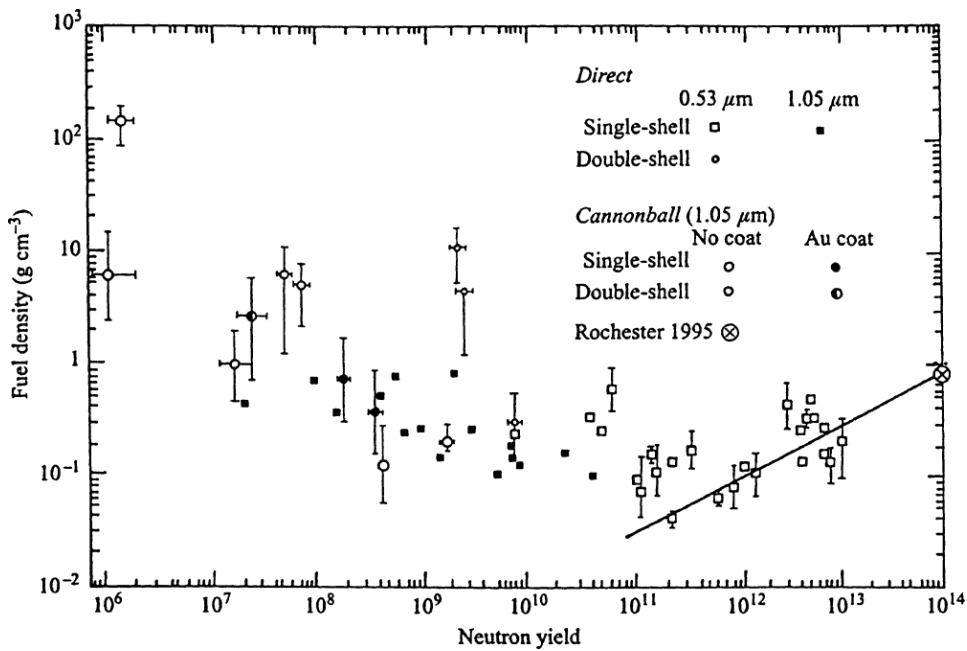

**Figure 9-2.** Measured DT maximum densities and neutron yields in laser fusion from the last four years, including an isothermal line (Hora 1991, Fig. 13.15) and the result from LLE Rochester (Soures et al. 1996).

## 9.3. HISTORICAL REMARKS ON VOLUME IGNITION

Following a very constructive request by earlier referees, we shall reflect on some historical developments that led to volume ignition. A difficulty arises owing to the fact that many of large-scale fusion reaction results were not published and not generally known, so we can draw some of our conclusions only indirectly.

Credit is given to Meyer-ter-Vehn (1982) for his analytical arguments for the isobaric scheme, although it has been pointed out (Teller and Nuckolls 1997) that his results were already well known—at least from numerical studies.

Volume ignition is preceded by volume compression and ignitionless burning of the nuclear fuel. We shall consider some of the relevant studies in more de- tail. What was calculated very early on (Hora 1964, 1971a) was the simple fusion burn in a sphere of radius $R_0$ and volume $V_0$ of DT fuel of a given uniform density $n_o$, resulting in a temperature $T_0$ after an energy $E_0$ had been deposited with an initial expansion velocity of zero. Subsequent self-similar adiabatic expansion with cooling resulted in the generation of nuclear fusion energy. The gain is then defined by

$$G = \text{(fusion energy)} / E_0. \tag{9-2a}$$

The justification for using the self-similarity model, as done by Basov and Krokhin (1964) and Dawson (1964), was clarified mathematically by Hora (1971b). The gain at fixed initial density $n_0$ depending on the input energy $E_0$ for a volume $V_0$ of the fuel sphere is shown in Fig. 9-3a (Hora 1964). The optimum gain $G$ is given by the asymptotic line (envelope) of the parabolic curves in Fig. 9-3a. Considering only this optimized gain gives

$$G = 1.47 \times 10^{-17} E_0^{1/3} n_0^{2/3} \tag{9-2b}$$

(where $E_0$ is in J and $n_0$ in cm$^{-3}$). This could be seen from the plots shown in Hora (1964) and was published in the form (9-2b) in Hora and Pfirsch (1970, 1972). It was noted by Hora (1964) that the gains calculated previously by Basov and Krokhin (1964) and Dawson (1964) were the same as those shown in Fig. 2—but on the parabolas at much lower values than the optimized values at the envelope. The maximum gains resulted only from the envelope.

On writing (2) with the energy $E_o$ normalized to the break-even energy $E_{BE}$ and the density of highest compression normalized to the solid-state DT density $n_s = 6 \times 10^{22}$ cm$^{-3}$ (corresponding to a density $\rho = 0.25$ g cm$^{-3}$), the following is obtained

$$G = \left(\frac{E_0}{E_{BE}}\right)^{1/3} \left(\frac{n_0}{n_s}\right)^{2/3}, \tag{9-3}$$

where the break-even energy $E_{BE} = 6.3$ MJ if compression and expansion are calculated and the optimum temperature $T_{opt} = 17$ keV, where a Maxwellian equilibrium has been chosen corresponding to the above-mentioned envelope lines. We mention further that the energy $E_0$ can be expressed as

$$E_0 = 3kT_0 n_0 \frac{4}{3}\pi R_0^3, \tag{9-4}$$

taking into account one electron to each nucleus, so that (3) reads

$$G = \text{const} \times n_0 R_0 \tag{9-5}$$

We see that this is algebraically identical with (9-3), where the constant in (9-3) has to be taken for the optimized temperature $T_{opt}$. It should be noted that the $nR$ criterion (or $\rho R$ when using the DT density $\rho$ instead of the particle density $n$) for inertial-confinement fusion as a substitute for the Lawson criterion for magnetic-confinement fusion was first published by Kidder (1974) and at nearly the same time by Fraley et al. (1974) as a result of extensive analysis and computation. As shown in Eqs. (9-3) to (9-5), this formulation is algebraically fully identical with the formulation (9-2b) published in Hora and Pfirsch (1970, 1972).

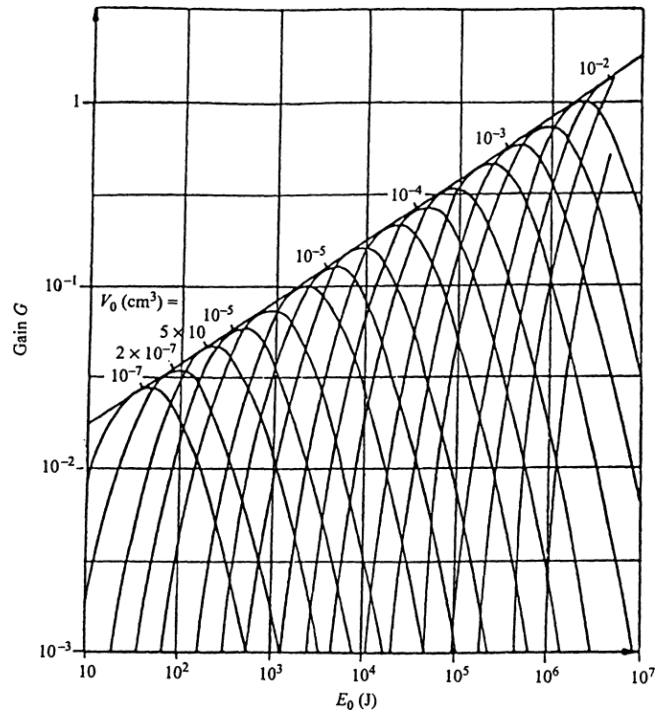

**Figure 9-3a.** Inertial confinement fusion gains G depending on the energy $E_0$ transferred into a DT plasma of density $N_0 = n_s = 6 \times 10^{22}$ cm$^{-3}$ (the solid-state density), with the spherical volume $V_0$ as parameter (Hora 1964).

The expression (9-3) for the gain $G$ has the advantage that it shows immediately how compression is favourable, for example, compression to $10^3$ times higher values requires $10^{-6}$ times the input energy $E_0$ if the same gain $G$ is to be reached. This usefulness of compression was also seen in the plots of Hora (1964), but it was not clear how laser irradiation by ablation of the plasma corona could compress the plasma interior over the full thickness of the volume and not only in thin layers as in shock waves. The ablation-compression mechanism of laser irradiation was first suggested by the results of numerical studies (Mulser 1970; Rehm 1970), and much more detailed results from the same time were published later (Nuckolls et al. 1972; Boyer 1973). It should be stressed that the advantage of compression for nuclear reactions was well known from unpublished work of Seth Neddemeyer, John von Neumann and Edward Teller 1942 (see Nuckolls et al. 1973). Nevertheless, it seems obvious that compression was a new aspect of a special improvement of the then undisclosed work in the Soviet Union in 1954 in contrast to earlier work. While the first exothermic fusion reactions were demonstrated by Edward Teller on 1$^{st}$ of November 1952 (see Teller 1987), Sakharov's H-bomb was ready in 1953 for use from an aircraft. It was reported that after this an enormous development in the efficiency of these devices was achieved within the few months to 1955 when compression was introduced, and theoretical contributions, with a long list of authors from E.N. Avrorin to A.N. Tikhonov, using self-similar solutions for "atomic compression" were especially highlighted (Goncharov 1996).

These results were indeed not known in the open literature in 1964, when the results of Hora were the work of a an individual, like those of Dawson, in contrast to the large-scale research and the later published work on laser fusion (Nuckolls et al. 1972, 1973; Fraley et al. 1974), where in one case more than 500 hours computations with a CD7600 were reported. The results for the gain (9-3), or for Kidder's $\rho R$, however, were based on very simplified assumptions. In the extensive computations by Nuckolls et al. (1972) and Fraley et al. (1974), as well as by Hora (1964), the following was neglected:
  (a) Losses by bremsstrahlung;
  (b) Depletion of fuel by the reaction;
  (c) Increased gain due to self-heating by reaction products: alpha particles and neutrons.

Problems of sufficiently fast equipartition for thermal equilibrium and double-layer effects were mentioned later. Despite these large-scale computations, the special effect of volume ignition was not noted, since attempts were to be made to reach very high laser fusion gains using a very sophisticated model of spark ignition with fusion detonation fronts. Nevertheless, it is important to mention the connection with the earlier results where the gain $G \propto E_0^{2/3}$ (Hora and Pfirsch 1970,1972) or Hora (1964, 1971a) and the $\rho R$ dependence were elaborated explicitly by Fraley et al. (1974).

The direct improvement in the gain computations of Hora (1964, 1971a and Hora and Pfirsch 1970, 1972) by including (a), (b) and (c) resulted in a remarkable numerical observation of volume ignition (see Fig. 9-3b). Starting from an appropriate initial density and volume, a small change in the energy $E_0$ leads to a very different gain $G$ (Hora and Ray 1978). An initial temperature of 1.07 keV results in a time dependence of the temperature in the form of a monotonic decay—indeed, with slower cooling, there is a little increased gain $G$ of 0.77 than without reheat (c). A small increase in the energy $E_0$ ca es an initially slow increase in the temperature $T$ (expressed by $E_{av}$ in Fig. 9.3b) and later a rapid growth to temperatures as high as 100 keV and more, with an abrupt decay when the plasma is adiabatically expanding very rapidly, resulting in a gain $G$ of 1900. A very slightly higher energy results in a faster rise of the temperature and faster expansion, causing a lower net gain. This temperature effect in volume ignition was essential also in the following work of Kirkpatrick and Wheeler (1981), and was most significant when too-rapid heating of the ions left the electrons at lower temperatures, as seen by Martinez-Val et al. (1994) and He and Li (1994).

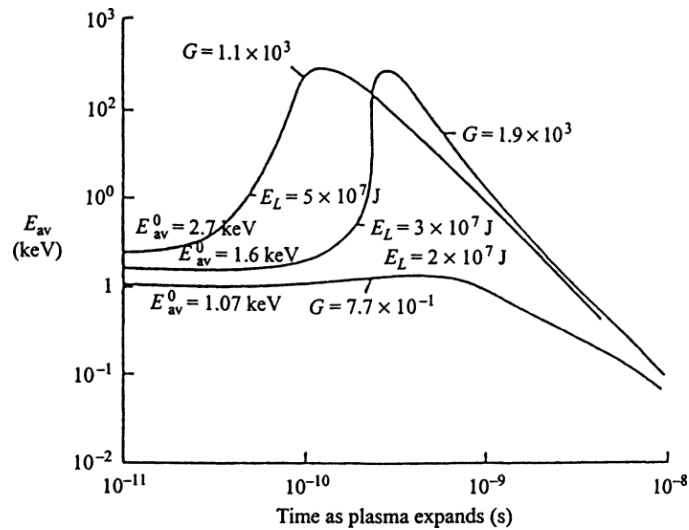

**Figure 9-3b.** Time dependence of the temperature (expressed by the internal average energy $E_{av}$) for a spherical DT plasma of volume $10^{-3}$ cm$^3$ at an initial compression to $5.8 \times 10^{25}$ cm$^{-3}$ ($10^3$ times the solid-state density) when an energy $E_{av}$ is put into the volume (equivalent to $E_o$ in the text) (Hora and Ray 1978). $G$ is the resulting core gain, Eq. (6) Superscript zero indicates initial values.

This temperature jump due to self-heating causes a self-produced strong addition of driver energy, added to that of the laser, and leads to high gains, comparable to that of spark ignition (see Subsection 9.4). The difference is that the compression dynamics follows the natural adiabatic parameters, avoiding entropy-producing shocks. This type of stagnation-free volume compression gave the highest fusion gains (Yamanaka et al. 1986b), with uniform X-ray emission, in contrast to the less-efficient cases with shocks. It was also underlined that this volume ignition is very robust compared with the dynamically very complicated compression for spark ignition (Lackner et al. 1994).

For historical considerations it is important to note that the most sophisticated computations (Nuckolls et al. 1972; Fraley et al. 1974) could well have shown the volume ignition effect of the strong rise of the temperature due to self-heating, but—according to the publications—it was not sought since the aim was to study spark ignition where indeed too-high temperatures in the ignition fronts had to be avoided. Apart from the previously ignored (and only derived later) effect of high temperatures on volume ignition (Hora and Ray 1978), the effect of partial reabsorption of bremsstrahlung included in these calculations (without the later introduced small modification due to relativistic effects on the optical constants; Scheffel et al. 1997; Kasotakis et al. 1997) was observed numerically by Fraley et al. (1974). At very high compression and/or sufficiently large volume, the initial temperature can be as low as 2.5 keV or even less (Hora and Ray 1978). The code for these computations was published later (Stening et al. 1992). This low- temperature ignition (LTE) due to partial reabsorption of bremsstrahlung common to the work of Fraley et al. (1974) and Hora and Ray (1978) was also noted by Caruso (1974), but again *the effect of heating to very high temperatures at volume ignition was not noted before the publication of Hora and Ray (1978)*, see Fig. 9-3 and Hora et al. (1979) We should mention that a more detailed discussion of LTE was recently made possible by an extension of the self-similarity model to non-adiabatic conditions with respect to radiation processes, where cases of fusion gains from volume ignition resulted that were even better than those for spark ignition (Murakami 1997; Johzati et al. 1998).

At this point, we should clarify some definitions. In this section we have used a gain based on the energy $E_0$ transferred from the laser into the reacting volume $V_0$. This is only 5–15% of the total interacting laser energy $E_L$. The remaining 85–95% of the energy $E_L$ is needed for the ablation of the plasma corona to produce the recoil to compress the reacting plasma. This depends on the hydrodynamic efficiency $\eta$. In contrast to gas-dynamic ablation, an efficiency of between 45% and 50% is possible if non-thermalizing ablation due to a nonlinear force (Hora 1969,1996) is used (Hora et al. 1996). In contrast to the gain $G$, Eq. (9-2a), used earlier now called the core gain, we define a total gain $G_{tot}$:

$$\text{core gain} \quad G = \frac{\text{fusion energy}}{E_0}, \tag{9-6}$$

$$\text{total gain} \quad G_{tot} = \frac{\text{fusion energy}}{E_L}. \tag{9-7}$$

In addition to the preliminary observations of Subsection 9.2, we note that the core gain in the experiments of the Osaka (Takabe et al. 1988) and Rochester (Soures et al. 1996) groups, with a core gain of energies of about a kilojoule, producing $10^{13}$ and $10^{14}$ neutrons respectively, correspond to the cases of good absorption that were mentioned as early as 1974 (Fraley et al. 1974).

## 9.4. VOLUME IGNITION COMPARED WITH SPARK IGNITION

In order to explain the conditions of fusion burn and ignition, we present here a comparison of spark ignition with volume ignition. (This has been done previously in Hora (1991, p. 357), but the treatment here is more detailed.) There were difficulties in understanding why in both cases the fusion gains were nearly at the same very high level although the volumes and densities were so different.

Let us first consider spark ignition, with the well-known case of Storm et al. (1988) (see also Fig. 13.13 of Hora 1991). Instead of the curved radial tem- perature and density profiles (Fig. 9-4), at the stage of highest compression we select an averaged inner spark region of 250 times the solid-state density $n_S$, and temperature 10 keV (maximum 12 keV), changing at the detonation-front radius $0.47 r_0$ into an outer region with average temperature 400 eV and density $1000 n_S$ (maximum $2300 n_S$). The plasma radius at highest compression is derived from gains based on an initial radius $r\ r\ 0_S = 3$ mm at a time before compression. The efficiency $f$ of the fusion detonation wave going through the outer high-density, low-temperature DT plasma after being triggered by the inner spark is rather low according to the Fraley–Linnebur–Mason–Morse formula (Hora 1964, 1971a)

$$f = \frac{H}{H + H_B} = 22\% \qquad (9\text{-}8)$$

using $H = \rho R = 2$ (see Fig. 4), and $H_B = 7$ g cm$^{-2}$. The total fuel depletion (fuel consumption) is then only 18.9%, giving a total gain $G_L = 100$ for 10 MJ laser energy, assuming $\eta = 10\%$ hydrodynamic efficiency, which is defined as the ratio of the energy $E_0$ that goes into the compressed DT plasma core of radius less than $r_0$ divided by the energy $E_L$ of the incident laser pulse; see Eq. (9-7). All of these values are reasonably close to the detailed computation of Storm et al. (1988), where, instead of our averaged values, the computed density and temperature profiles shown in Fig. 4 for the conditions of an isobaric compression were used.

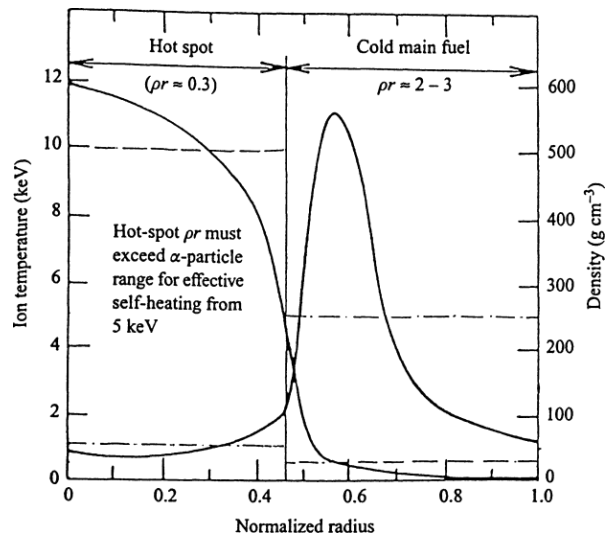

**Figure 9-4.** Radial profiles of the ion temperature and density in a DT pellet computed at highest compression for isobaric spark ignition according to Storm et al. (1988). The radius of the fusion detonation wave is at 0.47 of the actual plasma radius; the dashed lines are the average temperatures of the inner and outer regions and the dashed-dotted lines are the averages of the respective densities.

The comparison with volume ignition is based on the computations shown in Fig. 9-4 (Hora 1991, Fig. 13.6). The volume of the fuel at solid-state density before laser irradiation $V_{0s}$ is compressed adiabatically according to the isentropic self-similarity model (Hora 1964, 1971a) to a maximum density $n_o$, from which it starts to expand adiabatically. A typical case is shown by the dashed parabolas of Fig. 5. If the input energy $E_0$ is too low, the low maximum temperature permits only low core gains $G$ (9-1) or (9-6), owing to the low fusion cross-sections. If the input energy results in too high a maximum temperature then the adiabatic expansion is too fast and also results in low gains. At the optimum temperature of 17.1 keV the optimum gains are reached, which lie on the solid curves linking the peaks of all the parabolic curves as envelopes in Fig. 9-5 (as in Fig. 9-3a). The parabolas are symmetric only for core gains $G$ below 8, following the relation (9-3) given by Hora (1971b, 1991).

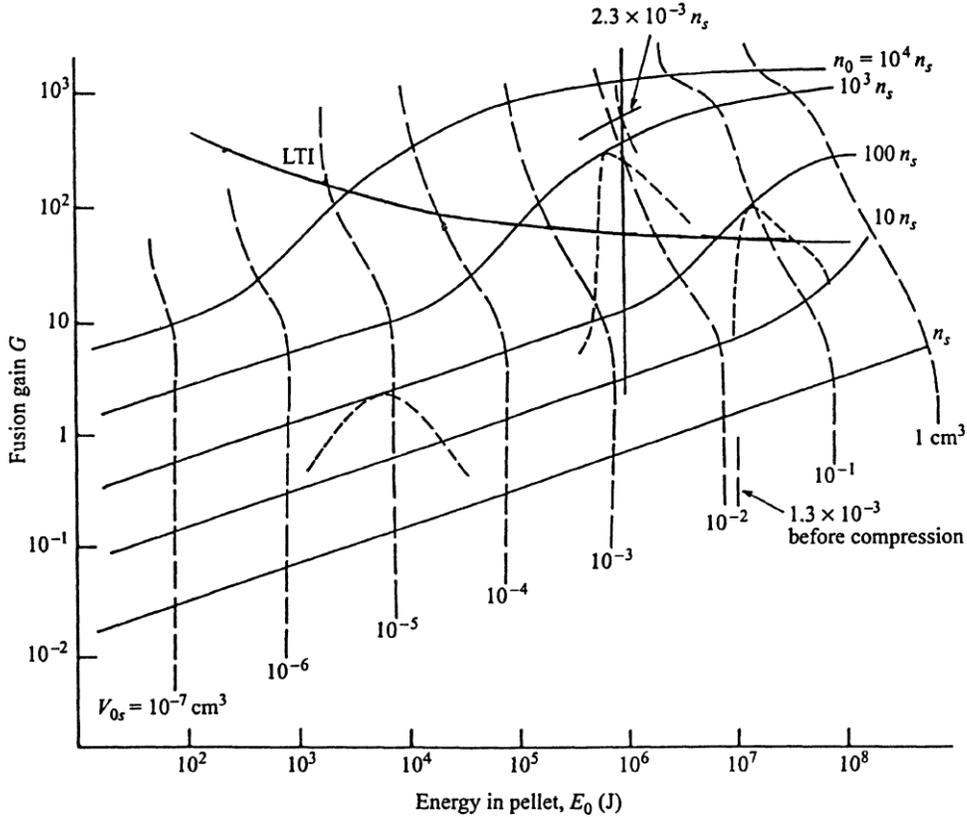

**Figure 9-5.** Optimized core gains $G$ for inertial fusion energy for varying maximum density $n_0$ of the uniformly compressed DT fuel in multiples of the solid-state density $n_s$ depending on the energy $E^0$ transferred from the laser pulse to the compressed core (Hora and Ray 1978; Hora 1991) using the computation code presented by Stening et al. (1992). For core gains $G$ larger than the LTI (low-temperature ignition) curve, the initial temperature $T_{opt}$ is less than 4.5 keV owing to strong reabsorption of bremsstrahlung.

The relation (9-3) holds only for $G < 8$, as can be seen from the bending up of the lines in Fig. 9-5 (Hora and Ray 1978), where in contrast to the preceding computations, the fuel depletion, the loss and partial recovery of bremsstrahlung, and the alpha-particle reheating were included. Figure 5 shows how the parabolic curves become distorted for $G > 8$; this is the ignition phenomenon. Steepjumps in the former parabolas appear at input energies above $G^* = G = 8$. Below this value, a simple burn and adiabatic expansion of the plasma occurs, and above $G^*$ (as shown in Hora and Ray 1978) the alpha-particle reheating and the reabsorption of bremsstrahlung heats up the ions, for example to temperatures above 100 keV, resulting in high gains and high fuel depletion (see Fig. 9-3).

For a comparison of spark ignition and volume ignition, we start from the same input laser energy of 10 MJ and the same maximum compression (uniform) density of $2300n_s$, and we assume a hydrodynamic efficiency of 10%. The core energy $E_0$ in the compressed plasma is then 1 MJ, the solid-state DT radius before compression $r_{0s} = 1.45$ mm and the volume before compression $V_{0s} = 1.3 \times 10^{-2}$ cm$^3$. The initial temperature is then $T_0 = 2.667$ keV, showing some reabsorption of bremsstrahlung (since this temperature is below 4.5 keV for the case without such reabsorption). The core gain taken from the diagram in Fig. 9-3b is then $G = 721$, showing an efficiency of 63%. The total laser fusion gain $G$ is then 72.1, which is only 28% less than the gain for the case of isobaric spark ignition.

We see that in these comparable cases for same laser input energy, the radius of strongest compression is much larger and the efficiency is much lower for spark ignition than for volume ignition. Volume ignition works like a diesel engine, and there is—apart from the favourable reabsorption of bremsstrahlung because of the very low temperature—a large amount of "additional driver energy" coming from self-produced alpha-particle reheating. Detailed calculations (see He and Li 1994; Martinez-Val et al. 1994) include additional self-heating by neutrons, and, instead of assuming LTE, they find the ions are much hotter at a maximum temperature, for example of 200 keV, with electrons having only 80 keV and the background blackbody radiation 8 keV. In this case (Martinez-Val et al. 1993b), a pulse of 6 GeV bismuth heavy ions as a driver of 1.6 MJ en- ergy and 10 ns duration produces 120 MJ of DT fusion energy. It should be men- tioned that the neutron reheating—not included in the case of Fig. 9-3b—improved volume ignition (gains may increase up to a factor of two), while it was shown by Nakao et al. (1996) that neutron reheating typically decreases the gain in spark ignition. This advantage of the lower electron temperature is important also for ion beam fusion (Roth et al. 2001) and resulted automatically at initiation of fusion flames (Fig. 10-5).

## 9.5. SELF-SIMILAR VOLUME COMPRESSION AGREES WITH MEASURED FUSION GAINS

Returning to the discussion started in Subsection 9.2 regarding the measured laser fusion gains and the agreement with the predicted line in Fig. 1, we now compare the measurements with the results for the fusion gains based on the isentropic compression model of volume compression. In Subection 9.3, we explained some of the main properties of this model and showed how laser fusion gains with volume ig-nition have similar high values as those from isobaric spark ignition. These model cases were for laser pulse energies in the megajoule range, while the discussed experiments are only about 10 kJ. The resulting core gains will be less than 8. Therefore we are in the regime below volume ignition in the range of simple fusion burn.

In all cases that follow, we assume—realistically—a hydrodynamic efficiency $h = 7\%$ and discuss the details of the new results using the same background core gain diagram for volume ignition as in Fig. 9-5 but concentrating now on the parameters of the following experimental results. In Fig. 9-6, with selected results of Fig. 3, we arrive at the measured point A for the Rochester result (Soures et al. 1996) with a laser energy of 30 kJ ($E_0 = 2.1$ kJ because $\eta = 0.07$) and a core gain of 13.4% according to $10^{14}$ DT neutrons. When we fit the standard parabola through A and the tangent to the line for the measured maximum density of $4n_s$ in the Rochester experiment, we derive a factor of 1.7 for the difference between A and the tangent point. This means that the temperature in the plasma is 1.7 times less than the optimum temperature of 17 keV of the straight lines, i.e., we conclude from the self-similar volume compression model that $T_0 = 10$ keV. This derived temperature agrees with the measured (Soures et al. 1996) temperature of 10 keV.

The same is done with measured gains from Osaka (Takabe et al. 1988), point B in Fig. 9-6, and with the measurements from Livermore (Storm 1986), point C, and from Arzamas-16 (Kochemasov 1996), point D. Table 9-2 shows how well the maximum temperatures derived from the isentropic-compression model agree with the measured temperatures. This agreement can be considered as very satisfactory in view of the very sophisticated techniques with which the experimental data were gained. It should be mentioned that these results are based on computations using the code published in Stening et al. (1992), where agreement between these and earlier results was confirmed using completely different computations of volume ignition (Atzeni 1995; Nakai and Takabe 1996).

It should be mentioned that the data from the Livermore experiment (Storm 1986) were available by private communication in greater detail and that the isentropic spherical model led to fusion gains that were 22% below the measured values. However, the reported one-dimensional simulation gains (Storm 1986) were too large by a factor of more than 3. In view of the more recent two- and three-dimensional simulations with better fits with the experiments,

C, and from Arzamas-16 (Kochemasov 1996), point D. Table 9-2 shows how well the maximum temperatures derived from the isentropic-compression model agree with the measured temperatures. This agreement can be consid- ered as very satisfactory in view of the very sophisticated techniques with which the experimental data were gained. It should be mentioned that these results are based on computations using the code published in Stening et al. (1992), where agreement between these and earlier results was confirmed using com- pletely different computations of volume ignition (Atzeni 1995; Nakai and Takabe 1996).

It should be mentioned that the data from the Livermore experiment (Storm 1986) were available by private communication in greater detail and that the isentropic spherical model led to fusion gains that were 22% below the mea- sured values. However, the reported one-dimensional simulation gains (Storm 1986) were too large by a factor of more than 3. In view of the more recent two- and three-dimensional simulations with better fits with the experiments,

**Figure 9-6.** Optimized core fusion gains $G$ (full lines) for the three-dimensional self-similar hydrodynamics of volume compression for simple burn ($G < 8$) as in Fig. 2 (sometimes called quenching: Atzeni 1995; Nakai and Takabe 1996) and volume ignition (for $G > 8$). The measurements from Rochester (Soures et al. 1996) (point A), Osaka (Takabe et al. 1988) (point B), Livermore (Storm et al. 1986) (point C) and Arzamas-16 (Kochemasov 1996) (point D) all agree with this isentropic volume burn model, while the earlier fast pusher (Kitagawa 1984) (point E), with strong entropy-producing shocks (Kitagawa et al. 1992), does not fit (see Table 9-2).

the discussion of this point is of minor importance, and is not followed up here.

As an example of how the isentropic model does not fit with experimental data, we use results from the Osaka fast pusher (Kitagawa et al. 1984): point E in Fig. 9-6 (see Table 9-2). It is obvious that this experiment does not fit the isentropic-compression model at all. The reason for this is evident: the fast pusher was mostly using shock-wave compression with very high entropy production, while it was essential for the very high laser fusion gains first achieved at ILE Osaka in November 1985 (Yamanaka et al. 1986; Yamanaka and Nakai 1986) that a shock-free and stagnation-free compression was used—just the conditions of the isentropic compression in the ideal self-similarity model.

**Table 9-2.** Reproduction of experimental laser fusion gains by the self-similar volume-compression model

| Experiment | Input Laser energy ($\eta = 7\%$) DT/$T_{opt}$ (kJ) | Ce DT temperature (keV) | M Core temperature (keV) | in neutrons | P isochore gain | F difference diagram | Rt parabola | Ci |
|---|---|---|---|---|---|---|---|---|
| Rochester 10 (1995) | 30 | 2.5 | $10^{14}$ | 0.14 | A Yes | 1.7 | 10 | |
| Osaka 10 (1989) | 10 | 0.7 | $10^{13}$ | 0.04 | B Yes | 2.4 | 7.3 | 8 |
| LLNL 23 (1986) | 23 | 1.7 | $2 \times 10^{13}$ | 0.025 | C Yes | 10 | 1.7 | 2 |
| Arzamas-16 2 (1993) | 0.135 | 0.010 | $9 \times 10^{9}$ | 0.0025 | D Yes | 10 | 1.7 | |
| Osaka, fast pusher (1994) | 0.125 2 | 0.009 (shocks) | $4 \times 10^{8}$ | $1.5 \times 10^{-4}$ | E No | ? | ? | |

## 9.6 THERMAL COMPRESSION AND VOLUME IGNITION OF BORON FUSION

The fusion reaction of the boron isotope $^{11}$B with protons was mentioned in Eq. (9-1b) as an exception of several alternatives, where primarily no neutrons are generated and this clean fusion generated less radioactivity per gained power than burning coal. This reaction for laser fusion was considered in an early stage (Hora, 1975a, see p. 85) and even the non-thermal interaction by purely electrodynamic driving of the plasma was envisaged about which the only the recent techniques with laser pulses of ps duration and exawatt power (Mourou et al 2013) may provide and access (Hora et al 2014). While these new (or very old!) schemes will be discussed in Section 10, it will be summarized within this section how far the possibilities have been evaluated by using the thermal compression and ignition with nanosecond laser pulses.

It is remarkable, that the need of compression of the plasma to 100,000 times solid density of HB11 is necessary (see Eq. 8.9 of Ref. Hora 1975). The very detailed computations for direct drive volume ignition (Stening et al. 1993, Khoda-Bakhsh 1993; Khoda Bakhsh et al. 2007) after a longer series of studies and publications using alpha reheat, partial bremsstrahlung re-absorption and collective stopping power, the same high compression was derived. Indeed there were a number of interesting details gained and the more precise evaluation of the fusion cross section in the resonance area (Nevins et al. 2000) arrived at some relaxation of the exotic conditions (Kouhi et al. 2011).

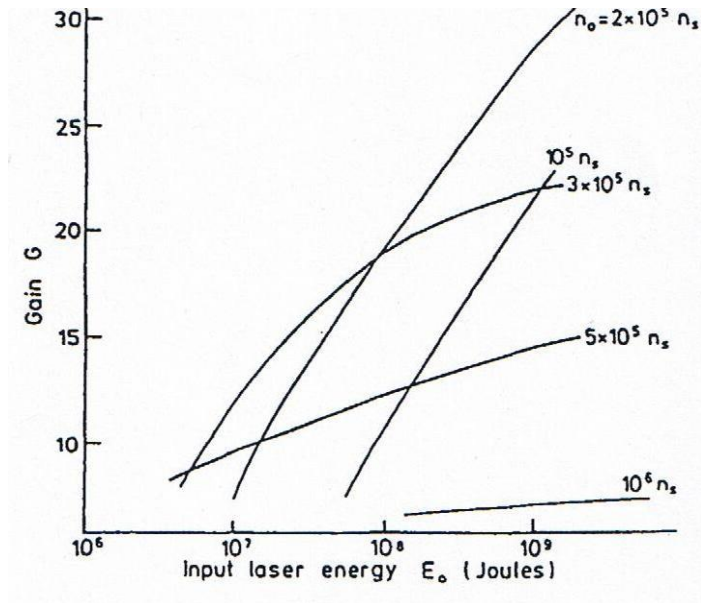

**Fig. 9-7.** Optimized HB11 fusion gains depending on the input energy $E_o$ into the reacting core at maximum ion compression densities $n_o$ expressed in multiples of the solid state fuel density $n_s$.

What is important to note, is that one still needs a compression to about 100,000 tims the solid state density as it was before without the special innovation of the detailed resonance in the nuclear fusion cross section of HB11. The of reduction the input energy from the range of GeV to nearly few MeV is the main advantage of the resonance property to reduce the difficulties by a factor of about 100. Before, the difficulty of the HB11 fusion was about 100,000 higher than the DT fusion and therefore prohibitive. The factor 100,000 came from the need that the compression had to be about 100 times higher, the input energy was about hundred times higher and the gain was about 10 time less.

With respect to the 100,000times higher compression, it was interesting to note the results of generation of ultrahigh den densities of deuterons in voids (Schottky defects) in solids. The generation of clusters with more than 100,000times normal solid state densities was concluded from voids in the surface with catalysts (Holmlid et al. 2009; Badiei et al. 2010) as inverted Rydberg states. Another kind of clusters within the volume of crystals was detected by SQUIDS in superconductive states (Lipson et al. 2005). Targets with such deuteron clusters showed an increase of fusion neutons (Yang 2011).

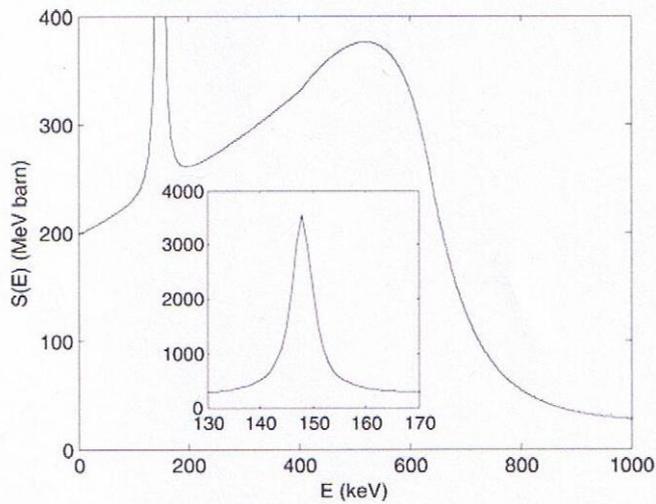

**Fig. 9-8.** Corrected resonace function of the HB11 fusion around the energy of 150 keV following Nevins et al. (2000).

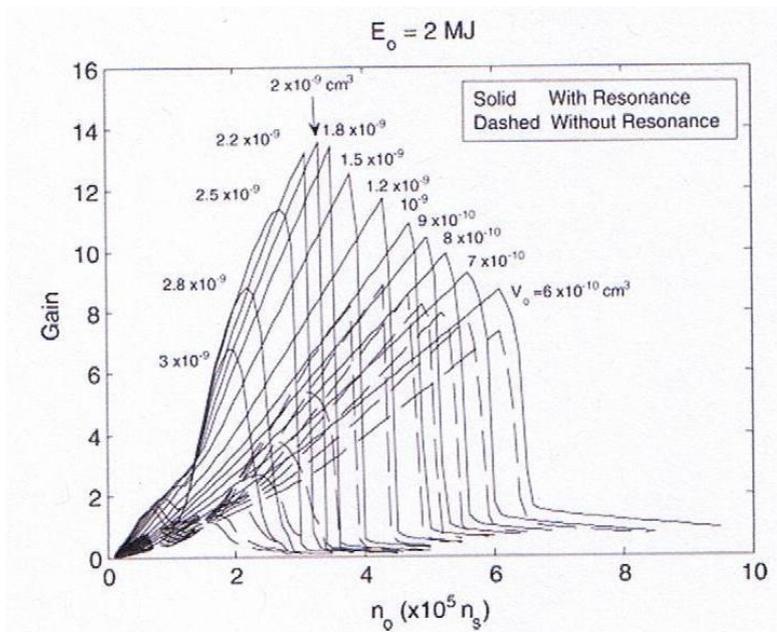

**Fig. 9-9.** HB11 fusion gainsdepending on the maximum compression density $n_o$ in multiples of the solid state density $n_s$ at an input energy $E_o$ of 2 MJ at different volume $V_o$ in the fusion plasma without and with the resonance of the cross section at 148 keV impact energy (Nevins et al. 2000) and with inclusion of the theory of Li et al. (2000, 2004) with a Schrödinger potential having an imaginary part.

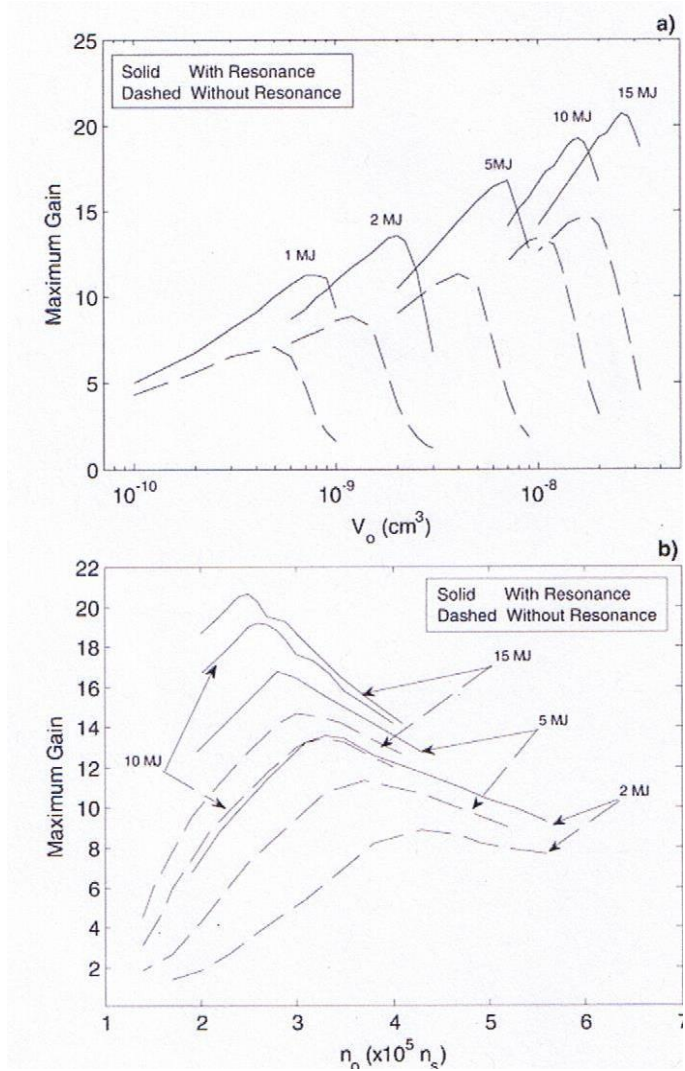

**Fig 9-10a&b.** Summary of measured HB11 fusion gains with a with and without the resonance according to Nevins et al. 2000 (Kouhi et.al. 2011).

One of the new results was the retrograde increase of the fusion gains on the increasing compression density (Pieruschka et al. 1992) which property could be understood (Scheffel et al. 1997). This can be seen in Fig. 9-7. The gain increases on the comprssion as normal when the density rises from 100,000 times solid state density. But above twice of this density, a further increase of the density results in lower fusion gains.

A significant change of the HB11 fusion happens when using the more accurate fusion cross reactions around the 150 keV resonance energy, Fig. 9-8. Due to the resonance, a strong change of the fusion gains appears as shown in Fig. 9-9 at the comparably very low input energy $E_o$ of only 2 MJ (not the usual GeV) resulting in comparably high gains up to 14. More examples at similar parameters were evaluated (Kouhi et al. 2011).

# 9.7. OUTLOOK FOR VOLUME IGNITION FOR NIF CONDITIONS

There is general agreement that volume ignition has the advantage of lower losses, lower ignition temperature, lower implosion velocity and lower sensitivity of the more robust capsule to small fluctuations and asymmetry in the drive system (Lackner et al. 1994), in contrast to spark ignition. For inertial fusion energy based on spark ignition only, the formation of an ignition DT volume in the centre of the implosion (hot spot) is possibly the most difficult aspect (Meyer-ter-Vehn 1996). There is a further problem of requiring extreme uniformity of the initiation of the fusion detonation wave (see the point $r = 0.47$ in Fig. 9-4): at a given value of the radius, the density and temperature must become uniform in all directions within a very short time interval. If ignition starts in one direction earlier, no detonation front will be possible. All these problems with the detonation front are avoided with volume ignition, because there is no detonation wave at all with the diesel-engine-like ignition occurring over the whole volume of the uniformly compressed plasma with volume ignition. Instead of the extravagant profiles of spark ignition (Fig.9-4), volume ignition only needs the natural adiabatic, compression. The problem of Rayleigh–Taylor instabilities and related mixing is relaxed by a factor of about three, since the compression front is most strongly concentrated outside the pellet with volume ignition, and not the inner part ($r = 0.47$ in Fig. 9-4) at the detonation front as with spark ignition.

Volume ignition applies for both direct and indirect drive. A very exotic condition for indirect-drive spark ignition can be seen in the example of Fig. 9-6 of Krauser et al. (1996) for the time dependence of the irradiated laser power. The simplest case with higher efficiencies is expected from direct drive if the problems of 20 ps stochastic temporal pulsation can be overcome by laser-beam smoothing. This pulsation and temporarily very high (phase) reflectivity may have been the most difficult obstacle for direct drive, but its suppression by smoothing, giving uninterruptedly low (mirror) reflectivity for the whole interaction time has been clarified by detailed numerical studies (Hora and Aydin 1992), see Figs. 6-21 to 6-24 (Kalal et al. 1987). In view of these general aims, it is interesting to see how volume ignition may be reached with the presently most advanced laser facilities.

We can conclude from the agreement of the highest measured fusion gains with the isentropic volume compression model as shown in Subsection 9.5 what needs to be done in order to reach ignition (a core gain above $G = 8$). As soon as ignition is reached the situation is simplified, since the optimum temperature decreases from 17 keV to 2 keV and even less. Without reaching volume ignition, one should preferably work always at the optimum temperature of 17 keV. If 17 keV is very difficult to achieve then one can work at 10 keV but with slightly higher values for the other parameters according to the gain diagram of Fig. 9-6. For the conditions of the Rochester experiment following a vertical line up from point A in Fig. 9-6 while keeping the maximum temperature of 17 keV means increasing the compression at the same input laser energy of 30 kJ and the same hydrodynamic efficiency of 7%. Volume ignition will then begin as soon as densities above $1200n_s$ are reached. The initial solid-state DT volume before compression is $1.05 \times 10^{-6}$ cm$^{-3}$, but at higher compression the initial volume increases and the maximum temperature decreases. For example, if a compression of $4000n_s$ is possible then the initial solid-state DT volume is $1.8 \times 10^{-6}$ cm$^3$ and the initial optimum temperature decreases to 11 keV.

The conditions for the basic demonstration of volume ignition change if laser pulses of 100 kJ are available, as initially envisaged by Mima et al. (1996), in which the merging of volume ignition computations using very sophisticated models with the earlier volume ignition calculations of Hora and Ray (1978) was demonstrated. Working with laser pulses of 100 kJ energy, hydrodynamic efficiency of 7% and, for example, a density of $4000n_s$, the core gain $G = 90$, the total laser gain $G_L = 6.3$, the initial temperature $T_o = 3.5$ keV and the initial solid state DT volume is $3.5 \times 10^{-5}$ cm$^3$, producing $3 \times 10^{17}$ DT neutrons.

Indeed the conditions change dramatically for driver energies of 2 MJ and higher as is the aim with the National Ignition Facility NIF (Cray and Campbell 1996; Campbell et al. 1997; Campbell 1989, 2006). As an example, we consider for this case a hydrody- namic efficiency of 7% and a compression to $3000n_s$. The core gain $G$ is then 600 and the total laser gain $G_L = 42$, the initial temperature is 3.1 keV and the initial solid-state DT volume is $1.8 \times 10^{-3}$ cm$^3$.

The question arises as to how the experiments discussed here and the theoretical agreement with gas-filled targets relate to the design of cryogenic capsules with DT shells. The example of the very detailed computation by Martinez-Val et al.,(1994), Hora and Pfirsch (1970, 1972) including non-LTE (in contrast to the simplified results of the diagrams of Fig. 9-3a based on LTE) use a cryogenic DT shell driven by gigaelectronvolt ions ablating an outer lithium shell. The resulting high gains are achieved with low-temperature (3 keV) volume ignition where the density profile and temperature profile at highest compression turned out to be very close to the ideal adiabatic self-similar case, in strong contrast to the profiles for spark ignition, as seen for example in Fig. 9-3a. Our projections in the preceding two paragraphs for volume ignition at megajoule driver energies, are therefore not necessarily restricted to a gas-filled uniform DT pellet but may well be applicable to the method of using cryogenic shells.

It needs not be argued whether isobaric spark ignition with very high deviation from the isentropic case may result in a better solution for the inertial fusion energy reactor. The difficulties of spark ignition were described by the initiator of the isobaric spark-ignition scheme himself as pointed out at the beginning of Section 9.7 (Meyer-ter-Vehn 1996). The results reported here, together with the very detailed cases in the megajoule range with cryogenic shells (Martinez-Val et al. 1994), indicate that the isentropic volume-ignition scheme could well offer an alternative and possibly very much simplified "robust" (see Lackner et al. 1994) method of nuclear energy generation by inertial confinement fusion, with interesting gains as elaborated here in Subsection 9.4.

This is the point to mention the rather realistic position about fusion energy as formulated as early as in 1996 by the leading Britsch expert, Sir William Mitchell (Butler 1996) or articulated by the former US-Minister of Energy (DoE) Nobel Laureate Steven Chu (2014), see Dorda (2014). Despite the progress on magnetic confinement fusion (Keilhacker 1999), the tokamak solution with DT fusion is confronted with the "hottest radioactive environment on earth" (Butler 1996). The less problematic DD-fusion is not possible by this tokamak solution, while the possibility with laser fusion using nanosecond laser pulses is not excluded. However, the DD fusion is much more difficult than DT fusion like the proton-$^{11}$B-fusion with nanosecond laser pulses (Section 9.6). But there may be a solution if extremely powerfull picosecond laser pulses are using a basical alternative physics principle by excluding thermal-pressure effects for HB11 thanks to alpha avalange by nonlinear force driven plasma block initiation as described in the following Section 10.

# CHAPTER 10

# LASER DRIVEN FUSION ENERGY WITH PICOSECOND PULSES FOR BLOCK IGNITION

___________________________________________________________

Laser and plasma physics with forces and the nonlinearity principle to be applied to controlled nuclear fusion energy generation, was considered in the two preceding sections 8 and 9. While Section 9 was presenting some aspects of the present very broad stream of experiments and theory using the traditional lasers with pulses in the nanosecond range this is interwoven with the crucially new physics of very short and non-thermal acceleration of plasma blocks to which Section 8 was introducing to achievements of the recent 15 to 20 years.

## 10.1 NEW ASPECTS BY SUB-PICOSECOND LASER PULSES

The push of developing the picosecond interaction indeed came from the developments of laser driven fusion thanks to the progress with the Chirped Pulse Amplification CPA (Section 8.1) offering sub-picosecond laser pulses above PW power. For the broad stream of laser fusion with ns pulses to heat and compress fusion fuel to more than 1000 times solid state densities for controlled igniting exothermal reactions, it was significant that densities of 2000 times the solid state were achieved (Azechi et al 1991). This compression and the generation of highest gain DT fusion neutrons was possible only thanks to the smoothing of the laser beams using the random phase plates discovered by Kato et al. 1984. This smoothing (Hora 2006) will enable in one immediate step to work with the second harmonics in the NIF project at LLNL in Livermore/CA instead of the usually applied third harmonics (Hora 2006) to increase the irradiated laserpulse-energy from 2 to 6 MJ (Roth 2014).

The record compression of 2000 times the solid state (Azechi et al 1991) however arrived at a problem. The measured maximum temperature of the plasma reached only 3 MeV while theoretically a much higher temperature was expected from the otherwise very successfully observed self-similarity compression (Fig. 2-8), called Yamanaka-compression as proved in purely empirical research at the record conditions for maximum numbers of fusion neutrons (Yamanaka et al. 1986) accompanied by detailed numerical confirmation (Yabe 1985). The result was based on ideal adiabatic compression as known from the self similarity model in volume ignition (Hora et al. 1978, Hora 2013). In this situation, Campbell (2006) had the idea before 1992 to achieve the missed higher temperatures by combining this with the extremely high power ps laser pulses of CPA (Section 8.1) discovering the scheme of "fast ignition" (Tabak et al 1994; Norreys 2005). This is described schematically in Fig. 10-1 where DT fuel within the complete outer circle is being compressed by a ns laser pulse to 1000 times solid state density and then a ps laser pulse entering from the left hand sice though a conical funnel is heating up the compressed fuel to ignition temperature. The preparation for this fast ignition experiment with establishing a 2 PW sub-picosecond laser pulse (Mourou 1994; Miley 1994; Perry et al 1994) led to target interaction with the very first convincing number of positrons from pair production, very intense gamma radiation to produce nuclear transmutations and numerous exotic results (Cowan et al. 1999) but before passing on to irradiate high density plasma from precompression with ns laser pulses, the experiment was decommissioned.

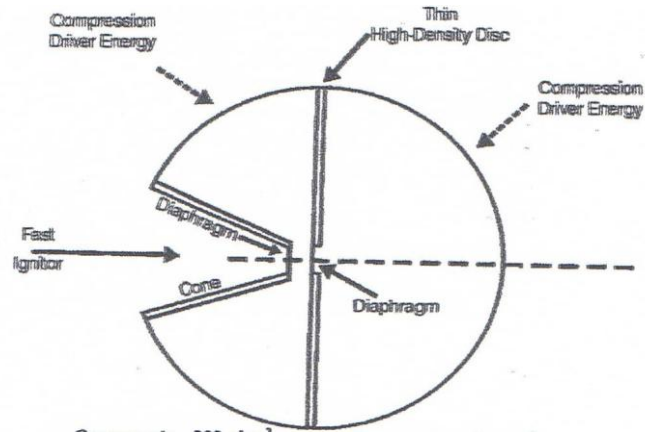

Figure 10-1. Fast ignition of DT fuel within the outer sphere compressed to 1000 times solid state denstity by a ns laser pulse with an additional ps fast ignitor pulse to heat up the compressed fuel to ignition temperature (Tabak et al. 1994).

Fast ignition experiments (Kodama et al.2001) used laser pulses up to petawatt power for the sub-picosecond fast ignitor generating up to $10^8$ fusion neutrons. The modified concept for ps laser pulse ignition of fusion with a modified fast ignition was elaborated by Nuckolls et al (2002) according to Fig. 10-1. The precompression by ns laser pulses has to be done only for the left half-sphere to densities of about 1000 times of solid DT. The right half sphere needs a precompression of the DT only to lower densities, in a good example only to 12 times the solid state. The fast ignitor beam from the left hand side produces an extremely intense beam of relativistic electrons having 5 MeV energy. These electrons ignite a fusion reaction with a gain of 10,000. It was postulated some time before by Nuckolls that such high gains are necessary for laser driven fusion power stations. This is a consequence of his leading role all the time in this field as expressed in the most valuable collected contributions by Velarde et al (2007) with nearly all leading centers, where by technical reasons only the contributions from France were not included like that from China or Poland and the Czech Republic.

An alternative option for fast ignition is the use of initiating the fusion flame for side-on ignition of the fusion fuel which was opened by Chu (1972 and Bobin (1974) as explained in Section 8.

## 10.2 NEW RESULTS FOR BORON FUSION

Attention was given from the beginning of the plasma block initiation of fusion flames to arrive at an application for the fusion of protons with the boron isotope 11, Eq. 9-1b (Hora 2002; Hora et al 2002a; Hora and Miley 2004), called HB11 fusion. The enormous advantages of this reaction is for energy production with less radioactivity than from burning coal, explained in Section 9.6, however, a reaction driven by ns laser pulses by the traditional thermal heating compression and ignition process is more than 10,000 times more difficult than the same process with DT, and this is only close to a breakthrough.

The more it was surprizing when the picosecond plasma block initiation of a fusion flame in DT was studied for HB11. Working only with the same kind of reactions as for DT and no secondary reations which are possible and most favourable for HB11 in contrast to DT, the difference between DT and HB11 was about a factor less than 4, not 10.000!

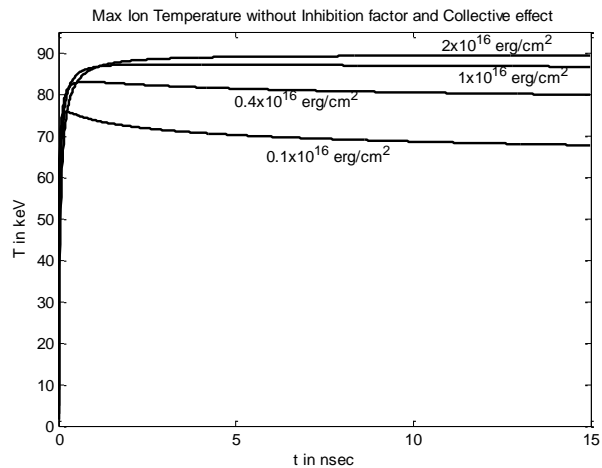

**Figure 10-2.** Fusion reaction diagrams (Hora et al. 2009) with reaction cross sections for HB11 showing the laser energy flux density of $E^* = 1.5 \times 10^9$ J/cm$^2$ with the same conditions as used for DT by Chu (1972) for comparison with Fig. 8-9.

This result (Hora 2009; Hora et al. 2009; 2010) in Fig. 10-2 shows the high electron temperature of the ignition threshold of 87keV in agreement with the fact, that HB11 fusion at equilibrium conditions without reabsorption of bremsstrahlung und without re-heat by the generated alpha particles as under the quasi two-dimensional conditions of the thin initiation volume of the fusion flame, has to be at a temperature above 60 keV. This temperature can only be reduced to lower values at the three dimensional conditions of volume ignition, see Section 9.6 in a similar way as the volume ignition for DT fusion can well happen below temperatures of 4 keV, if alpha-reheat and partial re-absorption of bremsstrahlung is included.

As a technical test only for comparison with the results in Section 9.6, the complete re-absorption of bremsstralung for the HB11 computation arrived at lower ignition temperatures. The final result is only that of Fig. 10-2 with the correct inclusion of the bremsstrahlung as it was done before for DT from the beginning by Chu (1972).

It needs to be noticed that the results of Fig. 10-2 can be improved further as used in the following estimations, if the secondary reaction of the alpha particles by elastic collisions with boron nuclei of the fuel have been neglected and only the reactions for comparision with DT were included in the computations as basis for the comparison with the work of Chu (Hora 1999).

## 10.3 FROM PLANE GEOMETRY TO USABLE IRRADIATION

Up to this stage, we have reported only for the case of plane geometry laser irradiation perpendicularly irradiating a plane fusion target. The results following the computations of Chu (1972) - reproducing, updating and applying the ultrahigh acceleration of plasma blocks produced by ps pulses of $10^8$ J/cm$^2$ energy flux density - could be based on the achievement and developing of PW-ps laser pulses (Mourou et al 2013). One important question is how to apply this plane geometry to laser pulses of beams with limited coss section. One cannot simply take out the aera of the beam hitting the target and assume that a cut out isolated cylindrical geometry carries the reaction as under infinite plane conditions, because there will be lateral losses by radiation and thermal conduction (Nuckolls 2009). This is indeed different to the scheme with relativistic electron beams according to Fig. 10-1 (Nuckolls et al 2002) where the 5 MeV electrons are interacting in three dimensions with the whole volume for the fusion reaction behind the diaphragma. In the case discussed up to this point, the plasma laser driven plasma block of few micrometer depth interacts as a kind of a two-dimensional layer with the target to initiate the fusion flame.

One way to solve the problem is to use spherical geometry of interaction where the problem of lateral losses is avoided completely. This case is considered in this sub-section 10.3 following Fig. 8-3 with a spherical fiber glass laser ICAN (Mourou et al 2013) where the irradiated fusion fuel 3 is in the center of the sphere of a radius r (Hora et al. 2013). We consider first the case of solid density fuel based on the before described computations.

We report on laser fusion of uncompressed solid density deuterium-tritium (DT) fuel at spherical irradiation where the properties can realistically followed up on numerous preceding computations. The use of solid density HB11 fuel resulted in similar gains when using the same presumptions as for DT (Hora 2009; Hora et al. 2009). This was not including the favourable secondary reactions of the generated alpha particles when hitting protons leading to an avalanche of reactions with an essentially much higher total fusion gain, even to higher values than with DT. One may note, that the HB11 reaction is then the first to realize a neutron-free reaction (Tahir et al. 1997; Labaune et al. 2013), which condition the highly supported DT fusion schemes cannot provide. The generated neutrons for the DT reaction decay with a half life of 12 minutes into protons and electrons to produce water but before reaching this, the neutrons can react with stable nuclei to produce large amounts of radioactivity. These problems are eliminated with the neutron-free HB11 reaction.

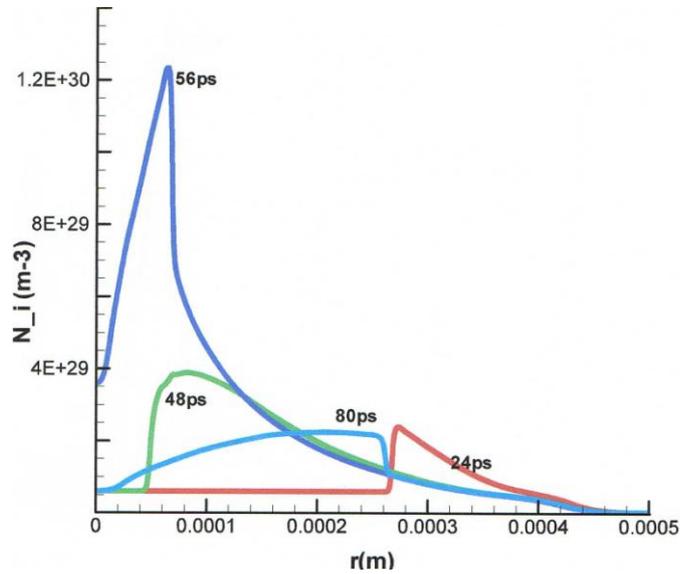

**Figure 10-3**. Initiation of a fusion flame in a solid DT sphere of 0.5 mm radius in the center of Fig. 8-3 irradiated by a 1 ps laser pulse of $10^{20}$ W/cm$^2$. The profile of the ion density is shown at different time after the initiating ps laser pulses deposited the energy to the driving plasma block.

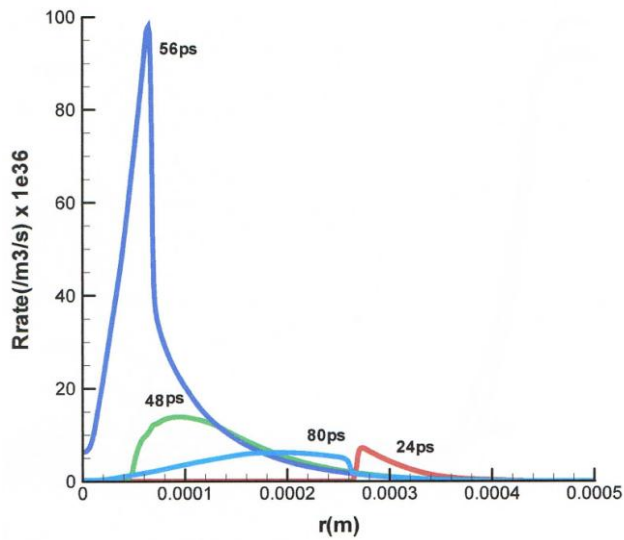

**Figure 10-4**. The reaction rate for the case of Fig. 10-3.

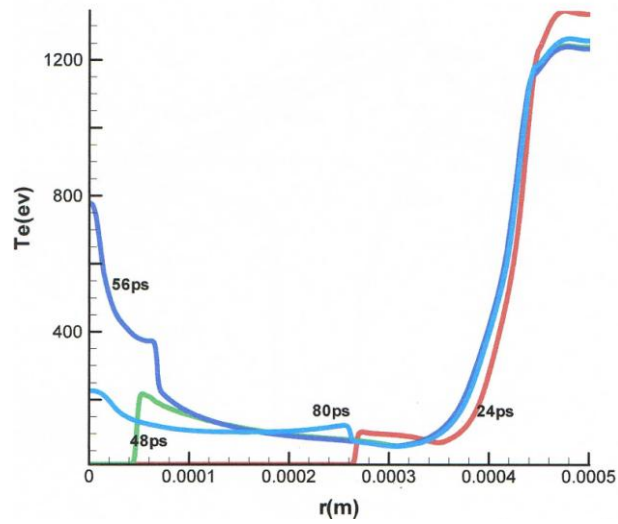

**Figure 10-5.** The electron temperature from the slowly thermalising of the directed ion energy of the nonlinear force accelerated plasma blocks

Without the availability of details of the essential secondary reactions of the alpha particles (Hora et al. 2012a), an evaluation of estimations for the fusion gain is possible for the conditions of irradiating spheres of solid density HB11 fuel by converging ICAN laser pulses (Fig. 8-3).

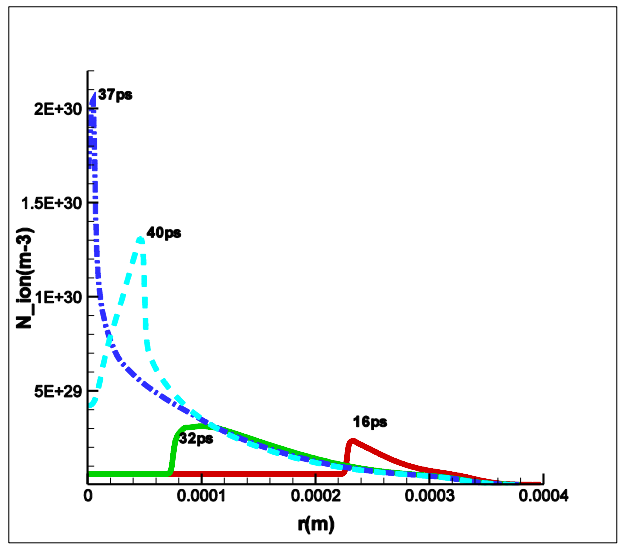

**Figure 10-6.** Spherical solid DT of 400 mm radius irradiated by ps $10^{20}$ W/cm$^2$ laser pulse of 248nm wave length. Radial dependence of the ion ion density $n_i$ at different times after ignition by the ps laser pulse.

The application to nuclear fusion was based on the Chu-Bobin scheme (Chu 1972) of side-on ignition. The application to nuclear fusion for spherical irradiation was based in similar way on the a fusion flame in uncompressed solid density deuterium-tritium where it was shown that the flame can be initiated by a ps energy input into the fuel having an energy flux above $10^8$ J/cm$^2$ (Hora 2009).

For a fusion power station, to eliminate problems of lateral energy losses of the plane irradiation this can be overcome by using spherical irradiation. ICAN fiber lasers offer a solution as can be seen can from Fig. 8-2 (Mourou et al. 2013). The output of the amplified beam before the focussing mirror 6 is from parallel fibers and has a plane wave front. In practical cases, the beam of 100 cm$^2$ cross section may be a pulse with 1kJ energy and of ps duration representing one PW. The wave front is of such a quality that focussing has been shown (Mourou et al 2013) to go to a diameter of 10 µm such that intensities at about $10^{21}$ W/cm$^2$ are reached. The fibers permit an exceptionally high single mode quality of the beam uniformity (Mourou 2013) where the usual undesired maxima in the beam profile with glass lasers are automatic eliminated due to the fiber optics quality.

A further advantage of the fiber optics is by avoiding the focussing mirror 6 of Fig. 8-2, when taking the axes at the ends of the fibers not parallel at the end. Instead of the plane wave fronts from parallel axes of the fibers the axes are directed radial to a spherical center 3 as shown in Fig. 8-3 such that a whole sphere 1 produces a spherical converging laser pulse front towards the center where a fusion fuel pellet 3 is located. The fiber ends should fill nearly completely the whole sphere 1. If the radius of 1 is one meter or more, a converging laser pulse with 1 ps duration and a power of more than Exawatt will be generated for hitting the fuel pellet 3. Based on kJ energy per 100cm$^2$ output from a sphere of at least of about 2 meter diameter, a spherical laser pulse of Exawatt and 1 ps duration would be needed.

In order to understand the mechanisms in the spherically irradiated solid density fuel, results from computations using the genuine two-fluid hydrodynamics are being shown. In this model case, the radius of the pellet is taken 0.5 mm similar to the cases evaluated before (Lalousis et al. 2013) with a laser intensity of $10^{20}$ W/cm$^2$ for the initiating ps laser pulse. Fig. 10-3 shows the ion density $n_i$ within the sphere at a sequence of times, Fig. 10-4 the fusion reaction rate and Fig. 10-5 the electron temperature. The main push is in the radial directed ions while the complete hydrodynamic code shows the highly delayed heating of the electrons in the collectively accelerated plasma blocks in Fig. 10-5. From Fig. 10-6 on we show a case for 0.4 mm fuel radius for comparison.

The conditions for the numerical evaluation of the initiation of the fusion flame by ultrahigh acceleration of nonlinear force driven plasma blocks by non-thermal direct conversion of laser energy into the collective driving process involves the result of Chu (1972) that the energy flux density E* threshold for the picosecond laser pulses of $4 \times 10^8$ J/cm$^2$ has to be reduced by a factor 20 (Hora et al. 2008) for DT to

$$E^* = 2 \times 10^7 \text{ J/cm}^2 \qquad (10\text{-}1)$$

The reason for this reduction is the fact that at the time of the work of Chu, the inhibition factor for the reduction of thermal conduction in inhomogeneous plasmas was not know and the collective stopping power of Denis Gabor (1952), see Section 8.4.3, was not used as needed for high density plasmas. The experimental confirmation of these processes was shown while the quantum correction of the collision frequency is important only for longer times than the ps interactions (Yazdani et al 2009; Sadighi-Bonabi et al 2010, 2010a; Hora et al. 2012).

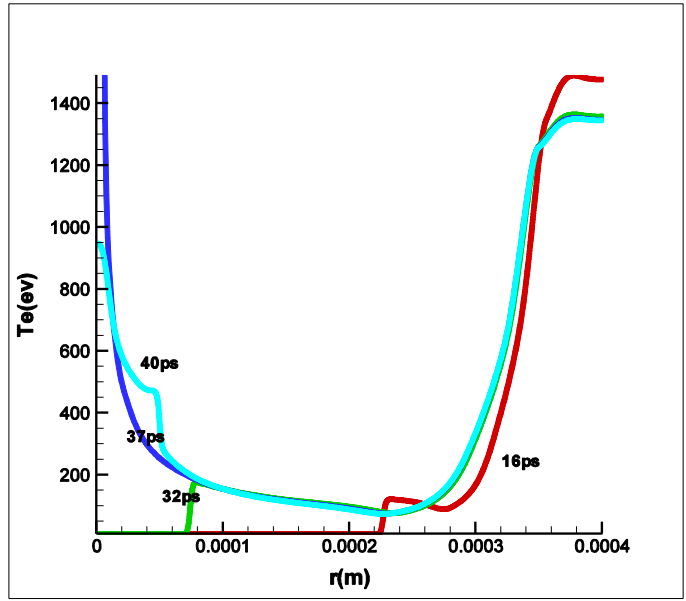

**Figure 10-7**. Same case as Fig. 10-6 showing the electron temperature.

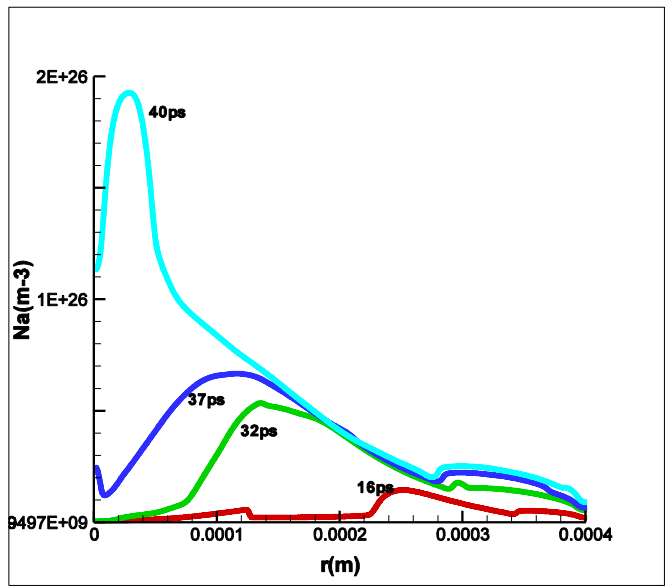

**Figure 10-8.** Alpha particle density created by the DT reactions at different times as parameters in the radial irradiated spherical plasma as in Fig. 10-6.

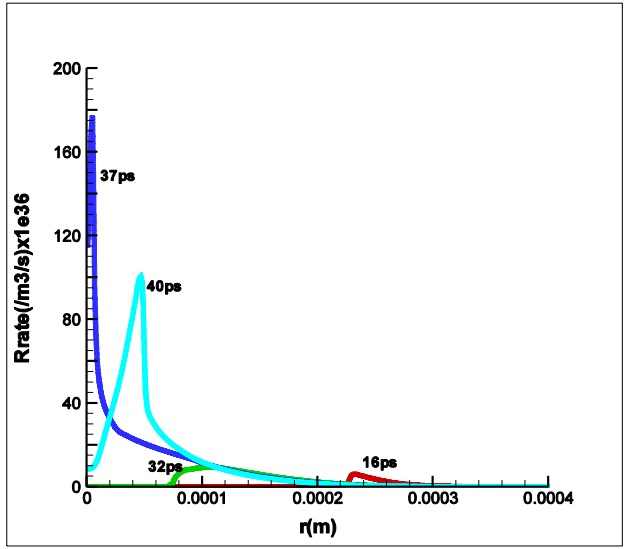

**Figure 10-9**. DT fusion reaction rate expressing the about 10,000 times lower alpha particle density $n_\alpha$ then the ion density comparing Fig. 10-8 with Fig. 10-6.

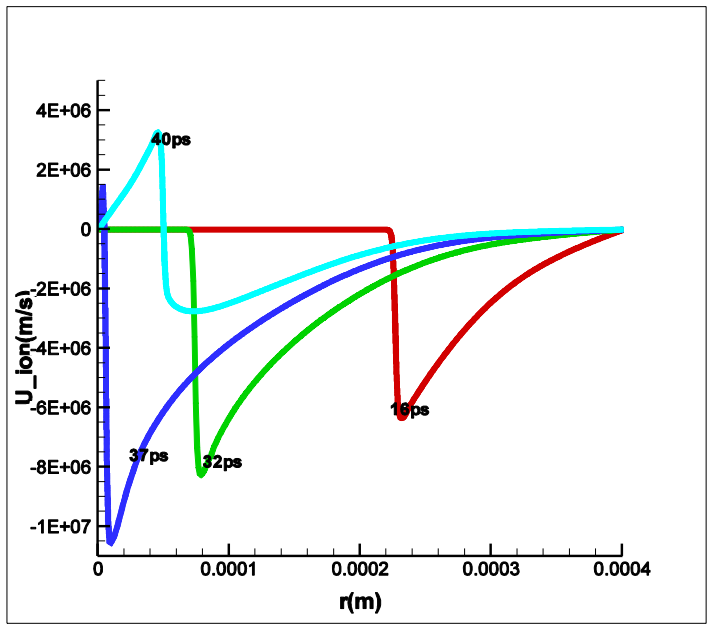

**Fig. 10-10.** Ion velocities $u_i$ depending on the radius r at different times for the case of Fig. 10-6.

For comparison, Figures 10-6 to 10-10 show the case for a solid DT fuel sphere with the radius 0.4mm (Hora et al 2014, 2014a; Lalousis et al 2014b).

The collective block acceleration could best be seen e.g. from the very low energy of the space charge neutralizing electrons between the directed motion of the blocks where the main energy is in the ions. The computations with the general hydrodynamic code (Lalousis 1983; Hora et al. 1984; Hora 2011; Lalousis et al. 2012; Hora et al. 2013a; Lalousis et al 2013, 2014a, 2014b; Hora et al. 2014; 2014a) include the separate temperatures of the electrons and of the ions based on the genuine two-fluid model with complete inclusion of collisions in contrast to the initial computations (Chu 1972). A study for spherical interaction following the earlier presumptions was published (Malekynia et al 2013). The stopping power for the generated charged nuclear products was a first option based on the same collective model as for DT (Hora 2009, Hora et al. 2009). This is in strong contrast to the fusion of HB11.

The result that the plane geometry side-on initiation of a fusion flame by ps laser pulses was only about four times more difficult for solid density HB11 than for DT (Hora 2009, Hora et al. 2009) was well a first important step, but it did not include a further strong improvement for the HB11 case: the alpha-avalange multiplications (AAM). Without them, the procees of the generated alpha particles was treated in the same way for both cases as the binary reactions, causing there the reheat of the plasma by the stopping length as in the solid density state of the fuel in the range not too far above 10 micrometers. This reheat process is well known from the crucial importance for the volume ignition with thermal fusion at nanosecond irradiation (Chapter 9) casing the ignition (Fig. 9-3b) and very high gains for direct drive of DT (Hora et al 1978, Lackner, Colgate et al 1993) apart from neutron reheat of about the same order of magnituted (He et al 1993; Martinez-Val et al. 1994). In similar way this re-heat is crucial to increase the gain at indirect drive at laser fusion with ns pulses (Hurricane et al 2014).

The difference from DT for the case of HB11 is the further strong increase of the fusion gain by secondary reactions of the alpha particles. When these collide with the boron-11 nuclei in the fuel within a range of about $60\mu m$ by central elastic collisions, an energy of about 600keV is transferred to the boron. Due to a most exceptionally high fusion cross section at 600keV, the boron nuclei react with protons and produce three new alphas (Eq. 9-1b) such that an avalanche multiplication of the reaction of the alpha particles can take place until the complete exhaustion of the fuel.

In order to estimate whether a sufficient fusion gain is possible with ps laser irradiation for initiating the fusion flame for the HB11 reaction, allowing the assumption of full burn by the alpha avalange, even optimistic assumptions about the initiating process of the fusion flame result in a gain of 69 only for a power of exawatt initiation power of a one picoseconds laser pulse (Hora et al 2014). This result is sobering if one could not consider a further considerable increase of the laser power.

## 10.4 SOLUTONS WITH ULTRAHIGH MAGNETIC FIELDS

The case of spherical geometry for the ps laser pulse initiation aiming to avoid the lateral losses following Nuckolls (2009) was one option, but another is to use a cylindrical geometry of the fusion volume and to confine the plasma by magnetic fields against the expansion in the direction of the cylindric radius. Use could be made with the magnetic field in the coil 2 from a pulsed discharge from a source 1 as shown in Fig. 10-11.

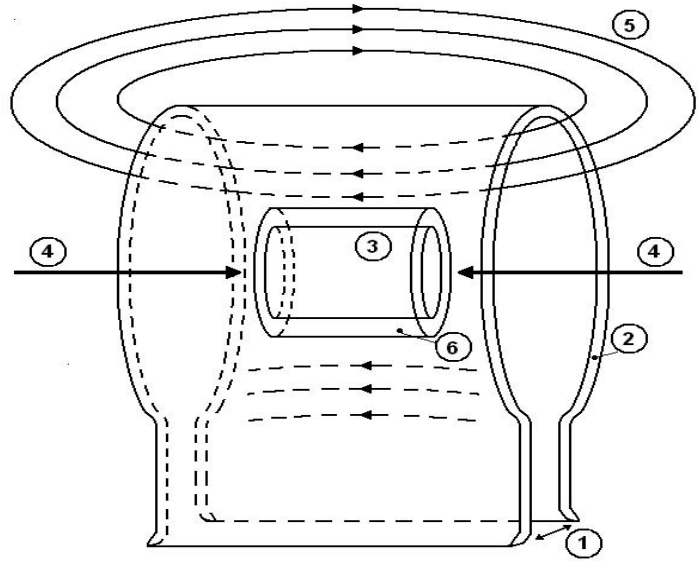

**Figure 10-11.** Generation of a pulsed magnetic field 5 of few nanosecond duration by a pulsed discharge 1 with the coil 2 for confining the cylindrical plasma 3 generated by ps laser pulses of interaction radius R = 1mm produced plasma blocks 4 from one or both sides. The cylindrical fusion fuel 6 has a radius larger than R.

When computations were done in 2010 with the then available highest magnetic fields of 100 Tesla, the reacting fusion plasma at the considerd conditions within the cylinder volume could not be confined. Fig. 10-11a shows the result where the plasma is expanding radially against the magnetic field with a loweering of the plasma density in the central parts of the cylinder and with an increasing of the magnetic field at thee boundary between the plasma and the applied magnetic field.

This situation changed since nearly 100 times higher magnetic field can be generated, again by using a short time laser pulse technique (Fukuoka et al 2013). Magnetic fields of 4.5 kiloTesla have been reached in experiments and it may not be impossible in the following computations to use fields of 10 kiloTesla. The loop of the coil is shown in Fig. 10-12.

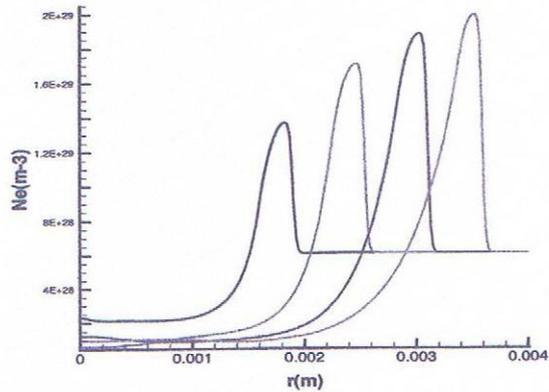

**Figure 10-11a.** Computation of electron density $N_e$ and axial cylindric magentic field $B_z$ depending on radial cylindric coordinate r at times (from the left of the plots) 0.4, 0.8, 1.2 and 1.6 nanoseconds of a cylindrical solid density DT plasma of 1 mm radius at time zero located in an axially parallel magnetic field of 100 Tesla. At time zero a $10^{20}$ W/cm$^2$ KrF laser pulse of one ps duration produced a plama block for igniting a fusion flame.

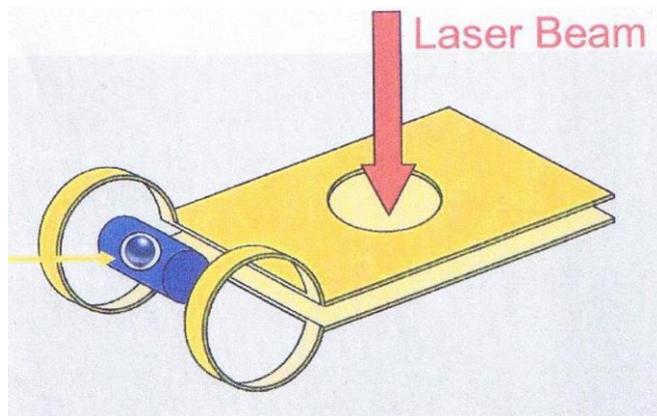

**Figure 10-12.** Modifying the coil with laser triggering of the pulsed current for generating of the magnetic field generated in the loop up to values of 4.5 kiloTesla (Fujioka et al. 2013; Moustaizis 2014)

The following computations are now reported with conditions from the following Figures 10-13 . We use a cylinder radius for the fuel cylinder of (part 6 of Fig. 10-11) 5 mm radius of solid HB11 irradiated from one side by 1ps, 248 nm wavelength laser pulses of intensity $10^{20}$ W/cm$^2$ on a coaxial cross section (Part 3 of Fig. 10-11) with 1 mm radius as described before. The fuel is imbedded in an axially parallel magnetic field of 10 kiloTesla which is assumed to be constant with the range of a nanosecond or longer, in contrast to the earlier cases which used 100 Tesla magnetic fields (Moustaizis et al 2013).

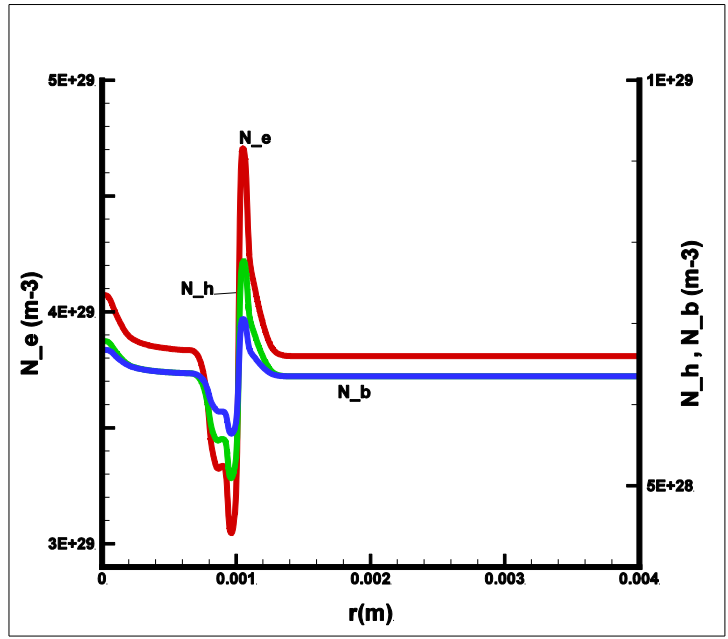

**Figure 10-13,** Solid HB11 cylinder of 5mm radius irradiated by a ps-$10^{20}$ W/cm$^2$ laser pulse of 248nm wavelength at concentric radius of R = 1 mm. After 100ps, the dependence on the radius r is shown for the density of electrons $N_e$, boron $N_b$ and hydrogen $N_h$ (sequence at R=0 from above) using a 10 kiloTesla magnetic field.

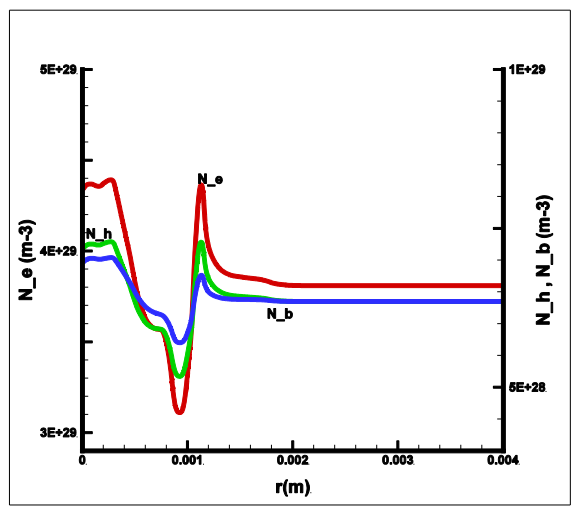

**Figure 10-14.** Same as Fig. 10-13 at time 500ps.

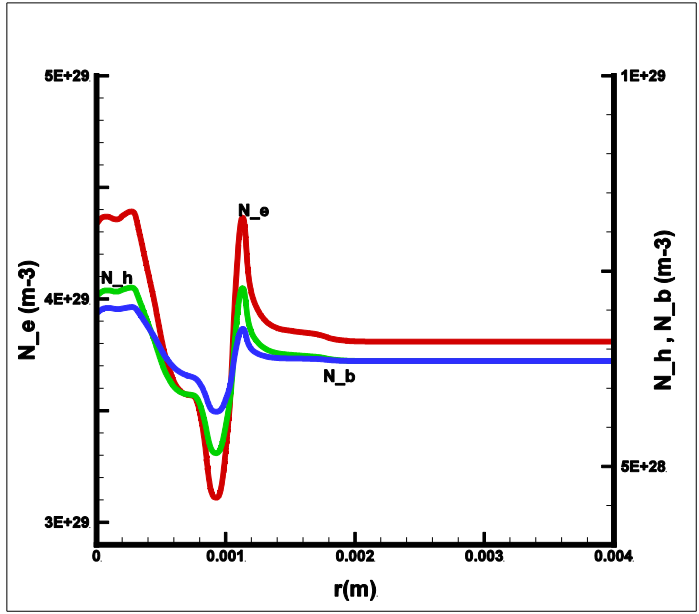

**Figure 10-15.** Same as Fig. 10-13 at time 1000ps

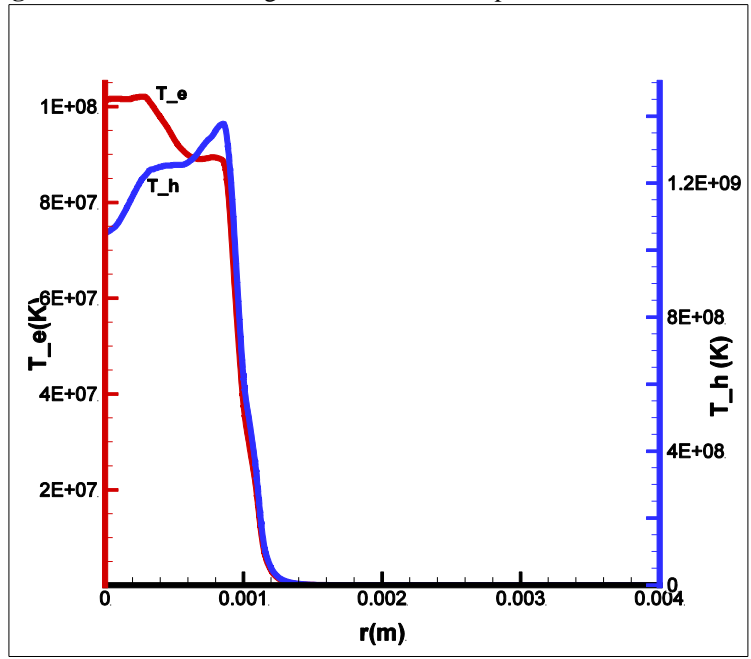

**Figure 10-16.** Same as Fig. 10-15 for 1000 ps with electron and ion temperature. At r=0 the higher curve refers to electron temperature $T_e$ with the left hand ordinate.

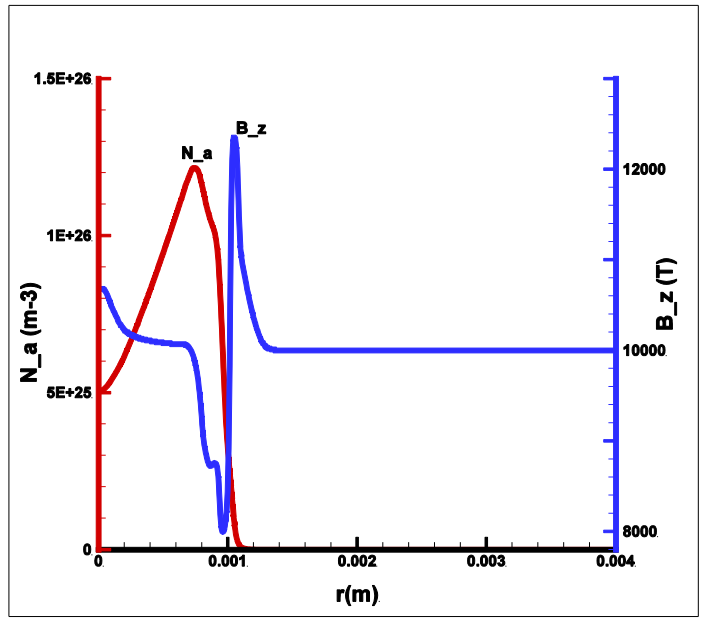

**Figure 10-17.** Radial dependence of the alpha particle density $n_\alpha$ and of the magnetic field B in axial direction in Tesla at 100ps.

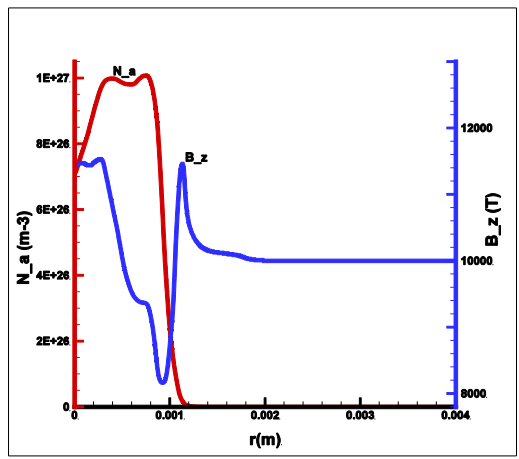

**Figure 10-18**. Same as Fig. 10-17 at time 1000ps.

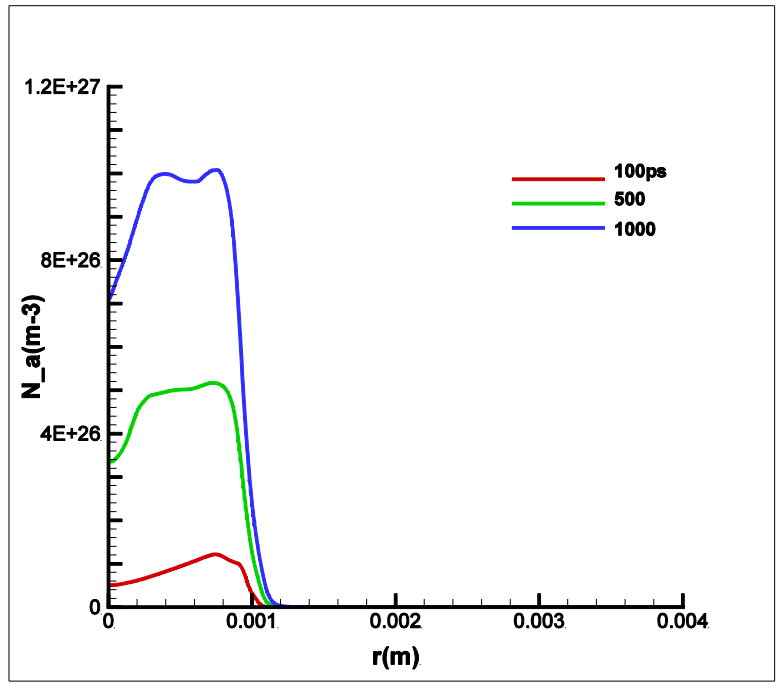

**Figure 10- 19**. Generated alpha paticle density at growing time from below.

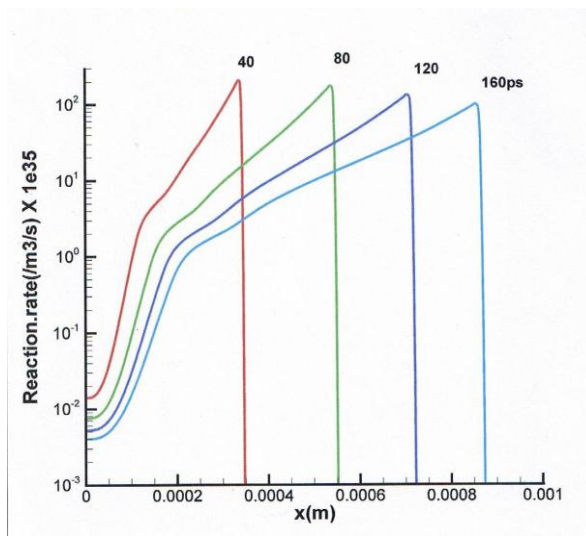

**Figure 10-20**. Binary eaction rate of HB11 along the axial propagation depth x parallel to the cylinder axis at different times without alpha avalange in 1-D computation parallel to the magnetic field, at times after the ps generation of the fusion flame initiated by $10^{20}$ W/cm$^2$, ps, 248-nm wave-length laser pulse.

fusion reaction rate at different depths x in 1-D computation parallel to the magnetic field, at times after the ps generation of the fusion flame initiated by $10^{20}$ W/cm$^2$, ps, 248-nm wave-length laser pulse.

The hydrodynamic computations are based on the initial genuine two-fluid model (Lalousis et al. 1983; Hora et al. 1984), where the electron and ion fluid use separate hydrodynamics with thermal equipartition and general collisions at laser interaction with electron heating, and using nonlinear (ponderomotive) force acceleration and the electric fields by the Poisson equation. This was the first general description of plasma hydrodynamics with the usually neglected (Kulsrud 1983) internal electric fields generated in inhomogeneous plasmas at the laser interaction and also known from ionosphere plasmas (Alfven 1981). These fields produced the inhibition of thermal conduction in inhomogeneous plasmas and included Gabor's collective stopping power for high plasma densities (Hora et al 2008; 2009). A clarification of the borderline between the Bethe-Bloch theory and that of Gabor may still be open. The details of the computations included the interaction of picoseconds-petawatt laser pulses with high density plasmas with extension to fusion produced alpha particles as further fluid as described before, see Section 6.4 for the initial conservation equations (Lalousis et al. 2012; 2013).

Fig. 10-13 shows the result of the radial dependences of the densities of electrons, boron and protons at 100ps after the ps laser pulse initiated the fusion flame when the cylindrical fuel was located in the 10 kiloTesla magnetic field parallel to the cylindrical axis. The plasma is well confined to its initial cylindrical radius in contrast to the radial expansion that occurs without a field or even with a 100 Tesla magnetic field see Fig. 10-11a (Moustaizis et al. 2013). The confinement for the 10 kiloTesla case can be seen in Fig. 10-13 showing the radial dependence of the ion temperature and electron temperature at the time 100ps. The electron temperature (left hand side ordinate) begins at R=0 with $9.7\times10^7$ K and the ion temperature (right hand side ordinate) with $9.7\times10^8$K. The results are shown for 500ps (Fig. 10-16) and at 1000ps are shown in Figure 10-17 and 4. First of all one can see that the plasma is still confined to the initial radius near 1 mm. What is remarkable is that even at longer times the electron temperature remains about ten times lower than the ion temperature, Fig. 10-16.

The positive result of ignition can be seen also from the figure 10-17 at 100ps and figure 10-18 at 1000ps for the radial dependence of the alpha particles (obviously less than 60 times of the ion density at 1000 ps). The magnetic field shows some increase above 10 kiloTesla due to some dynamics at the 1 mm border of the plasma. When looking to the alpha density in Fig. 10-19, this is about at $1\times10^{26}$ m$^{-3}$ at 100 ps growing up to $9\times10^{26}$ m$^{-3}$ ten times later at 1000ps. Clearly these results confirm well the ignition process.

The development of the fusion reactions of HB11 based on the primary binary reactions – as in the DT cases - is shown in Fig. 10-20 at different times depending on the depth x of the cylindrical axis. Though the irradiation from the ps laser pulse has long gone, the fusion reaction continues nearly unchanged apart from a minor relaxation.

The gyroradius of alpha particles from the HB11 reaction in a 10 kiloTesla field is 42.5 μm, which is sufficiently small within the cylinder of 1 mm radius for the magnetic field confined volume of the reaction. Hydrodynamics is resulting in propagation of the fusion flame by a length x = 0.543 cm within one nanosecond. The emitted cyclotron radiation (Eq. 33 of Ref. Gulkis 1987) is less than 3 kJ per ns from the cylindrical reaction volume thanks to the low electron energy of less than 10 keV (Fig. 10-16). Thus cyclotron radiation can be neglected. In contrast to HB11, the complications of the neutron production by the usual DT fusion reactor associated with their 12 second average life time with neutron damage to materials, and induced radioactivity in blanket structures were elaborated in a study by Tahir and Hoffmann (1997).

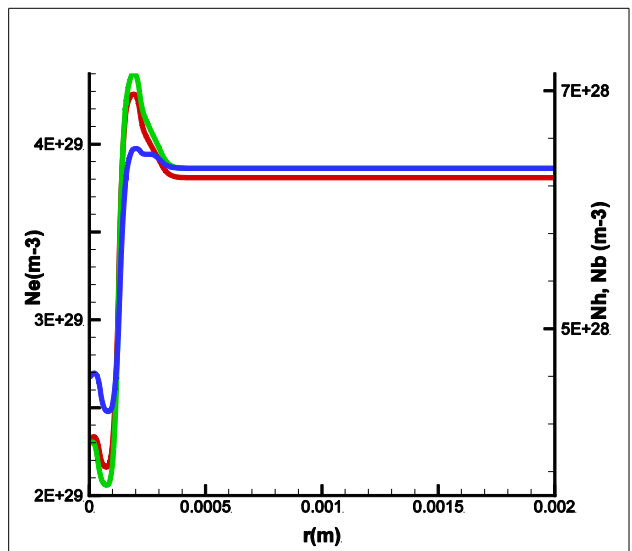

**Figure 10-21.** Dependence on the radius r of the electron density $N_e$, the proton density $N_h$ (highest maximum), and the $^{11}$B nuclear density $N_b$ (lowest maximum) at time 100 ps for solid state HB11 fuel at an irradiation radius r of 0.1 mm of a 248 wave length laser pulse of 1 ps duration and $10^{20}$W/cm$^2$ intensity.

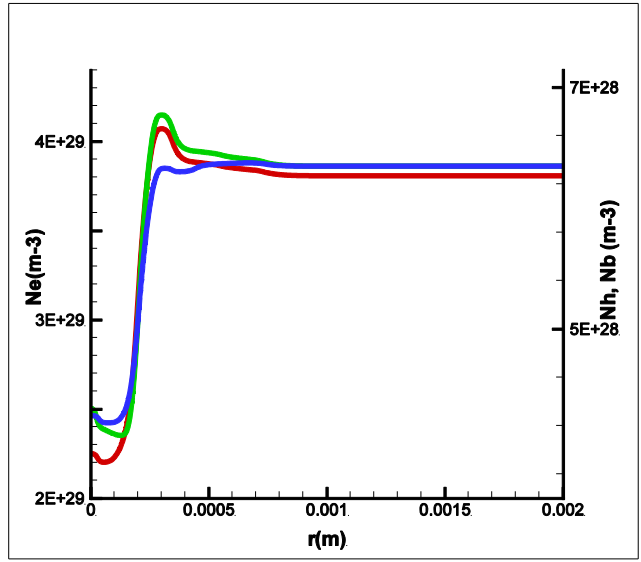

**Figure 10-22.** Same as Fig.10-21 at time 1000 ps.

The reported computation results included only the fusion energy of binary HB11 reaction given by the cross sections similar to our results for DT. However, these results did not include the subsequent secondary avalanche reactions (Hora et al 2014) by the alphas which transfer about 600 keV energy to $^{11}$B nuclei. This very important effect – as mentioned before - is due to the exceptionally high cross section at this energy for 11B reactions with a proton. The associated avalanche reactions result in a very high degree of fusion within the time in the range of ns or less. The resulting advantage of HB11 fusion compared with the DT reaction can be estimated in the same way as done for a spherical reaction of solid density HB11 initiated by an exawatt-ps laser pulse (Hora et al 2014). For a reaction of one ns in the confined cylinder, a gain of 300 times above the laser pulse energy for initiating the fusion flame occurs within the first ns. If this can be continued to 2 ns, the estimation arrives at a gain above 500. This is an attractive gain for inertial fusion energy IFE (Nuckolls. See Fig. 10-1) to provide an economical GW power station. The next more precise evaluation may well reduce and relax the conditions for the nuclear energy production of the clean boron-11 hydrogen fusion with elimination of neutron and tritium issues associated with conventional DT fusion power plants.

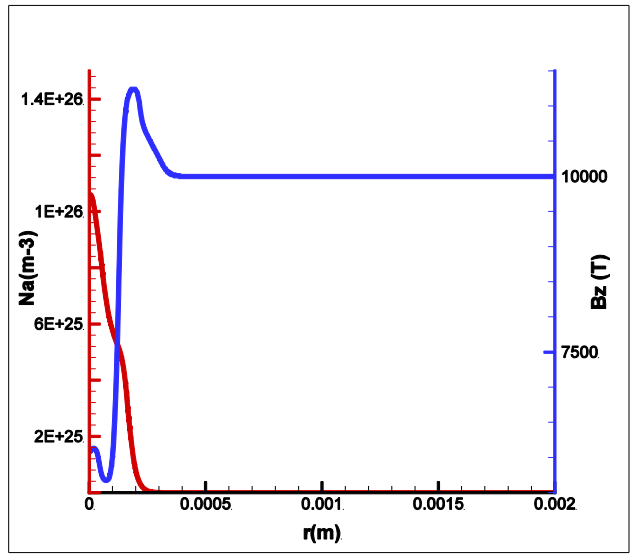

**Figure 10-23.** Same case of Fig. 10-21 with the magnetic field $B_z$ parallel to the cylindrical axis z merging into 10 kiloTesla above the radius 0.5 mm, with the curve at the left showing the density of the generated alpha particles $N_a$.

The well-known inertial confinement fusion schemes with DT are based on fusion gains up to a few hundred. In the present case of fusion with the measured 4.5 kiloTesla magnetic fields at each reaction onc can reach much higher fusion gains by using the needed the laser controlled magnetic field generated by the coils, Fig. 10-12 (Fujioka et al 2013) using an electromagnetic flux compression (Yoneda et al. 2012) for simplifying a shock methoed (Chang et al. 2011) such that each laser shot uses a target which may cost a few dollars. This situation then requires very high gains. Such high gain operation (up to 10,000) was postulated by Nuckolls (see Fig. 10-1) by use of very intense beams of electrons of 5 MeV energy focused on the target to achieve a compression about 12 times solid density or even less for DT fusion. However, conventional laser compression results in much lower gains (Section 9). The higher gain than usually attributed to HB11 fusion becomes evident by noting that in the present case, a 3 MJ-ps laser pulse on HB11 can produce >GJ energy gain. One GJ is 277 kWh energy. If a power station with a pulse sequence of 1 Hz would produce nearly the same amount of energy due to highest efficient electrostatic conversion of alpha particle energy into electricity, the cost of few dollars for generating each of a solid density HB11 fuel cylinder with the 10 kiloTesla magnetic field equipmen appears to be feasible.

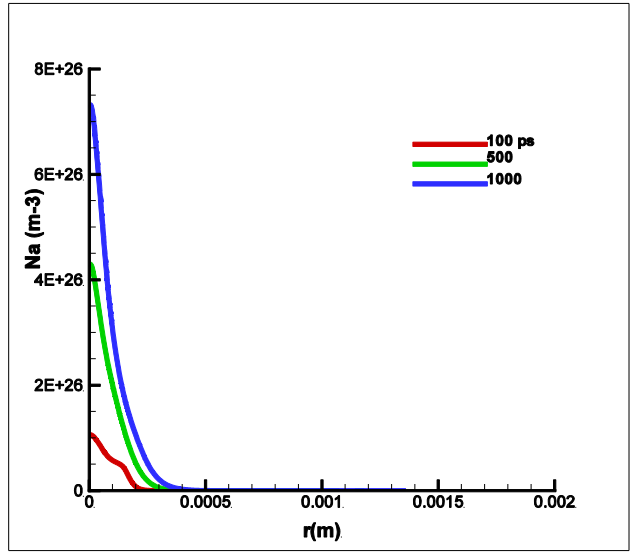

**Figure 10-24.** Alpha density $N_a$ depending on the radius r at different times (from lowest to highest curves for 100ps, 500ps and 1000ps respectively) showing ignition from the increase of the curves on time.

As next step, a further improvement was studied by reducing the cross section of the laser irradiated cylinder 3 in Fig. 10-11 from the radius of 1 mm to smaller values. These computations were done (Lalousis 2014a) to values 0.3 mm and 0.1 mm (Lalousis et al. 2014). The results are similar to those for r = 1mm in Fig. 10-13 to Fig. 10-20.

From a series of cases with each irradiation of ps laser pulses of 248 nm wave length and $10^{20}$W/cm$^2$ intensity at an interaction diameter of 0.1mm radius on solid density HB11 for initiation of fusion, Fig. 10-21 and 10-22 show the resulting densities of the electrons $N_e$, protons $N_h$ and boron nuclei $N_b$ along the radial coordinate r at times 100ps and 1000ps respectively of the two figures. From the initial radius of 0.1mm, the plasma has expanded showing a depletion at the cylinder axis up to the radius of 0.25 mmbut with some compression at higher radius. Above 0.4mm the plasma has the untouched densities. Fig. 10-23 shows at 100ps the magnetic field which is 10 kiloTesla in the axis-parallel z-direction while the curve on the left hand side is the density $N_a$ of the generated alpha particles centred inside the plasma. The fact of the ignition can be seen in Fig. 10-24 where the density of the generated alpha particles is shown increasing on time with a rdius growing to 0.3 mm at 1000 ps.

All these calculations are similar to the DT fusion using binary reactions without the secondary alpha avalange reactions. The secondary reactions of the 2.9 MeV alphas when hitting a boron nucleus and transferring about 600 MeV energy at central collision are not included in the computations (Hora et al. 2012a). The gyro radius of the alpha particles at 10 kiloTesla magnetic fields is 42.5 μm and their mean free pass for collective stopping at solid state density is nearly independent on the electron temperature in the range of 60 μm at solid state density such that an avalanche multiplication is resulting in an exponential increase of the fusion gain until fuel depletion. Similar estimations as for spherical geometry (Hora et al. 2014) show how a ps-30PW laser energy input into the block for the initiation of the flame of 30 kJ can produce alpha energy of >1 GJ. By this way, the requested fusion gain for DT of 10000 postulated by Nuckolls et al. for a power station (Nuckolls et al 2002) will be fulfilled. The aim to produce more than 100 MJ fusion energy per pulsed fusion shot was also underlined by Feder (2014) mentioning Dawns Flicker (2014). Her understanding with respect to the costs for a fusion power station is evident. The scheme of Nuckolls et al (2002) using relativistic electron beams for fast ignition arrives at comparable values for HB11. It is remarkable that the alpha-avalanche process is arriving at comparable values with clean HB11 fusion above those with DT.

## 10.5 CONCLUDING REMARKS

The book encircles a selection of topics of laser-plasma intraction which may begin with long settled facts kept in the first edition and followed by opening views of potential progress gained from broad stream reseach during more recent developments. This does not only relate to laser fusion work with nanosecond laser pulses in Section 9 while well understanding the option of direct drive with volume ignition and the alternaitve of indirct drive with spark ignition, now focusing both on the role of alpha particle reheat. The alternative direction with picoseconds laser pulses of Tabak's et al. (1994) fast ignitor opened new options for sub-picosecond laser interaction and power level much above petwatt. The basically and crucial difference of the picoseconds case for the physics of nonlinear (ponderomotive) plasma acceleration known since 1978 is shown in Figures 4-14 and 4-15 with eliminating thermal mechanisms in contrast to the thermal-gasdynamic pressure mechanisms of nanosecond laser pulse interaction.This is essential and presented in the sections 8 and 10.

Even within the broad stream of sub-picosecond interaction, there is a very small section only seen from transparent conditions with the measurement of the ultrafast acceleration of plasma blocks – called only in some sense pistons - (Sauerbrey 1996; Földes et al 2000) using extreme high quality single mode beams. This was clarified by the elimination of relativistic self-focusing shown by Jie Zhang (Zhang et al. 1998). This discovery was based on the attention and ingenious clarification of very rare facts of x-ray emission which were usually ignored by the borad stream of research. This was realised also by the most unusual and unorthodox measurements by Badziak et al (1999) pioneering the fact of ultrahigh acceleration of the dielectrically swelling-increased depth of the plasma skin layers (Hora et al 2002). This led to understand the ultrahigh acceleration of plasma blocks including also ultrahigh space charge neutralized ion current densities far higher than million times of accelertors. Credit should be given also to journals against the broad stream to publish these results. Only following these facts (Hora et al 2007), Chu's (1972) discovery of ps very high energy flux density laser pulses for initiation of a fusion flame (Bobin 1974) led to the side-on ignition of solid density fusion fuel (Hora et al 2002; Hora and Miley 2004; Hora et al 2007; Hora 2009).

The exceptional conditions were realized also through the extreme nonlinear absorption in nanometer thin diamond layers (Steinke et al. 2010) as consequence of very special experimental instrumentation (Kalashnikov, Nickles et al 1994) and understood by unusual approaches (Hora 2012) to the result of laser powers above the petawatt and ongoing much higher level new options.

These developments all began as fundamental new direction of picosecond pulses opened with Tabak's et al (1994) fast ignition inspired since 1991by Mike Campbell (2006). This is the motivation why from the very broad and extremely diversified stream of sub-picosecond laser acceleration of plasmas, the topics of this book were selected. This may be summarized by mentioning the following key points:

1) We should not forget how the nonlinearity became visible with the experiment by Linlor (1963). When irradiating Hellwarth's (McClung et al. 1963) several nanosecond long Q-switched laser pulses of few MW power, the irradiated target led to the observation of an extreme anomaly of the emitted ions. Instead of classical-thermal generation of ions with few eV energy, known from irradiation with less than MW laser pulses, it was a surprise that keV ions appeared whose energy was separated linearly on the ion number Z. This was an electrodynamically determined non-thermal process (Hora et al. 1967) needing to include ponderomotive self-focusing (Hora 1969a). The involved forces were related to Kelvin's (Thomson 1845) discovery how mechanical forces can be generated on uncharged, electrically neutral bodies by quadratic expressions of the force quantities of electric fields. This ponderomotion was explained for plasma (Hora et al. 1967; Hora 1969;1985) by using the optical response given by the dielectric constants of the plasma to modify Maxwell's stress tensor. Another relativistic self-focusing is due to quiver motion of electrons if their quiver energy in the laser field exceeds $mc^2$ (Hora 1975).

2) Evaluation of the Boreham et al. (1979) experiment arrived at the fact (Section 12.3 of Ref. Hora 1981) that laser beams apart from the main transversal fields have tiny longitudinal fields as exact solution of Maxwells' equations. Only with these tiny fields, Boreham's measurements of the polarization independence of electron accelertion could be described correctly, see general case needing all components of Maxwlls stress tensor (Ciccitelli et al 1990). Neglection of the tiny fields arrived at a totally wrong result. This reality showed that nonlinear theory can be necessary while linear theory completely failed. This led to a nonlinearity principle for opening new physics research with predicting or explaining completely unexpected new phenomena (see Section 6.3 of this book).

3) The nonlinear (ponderomotive) forces opened new conditions when studying processes of shorter than picoseconds (ps) duration where laser pulse energy is converted in a non-thermal way directly into mechanical motion of macroscopic plasma blocks in drastic contrast to thermal heating and gasdynamic presure at interaction by nanosecond (ns) laser pulses. Laser pulses with 2 petawatt power (Cowan et al. 1999) were produced by chirped pulse amplification based on Mourou's (2010) crucial discovery of 1985 for solving the problem of the stretcher-compressor element (Strickland et al 1985). Pulses as short as attoseonds have been achieved and zeptosecond ($10^{-21}$s) pulses are on the way (Krausz et al 2009).

4) Ultrahigh acceleration of plasma blocks above $10^{20}$ cm/s$^2$ has been measured (Sauerbrey 1996) and was reproduced (Földes et al 2000) in agreement with predictions by theory and computations of 1978 (see Fig. 10.18b of Ref. Hora 1981; Hora et al. 2007). These accelerations are more than 10000 times higher than measured before in laboratories. Ultrahigh accelerated plasma blocks reach the conditions of energy flux densities of at least $10^8$ J/cm$^2$ at ps interaction for initiating nuclear fusion flames in deuterium-tritium and even in proton-boron(11). This was following the side-on ignition (Chu 1972) of uncompressed solid density fuel after updating the theory (Hora 2009).

5) These fusion reactions by laser beams of limited cross section can be spherically confined (Hora et al 2014) or alternatively in cylindrical fuel geometry using laser operated cylindrical magnetic fields of 4 kiloTesla (Chang et al. 2011) and more using electromagnetic flux compression (Fujioka et al. 2013). Differing from DT fusion, the HB11 fusion should lead at high gain by a secondary alpha avalange processes. First estimations about the secondary processes arrive at the possibility that ps laser pulses of 30 PW power may produce >GJ alpha particle energy from HB11 (Lalousis et al 2014a) – taking into account that 3 PW have been measured (Li Ruxin 2013). This kind of fusion leads to less generation of radioactivity than from burning coal (therefore being negligible) about which the Edward Teller medaillist Steven Haan (2010) commented that "This has the potential to be the best route to fusion energy".

# APPENDIX 1
# COLLECTIVE ELECTRON INTERACTION AT ULTRAFAST ACCELERATION OF PLASMA BLOCKS*

**ABSTRACT**

A fundamental difference between interaction of laser pulses of less than picosecond duration and power in the range of and above Petawatt appears in contrast to pulses of nanosecond duration. This is due to the basic property that the long pulse interaction is based on thermal effects with inefficient delays of chaotic microscopic thermal motion while the short pulses avoid these complications and the interacting plasma reacts as a macroscopic collective known from atomic physics. Optical energy is converted into mechanical motion with high efficiency and nearly no thermal losses. These developments cover a long history of laser developments leading now into a new era of nonlinear physics combined with quantum properties. One of the applications is laser driven fusion energy

## 1. INTRODUCTION

This topic is of interest for considering a new and unexpected phenomenon appearing in the range of femto- and atto-second (fs & as) laser pulses with powers in the range of about and far beyond Petawatt (PW). This is leading to very unusual results for example with the generation of plasma blocks by an ultrahigh acceleration in the range of $10^{20}$ cm/s$^2$. When this was first experimentally verified from directly visible Doppler line shifts [1], it was underlined that these ultrahigh accelerations are far beyond - more than 100,000 times – to all what was known before from measured accelerations of macroscopic bodies in laboratories. The exotic acceleration was comparable to concluded values at neutron stars or black holes only. Nevertheless, not much attention was given immediately to this result – though underlined in the abstract - while these signals were a breakthrough into the new dimension of physics which was opened thanks to the discovery of the Chirped Pulse Amplification CPA of laser pulses [2][3] with applications to many new phenomena [4]. Last not least one application was given to a new scheme of low cost energy production from fusion by lasers [5] with less generation of dangerous nuclear radiation than from burning coal [6]. This would be a first step to profit from the 10 Million times more efficient gain of nuclear energy in an absolute clean way compared with chemical processes.

The roots for these developments go back to the years before 1970 of laser research where the basic new concept of nonlinearity was opened. It is no surprise that the ultrahigh acceleration of directed plasma blocks by $10^{20}$ cm/s$^2$ were the result of the theory as seen in numerical outputs in 1978, see the figure on p. 179 of Ref. [7], well agreeing [4] with the later measured ultrahigh accelerations [1] using laser pulses of 300 fs duration and more than TW power. A necessary condition for the experiments was to avoid relativistic self-focusing. The theory was based on the correct inclusion of the optical properties of plasma in Maxwell's stress tensor leading to the nonlinear force $f_{NL}$ of which under simplified conditions the ponderomotive force is a long known approximation.

The femtosecond pulse interaction is a direct conversion of optical energy to the collective macroscopic motion of particles without all thermal complications of long lasting delays for generation and conversion of electron heat into ion pressure with instabilities and inefficiencies for macroscopic motion as known for longer laser pulse interaction. Instead of chaotic thermal mechanisms known from all linear macroscopic physics, the short pulse interaction provides a change of microscopic properties to macroscopic systems in a collective way, what usually is known only to the microscopic states in atomic physics. A glimpse of connection was seen before with the Mössbauer effect where the thermal properties of a crystal were frozen and the interaction of gamma radiation takes place with thermally undisturbed nuclei.

In view of the here presented example for a breakthrough between macroscopic physics in contrast to atom physics with the present laser pulses of fs duration and beyond PW power, it is interesting how these new results open doors for new physics. It is therefore indicated that some of the laser developments of all these years have to be explained where a lot of unexpected hurdles had to be overcome.

## 2. NONLINEARITIES WITH LASERS

The crucial turning point from linear to nonlinear physics by lasers was experienced in a very early stage. When laser pulses with a power of up to about 1 MW were hitting a solid target in vacuum, the interaction was fully classical. A plasma was produced with temperatures of about 20,000 K (about 2 eV) and the expanding plasma was expanding following classical hydrodynamic equations. The energy of the emitted ions could be measured from the time of flight by an oscilloscope showing energies of few eV and the charge number Z of ionization was close to unity. The emission of electrons followed the Langmuir-Child's law with a limit of an electron emission density in the range of the well known space charge limitation of 0.1 Amps/cm$^2$. These results for industrial applications of lasers were summarized by Ready [8] and were expected from the preceding knowledge.

Hellwarth [9] discovered the Q-switch for laser pulses of few ns duration with fully reproducible profiles of the time dependence of the pulse power for a maxium at or above 10 MW. When these pulses irradiated a target, Linlor made a most unexpected observation [10]. The energy of the emitted ions were not longer in the range of few eV but were (!!) 1000times higher. These keV ions had a high ionization number Z. Their ion energy was not uniform as expected from thermal equilibrium in the plasma and instead it was linearly increasing on the charge number Z [11]. The number of the keV ions was very high (up to about $10^{13}$). The linear dependence on Z indicated a non-thermal emission process, therefore the ion emission had to be driven by electrodynamic forces. The suggestion could be forgotten that the electric fields in the plasma surface produce the charge separation of the ions, because it was estimated, that then only about $10^8$ ions could have been accelerated which number was by orders of magnitude smaller than measured. The small number of the emitted keV ions due to the ambipolar surface fields could separately be measured later [12] to be close to the earlier estimated value of $10^8$.

The first result about the plasma acceleration based on laser driven forces due to dielectric response [15] Fig. 1a was using the decrease below the vacuum value unity of the optical refractive index *n* in a plama slab causing an increase of the electrical laser field amplitude $E_v/n^{1/2}$ above its vacuum value $E_v$. This was resulting in an equation of plasma motion by using the solution of Euler's differential equation for an initial Rayleigh profile of the plasma density *N* on the depth x (see last part of Section 3.4 of this present book). The equation of motion, expressed by the Lorentz force parallel to a stress tensor formlation arrived at the forces shown by arrows in Fig. 1a tearing the plasma into a block moving against the laser light and another moving into the direction of the laser beam.

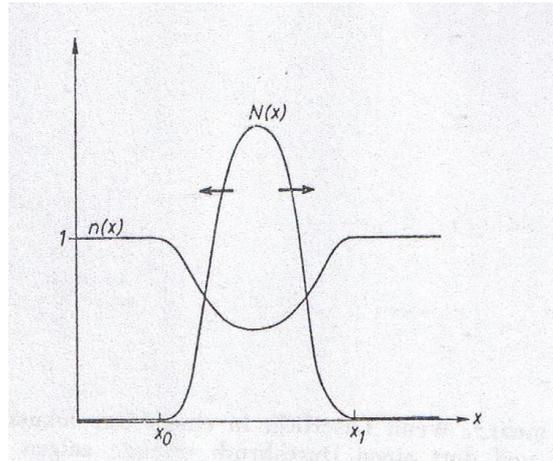

Fig. 1a. A laser pulse arriving from the left hand side on a plasma slab of density *N(x* with the schematic drawn curve for the optical constant *n(x)* within the plasma desreasing below the value unity causes nonlinear p(onderomotive) forces (arrows) in the plasma equation of motion tearing the slap in a part moving against and another with the laser beam [15].

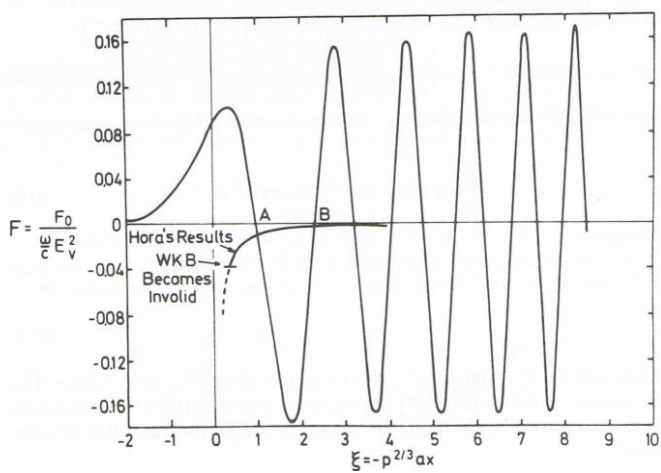

Fig. 1b. Analytic-numerical evaluation with Airy functions by Lindl and Kaw [21] in order to check the appearence of the nonlinear (ponderomoritve) force following [17])

The way for a correct explanation of the keV ions was the result of combining the following three facts. First the optical constants for the optical dispersion due to the longitudinal plasma oscillations (Langmuir waves) of plasmas had to be calculated with inclusion of damping. This could be based on the theory of electron-ion collisions of frequency $\nu$ which could be calculated with Spitzer's classical plasma collision frequency [13]. Interestingly this result arrived at nearly the same values as the quantum-electrodynamic (Dirac) theory with the Gaunt factors as summarized [14] including plasma densities above the critical value. The complex optical constant **n** in a plasma is then with the laser frequency $\omega$, the plasma frequency $\omega_p$, and with the charge e and the rest mass m of the electron

$$\mathbf{n} = 1 - (\omega_p/\omega)^2/(1 - i\nu/\omega); \quad \omega_p^2 = 4\pi e^2 n_e/m \tag{1}$$

With this, the second fact was the combination with optically modified averaged electric laser fields **E** in the Maxwellien stress tensor for describing forces. In a first step it was evident that this dielectric modification of the laser fields in plasmas could produce an acceleration determined by a ponderomotive force given be the negative gradient of $\mathbf{E}^2$ [15]. This could show the acceleration of the keV ions and their energy separation linear on Z. The third ingredient was how the sufficient high laser intensities could be produced. This was solved by the ponderomotive self-focusing [16] where the resulting filament of the laser beam arrived at the experimental confirmation of the diameter resulting in the threshold power of 1 MJ laser pulses reproducing the observed threshold separating the thermal eV ions generation from the keV nonthermal ions at higher powers.

Summarizing, the force density **f** in a plasma is given by the classical thermal gasdynamic pressure $p = 3n_p kT/2$ where $n_p$ is particle density, k is Boltzmann's constant and T the temperature, and given by the nonlinear force $\mathbf{f}_{NL}$ due to electrodynamic interaction

$$\mathbf{f} = -\nabla p + \mathbf{f}_{NL} \tag{2}$$

The nonlinear (optical corrected) nonlinear force of the simplified initial study [15] was derived [17] for the stationary condition without the temporal derivation in the stress tensor using the unity tensor **1** and the magnetic field **H** of the laser

$$\mathbf{f}_{NL} = \nabla \cdot [\mathbf{EE} + \mathbf{HH} - 0.5(\mathbf{E}^2 + \mathbf{H}^2)\mathbf{1} + (1+(\partial/\partial t)/\omega)(\mathbf{n}^2-1)\mathbf{EE}]/(4\pi)$$
$$- (\partial/\partial t)\mathbf{E} \times \mathbf{H}/(4\pi c) \tag{3}$$

The non-stationary case led to six different formulations between each of the six different research centers different from Eq. (3). The final result of Eq. (3) was derived by the addition of a missing small logarithmic term [18] which could not have been recognized before in the otherwise most advanced approximation by Zeidler et al. [19]). Only Eq. (3) is the complete and general nonlinear force density in plasma. This and only this was shown to be gauge and Lorentz invariant (Eqs. 8.87 and 8.88 of Ref. [20]).

For simplified geometry with plane wave laser interaction, Eq. (3) can be reduced to

$$\mathbf{f}_{NL} = -(\partial/\partial x)(\mathbf{E}^2+\mathbf{H}^2)/(8\pi) = -(\omega_p/\omega)^2(\partial/\partial x)(E_v^2/\mathbf{n})/(16\pi) \qquad (4)$$

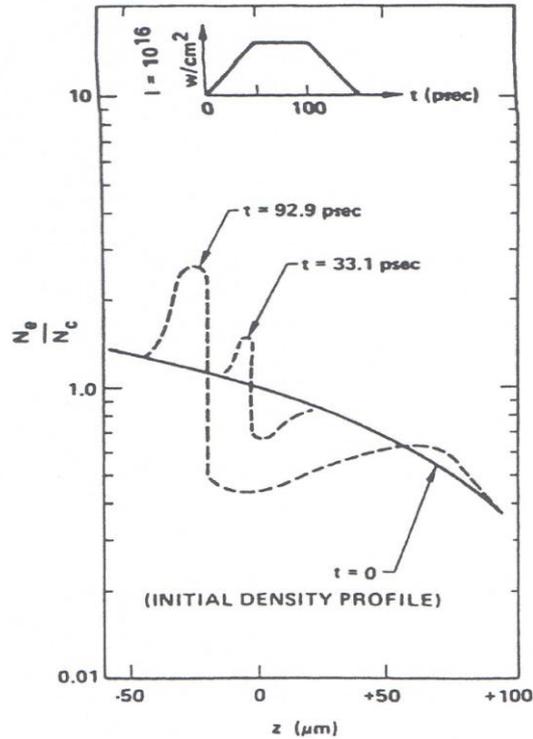

Fig. 2a. plasma-hydrodynamic WAZER computation of a the temporal changing plasma density by a laser intensity with dominating nonlinear force creating profile steepening and a density minimum (caviton) by Shearer, Kidder and Zink [22].

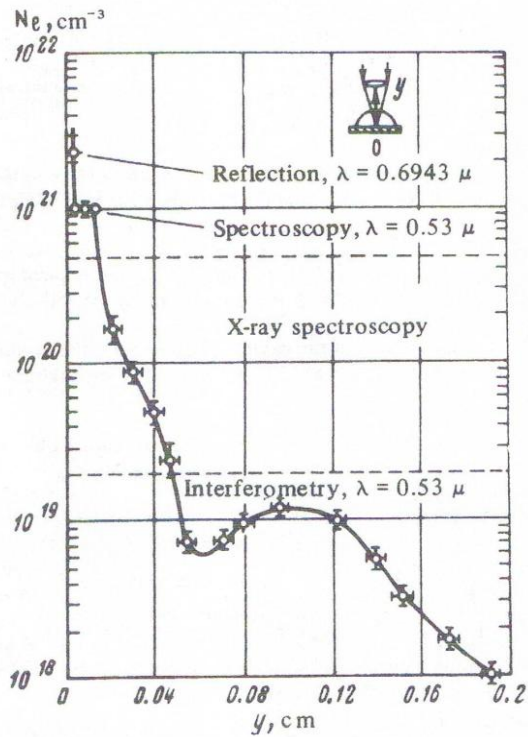

Fig. 2b. After numerical discovery [22], measured cavtion by Zhakharenko et al. [23].

The first expression is the gradient of the electromagnetic energy density and the second expression is related to the ponderomotive force which was derived for electrostatics without magnetic fields and not for plasma properties.

Following Eq. (2), the nonlinear force dominates over the thermal pressure, when the non-relativistic oscillation (quiver) energy of the electrons in the laser field expressed by the critical electron density $n_{ec}$ where in Eq. (1) the plasma frequency is equal to the laser frequency

$$\varepsilon_{osc} = \mathbf{E}_v^2/(8\pi n_{ec}) > (3/2)nkT \qquad (5)$$

is higher than the energy of thermal motion given by the temperature T. A first numerical comparison was seen with the action of the nonlinear force, Fig. 1b [21], where a plane geometry dynamic computation by Shearer, Kidder and Zink [22] with dominating quiver energy showed the temporal development of an initially smooth plasma density profile under laser irradiation. The nonlinear force resulted in a profile steepening with generating a density minimum, Fig. 2a, later called a caviton. The measuring of the caviton, Fig. 2b, within the very short times and the microscopic lengths was proof of the dominance of the nonlinear force. This proof was reproduced, even by sophisticated three dimensional measurements of the caviton (Azechi et al. [23]).

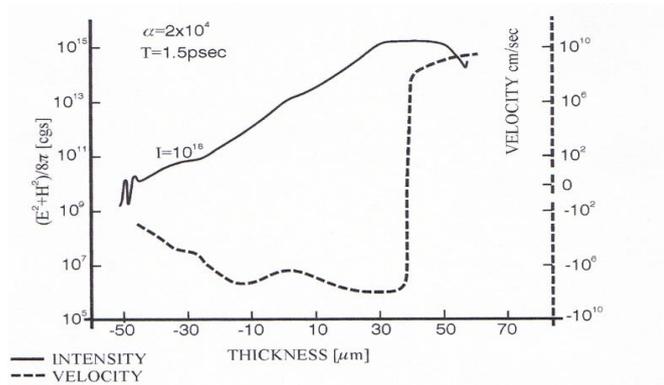

Fig. 3. $10^{18}$ W/cm$^2$ neodymium laser incident from the right hand side on an initially 100 eV hot very low reflecting bi-Raleigh plasma profile showing after 1.5 ps interaction the electromagnetic energy density [Eq. (4)]. The dynamic development had accelerated the plasma block of 20 vacuum wave length thickness moving against the laser and another into the plasma (combining results from p. 178 & 179 of Ref. [7]) with ultrahigh >$10^{20}$cm/s$^2$ ultrahigh acceleration.

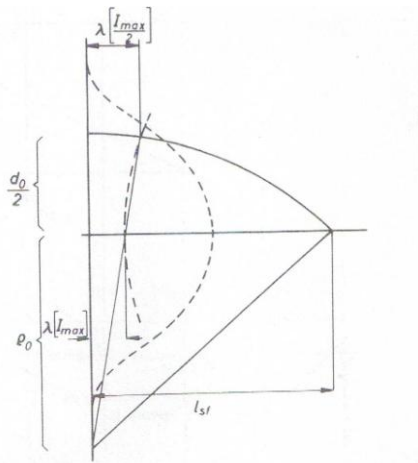

Fig. 4. Relativistic self-focusing of a laser beam of diameter $d_o$ (dashed intensity) hits a plane plasma with shorter effective wave length at higher intensity than at lower such that the plane laser wave front is bending concavely with shrinking the beam to less than wave length diameter resulting in very high intensity [25].

Irradiation on deuterium plasma with an initially very low reflecting Rayleigh profile of 100 vacuum wave lengths thickness (Fig. 3) ([7]: p.179; [24]) showed numerically in 1978 that a neodymium glass laser pulse of 1.5 ps duration and $10^{18}$ W/cm$^2$ produced a plasma block of nearly 20 vacuum wavelength thickness with velocities above $10^9$ cm/s. The acceleration is then $6\times10^{20}$ cm/s$^2$.

## 3. RELATIVISTIC SELF-FOCUSING

After the results with the dominating nonlinear force in plasmas at laser interaction with the ultrahig acceleration of plasma blocks by computations was shown, the question was why have these accelerations not been measured? One simple answer is that ps laser pulses of this intensity were not available in 1978. But there was another serious obstacle due to the then discovered relativistic self-focusing [25].

The velocity **v** of an electron oscillating (quivering) in the presence of a laser beam with an electric field and magnetic field **E** an **H** respectively is given by the equation of motion

$$(d/dt)m\mathbf{v} = e\mathbf{E} + (e/c)\mathbf{v} \times \mathbf{H} \qquad (6)$$

For linear plarized laser radiation with **E** in vacuum in the Cartesian y-direction and **H** in z-direction the only velocity components $v_x$ and $v_y$ in the x- and y-directions are

$$(d/dt)mv_y/[1-(v_x^2+ v_y^2)/c^2]^{1/2} = eE_y \cos(\omega t) - (e/c)v_x H_z \cos(\omega t + \phi) \qquad (7)$$

$$(d/dt)mv_x/[1-(v_x^2+ v_y^2)/c^2]^{1/2} = (e/c)v_x H_z \cos(\omega t + \phi) \qquad (8)$$

where $\phi$ is the phase between **E** and **H**. If v is much smaller than the speed of light c, the Eq. (7) remains for an elementary solution of the oscillation equation resulting in a maximum quiver velocity $v_{max}$ determining in Eq. (5) the oscillation (quiver) energy $\varepsilon_{osc} = mv_{max}^2/2$.

For relativistic laser intensities, the solution of the two coupled nonlinear differential equations (7) and (8) are a complicate mathematical problem, for which initially only approximations for the range of sufficiently low sub-relativistic conditions were used [26]. However for determining the quiver energy for the whole sub-relativistic and relativistic range, an exact solution was possible [27]. Using the Lorentz factor (in the detailed derivation, Section 6.5 of Ref. [7], the reciprocale value was used)

$$\gamma = 1/[1-(v_x^2+ v_y^2)/c^2]^{-1/2} = 1/[1-\mathbf{v}^2/c^2]^{-1/2} \qquad (9)$$

and neglecting the second term in Eq. (7) – a correction factor [28] results in a deviation of the final oscillation energy of the electrons of less than 3% with a maximum only for very high laser intensities – Eq. (7) can be integrated using the wave number $k = \omega/c$

$$v_y = (eE_y/m\omega\gamma) \sin(-kx + \omega t) \qquad (10)$$

Using Eqs. (9) and (10), in Eq. (8) leads [26] to the upper limit of

$$\gamma = [1 + e^2\mathbf{E}^2/(m^2c^2\omega^2)]^{1/2} \qquad (11)$$

where **E** is the electric field of the laser in the plasma given by its vacuum value $\mathbf{E}_v$ using the refractive index **n**, ($\mathbf{E}=\mathbf{E}_v/\mathbf{n}$). The oscillation energy of the electron is then

$$\varepsilon_{osc} = (\gamma - 1) = mc^2[(1 + 3I/I_{rel})^{1/2} - 1] = mc^2[(1 + a_L^2)^{1/2} - 1] \quad (12)$$

I is the laser Intensity and $I_{rel}$ the relativistic threshold intensity

$$I_{rel} = (3c/8\pi)(mc\omega/e)^2 \quad (13)$$

where the oscillation energy of the electron is equal to its rest mass m, and the abbreviation $a_L$ for defining the laser intensity is given by

$$a_L^2 = 3I/I_{rel} \quad (14)$$

The detailed *derivation of the important relativistic oscillation energy $\varepsilon_{osc}$ (12) [27]* from the coupled two nonlinear differential equations (7) and (8) for the velocity components $v_x$ and $v_y$ can be seen in Capter 6.5 of Ref. [7] where, however, the usual Lorentz-factor (11) was used in inverted form.

The effective wave length λ is modified relativistic by γ resulting in a shorter wave length at higher laser intensity than at lower such that (Fig. 4) the plane laser wave front is bent [25] to produce a beam diameter of less than a vacuum wave length (Fig. 5) at a focal length $l_{sf}$ in relation of the laser beam diameter $d_o$ before the interaction with the plasma. This can be expressed by the refractive index n at the beam center to that at half maximum as

$$l_{SF}/d_o = 0.5(n_{max} + n_{max/2})/(n_{max} - n_{max/2}) \quad (15)$$

The shrinking of the laser beam results in very high laser intensities with high ponderomotive acceleration gradients for emission of ions with high charge number Z and MeV to GeV ion energies [29] going into all directions [30] in agreement with the theory. This was confirmed [31] while it was difficult for leading plasma physicists to accept that nanosecond laser pulses of few Joules energy could produce phosphorus ions above 10 MeV [32]. In the mean time, the 100 MeV ions in agreement with theory [25] are usual and accepted and no other kind of observations were known.

For completeness it should be mentioned that the observation of the very high energies of the emitted ions could be a lot more complex than the just mentioned fastest energetic ions. When evaluating the oscillograms of time-of-flight measurements, there were not only the fastest peaks separated by the linear energy dependence of the ions on the charge number Z in agreement with the relativistic self-focusing theory. There was a slower group of ions well separated with another linear dependence of the energy on Z [32] before the non-differentiated spectra of

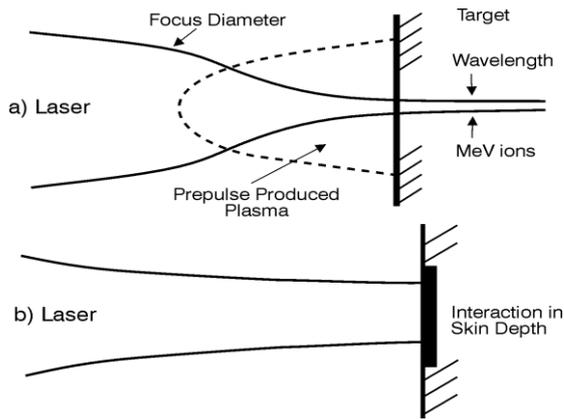

Fig. 5 Laser irradiation of a plane target. (a) The pre-pulse of the laser beam produces a plasma plume in front of the target causing relativistic self-focusing relativistic self-focussing of the laser beam, while (b) without pre-pulse, plane wave interaction takes place.

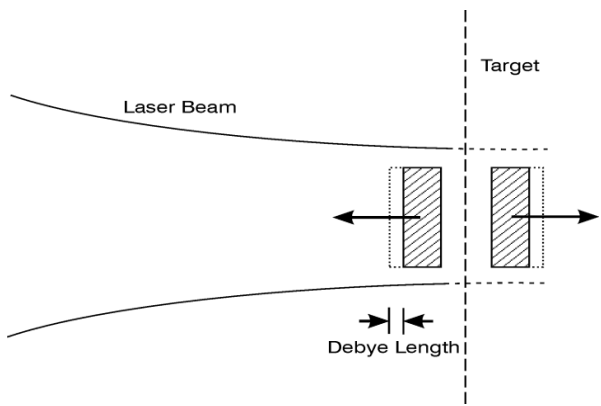

Fig. 6 laser beam without pre-pulse (Witte et al. [41]) hits a target producing two blocks of plasma, one moving against the laser light and one in the direction of the laser light [25][37][38].

the slowest thermal part of ions were appearing. At the very high laser intensities, there was a further mechanism with hot electrons discovered as clarified by Gitomer [33] and analysed [34]. While the ambipolar electric acceleration and Z-separation was not explaining [12] the large number of ions in the early experiments but this was different at much higher intensities. The very energetic quivering electrons in the high laser fields were partially thermalising by collisions resulting in hot electrons. In these surface areas there was well a strong Z-dependent ambipolar acceleration with parallel x-ray emission but with remarkable lower maximum ion energies [32]. A necessary condition is that the (in many cases only partial) themalization of the high energy quivering electrons could be explained only [34] by the quantum branch of collisions [35] and not by classical Spitzer [13] collisions.

At the very high laser intensities in the filament of relativistic self-focusing, very hard, short wave length x-rays are produced of very high intensity. In contrast to all these usual observations, cases with very low intensity and soft, long wave length x-rays were occasionally measured. It was the merit of Jie Zhang et al [36] to give attention to this fact and to find out what the reason was. At this time it was necessary for measurements with sub-picoseond laser pulses of more than TW power, that – for other experimental reasons – the pre-pulse before the main pulse had to be cut-off to a very high degree (contrast ratio) of up to $10^8$ or more [37]. Zhang et al. irradiated then similar short pulses at times t before the main pulse. When t was long enough (70ps) the generation of the high x-ray emission appeared as usually measured. During this time for a 30 wave length diameter laser pulse (case b in Fig. 5), a plasma plume was of about 60 wave length depth was created, just enough to produce relativistic self-focusing [38]. This explanation for suppressing relativistic self-focusing was further confirmed for the very high contrast experiments by Badziak et al. [39] where instead of fast ion emission into all directions, a very directed emission of ions were observed generated in the dielectrically expanded skin layer of the target [38].

## 4. SAUERBREY'S FIRST MEASUREMENT OF ULTRAFAST PLASMA ACCELERATION

After realizing the preceding description of facts, it will be understood what a turning point was arrived by the first measurement of the ultrahigh acceleration of plasma fronts [1]. A first indication of an unusual effect at interaction of very clean (high contrast) sub-picosecond laser pulses of power above TW was noticed [40] and a hint was given that this is due to the dominance of the nonlinear (ponderomotive) force. Experiments at this time needed the suppression of pre-pulses (very high contrast ratio) in order to avoid the superradiance in the lasing process. This was especially developed in the Max-Planck-Institute of Hans-Peter Schäfer in Göttingen with most advanced KrF lasers [41]. Thanks to this situation, Sauerbry [1] could measure under the precise conditions of plane geometry (case b of Fig. 5) where laser pulses of about 30 wave length diameter were hitting the surface of a carbon target without any plasma plume generated by prepulses. This was precisely avoiding relativistic self-focusing in a similar way as this was necessary at the discovery of Jie Zhang et al [36] why only low x-ray emission could exceptionally be measured from targets. A crucial clarification of the problems with the high contrast was given by Witte et al [41]

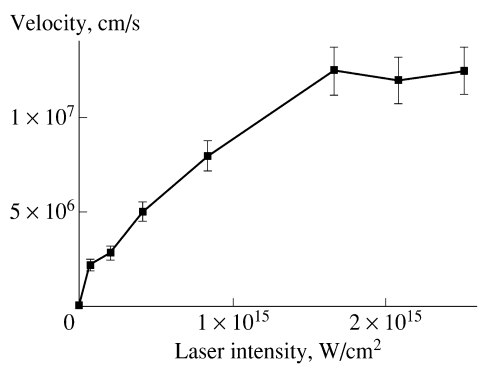

Fig.7. Intensity dependence of the velocity of the plasma front from the Dobppler shift of reflected 700fs KrF laser pulses from an Al target (Földes et al. [43]).

The next problem for Sauerbrey after having the real plane geometry, was to find a very sophisticated solution for measuring the Doppler effect from the plasma block emitted against the laser light as expected theoretically (see Fig. 3). Only these two facts - the clean laser pulses and the special method for the Doppler effect - arrived at the significant result [1], Fig. 6. The agreement with the theory of the nonlinear force acceleration [4] involved the necessary dielectric swelling of the laser pulse in the target where the factor of 3 was fitting with comparable experiments [38]. The repetition of the measurement ofthe ultrahigh acceleration was complicate due to scientific most questionable decisions, that the activities of Schäfer were decommissioned. Sauerbrey was then trying to repeat his experiment at the very advanced and well developed KrF laser with most perfect diagnostics and a long experienced staff at the Rutherford Appleton Laboratory in the UK. The earlier result of the Doppler effect could not be repeated [42]. Sauerbrey was then receiving an exceptionally high support to repeat the effect with solid state lasers at the University of Jena. Numerous very important results were discovered with several publications in top-class journals, but the Doppler effect could not be reproduced. Obviously the necessary condition of plane geometry interaction at very high contrast ratio was not sufficiently fulfilled.

Szatmari knew the special properties of the KrF laser from Göttingen, and with a new team in Szeged/Hungary, he could build the laser such that Földes et al [43] could fully reproduce the measurement of the ultrahigh acceleration (Fig. 7). The directly shown acceleration near $10^{19}$cm/s$^2$ could be understood by the theory of the nonlinear force acceleration due to the lower laser power than in the experiment of Sauerbrey [1].

These developments were supported from another experiment with extremely high contrast ratio similar to Fig. 6, by Badziak et al. [38] based on ion emission experiments. Against of all common knowledge with relativistic self-focusing, maximum energies in the range of or above dozens of MeV to be emitted into all directions, the fast ions were highly directed against the laser light and the maximum ion energy was 0.5 MeV while relativistic focusing under the experimental conditions should have had to be 22 MeV. Against all usual cases, the ion energy increased linearly on the laser power. The clarification was very easy: there was no relativistic self-focusing and there was the plasma block acceleration as expected (Fig. 3) from a skin layer (Fig. 5b) whose fixed depth was remarkably enlarged by dielectric swelling. This nonlinear force driven skin layer acceleration (Fig. 6) derived from the numerical results of Fig. 3 was reproduced also from Particle in Cell (PIC) computations [44]. While further PIC computations agree [45] with the plasma-hydrodynamic results of the nonlinear force [4] other experimental results [46] showed some difference to this PIC calculated "Transverse Normal Sheath Acceleration" TNSA. This difference is in the range of possible dielectric effects (swelling) which are not included in the PIC computations as it is automatically achieved in the hydrodynamic compoutations by using their electrically modified Maxwellian stress tensor in the hydrodynamics with the nonlinear force following Eq. (3).

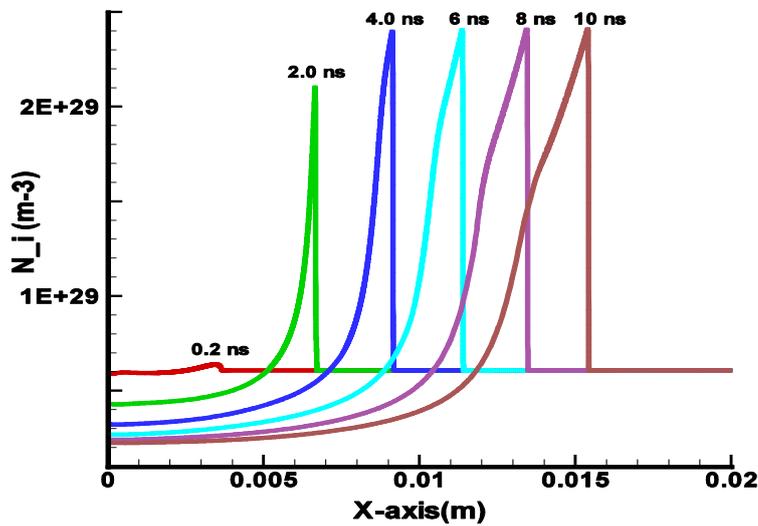

Fig.8. A ps KrF laser pulse of an energy flux $E^* = 3 \times 10^8$ J/cm$^2$ ignited a fusion reaction flame in solid DT. The ion density $N_i$ depending of the depth X at different times of the flame propagation shows a compression of a shock wave [68], however only after 2 ns.

## 5. ULTRAHIGH HIGH ION CURRENT DENSITIES APPLIED FOR LASER FUSION

The ultrahigh acceleration of plasma blocks by the nonlinear force of laser interaction arrived at a further significant aciewement following the results of the preceding Sections. The achievement is that petawatt-ps laser pulses are generating ultrahigh ion current densities in the plasma blocks as results of the dominating nonlinear force. This was implicitely included in the results of 1978 (Fig. 3) [24] and was fully reproduced by the PIC codes by Wilks et al. (1992, 2001) as target normal sheath acceleration (TNSA). The surprise was that ion current density in the space-charge neutralized plasma block corresponded to electron densities were in the range of

$$j = 10^{12} \text{ Amps/cm}^2 \qquad (16)$$

These values are about $10^6$ times higher than accelerators can produce. Apart from an approach of the particle beam ignition scheme for laser fusion, the new result (16) permitted a comeback [49][50] of the generation of a fusion flame in uncompressed solid density deuterium-tritium (DT). This scheme was studied by Chu [51] and Bobin [52] with the very disappointing result that such a fusion flame can be produced by pulses with an energy flux density E* irradiated within about picoseconds times on the solid DT needing

$$E^* > 10^8 \text{ J/cm}^2 \qquad (17)$$

This was completely out of consideration at the time of 1971 and from then on the scheme of spherical compression of DT fusion fuel by lasers to densities at or above 1000 times solid state density $n_s$ with subsequent thermal ignition was followed up culminating in the establishing of laser technology like the NIF
laser at the costs of more than $3.5Billion [53].

This scheme for producing fusion energy by lasers is based on thermal-pressure mechanisms using laser with nanosecond pulse duration where the mechanism of converting the energy of the laser pulses are involving numerous inefficiencies by thermal losses, instabilities and bremsstrahlung emission, delays of energy absorption by electrons and their delays by conversion to ions for the necessary thermal pressures of the plasma dynamics and many other typical problems of statistical mechanics, theory of heat and plasma physics. This can be changed in a fundamental and crucial way by using the nonlinear force driven plasma blocks by the PW-ps laser pulses where the laser energy is converted with nearly no thermal losses directly into mechanical motion of the blocks.

It has to be underlined that the thermal-compression and ignition of nuclear fusion with nanoseond laser pulses [53] may well succeed [54] to arrive at the needed high nuclear fusion gains. At present, the use of indirect drive spark ignition [53][55] is producing $10^{15}$ DT neutrons with 2 MJ laser pulses while $10^{18}$ or more neutrons are needed for higher than break-even gains, see Fig. 1 of Ref. [56]. This can well be achieved by direct drive and the optimized thermal compression and volume ignition [57][58] which direction is now being aimed in experiments [54]. The study of spark ignition [59] producing 1000 times lower gains than volume ignition in the past with less energetic laser pulses, is well important for other applications. More about fusion by laser pulses of nanosecond duration is discussed in Appendix 1.

The use of the PW-ps laser pulses for reaching efficient conditions according to Eq. (17) for igniting fusion flames in uncompressed solid DT was well discussed [4][49][50][60]. A repetition of the computations of Chu [51] needed some updation about effects which were not known or used in 1971, e.g. the inhibition of thermal conduction in the electric double layers of highly inhomogeneous plasmas, and Gabor's collective stopping power in contrast to the Bethe-Bloch individual binary collision theory. This led to a limit instead of Eq. (17) to [61]

$$E^* > 3 \times 10^7 \text{ J/cm}^2 \qquad (18)$$

by using the same plasma one-fluid computation as it was done by Chu [51] and extensive further computations summarised in Ref. [5].

A remarkable result appeared when the Chu-Bobin side-on ignition of solid density fusion fuel was applied for igniting the reaction of protons and the boron isotope 11 (HB11). This reaction was from the beginning very interesting because it resulted only in alpha particles and no neutrons and the radioactivity by secondary reactions is less than from burning coal per generated energy. Using thermal compression and ignition, fusion of HB11 was about 100,000 times more difficult than DT fusion as seen for temperatures above 60 keV for HB11 instead of the 4 keV limit for DT, which limits however could well be reduced by re-absorption of bremsstrahlung and and by re-heat within the spherical reaction plasma. Using PW-ps block ignition, the HB11 reaction was only less than ten times more difficult than for DT, therefore very interesting to produce nuclear energy with less radioactivity per generated energy than from burning coal due to its 2 ppm contents of uranium [5][62]. Detailed computations showed well temporally and specially limited temperatures in the 70 keV range.

A new approach for studying the block ignition mechanism led to see the generation of ion compression at the flame front when propagating through the cold fuel [63] where the genuine two-fluid hydrodynamics were used which initially showed for the very first time how the very high longitudinal electric fields are appearing in inhomogeneous plasmas [64] as appearing in electric double layers [65] and a new kind of instabilities [66]. At the fusion flame ignited by the ps laser pulse, the propagating compression was by a factor 4, exactly as predicted by the simplified analytical Rankine-Hugoniot theory of shock waves. The evaluation of the very general hydrodynamic computation resulted in more details of the thermal dissipation processes which led to the broadening of the compression front [67] after 1000 ps in the reaction. It was remarkable, that the highly non-equilibrium processes at the one picoseconds initiation, developed very slow into the shock. Only between 200 and 2000 ps the shock had been developed (Fig. 8) and more results were produced showing velocities of the flame front of more than 2000 km/s [67][68]. About fusion reactions, production of MeV nuclear reaction products and gains, more is reported in the parallel SPIE conference [69].

It should be noticed that the fusion energy scheme with PW-ps nonlinear force ultrafast accelerated plasma blocks is well in an initial state only, as an Editor of the Royal Society of Chemistry in London [70] with reference to [6] expressed about this scheme:"'This has the potential to be the best route to fusion energy,' says Steve Haan, an expert in nuclear fusion at Lawrence Livermore National Laboratory in California. However, he also points out that it is still only potential at this point, 'there's a fair amount of work to do before this technology is at hand.'"

One of the problems for generating fusion power stations is the specification of the infinite plane geometry result which has to be reduced to a section of cylindrical geometry and how to control and reduce the radial losses. For the very short time of the fusion flame reaction of about 10 ns, it may be possible to use very high magnetic confining fields [71] based on related results [72] with cylindrical geometry [73]. Another option is to change the two-dimensional interaction area for the PW-ps laser pulse from the plane interaction front into spherical geometry [74]. These developments were based on an ion stopping process for HB11 [75] including reactions of the generated alpha particles which were the results of ion stopping.

Another research is to apply PIC computations to the problem of the generation of the fusion flame by ps or shorter laser pulses. A very encouraging step was for a similar problem of the usual fast ignition where a similar energy flux E* of $10^9$ J/cm$^2$ resulted in ignition of DT [76] is the same as for using the nonlinear force driven plasma blocks as piston for ion beam fast ignition [77].

# 6. PW-ps INTERACTION CONVERTS ATOMIC INTO MACROSCOPIC PHYSICS

The question remains about the basic properties for distinguishing between the sub-picosecond laser interaction with targets against to that of nanosecond at very high intensities. This question was well articulated [39][75] before this became significant by the definite measurements of ultrahigh acceleration of plasma bocks [1][43] after this was theoretically expected (Fig. 3) from the dominance of the nonlinear force in contrast to thermal pressures, Eq. (2). In the same way as a laser beam is nearly an ideally atomistic ordered entity without thermal chaos properties, in the same way is the nonlinear force driven plasma block after ultrahigh acceleration an entity nearly free of chaos. Just the results of the building up of the shock-like plasma front at block ignition of solid density fusion fuel [68], Fig. 8, demonstrated how it needed a comparably long time after the picosecond igniting push by the ultrafast accelerated plasma block before the shock properties were developed after nearly 1000 times longer duration.

   Historical aspects may be interesting. After the basic physics laws were discovered by Newton with the help of mathematics. The sbsequent mathematics formulations of fields or of Lagrangeans or Hamiltoneans developed to an over-determinism of mathematical rigidity based on Cauchy's complex analysis of holomorphic functions, based on Maxwellian fields and differential equations including a model of hydrodynamics up to the end of the nineteenth century. Only the statistics of particles for chaotic motion for explaining thermal processes opened up, while the later discovery of the quantum structure was again another aspect of rigid quantities without thermo-statistic chaos. In a simplified way it may be seen, that the plasma irradiation by nanosecond laser pulse irradiation is so drastically different in contrast to the PW-ps laser pulse interaction where the chaos properties with temperatures are mostly excluded. Nevertheless the plasma block laser ignition is then mixed with the usual temperature determined processes when fusion flames have to be followed up to ns times and longer. The new knowledge about complex systems is then leading to some ordering conditions.

   Theses difficulties in plasma physics were well addressed by Edward Teller [78] when he is referring to his situation in 1952. "Research on controlled fusion means dealing with the hydrodynamics of a plasma. I had a thorough respect for the fearsome nature of hydrodynamics, where every little volume does its own thing. Plasma does not consist of molecules, like a gas, but of ions – heavy slow moving positive ions – and light fast moving electrons Those, in turn, create and are coupled with electric and magnetic fields. For each little volume of plasma, several questions have to be answered: How many positive ions? How many electrons? How fast does each move on the average? What is the electric force, and what is the magnetic force acting on them?

   Mathematicians can predict the flow of matter as the volumes involved move in an orderly way. But even hydrodyanmics of air was (and to some extend is – see weather forecasts) beyond the grasp of mathematics. Theoretician of the nineteenth century proved that flying was impossible! In the twentieth century, they retreated to the statement that flying is impossible unless the air flow is confused and disordered (turbulent). Hydrodynamics as a science remains uncharted water.

   The same complications occure in planning a thermonuclear explosion. But an explosion occurs in a so short
time that many of the complicated phenomena have no chance to develop. Even so it took a decade from Fermi's first suggestion of a thermonuclear reaction to the point (which occurred after the first full-scale demonstration of fusion) that the theoretical calculations of the explosions were reasonably complete. *I had no doubt that demonstrating controlled fusion would be even more difficult.*"

This problem with plasma physics and hydrodynamics is still not fully solved now 60 years later, though a lot has been learnt. Only a research with thorough depth can lead us forward, as seen from the eminent achievements of Lord May of Oxford [79] who ingeniously changed the initial insufficient approaches form the nineteenth century in theoretical physics now to master complex systems. Solutions are based on realizing "will a large complex system be stable?" and applying this to population systems in zoology and finally to "systemic risk in banking systems" [80] with many groundbreaking results and insights.

A laser pulse with a nearly ideally directed collective of photons without thermally chaotic disturbance is turning out to result in the collective behavior of the block of plasma particles moving all into one direction with (nearly) no thermal disturbance, when there is the ultrahigh acceleration being dominated by the nonlinear (ponderomotive) force by the laser sub-picosecond laser pulses in the energy range of petawatts. This is in essential contrast what all is known from longer pulses with all the details gained from nanosecond laser pulse interaction, especially for the application to gain nuclear fusion energy using lasers [53].

One of the not fully solved problems is how high-contrast 45 fs 1.2 Joule Ti:sapphire-laser pulses interact with thin diamond-like layers [81]. If the layers have a thickness of only 2% of the vacuum laser wave length, the light is completely absorbed. No tunneling occurs, full ionization of carbon occurs within much shorter time and the acceleration of materials is due to $10^{19}$cm/s$^2$. Without having all usual parameters under control, the properties of block acceleration appear [82], however with a lot of questions to a complete explanation. This may even be more complex when realizing that the ohmic resitivity R of quartz changes within 40 attoseconds by 18 orders of magnitude [83] with a rate of

$$dR/dt = 10^{34} \text{Ohms/s}$$

in which range of attoseconds [84] no chaotic properties of temperature are possible.

## 7. CONCLUDING REMARKS

The studies of the ultrahigh acceleration turned out to need sophisticated experimental approaches as it was shown in the steps from the measurements by Sauerbrey [1] to that of Földes et al. [43]. Even the first experimental approaches to measure the predominance of the nonlinear force against thermokinetics [85] gave a negative, inverse result despite all the theoretical background of theory [17][20] and numerical (Fig. 3) results. This induced the next step of first suggested indications by Kalashnikov et al. [39] leading finally to the significant and transparent measurements of Sauerbrey [1] clarifying completely the agreement with earlier predictions by the nonlinear force [4] (Fig. 3).

Even in the theory of the particle acceleration, surprising contradictions were experienced. In contrast to the usual assumption that optical fields are only transversal (correct only for infinite plane or spherical geometry) it had to be realized that the optical beams are Maxwellian exact only with *longitudinal* components. At least this was the result ([7], see Chapter 12.3) about the polarization dependence of the forces [86] which enforced the use of these exact electro-dynamic fields and the use of the complete Maxwellian stress tensor for correct theoretical clarification leading to the nonlinearity principle [87]. These usually very small longitudinal field components of the laser beam appeared to be absolute crucial. Neglecting these small nonlinear parts cause a change from "yes" into "no", from fully polarization in the linear case to non-polarization in the correct nonlinear case as measured (Chapter 12.3 of [7])[86][87]. Richard Feynman [88] critically argued that nonlinearity was well known for a long time. But these were approximate extensions. The examples [87] show that neglecting very tiny quantities can change theory from right to totally wrong. For electron acceleration by lasers in vacuum, the longitudinal components were initially assumed that this may cause an increase of the acceleration of electrons in vacuum [89]. When this was studied exactly, it turned out against all the expectations that the phases of the longitudinal components of the fields in laser beams are even causing a little reduction in the acceleration [90].

Similar results against the initial assumptions are known with respect of the particle acceleration at the conditions near black holes. The comparison of the fields from the Hawkings and the Unruh radiation [91] with the optical fields in the range of the Heisenberg-Schwinger pair production in vacuum [91] led to the result of a difference between the these radiations [92][93]. **The final conclusion is then that there is a need to study the ultrahigh acceleration which is one of the topics to be given attention next, how Petawatt to Exawatt and to Zetawatt laser pulses** are an option for producing particles with PeV energy [94] as the possibility to overcome the limits with the size of usual accelerators.

# APPENDIX 2
# SUMMARY FOR UNDERSTANDING A BREAKTHROUGH FOR LASER DRIVEN FUSION ENERGY REPRODUCED FROM AN INTRODUCTION ABOUT CLIMATIC CHANGES DUE TO POLLUTING THE ATMOSPHERE

Appendix 1 was a summary of the phenomenological description of plasma acceleration by lasers without mathematical details of the theory and with reviewing fusion energy and recent results on reaching the clean hydrogen-boron11 nuclear reaction. For readers with primary interests on environmental problems, the following appendix is an introduction again with some repetition of results of the main text but more oriented to a general information. The source of the used book is mostly unchanged for reasons of the originality. Some parts are added where the citations are given by letters [a], [b] etc..

**Translation from Book** Heinrich Hora *Klimaprobleme und Lösungswege (Climatic Problems and Ways of Solutions)* S. Roderer-Verlag Regensburg 2010, 256 pages ISBN-10: 3897837153. ISBN-13: 978-3897837157, Subsection 2 of Section D.
Citation on the page before end of the text is with permission of the Royal Society of Chemistry, London for the reference book.

**D) How to overcome Climate Problems**

**2. Possible low cost fusion energy by Lasers: nuclear energy without radiation hazard**

When the New Zealander Ernest Rutherford (later Lord Rutherford) began around 1903 to discover nuclear physics in the Cavendish Laboratory of the Cambridge University in England, it was characteristic that the exchanged energy is *ten million times* higher than in chemical reactions. The desire to use this for producing energy was immediately clear. In view of the result explained in Section B) that the energy source from fossil carbon sources (coal, petrol and natural gas) can be used only up to the level of 1960 [1] if a catastrophic damage of the earth atmosphere with a change of the climate has to be avoided. Nuclear energy came into the focus of interest. This section is describing very recent results, how lasers may solve the generation of nuclear energy by low cost and unlimited available sources of fusion energy with less emission of radiation than from burning coal.

    The cardinal disadvantage of nuclear energy was immediately recognized due to the danger of nuclear radiation. In very small doses and in controlled way, nuclear energy led to enormous advantages for example in medical techniques. A first indication for healing by very low doses was developed at hospitals in the so called "Radium-Bad" Brambach in the German Erzgebirge near St. Joachimstal. In an extreme contrast, the big danger of radiation from atomic bomb explosions is well known, and the control of the large scale radiation problems from the waste of nuclear power stations is a problem where rather perfectly developed methods are used for nuclear power generation of electricity which is now the second largest source next to the still mostly generated energy from burning fossil resources despite of the growing fatal effect on the climate due to the emission of carbon dioxide $CO_2$ polluting the atmosphere.

    Ways for using nuclear energy sources were discussed in Section C) where the most developed source is to use uranium. This is apart from many possible improvements in the future by using thorium as source for nuclear fission reactors, or by the absolute safe small underground reactors being at the stage close to the market, or by similar techniques. The example of France is a reality where 85% of electricity is produced nuclear, 10% from water power, or Sweden where not only any fossil fuel is producing electric power on a low scale (5%), the rest using nuclear and water power is at nearly same amounts. The storage of the nuclear waste with its most dangerous nuclear radiation is completely under absolute control by digging 3000 meter deep holes into the Swedish granite mountains for deposits.

    The environmental pioneer James Lovelock [2] had already described the now unstoppable and only possibly delaying of the development, indicating that in not very too

far distant future the increase of the sea level by 60 meters is possible as it has happened in times more than one million years ago [a], where the concentration of carbon dioxide $CO_2$ was at a level which may be close to the expected slowly rising level after 2050 if the emission from fossil energy use will not be radically reduced to the level of 1960. To intervene quickly with nuclear fission reactors was mentioned to be needed, but that should be later solved in more ideal way with nuclear fusion energy.

As is described in more detail in Section C.5, research follows two basically different approaches. One is to work with the necessary plasma of heavy and super-heavy hydrogen isotopes, deuterium D and tritium T, for reactions within a plasma of several dozens of million degrees centigrade temperature confined with magnetic fields. The leading project is the up to 20 Billion Dollar ITER for demonstration of the feasibility, scheduled for 2026, or another low current stellarator project Wendelstein. The high current ITER project suffers from wall erosion as discovered by the Razumova disruption instabilities [b]. These difficulties may be overcome by very intense neutral beam injection [56] which is converting the fusion scheme into a beam fusion following the nonlinear principle, see preceding section D1) with reference to [20] (page 129).

The other approach for fusion energy is to work with extremely high-intensity and high energy laser beams for interaction with the DT fusion fuel fast enough for igniting an exothermic energy gain by inertial confinement of the plasma, see sections C5.2 to 5.4. Experiments are on the way, to use the largest laser in the world NIF (Fig. 11) with laser pulses of nanosecond duration for igniting the fusion reaction by heating highly compressed plasma. Alternatively, a basically new method by using extremely intense laser pulses of ps or shorter duration was elaborated after new laser technology was discovered by Strickland and Mourou [c] in 1985 as a most significant turning point in laser development [d]. Fig. 15a [e] shows how this may lead to acceleration of particles to 1000 times higher energies (>PeV) than the largest accelerator LHC at CERN in Geneva/Switzerland can generate. This LHC limit can in no way be much more increased by classical accelerators than only by lasers [e1]. For the application in laser generation of fusion energy, these developments open a basically new approach described in this chapter. This was realized immediately by the leaders for laser fusion [e].

The novelty consists not only in a new approach for the DT fusion [f], but it initiated a new ignition scheme for fusion energy [g] where the reaction of hydrogen with the Boron-11 isotope (HB11 reaction) may be possible now against the earlier view that this is 100,000 times more difficult than the DT igniton (see last two sentences before the Section D). It is to realize that this HB11 reaction results in less nuclear radiation per generated energy than from burning coal what value was never considered as dangerous. For this purpose some points of the present state of laser driven fusion may be recapitulated. The key question is that the ps or shorter laser pulses permit a direct conversion of the laser energy into mechanical motion of plasma with hundred-thousand times higher acceleration than by thermal mechanisms.

The use for laser driven fusion energy has reached a highly developed state but has not yet reached the solution. Many of the newly emerged problems had to be identified and explored. The solution with megajoule nanosecond laser pulses at volume ignition of a thousand times of solid density compressed DT fusion fuel seemed to be the safest solution to an expected breakthrough before 2015 [67]. Its success is also fully supported by experimental measurements of underground nuclear explosions following W J Broad [60] when intense X-rays in underground explosion were used instead of lasers and where comparable compression and thermal ignition of the DT-fuel resulted in high gains of fusion energy.

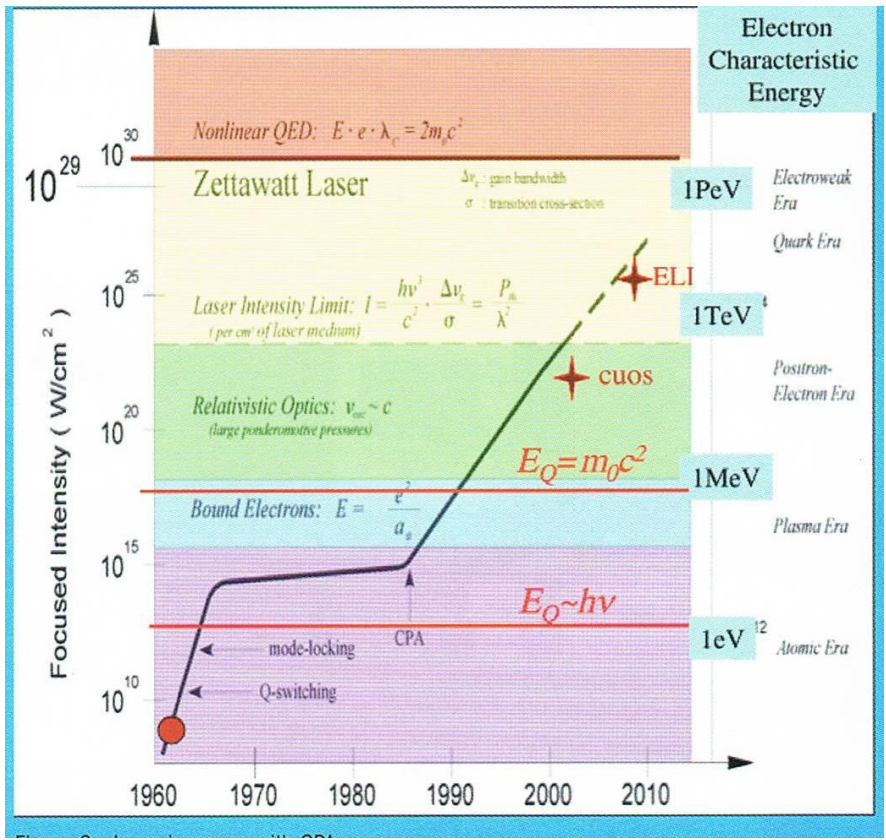

*Fig. 15a. Gerard Mourou diagram: Development of lasers with the turning point at 1985 [e] with the discovery of the Chirped Pulse Amplification CPA [d].*

Compared with this normal development as indicated in the following, there is a completely new result: What advantage will it be, if nuclear power can be generated with less radioactivity per gained energy compared with burning of coal, which contains 2 parts per million of uranium which was considered what as insignificant and could always be ignored.

According to the research results since 2008 there should be "nuclear power without radioactive radiation hazard" possible with lasers, based on the ignition of the neutron-free fusion reaction of ordinary light hydrogen with the boron isotop 11, equation (16, page 112). The contrast is with respect to the exploration of the many newly discovered phenomena in fusion with the manageable methods of the giant laser of Figure 11, where there is a very sophisticated solution close [67], including the option of an economical fusion power and power plant [69] with the now dominated nanosecond-megajoule technology. However, for the direction of the laser fusion research another option with laser pulses of picosecond or shorter duration has been developed. These results, however, are still at the stage of some not yet solved scientific problems, so that predictions are given in the following are to be taken with precautions for a laser fusion reactor compared with the other case with the giant lasers. But there is a very likely prospect that perhaps fusion energy will be very far cheaper than by all other existing sources of energy.

The novelty of these results is that nuclear fusion is not ignited by thermal means, but by direct electrodynamic conversion of optical energy into mechanical laser plasma acceleration. This is theoretically known since 1969 [78], based on mechanical conservation laws known, and what has been confirmed experimentally in detail [20] [49:

Section 10.4] and led to the closed equation of motion of plasmas with electromagnetic fields [49: Equations 8.87 and 8.88] in accordance with Lorentz and gauge invariance. Furthermore it was essential that the optical properties of plasma where refractive index and absorption were included in a general way. It led to the formulation of the nonlinear force $f_{NL}$ [49] for general electro-magnetic fields. Numerical evaluations resulted in acceleration of plasma blocks above $10^{20}$ cm/s$^2$, see Fig. 10.18b on page 179 of Ref. [77].

For simplified geometry it was known already since 1846 [see Eq. (22)] for constant electrostatic fields, that there is an acceleration of not electric charged particles by a ponderomotive force in contrast to the electric Coulomb force between charges. This result was confirmed directly from Maxwell's electrodynamics due to Maxwell's stress tensor where for plasmas the inclusion of the dielectric optical properties was an essential ingredient [78].

When the laser was discovered in 1960, it was obvious that the enormous high radiation density can reach values that were known by the preceding explosive fusion reactions. In this case, with cavity radiation (Planck radiation) worked in accordance with temperatures of several million degrees. The radiation density at ten million degrees black body radiation corresponds to a laser intensity of about $10^{17}$ Watts per cm$^2$, a value that was achieved with lasers since a long time and has been surpassed by more than hundred thousand times.

The original question was whether and under what conditions one can simply use laser pulses on an uncompressed, solid state dense deuterium-tritium target to drive a reaction in the form of a fusion flame. More detailed calculations were known around 1972, where Chu had found [79], confirmed by Jean-Louis Bobin [79], that a very incredibly high energy flux energy density

$$E^* > 4 \times 10^8 \text{ J/cm}^2 = 450 \text{ Megajoules/cm}^2 \qquad (23)$$

is necessary, or that the action of ion beams of optimum ion energy of 80 kilovolts will be needed with an ion-current density

$$j > 10^{10} \text{ A/cm}^2 = 10000 \text{ Mega-Ampere/cm}^2 \qquad (24)$$

to act on uncompressed solid DT within a period of about a picosecond.

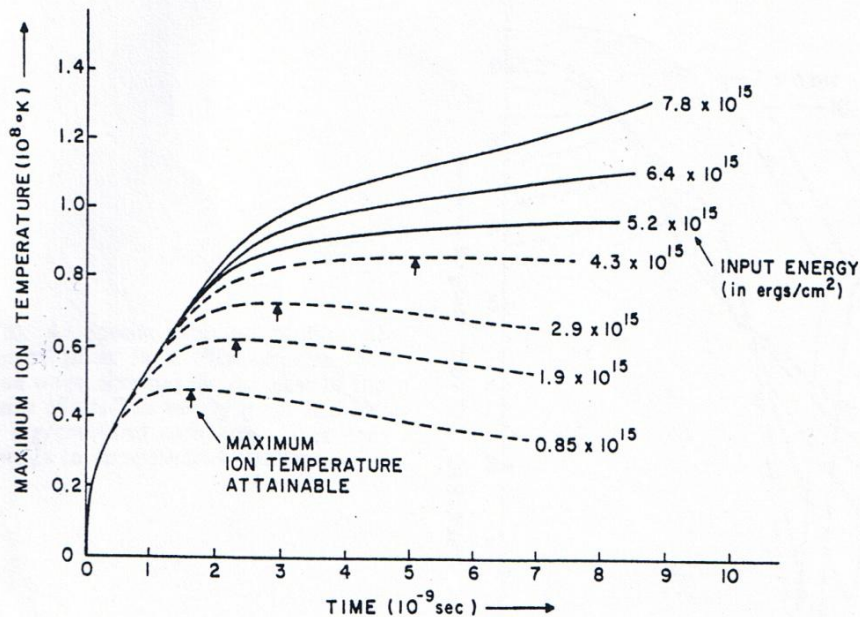

*Fig. 16 Hydrodynamic computations of the maximum ion temperature T in a fusion flame in solid state deuterium-tritium DT depending on time t for after the picosecond duration of different energy flux densities (input energy) (from Fig. 2 of citation Chu [79]).*

This result can be seen from Figure 16. Chu [79] calculated the temperature T of the thin fusion flame in the solid DT as a function of time t after an approximately picosecond long pulse of input energy flux E* as parameter had acted on the solid state. If the curve to a maximum at lower temperatures was decreasing on time (dashed curves), no ignition takes place. If the curves continue to increase on time (fully drawn curves), there was ignition. The minimum energy flux density is the boundary between the two sets of curves resulting in the ignition condition of Eq. (23).

In 1972, picosecond laser pulses with energy flux densities E* of Eq. (23) were an utopia while ion current densities of accelerators were over a million times smaller than (24). This has now been changed thanks to the laser development, see Fig. 15a. It is now possible thanks to the ultgra-high acceleration of plasma blocks for ignition of fusion with laser pulses achieved today having petawatt power (PW) and picosecond (ps) pulse lengths. On the level of 1972, the condition (23) could be reached only theoretically by a very tricky central ignition scheme, the central spark ignition [66] which is very difficult to achieve due to "Rayleigh-Taylor" and other instabilities and asymmetries of laser radiation. After all, with this "indirect drive" in Figure 12, the hohlraum has reached black body radiation of about 3 million degrees [68].

In order to find an easier alternative to this indirect drive spark ignition, the motivation for the development of the PW-ps laser pulses was based on the discovery of the kink in Fig. 15a [d] in 1986, that this laser technology was developed from 1988 on (see p. 13 of Ref. [f]). At this stage the high compression of plasma was still included in contrast to the new scheme without compression described in the following as a further important simplification. Before this very new scheme, compression was of interest for the usual laser fusion with giant lasers for spherical irradiation. It has always been important that the DT fusion fuel should be compressed to more than a thousand times more dense than the normal solid-state density. This very difficult task was achieved with lasers in 1991 in Japan after the pioneering work of Chiyoe Yamanaka [58], where the laser-driven blowing off (ablation) of the outer layer by irradiation of a tiny spherical target into compression to the inside leads and the already mentioned densities of 2000 times solid density in carbon polymers.

What was disappointing is that the temperature after compression reached only about 3 million degrees. A much higher temperature was expected. To overcome this problem it was suggested by Dr. Michael Campbell, at the Lawrence Livermore National Laboratory near San Francisco from 1988 on [f], that after the mentioned compression was reached, one could increase the temperature by a second irradiation using an additional irradiation by the new laser pulses of picosecond duration and petawatt power into the compressed DT. This system of heating is called "Fast Igniter" and was studied intensively by Tabak et al. [80].

In this context it should be mentioned that too low temperatures at very high density compression with lasers can even be tolerated at special conditions. It has been shown for volume ignition of DT (details see Chapter 5.4, Fig. 13 with references [64][65][66]) that this is possible at comparatively very low temperature of only 5 million degrees with large laser pulses if densities of few thousand times the solid state can be reached without any need further adding heat [66][81]. It is noteworthy that this volume ignition after long neglect, perhaps will be interesting [82].

In the study of fast igniters - however quite unexpectedly with the petawatt-picosecond laser pulses – several other phenomena were observed instead of the desired simple reheating. These resulted in many novel relativistic effects, as very intense electron beams of several hundred MeV (million electron volts) of energy, and as the long predicted intensive laser production of positrons (antimatter: pair production of electrons) were generated, gamma rays of several million electron volts of energy with subsequent nuclear transmutations, highly-ionized heavy ions of GeV energy, and other phenomena [99].

From the broad stream of these experiments, some abnormal phenomena were recorded especially when "clean" laser pulses were used where pre-pulses of the laser were cut off by a high contrast ratio before the picosecond main pulse was arriving. Contrast ratios had to be better than hundreds of millions for the elimination of prepulses up to about one ps before the main pulse arrived. In all the many standard experiments, this contrast ratio was usually much smaller than $10^8$ for the suppression of the pre-pulse.

It is very exclusive and exceptional within the wide stream of laser-plasma interaction studies, what was realized by the cooperation between Fritz-Peter Schäfer [83] and Roland Sauerbrey [84]. Schäfer as discoverer of the dye-laser became Head of a laser center at a Max-Planck-Institute in Göttingen where he also was involved with the development of excimer lasers, especially with krypton-fluoride KrF with an ultraviolet wave length of 249 nm. These lasers were pumped by gas discharges and produced pulses in the range of few dozens of nanometers. To get shorter pulses, dye laser pulses of less than a pico-second duration were sent through an up-pumped KrF laser where the ps-pulse was loading up the pumped states by Einstein's stimulated emssion and produced laser pulses of terawatt power and about 0.5 ps duration. A contrast ratio above $10^8$ could be reached.

When Sauerbrey [84] irradiated targets with carbon, he could measure nearly plane plasma blocks being accelerated from the targets whose acceleration had reached more than $10^{20}$ cm/s$^2$. Sauerbrey underlined that these ultra-high accelerations were much higher than the astronomically high accelerations at neutron stars and were more than hundred-thousand times higher than any ever measured acceleration by gas-dynamic processes using very powerful laser pulses generated by laser pulses of nanosecond duration [85] showing accelerations of less than $10^{16}$ cm/s$^2$ even when using the huge NIF laser (Fig. 11). Even with these most powerful ns laser pulses of 2012, Sauerbrey's ultrahigh accelerations are many orders of magnitudes higher.

As mentioned before, these ultrahigh accelerations were predicted with same high intensity ps duration laser pulses in numerical computations of plane geometry plasma block generation (Fig. 10.18a of [77]) in 1978. The basic difference between the ultrahigh accelerations at ps pulse interaction in contrast to the interactions with ns pulses is of fundamental importance and was reached thanks to the kink of 1986 in Figure 15a as summarized by Mourou and Tajima [86]: in the ps case there is a direct conversion of laser energy into mechanical plasma motion by the nonlinear force and in the ns case there are thermal processes involved with well known loses and temporal delays. Characteristic for this difference is that quiver motion of the electrons in the laser field is remarkably higher than the thermal motion (see Chapter 9 of [49]). This is being explained in more details in the further parts of theis section.

The essential reason for the success in the measurement of the ultrahigh acceleration [84] is that the very high contrast ratio prevented relativistic self-focusing and provided for the first time the conditions of plane geometry as presumed in the computations of 1978. For understanding these facts, we have first to explain the long history of these discoveries in the following. It should be mentioned that first indications of the dominance of the nonlinear force with very high contrast ps laser pulse interaction was indicated by the experiments of Kalashnikov et al [h], but the significant breakthrough came by the Doppler measurements of Sauerbrey [84].

The repetition of the Doppler measurement for the plasma blocks was not easy. When Sauerbrey together with most experienced colleagues was using the highly developed KrF laser at the Rutherford Laboratory in Appleton/England [j], the effect was not reported. One competing colleague in Japan triumphantly commented: "non-reproducible effect". A repetition of the measurement was possible by Földes et al. [k] using a KrF laser with beam contrast of $10^9$ with some lower power following the design of the laser built by Szatmari in Göttingen. The block velocities published in 2000 show immediately an acceleration of more than $10^{19}$ cm/s$^2$. The measurement is reproduced as Fig. 1 in a recent publication [l].

The starting point for the new situation of nonlinearity of laser-plasma interaction as it by was    initiated by measurements of Linlor in 1963 [87] studying laser driven emission of ions from targets in a vacuum. When the laser pulse energy was below one megawatt MW, the plasma reacted completely classical. The target was heated to temperatures T well over few thousand degrees and turned into a plasma. At this dynamic expansion, plasma electrons and ions are emitted, where their energy was determined by classical time of flight measurements were of a few eV had the expected values according to the temperature    T    of    a    few    tens    of    thousands    of    degrees.

When William Linlor [87] in the Hughes Aircraft Laboratory in Malibu, near Los Angeles, could just apply Hellwarth's discovered Q-switch technique, with well reproducible nanosecond laser pulses of about 1 to 10 joules of energy onto targets in vacuum for measurement of the energy of emitted ions, something was completely incomprehensible. Instead of the cases with laser powers of MW with completely classical few eV energy ions, the ions had *thousand times higher ion energy* (up to 10 keV) at the only ten times higher 10 MW laser power. Linlor was immediately promoted to the U.S. Atomic Energy Commission in Washington, because these ion energies could ignite fusion reactions. But it turned out, unfortunately, that the ions were not thermal in nature and were not (yet) suitable for thermonuclear reactions. This Linlor-effect with up to 10 keV ion energy was immediately confirmed and reproduced by Opower at the institute of Wolfgang Kaiser at the Technical University of Munich [88] and others (Chapter 1.4 of [77]).

Had it been a thermally determined process, the ions, would all have similar energy distributions, regardless of the number Z of their ionization. But that was not the case, since the ion energy increased linearly on Z. This indicated that there was basically an electro-dynamic process is typical for non-linear effects with lasers. After a first result to calculate such a force in the plasma with the just gained derivations of the optical properties of a Rayleigh profile of optical constants [75] the nonlinear force was derived $f_{NL}$ for this special case. For general geometry, the consequence was that additional nonlinear terms had to be added to Schlüter's equation of motion for space charge neutral plasmas [78] derived from momentum conservation for stationary conditions. The general time-dependent equation of the nonlinear force (Equation 8.88 in [49]), was derived after a long controversy between a number of research centers was solved by using symmetry consideration [89] arriving at the final and full general solution of the Maxwellian stress tensor. This was proved by Lorentz and gauge invariance that this and only this is the correct formulation.

To clarify the measured 10 keV ions of Linlor-effect, another consideration was needed in addition to the above fundamental theoretical investigations [78]. How was the result possible with of the high ion energies by irradiation with laser intensities from such small lasers? For this purpose, quantitatively the ponderomotive self-focusing of the laser was derived in the plasma [90] (see chapter 12.1 of [77]). The essential mechanism is that the laser beam pushes away plasma radially from its axis by means of the nonlinear force $f_{NL}$ due to the gradient of the radially decaying laser field of the laser beam, see Eq. (22) where an equilibrium with the thermal pressure is given due to temperature and the generated density profile. This determines the first condition together with the second condition of refraction and total reflection for the laser beam, and as third condition the diffraction limit. These three equations resulted in the intensity threshold for self-focusing of about MW laser power, showing exactly the value separating classical conditions from the observation of the nonlinearities observed at higher laser powers. This process of emptying of the plasma from the beam center was experimentally confirmed by Martin Richardson (see section 12.1 of [77]). The threshold of the self-focusing of about 1 MW laser power of the pulses of ns duration proved why at these lower laser powers the interaction by classical classical dynamics ended and the the self-focusing of the laser beam began with producing the channels of a few wavelengths diameter exactly as measured with microscopes by Korobkin and Alcock (see [77]). In these self-focussing channels, the nonlinear force was exceeding the thermal pressure resulting in the non-linear force acceleration of the ions. This force resulted in ion energy energies increasing linear on the ion charge Z arriving at the measured keV energies. The very precise measurements of ion energies and diameters of the self-focusing channel by Korobkin and Alcock (see [90]) have all been reproduced theoretically and that explained the Linlor-effect completely.

As further steps, this allowed the theory that nonlinear force generated that ultra-high accelerations dominated by $f_{NL}$ in very general one-dimensional hydrodynamic numerical calculations for plasma flat plasma geometry. An example published in 1978 [91] (see Figures 10.18a and 18b [49] [77]), shows how deuterium plasma of the electron density $10^{19}$ cm$^{-1}$ with neodymium-glass laser pulses of $10^{18}$ W/cm$^2$ within 1.5 ps reached velocitdies of more than $10^9$ cm/s corresponding to the ultrahigh acceleration above $10^{20}$ cm/s$^2$. This corresponds approximately to the ultrahigh acceleration measured very much later, by Sauerbrey [84] measured and evaluated in details for these conditions for the accelerated carbon ions in complete agreement [92].

Why has it taken so many years between the calculation [91] and the verification of ultrahigh acceleration by the nonlinear force, which was proved experimentally not before 1996? The obstacle was that the self-focusing of the laser beam destroyed the plane geometry which was a condition for the plasma block generation. This obstacle even was more severe with the discovery of relativistic self-focusing in 1975 [93], resulting in subsequent measurements of ion energies confirmed exactly in the above MeV range. Basically, however, at that time the properties of the nonlinear force could be proved in details from direct experiments (Section 10.4 of [49]) as well as by the ponderomotive self-focusing and the Linlor-effect.

The process of relativistic self-focusing was on the change of the quiver motion of electrons in laser fields where the relativistic change of the electron mass has to be taken into account. For studying the electron pair production at very high laser fields, the differential equation for the motion in these fields had to be solved. For subrelativistic conditions, this is very easy. For the relativistic case for linearly polarized laser fields there were two coupled differential equations for the two electron velocities, one parallel to the electric E-field of the laser and one for the velocities in axial direction of the laser beams. The difficulty is that these were nonlinear differential equations. An exact solution for the electron quiver energy $\varepsilon_e$ valid for all laser intensities (neglecting only radiation damping at extremely high intensities) was simplified for very lowe density plasmas given by [i][93]

$$\varepsilon_e = mc^2[(1 + 3I_r/I)^{1/2} - 1] \qquad (24a)$$

Where I is the laser intensity and $I_r$ is the relativistic threshold intensity at which $\varepsilon_e = mc^2$ and whose value is $I_r = 3.66\times10^{18}$ W/cm$^2$ for the neodymium glass laser wave length which is proportional to the square of the critical plasma density. This familiar formula is generally valid and is widely used but reference is not given to [i] but to authors from the same time who were able to work only with the very low intensity approximation for the solutions of the nonlinear differential equations.

This relativistic solution resulted in optical dielectric constants for plasmas showing that the effective wave length is shorter for higher than for lower intensities. Following Fig. 17a, if a plane wave front of a Gaussian laser beam is irradiating a plane target, the usually unavoidable pre-pulse will produce a pre-plasma in front of the target. The plane wave front will then be bent concavely into spherical wave fronts such that the beam is shrinking to a little less than one wave length diameter containing then an enormously high laser intensity. These very high laser intensities led to the emission of highly charged ions with MeV [94][95] energy up to GeV and very intense and short wavelength X-rays are emitted.

This generation of MeV was very unexpected and it took some time that this fact was accepted. But it the it happened everywhere and one had to accept this. Only when the pre-pulse was cut away and no pre-plasma was appearing, Fig. 17b, plane geometry interaction occurred and the plasma block received the ultrahigh acceleration as first measured by Sauerbrey [84].

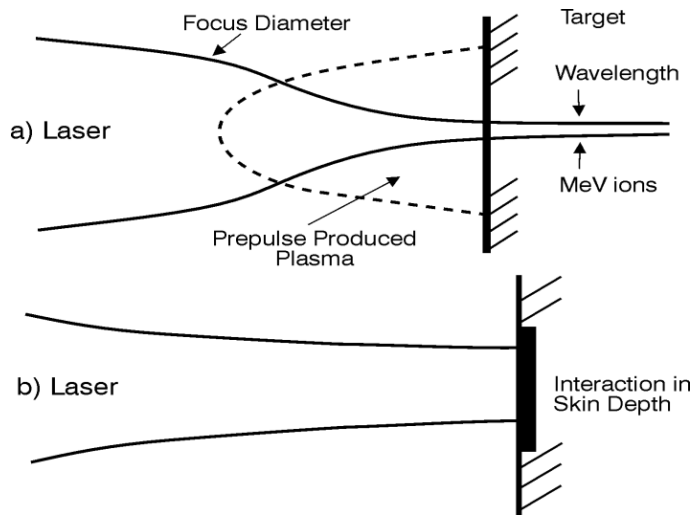

*Fig. 17. An optically focuesed laser beam of 20 to 30 vacuum wave length diameter hits a target in vacuum. Usually (a), the laser pre-pulse produces the dashed drawn pre-plasma where the relativistic self-focusing shrinks the beams to less than wave length diameter whose extremely high intensities emit ions by the nonlinear force to MeV and very much higher energies. Ideal is case (b) where self-focusing is avoided by cutting off the pre-pulse by using an extremely high contrast ratio as in the experiment of Sauerbrey [84].*

How difficult it was to accept that MeV ions were produced exactly following the theoretically expected values [93] (see sections 12.2 and 12.6 of [49]) by relativistic self-focusing, can be seen even from a case several years later. The problems were with the PhD thesis of Andrei Rode in the Moscow Lebedev Institute of the Academy of Sciences. Rode worked with nanosecond laser pulses of only a few joules of energy irradiating phosphorus and detected almost completely ionized phosphorus ions of 11 MeV energy. This was measured not only as in all other experiments worldwide using the time of flight of the emitted ions, but additionally Rode used a unique a technique of Doppler shift of spectral lines in the X-ray range below nm wave length using the uniquely developed instruments of Gleb Sklizkov. Rode was able to see from the Doppler shift of the wavelengths that the ions received their energy far above MeV in situ, i.e., at their origin of acceleration in the target within the relativistic self-focusing channel of wavelength diameter. By this way he could even measure the ions with deviating from a spherical distribution into rotation ellipsoids with increasing deviations for higher ion charges Z. This could subsequently be theoretically reproduced by application of soliton relations [96]. A very senior examiner of Rode's PhD thesis in about 1985 at the Kurtschatov Institute could absolutely not believe how these pulses from small laser could produce a few joules of energy phosphorus ions with an energy of 11 MeV. At an occasional visit to Moscow by the author who as discoverer was familiar with the formulas of relativistic self-focusing it could be calculated very quickly how Rode's measurements were exactly agreeing with the theory [93]. What a headache was then involved with the publication of the results in the Journal of Experimental and Theoretical Physics (Soviet Physics JETP) in Moscow, can be seen from the very exceptionally long two-years period between submission and publication of the work, even though the first author was Nobel Laureate Nikolai Basov [97].

Relativistic self-focusing became therefore quite familiar for these measurements of ions with energies in some cases exceeding far more than 100 million electron volts, as shown in the case in Figure 17 (a) schematically. It is the suppression of the prepulse laser intenisty (case b) by a factor of at least hundred millionen times smaller than the intenisty of the entire laser pulse arrived (contrast ratio, $10^8$). Only such contrast for a time of few picoseconds before the main pulse enabled Sauerbrey [84] that the interaction occurred in a planar geometry as shown in Fig. 17 (b) and as it was used in the calculations with the nonlinear force [92] with the ultrahigh acceleration of $2\times10^{20}$ cm/sec$^2$.

One can easily see how only a hundred thousand times smaller acceleration at nanosecond interaction of the NIF experiment [85] can be explained. The pulses for the acceleration by the nonlinear force were a thousand times shorter and the plasma reached velocities of more than $10^9$ cm/s which were with the nonlinear force a hundred times higher than in the thermal case.

The anomaly that Sauerbrey discovered in agreement with the preceding non-linear force theory was also confirmed by Jie Zhang et al. [98] by another measurement of the importance of the very high contrast. A solid-state laser pulse of 0.3 ps duration was used instead of a KrF laser with a similar and equally high contrast ratio of $10^8$. What happened in drastic difference to all the usual emission of very high intensity and short wave length X-rays after relativistic self-focusing, there was only a low intensity X-ray emission and no hard X-rays were detected. For clarification of this result, the trick was used that a prepulse was cut out from the main pulse and irradiated to the target at times τ before the main pulse arrived. If τ was 10 ps and further increasing, no change of the low X-ray emission was observed until τ was 70 ps, when the usual high intensity short wave length X-rays appeared. This was a proof that the laser pulse of about 30 wave length diameter at the target had not produced a pre-plasma plume at smaller τ and no relativistic self-focusing had happened until the time τ was 70 ps. Hydrodynamic estimations explained that with the 70 ps prepulse, a pre-plasma of about 60 wave lengths depth had been established, just sufficient for relativistic self-focusing for changing the conditions in Fig. 17 from (b) to (a).

Another experiment with very high contrast ratio confirmed the plane geometry laser plasma interaction without relativistic self-focussing. This was based on exceptional high contrast plane geometry interaction by Badziak et al. [100] confirming the abnormally ultrahigh plasma block acceleration by the nonlinear force $f_{NL}$ smilar to [84] and [98] terawatt-ps laser pulses experiments. In this case, the diagnostics was based not on the Doppler acceleration, or the X-ray emission, but on the otherwise generally conventional ion emission from time of flight measurements. The fact that there was the interaction in a planar geometry as in (b), Figure 17, had been achieved without relativistic self-focusing. This could be seen from the remarkably low energies of the fastest ions of 0.5 MeV while relativisitic self-focusing would have led to 22 MeV [101]. The 0.5 MeV ions resulted immediately from the nonlinear force in a planar geometry when avoiding self-focusing. The direction of the fast ions was nearly parallel while the ions from relativistic self-focusing were going into all directions [96][97].

What was also a sharp contrast to all the usual measurements, is that the number of fast ions remained constant while changing the laser power by a factor of 30 with linear increasing of the ion energy on the laser power. This led to the conclusion [101] [102] that the acceleration must have come from the constant volume of the skin depth (skin effect for electromagnetic radiation) in Figure 17 (b) if the non-linear force acts at a laser plasma block flying against the laser beam. It should be noted that the plasma blocks were not only of a depth of a few wavelengths, but could have after the computations of 1978 ([91] and Fig. 10.18a of [77]) depths of 10 to 20 vacuum wavelengths of laser light. This was caused by the refractive index of optical properties of the plasma and was important for the explanation of the calculated results of 1978 [91] following a lecture by the author in 1979 in Livermore in discussions with John Nuckolls and John Emmett.

As next steps after these findings, it should be noted that discussions with Lang Wong, as co-author of [98], at a summit in Islamabad, contributed February 2001, led to the explanation of this experiment as being caused by the suppression of relativistic self-focusing. This was evaluated quantitatively and presented at the conference. It should be mentioned here, what difficulties were experienced with referees at the subsequent publication of such an unusual result to explain the anomaly [100].

Following these results, experimental details were measured by Badziak et al. about the mechanisms involved confirming the acceleration of the block acceleration by the nonlinear forces [102]. The dielectric swelling of the laser intensity within the plasma block was found to have a value of about three [101] such that it could be concluded, that there were two plasma blocks, one flying against the laser light and one along the laser beam directed into the target (see Fig. 9.1 [77]). This second block was subsequently measured with very thin targets [102]. It was also confirmed [101] that the blocks were made of space-charge-free ionic currents with ion current densities of

$$j \geq 10^{11} \text{ Ampere/cm}^2 \qquad (25)$$

This was the argument that the ultrafast acceleration can be connected to the side-on ignition of solid density DT with a ps duration energy flux density E* of the theory of Chu and Bobin [79] This ignition of a fusion flame in uncompressed deuterium-tritium DT by lateral laser radiation found then a comeback, as described in an initially classified patent application [103] which was permitted for publication after a very similar ignition of a low (up to 12 times the density of solids) compressed DT using 10 PW-ps laser pulses produced by 5 MeV electron beams of Nuckolls et al [104] was disclosed for controlled nuclear fusion. This is known for power plants with yields of 10,000 times more energy than the laser has to inject. Initial estimates for this and the equally efficient ignition with the nonlinear force driven by ion blocks have been published [103] (Hora and Miley, see [104] p. 14) and summarized [92][g].

After the problems of the nonlinear force are resolved safely for the application described here, the laser ignition of nuclear fusion using ultra-high accelerations by direct transfer of laser energy to mechanical plasma motion without heating process, it was the next task, to study the process how nonlinear force driven plasma blocks can produce a fusion flame in uncompressed fusion fuel of solid state density.

The first task was to repeat the plasma-hydrodynamic calculations by Chu [79] and to amend this by inclusion of several new basic discoveries gained since. This work was supported by the International Atomic Energy Agency in Vienna in projects initiated at the ICTP (International Centre of Theoretical Physics) in Trieste/Italy. This center was established by Nobel Laureate Abdus Salam from the Imperial College in London, who was born as Pakistani and aimed to support theoretical physics in developing countries. The support by the UN International Atomic Energy Agency (IAEA) in Vienna was linked with Coordinated Research Projects (CRP).

First, the calculations by Chu were repeated busing his former conditions and presumptions for checking agreement and to recognize conformity with Figure 16, when a stream of energy flux E* in Joules (or equal $10^7$ more erg) per cm$^2$ is incident within a picosecond on uncompressed DT fuel to produce a fusion flame. The flame front propagates as a localized very thin front into the fusion fuel. The reactions cause heating to a temperature T. In pursuit of T over nanoseconds for the deuterium-tritium increases significantly above 4 keV at which losses by bremsstrahlung are balanced by nuclear fusion energy gain if there is no re-absoprtion in the plasma as known from volume ignition (Section 5.2). When T drops after a few nanoseconds, there is no ignition (dashed curves in Fig. 16). At sufficiently high E * will be no drop in temperature and the ignition is perfect. The limit of ignition for solid DT was found at 450 MJ/cm$^2$ by Chu [73].

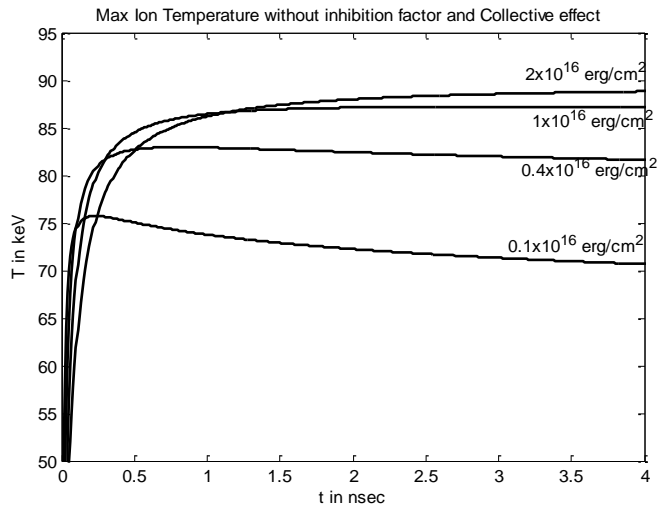

*Fig. 18. Characteristic curves of the dependence of the Temperature T of the fusions flames depending on time t for the fusion flame after the ignition by a picosecond irradiation of an energy flux desnityin erg/cm$^2$ for the fusion of hydrogen with boron11 (HB11 resulting in and ignition threshold of $10^{16}$ erg/cm$^2$ = $10^9$ J/cm$^2$ at a temperature of 87keV.*

The calculations of Chu required a correction by the later discovered in "inhibition" factor of the strong reduction of the heat conduction from the hot fusion plasma in the cold plasma of the fuel. Furthermore, the stopping length of the resulting alpha particles had to be corrected, since measurements have shown since 1974 that the collective theory of Denis Gabor is dominating against the Bethe-Bloch theory of binary collisions [105], at the plasma densities considered as summarized in [70]. The result was a reduction for E* by a factor up to 20 [106].

Under these conditions, with nonlinear force driven plasma blocks by picosecond clean laser pulses, it can be estimated that 10 to 30 PW power laser pulses frozen irradiating uncompressed DT in a power plant may produce more than 1000 times more energy than had to be in the laser pulse [70]. The advantages over the alternative solution to the ignition of DT by thermal processes at high compression with nano-second laser pulses (section C.5.4) are evident alone from the fact that the most complicate compression of the fuel to 1000times the solid state ist not necessary for the picosecond nonlinear force driven reaction in the solid state. But still the modified and advanced process the fusion flame after Chu and Bobin [79] have to be further explored for the ignition. The presented analysis of the block ignition has been treated initially only on the basis of hydrodynamics and further non-equilibrium conditions [n] have to be included.

A special surprise happened [71] when a the block ignition of DT was generalized for the case with fusion fuel HB11, equation (16). Igniting of a fusion flame in HB11 is only less than ten times more difficult than in DT, as the characteristics in Figure 18 showed as compared with DT, Figure 16 with an area of the ignition and an area showing no ignition. This is in stark contrast to the thermal ignition by very high density laser-compression, see page 139, where the result for HB11 showed that this is 100,000 times more difficult than for DT. The estimation for energy production in a power station [70] with DT, can be extented for HB11 to be in the range of 100 PW laser pulses with very high contrast ratio and about ps duration at fixed HB11. This is less than two orders of magnitude above that present technology. The generated radioactivity for the HB11 fuel per gain energy is less than for the case of burning coal [71], see also last paragraphs of Section D.5.4. A very minor question of the difference between the uranium in the ash of coal compared with gaseous output for HB11 is discussed in Appendix 1). Apart from this very marginal radiation problem of the coal and for HB11, not any pollution of the atmosphere with the billion of tons of $CO_2$ per year is the result of the laser driven nuclear fusion of HB11.

Under the assumption that the fusion flame can be described by the updated hydrodynamic theory of Chu [75] [70] [71] [106], the result of the calculation for HB11 is shown in Figure 18 with the time parameters as used before for the DT reaction. The necessary energy flux density E* is $10^9$ J/cm$^2$. Further studies used the detailed multi-fluid hydrodynamic model [o] with separate fluids for electron and ions including the electric fields inside the inhomogeneous plasmas in contrast to the usual plasma hydrodynamics [l]. The generated alpha particles could then be computed as an additional fluid. The results were basically similar to the earlier results [70], but it could be shown how the fusion flame is generated on time over nanoseconds in the solid state fuel after the picosecond initiation of the flame was initiated by the laser pulse. It took a considerably long time of more than 100 ps before a kind of shock wave was generated with the ion density up to four times the fuel density. This factor four is well known from the simplified analytical Rankine-Hugoniot theory. But the temporally growing thickness of the compressed plasma could only be reproduced numerically by the very general code including friction and thermal mechanisms. The shock velocity was seen of more than about 1000 km/s as usual. The slow building-up of the shock generation indicates the basic difference of the ps processes compared with the mechanism of "impact fusion" using nanosecond laser pulses [107][108].

The crucial aspect of the ps-laser pulse side-on ignition of the fusion reaction in contrast to the nanosecond option is the basic difference by nonlinear physics. This is a visible consequence of the general new dimension into which physics research is developing beyonf the views by Feynman, Hawkings and von Weizsäcker (see the preceding section). The nonlinear force with inclusion of the optical plasma properties [78] is an example providing from the very beginning the direct conversion of optical energy into macroscopic plasma motion with negligible heating and delaying thermal effects [n] and without losses including instabilities. These thermal processes are known from complex systems following the basic research by Lord May [p] about which Edward Teller was well aware in 1952 when judging about controlled energy generation by nuclear fusion [q]. This new aspect was well realized by the theory [78] and the numerical result of ultrahigh acceleration [91] (see Fig. 10.18a of Ref. [77]) and finally the measurement by Sauerbrey [84] thanks to the kink of the curve at 1986 in Fig. 15a by Mourou et al [d][e]. This confirms how to avoid the chaotic thermal effects. Though this basically is all well konnw a for a long time, the present turning point [r] can not strongly enough be articulated.

**Highlights in Chemical Technology**

Chemical technology news from across RSC Publishing.

# Nuclear power without radioactivity

24 March 2010

Radiation-free nuclear fusion could be possible in the future claim a team of international scientists. This could lead to development of clean and sustainable electricity production.

Despite the myriad of solutions to the energy crisis being developed, nuclear fusion remains the ultimate goal as it has the potential to provide vast quantities of sustainable and clean electricity. But nuclear energy currently comes with a serious environmental and health hazard side effect - radiation. For fusion to gain widespread acceptance, it must be able to produce radiation-free energy but the key to this has so far remained elusive, explains Heinrich Hora at the University of New South Wales in Sydney, Australia.

Conventionally, the fusion process occurs with deuterium and tritium as fuel. The fuel is spherically compressed - meaning compression occurs from all directions - with laser irradiation to 1000 times its solid state density. This ignites the fuel, producing helium atoms, energy and neutrons which cause radiation. Fusion is also possible with hydrogen and boron-11, and this could produce cleaner energy as it does not release neutrons, explains Hora. But this fuel requires much greater amounts of energy to initiate and so has remained unpopular.

Now, a team led by Hora has carried out computational studies to demonstrate that new laser technology capable of producing short but high energy pulses could be used to ignite hydrogen/boron-11 fuel using side-on ignition. The high energy laser pulses can be used to create a plasma block that generates a high density ion beam, which ignites the fuel without it needing to be compressed, explains Hora. Without compression, much lower energy demands than previously thought are needed. 'It was a surprise when we used hydrogen-boron instead of deuterium-tritium. It was not 100 000 times more difficult, it was only ten times,' says Hora.

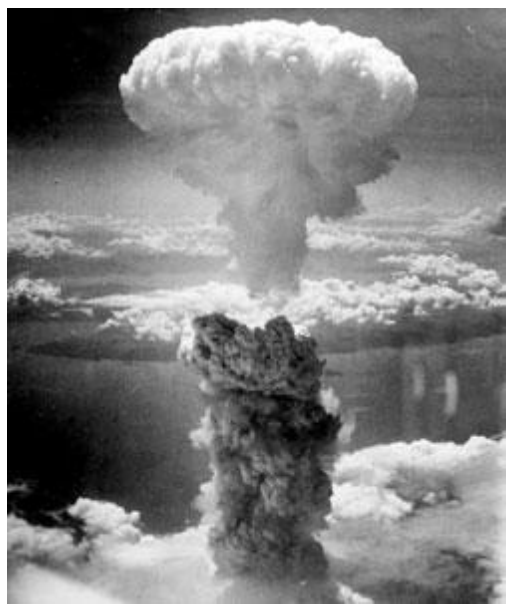

The power of nuclear fusion has yet to be tamed

'This has the potential to be the best route to fusion energy,' says Steve Haan, an expert in nuclear fusion at Lawrence Livermore National Laboratory in California. However, he also points out that it is still only potential at this point, 'there's a fair amount of work to do before this technology is at hand.'

Hora agrees that much more work is needed to fully understand this radical new approach. Its achievement will depend on continued advances in laser optics, target physics and power conversion technology, he concludes.

*Yuandi Li*

*Editor, Royal Society of Chemistry, London.*

**Energy Environ. Sci. 2010, 3, 479-486**

For the interesting case of power plants for energy generation, the comparable fusion gains of up to 10,000 are of interest under the conditions of the side-on ignition with relativistic electron beams similar to the results by Nuckolls and Wood [104]. This is in contrast to the spherical compression ignition with lasers in which case the yields can never be above a few hundred du to the limited amount of fusion fuel in the irradiated sphere.

If this solution is achieved, the dream of a completely safe nuclear energy with a ten milli-million times better nuclear energy efficiency compared to the combustion of fossil fuels would be essential. Yuandi Li, Editor, Royal Society of Chemistry, London, highlighted the results based on several interviews about HB11 [71] as a special "highlight", see the citation at the end of this section where one of the leaders of the NIF laser fusion experiment with spherical compression and thermal ignition by nanoesond laser pulses, Steven Haan said, that the new solution by side-on igntiontion with ps laser pulses has the potential to be the best way to fusion energy, which however still needs a lot to work on solving the problems. The publication [74] was downloaded electronically from the journal "Energy and Environmental Science" in the first 5 weeks about 600 times, which journal has the ranking (impact factor) of all scientific journals of the unusually high value 9.1.

(Note:) Gerard Mourou (see here on p. 3, Fig. 15a) became Director of the European Research Project IZEST-ICAN with a volume of more than One Billion Euro. About collaboration for application to laser driven fusion energy, see references [o].

[o] Paraskevas Lalousis, Heinrich Hora, Shalom Eliezer, Jose-Maria Martinez-Val, Stavros Moustaizis , George H. Miley & Gerard Mourou. Shock mechanisms by ultrahigh laser accelerated plasma blocks in solid density targets for fusion. Physics Letters A, 377, 885-888 (2013).

## References [1] to [108] from the book "Climatic Problems and Ways to Solution":

**Subject index**